\documentclass[onecolumn,preprintnumbers,floatfix,amsmath,amssymb,superscriptaddress,longbibliography]{revtex4}
\usepackage{graphicx}
\usepackage{color}
\usepackage{dcolumn}% Align table columns on decimal point

\newcommand{\be}{\begin{equation}}
\newcommand{\ee}{\end{equation}}
\newcommand{\ba}{\begin{eqnarray}}
\newcommand{\ea}{\end{eqnarray}}
\begin{document}

\title{Breakup of finite-size liquid filaments: Transition from no-breakup to breakup
including substrate effects}

\author{A. Dziedzic}
\affiliation{Department of Mathematical Sciences, 
             New Jersey Institute of Technology, Newark, NJ 07102 USA}
             \author{M. Nakrani}
\affiliation{Department of Mathematical Sciences, 
             New Jersey Institute of Technology, Newark, NJ 07102 USA}
             \author{B. Ezra}
\affiliation{Department of Mathematical Sciences, 
             New Jersey Institute of Technology, Newark, NJ 07102 USA}
             \author{M. Syed}
\affiliation{Department of Mathematical Sciences, 
             New Jersey Institute of Technology, Newark, NJ 07102 USA}
\author{S. Popinet}
\affiliation{Sorbonne Universit\'{e}, Centre National de la Recherche Scientifique, 
Institut Jean le Rond $\partial$'Alembert, F-75005, Paris, France} 
\author{S. Afkhami}
\affiliation{Department of Mathematical Sciences, 
             New Jersey Institute of Technology, Newark, NJ 07102 USA}

%\date{\today}

%\pacs{47.11.-j,68.08.-p,81.16.Rf,68.55.-a}

%\noindent{\texttt{FOR THE FULL-LENGTH PAPER}} \\

\begin{abstract}

This work studies the breakup of finite-size liquid filaments, 
when also including substrate effects, using direct numerical simulations.
The study focuses on the effects of three parameters: Ohnesorge number, the
ratio of the viscous forces to inertial and surface tension forces, the liquid filament
aspect ratio, and where there is a substrate, a measure of the fluid slip on the substrate, 
i.e.~slip length. Through these parameters, it is determined whether a liquid filament 
breaks up during the evolution toward its final equilibrium state. Three scenarios are identified:
a collapse into a single droplet, the breakup into one or multiple droplets,
and recoalescence into a single droplet after the breakup (or even possibly 
another breakup after recoalescence). The
results are compared with the ones available in the literature  
for free-standing liquid filaments. The findings show that
the presence of the substrate promotes breakup of the filament. The effect of
the degree of slip on the breakup is also discussed. 
The parameter domain regions are comprehensively explored
when including the slip effects. An experimental case is also carried out
to illustrate the collapse and breakup of a finite-size silicon oil filament supported on a 
substrate, showcasing a critical length of the breakup in a physical configuration.
Finally,  direct numerical simulations reveal striking new details 
into the breakup pattern for low Ohnesorge numbers, where
the dynamics are fast and the experimental imaging 
is not available; our results therefore significantly extend the range of
Ohnesorge number over which filament breakup has been considered.

\end{abstract}

\maketitle
\section{Introduction}
\label{sec:intro}

The breakup of liquid jets has been a subject of
extensive research \cite{lister,Eggers,eggers_rmp97,basaran,notz_jfm04}, with applications, for
instance, ranging from droplet generation in microfluidics, paint spraying, and 
ink-jet printers, to biological and geological systems. 
Theoretical, computational and experimental studies of the breakup of finite-size free-standing
filaments (of Newtonian liquids) exist in the literature; see e.g.~\cite{lister,notz_jfm04,_pita,pita,Driessen2013}.
Studies of the breakup of finite-size liquid filaments involving substrate effects, however, have
so far been very limited; see e.g.~\cite{Hartnett2015,Feng2013,Javier2017}. This is mainly due to difficulties
associated with the presence of a substrate. 
The study in \cite{Feng2013} reports on simulations of liquid filaments breaking up into
droplets on partially wetting substrates, however ignoring the inertial effects. 
The study in \cite{Javier2017} reports on the retraction of liquid filaments on substrates
under partially wetting conditions, the study however is limited to the evolution just before the
breakup. Finally, the work in \cite{Hartnett2015} reports on experimental and numerical
studies on the breakup of finite-size, nano- and microscale, liquid metal filaments 
on a substrate. The work however is limited to a small range of parameter space.
Our work here differs from previous studies in that we explore the parameter domain 
regions comprehensively, including the substrate slip effects, by solving the full three-dimensional
Naiver-Stokes equations based on a Volume-of-Fluid interface tracking method
that can accurately and efficiently (adaptively) model the breakup and coalescence 
of fluid interfaces \cite{popinetGerris}.

The experimental work of Pita et al.~\cite{pita} is the main motivation of the computational
study carried out in this paper. In \cite{pita}, the authors experimentally show that a finite-size 
liquid filament can collapse into one droplet or break up to multiple droplets, depending on
the viscosity and the initial dimensions of the filament. This phenomenon can be explained
as the competition between surface tension, that drives either the pinching-off or
shortening of the filament, and viscous forces, resisting the deformation \cite{lister}.
In the work of Pita et al.~\cite{pita} and Notz and Basaran \cite{notz_jfm04}, two
dimensionless quantities are identified as the relevant parameters controlling the
final outcome: the filament aspect ratio and the Ohnesorge number.
The Ohnesorge
number, denoted by Oh, represents the significance of surface tension and viscosity
(Oh = $\frac{\mu}{\sqrt{\rho\sigma R_0}}$) where the liquid viscosity, density, and surface
tension are denoted by $\mu$, $\rho$, and $\sigma$, respectively, and $R_0$ is the filament
initial radius. The aspect ratio, AR $=\frac{{L_0}}{R_0}$, represents the ratio between the half 
of the initial length, denoted by $L_0$ here, and the initial radius of the filament, $R_0$.

The experimental study in \cite{pita} explores the breakup of finite-size free-standing liquid 
filaments over the range $0.003 <$ Oh $< 10$, for filaments with $1 <$ AR $< 70$.
The results in  \cite{pita} show, in agreement with the prior computational investigation in 
\cite{notz_jfm04}, that within the range of considered Oh and AR, the filament does not
break up for AR$<6$, regardless of the Oh number, and that no
filament breakup is occurred for Oh$> 1$, regardless of AR values.
The work in \cite{pita} also identifies the critical filament aspect ratios,
above which the filament breaks up, for a range of 
$0.003<\mbox{Oh}<10$.

\begin{figure}[thb]
\centering
\begin{tabular}{c}
  \includegraphics[width=0.8\textwidth,trim=0 40mm 0 0]{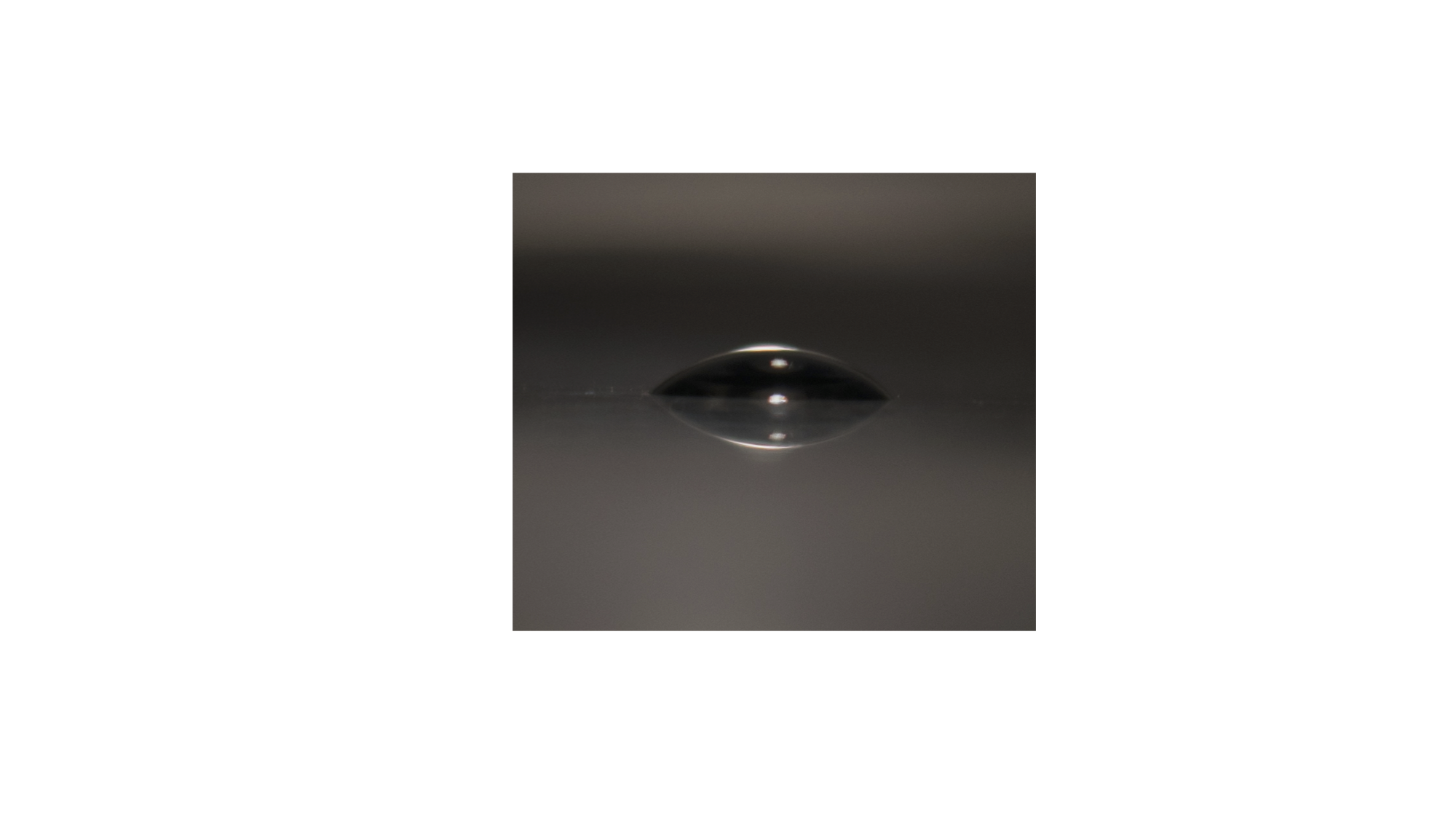}
\end{tabular}
\caption{The contact angle is measured from photographic images of droplets 
formed on the substrate using MATLAB\textregistered\, image processing toolbox.
The measured contact angle is about $30^\circ$. Photographs are taken after the filament
has broken up into droplets and the resulting droplets reached an equilibrium state.}
\label{fig:CA}
\end{figure}
\begin{figure}[thb]
\centering
\begin{tabular}{c}
  \includegraphics[width=0.9\textwidth,trim=0 40mm 0 0]{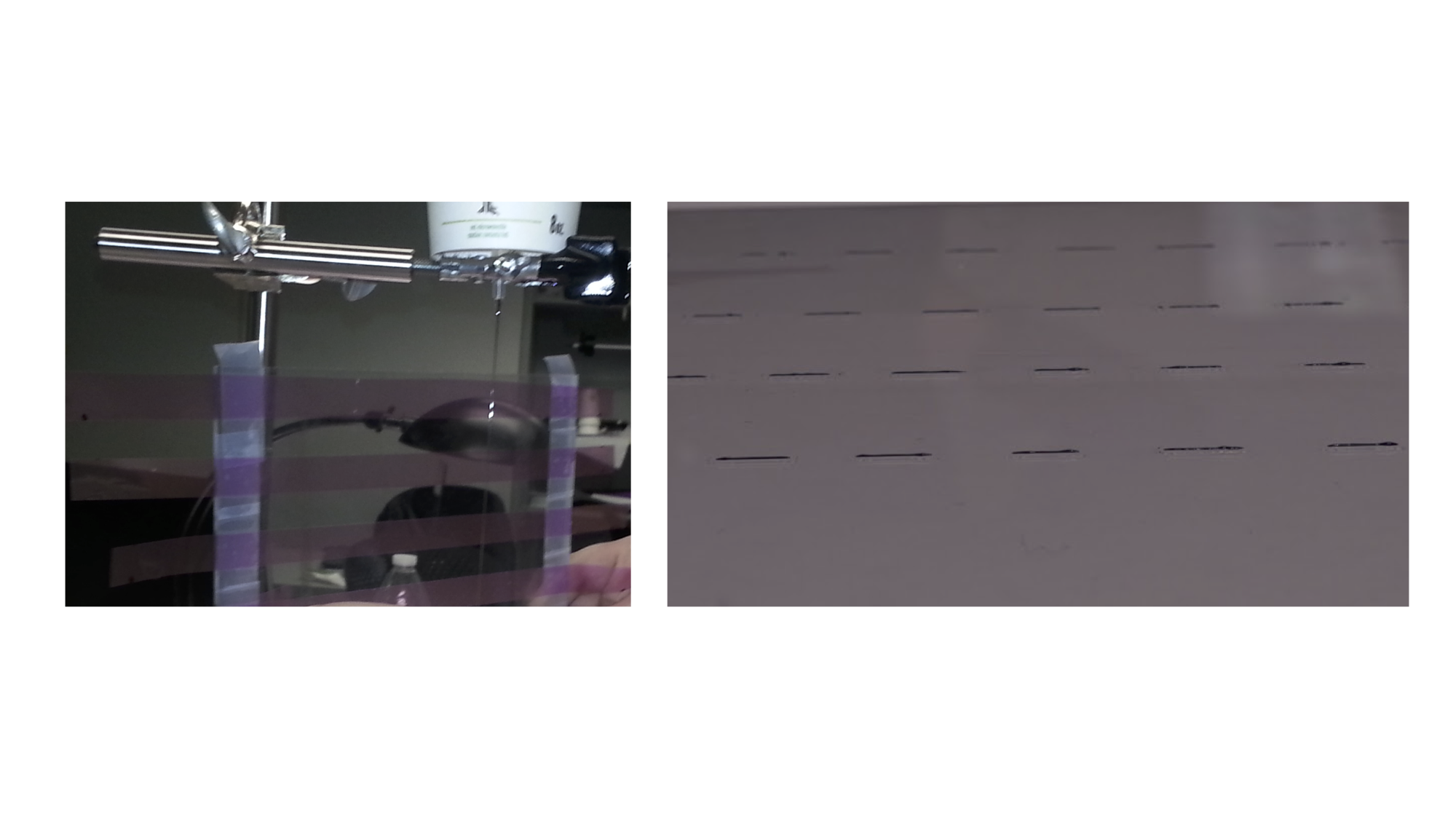}
\end{tabular}
\caption{Illustration of the experimental setup: filaments are captured 
by swiping the substrate through the liquid stream flowing out of the 
nozzle attached to a container filled with polydimethylsiloxane up to a certain height (left).
Cellophane (pink) strips are taped to the plate to produce filaments of
various lengths. The substrate is then immediately placed on a 
horizontal surface and cellophane strips are removed (right).
The initial width of the filaments is about $1$ mm.
}
\label{fig:Experiments1}
\end{figure}

\begin{figure}[tbh]
\begin{tabular}{c}
\centering
  \includegraphics[width=1.2\textwidth,trim=20mm 0 0 0]{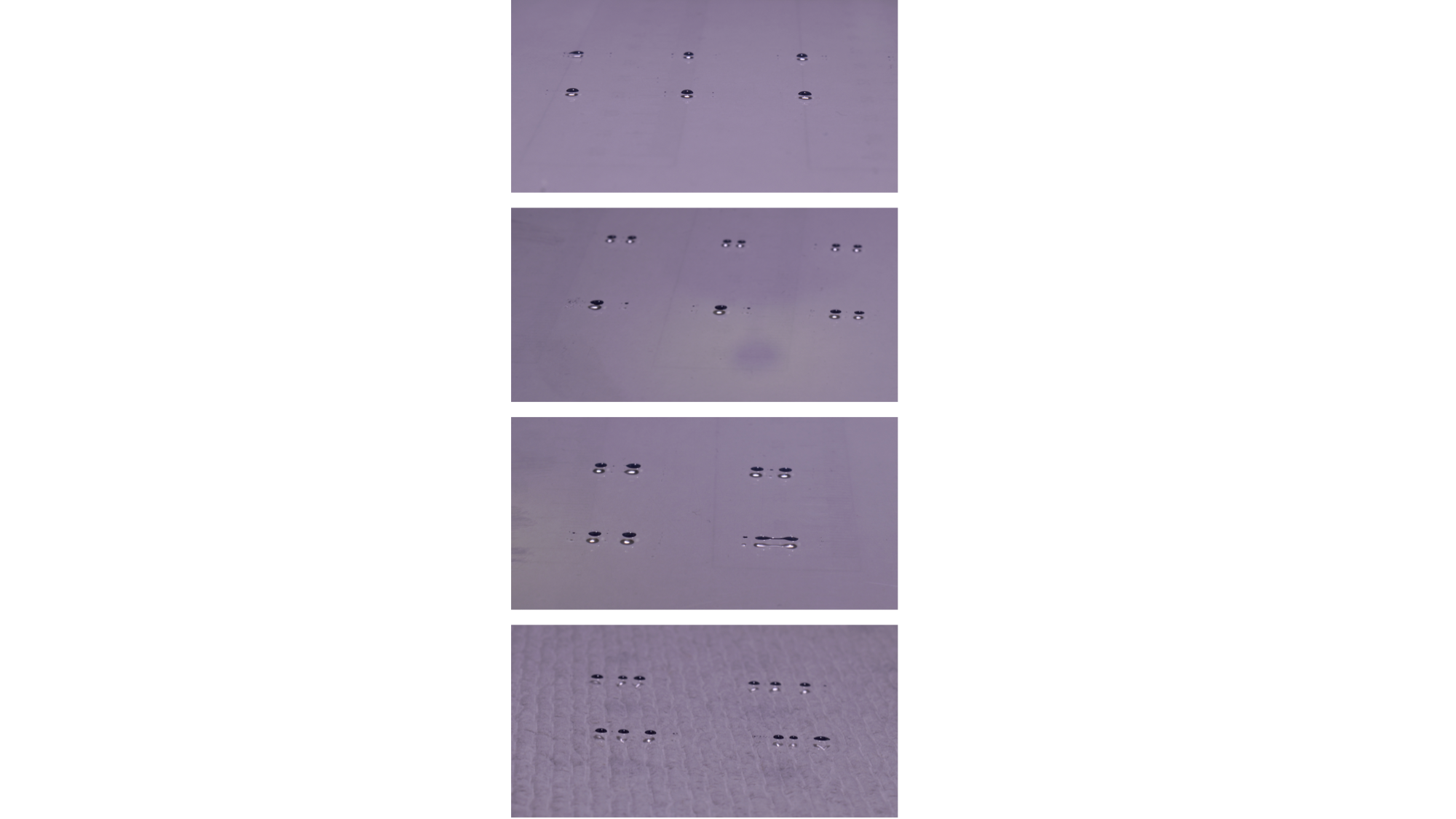}
\end{tabular}
\caption{Photographic images of the polydimethylsiloxane filaments 
increasing in length, $2L_0\approx$ $9$ mm, $11$ mm, $15$ mm and $17$ mm from bottom to top.
The initial width of the filaments is about $1$ mm.
While the filament of length $9$ mm collapses into single drops,
filaments of length $11$ mm and above break up into multiple droplets.
Oh $\approx 17$, $R_0\approx0.17$ mm.}
\label{fig:Experiments2}
\end{figure}

In this paper, we carry out an investigation of the collapse and breakup of finite, 
millimeter size, liquid filaments on a substrate. We first motivate our study
by conducting experiments to show the unstable evolution of liquid filaments
deposited on a substrate that leads to either the collapse to a single droplet or 
the breakup. We show this transition by increasing the length of the filament
deposited on a substrate, illustrating the existence of a threshold, 
above which the filament breaks up into multiple droplets. 
A similar study
has been carried out previously in the context of pulsed laser-induced dewetting
of nanometer size geometries \cite{Hartnett2015}. 

The experiments are conducted using $28$,$000$ molecular weight polydimethylsiloxane
deposited on a plastic substrate coated with Rust-Oleum\textregistered\, NeverWet\textregistered\,
treatment to make the substrate partially wetting. We ensured the 
wetting reproducibility by the treatment process through measuring the contact angle
by taking photographic images of the resulting droplets on the substrate (see Fig.~\ref{fig:CA}), 
using MATLAB\textregistered\, image processing toolbox.  
We adopt an experimental procedure similar to the one reported in \cite{Diez_04},
where filaments are captured from a jet flowing out of a nozzle attached to a container 
filled with polydimethylsiloxane up to a certain height. In order to produce filaments of various
lengths, we tape the substrate with cellophane strips (see Fig.~\ref{fig:Experiments1}). 
The substrate is then swiped through the
liquid stream to create an even, straight line.  The substrate is then immediately placed on a 
horizontal surface, after which
the cellophane strips are removed (see Fig.~\ref{fig:Experiments1}). 
The filaments are then left undisturbed to retract to form droplets (see Fig.~\ref{fig:Experiments2}).
This process is repeated to ensure the reproducibility of the results, 
as well as the controllability of the length.
We carry out the above procedure for 9 mm $\le 2L_0\le$ 17 mm and show
the results in Fig.~\ref{fig:Experiments2}. As shown, there is a critical length below which
the filament does not break up but rather collapses to a single
droplet. The transition from collapse to breakup can be
described as a competition between the capillary driven end retraction, 
the Rayleigh--Plateau type instability mechanism \cite{Rayleigh1878}, and the viscous dissipation due
to the presence of the substrate \cite{Feng2013,Hartnett2015}. Assuming typical physical properties of the
polydimethylsiloxane at room temperature, 
e.g.~$\mu=1$ Pa s, $\rho=971$ kg/m$^3$, $\sigma=0.02$ N/m,
we estimate Oh $\approx 17$, where we calculate $R_0\approx0.17$ mm
based on the initial filament thickness of approximately $1$ mm and the
contact angle of $30^\circ$ (i.e.~$R_0$ is the radius of a cylinder whose
cross sectional area is equivalent to that of the filament resting on the
substrate with a given width and contact angle). 
This result is interesting because the experimental evidence in \cite{pita}
suggests that no filament breakup can be observed for Oh $> 1$, 
regardless of the filament initial aspect ratio. Our experiments  however
show that the presence of the substrate can promote breakup leading to 
a much larger Oh for which the breakup can occur.  

In what follows, we will focus on the effect of the interaction with the 
substrate on the filament breakup. We explore
the boundary between breakup and no-breakup regimes as a function
of Oh and AR for various substrates characterized by different slip lengths. 
We report the results of numerical simulations, extending the range of
Oh that has been so far considered in the literature, as well as 
revealing peculiar breakup patterns for low Oh. 
For our study, we carry out direct numerical computations of the full 
Navier--Stokes equations, using Gerris code \cite{popinetGerris}.
We explore the range of $2\le$ AR $\le60$ and $10^{-3}\le$ Oh $\le14$,
focusing on the effect of the degree of slip on the results. 
Our study here mainly focuses on generating a large numerical dataset, 
by carrying out an exhaustive and systematic investigation.
Additionally, our study enhances the current numerical results 
regarding the freely standing finite size filaments.
We hope that this work will provide a basis for future theoretical analysis,
when including the substrate effects.

\begin{figure}[tb]
\begin{tabular}{c}
\centering
  \includegraphics[width=0.8\textwidth]{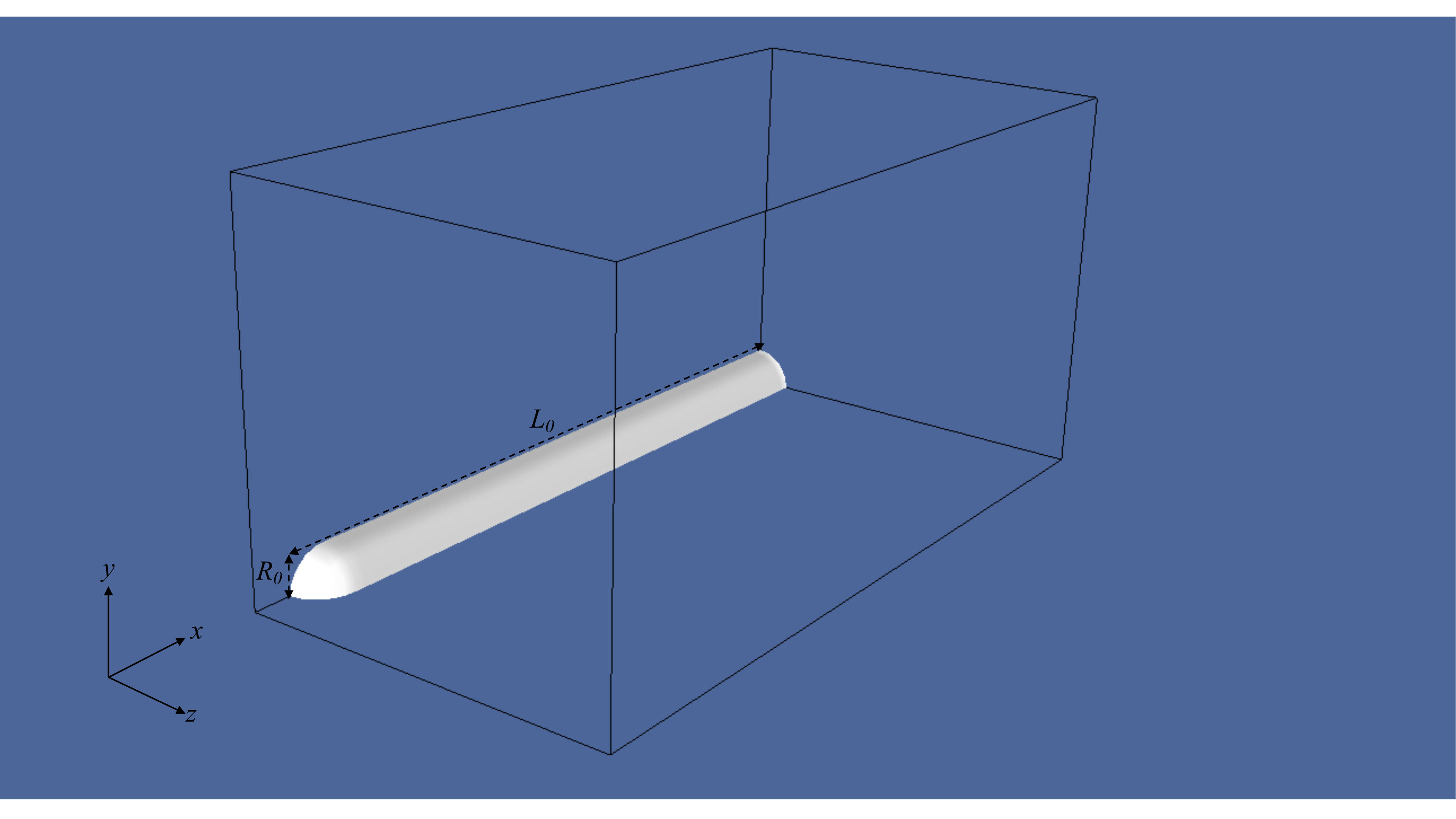}
\end{tabular}
\caption{Computational geometry, where
$R_0$ is the radius and $L_0$ is the half length of the filament with spherical end caps. 
For a freely standing filament, we use a symmetry boundary condition on all computational 
boundaries. For a filament on the substrate,  
we use a symmetry boundary condition on all computational  boundaries
except on the bottom boundaries where the substrate is modeled.}
\label{fig:SimSetup}
\end{figure}
\section {Computational Setup}  
We numerically investigate the dependence of the 
the breakup of a filament, either freely standing or being deposited on a substrate.
Initially, a liquid filament of radius $R_0$ and length $2L_0$ is considered. We take
advantage of the symmetries that exist in the problem and model only $1/8$ of the 
freely standing filament or $1/4$ of the filament on the substrate;
see Fig.~\ref{fig:SimSetup}.

\begin{figure}[tb]
\begin{tabular}{c}
\centering
  \includegraphics[width=0.5\textwidth,angle=-90,trim=0 0mm 31mm 0,clip=true]{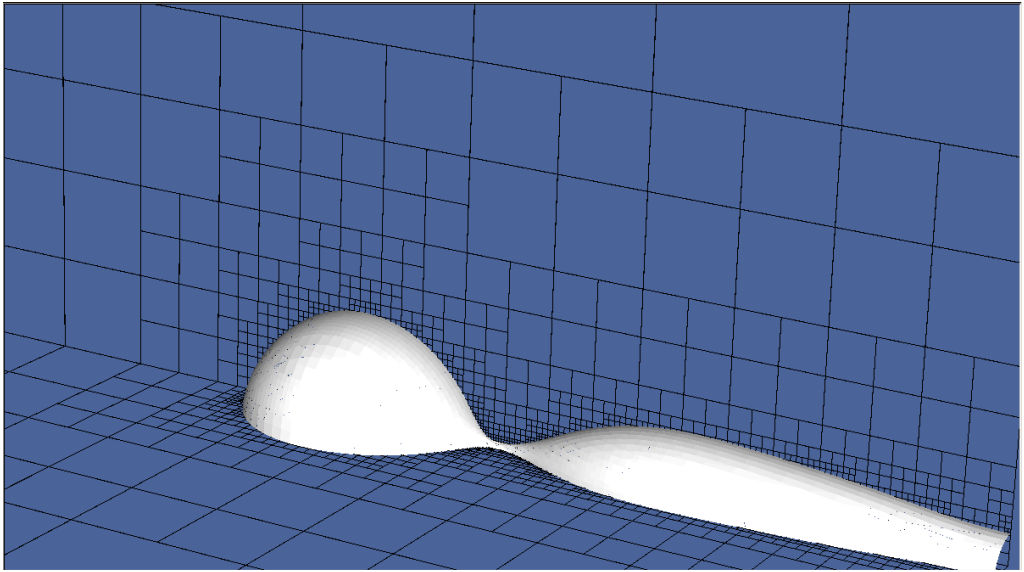}
\end{tabular}
\caption{Details of the adaptive mesh refinement where the
resolution is adjusted so that the size of any cell containing the interface
is $\Delta=1/2^8$.}
\label{fig:mesh}
\end{figure}

For a freely standing filament, we use a symmetry boundary condition 
for all variables on all computational boundaries. 
The code for the computational setup is derived from the ``Savart-Plateau-Rayleigh''
example, which is one of the examples from the Gerris open source site \cite{TestGerrisRP}.
When there is a substrate,  
we use a symmetry boundary condition on all computational  boundaries
except on the bottom boundaries where the substrate is modeled
as either a no-slip boundary or if partial slip is allowed, we then
impose the Navier slip boundary condition \cite{afkhami_jcp09} at the substrate ($y=0$), 
\begin{equation}
\label{eq:slip}
(u,w){|}_{y=0}=\lambda {\partial (u,w)/\partial y}{|}_{y=0},
\end{equation}
where $\lambda$ is the slip length and $(u,w)$  are the components of the velocity field
tangential to the solid boundary. 
For issues regarding the regularization 
of the viscous stress singularity at the contact line, the reader is referred to \cite{MCL}.
When the filament is placed on a solid surface, we impose a contact angle of $90^\circ$;
this is due to the limitation of the code for having an arbitrary contact angle 
and the simplicity in imposing a $90^\circ$ contact angle. For the details of the
numerical methodologies, including the Volume-of-Fluid 
method for tracking the interface and for the computation of the surface tension force,
the interested reader is referred to \cite{Popinet03,Popinet2009,PopinetARFM}. 
A computational domain of $[0$,$2]$$\times$$[0$,$1]$$\times$$[0$,$1]$ is considered. We vary 
$R_0$ and $L_0$ in order to have $2\le$ AR $\le60$.  
We set the surface tension $\sigma=1$ and the filament liquid density $\rho=1$,
and vary the filament liquid viscosity according to $\mu=\sqrt{R_0}$Oh. 
We set the density and viscosity ratio to approximately 
$800$ and $50$, respectively (corresponding for example to air/water physical parameters).
Simulations are then performed for various Oh numbers ($10^{-3}\le$ Oh $\le14$) and slip 
length values ($0\le\lambda\le 1$). We use dynamic adaptive mesh refinement as in \cite{Popinet2009},
particularly around the breakup point, see Fig.~\ref{fig:mesh}, and the maximum level of 
refinement is arbitrarily set to $8$, unless stated otherwise; 
using an octree mesh (in 3D), the root cell is at level zero and each 
successive generation increases the cell level by one, i.e.~ the smallest mesh size $\Delta=1/2^8$.
We note that for the substrate-supported filaments, Oh number is also defined 
using $R_0$. This is because the presence of a substrate only slightly modifies the Rayleigh--Plateau analysis,
see e.g.~\cite{Mahady2015,Hartnett2015}, and therefore
the relevant length scale should not change.

\begin{figure}[thb]
\centering
\begin{tabular}{ccc}
  \includegraphics[width=.25\textwidth]{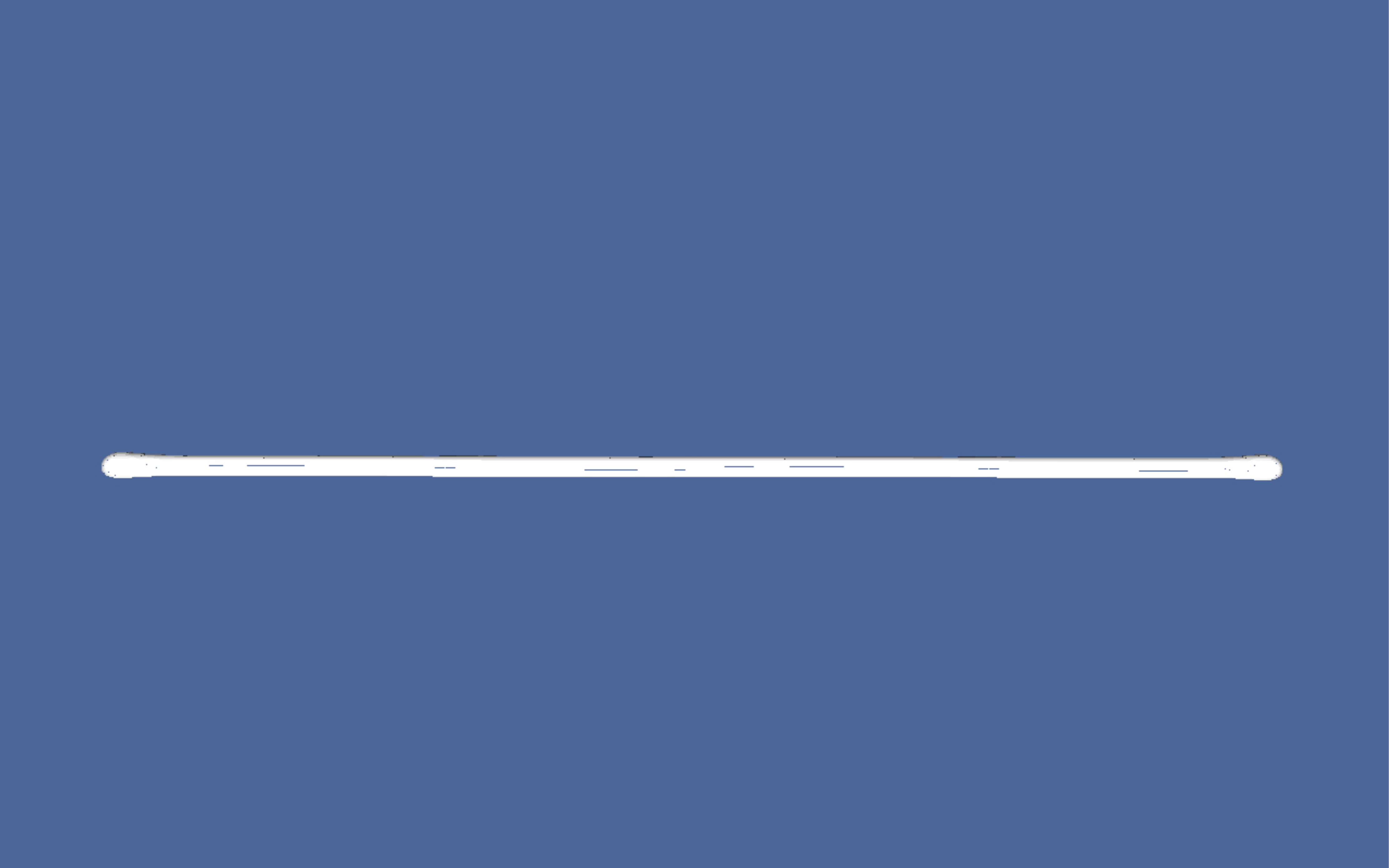}&
  \includegraphics[width=.25\textwidth]{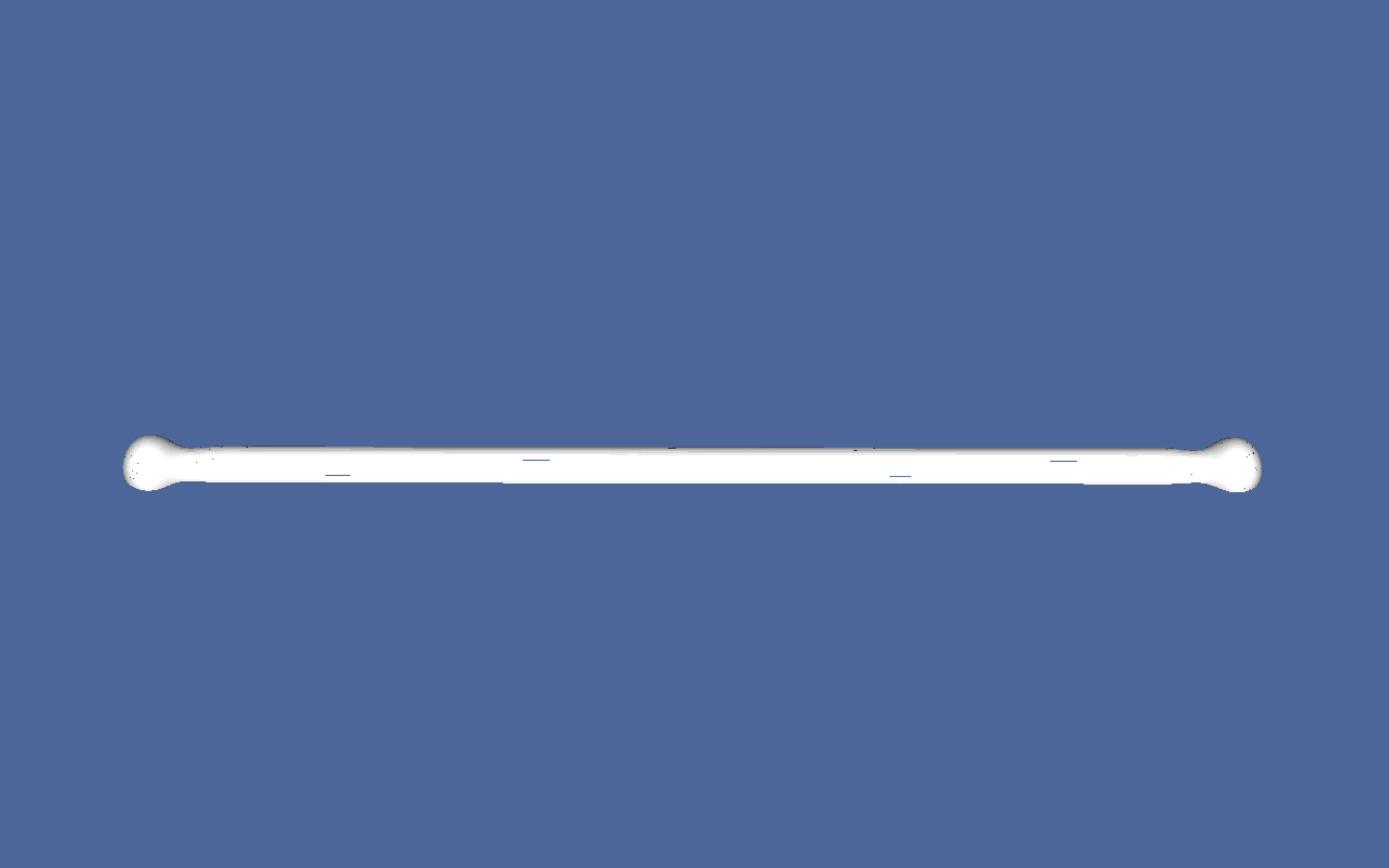}&
  \includegraphics[width=.25\textwidth]{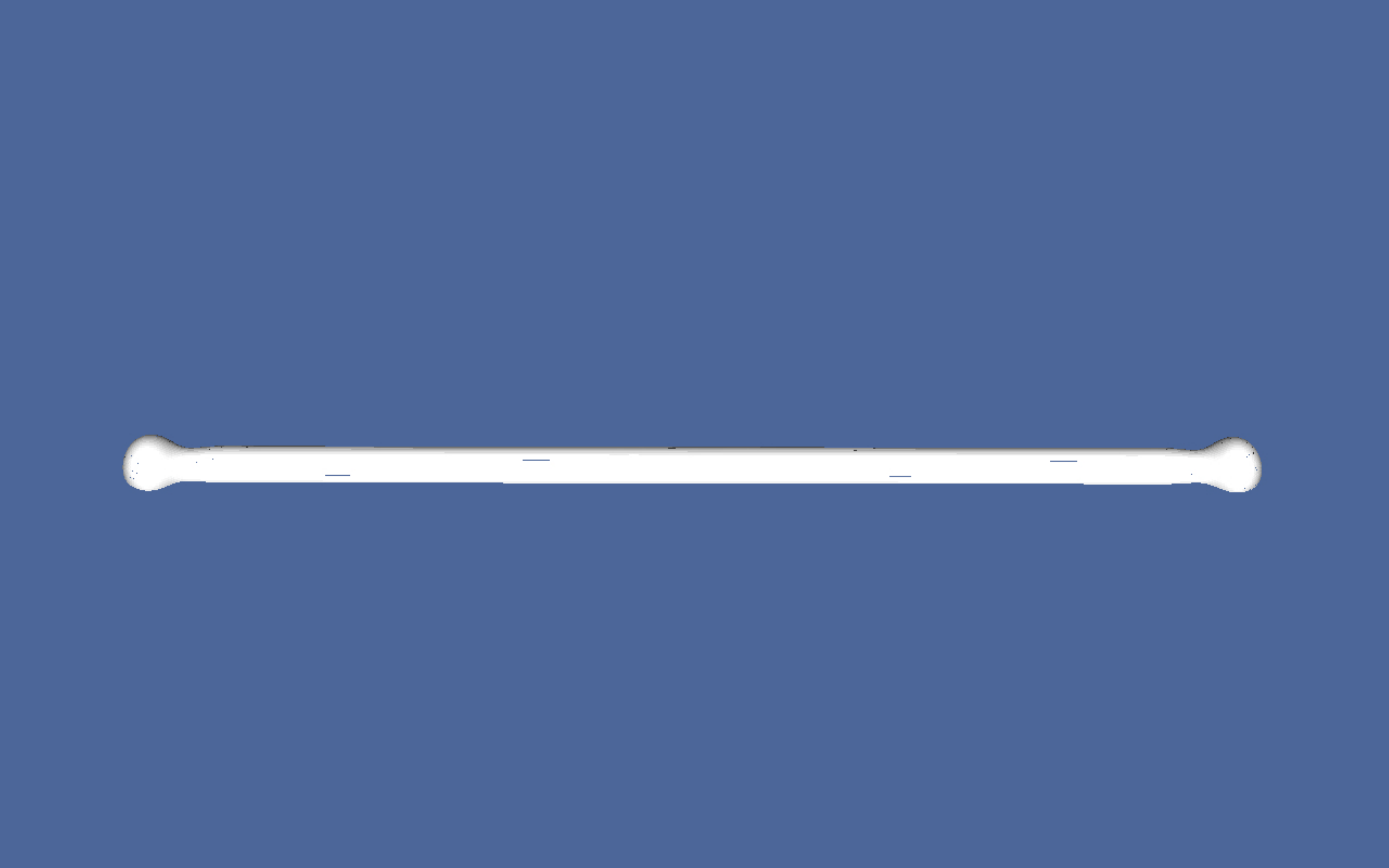}\\
  \includegraphics[width=.25\textwidth]{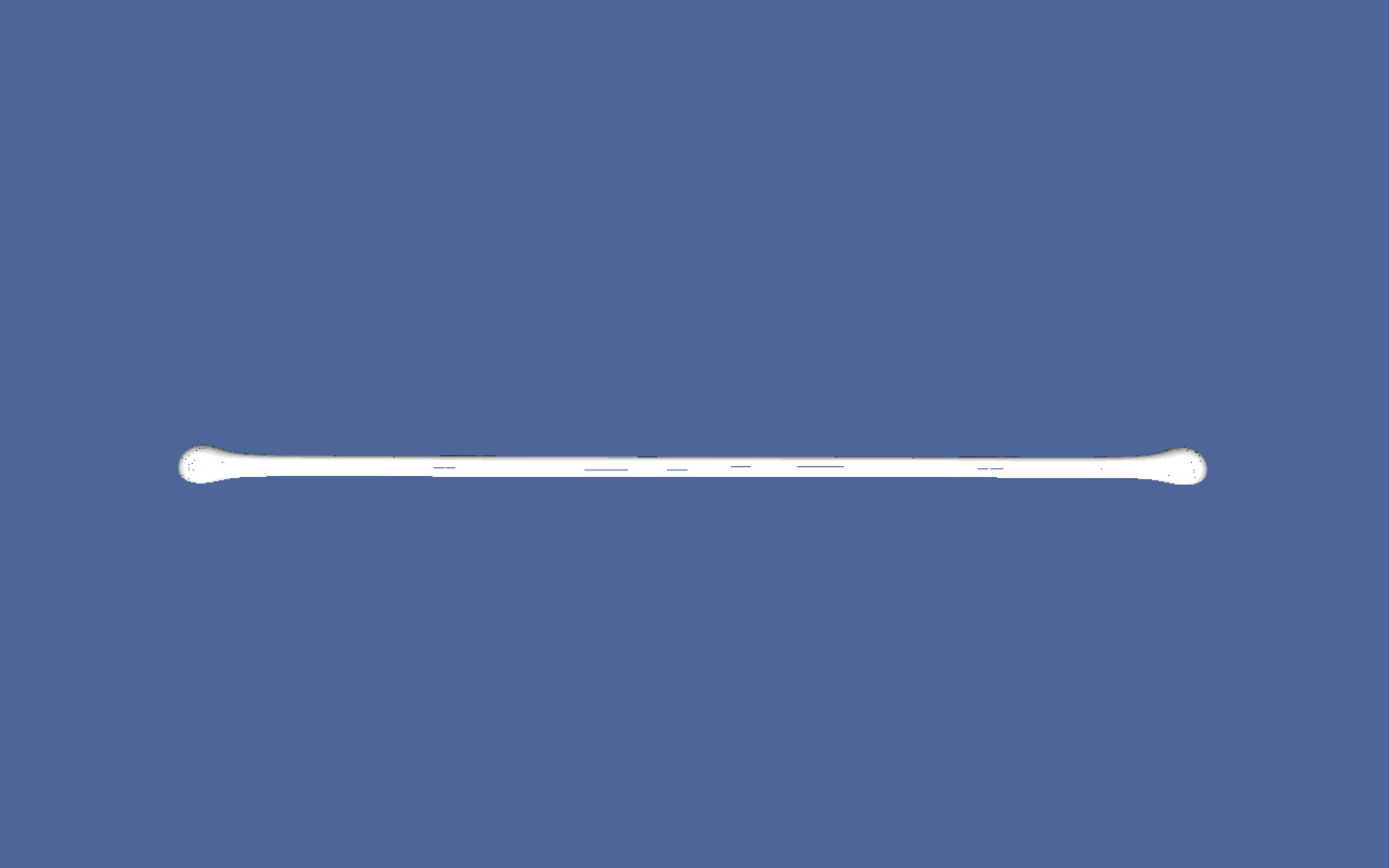}&
  \includegraphics[width=.25\textwidth]{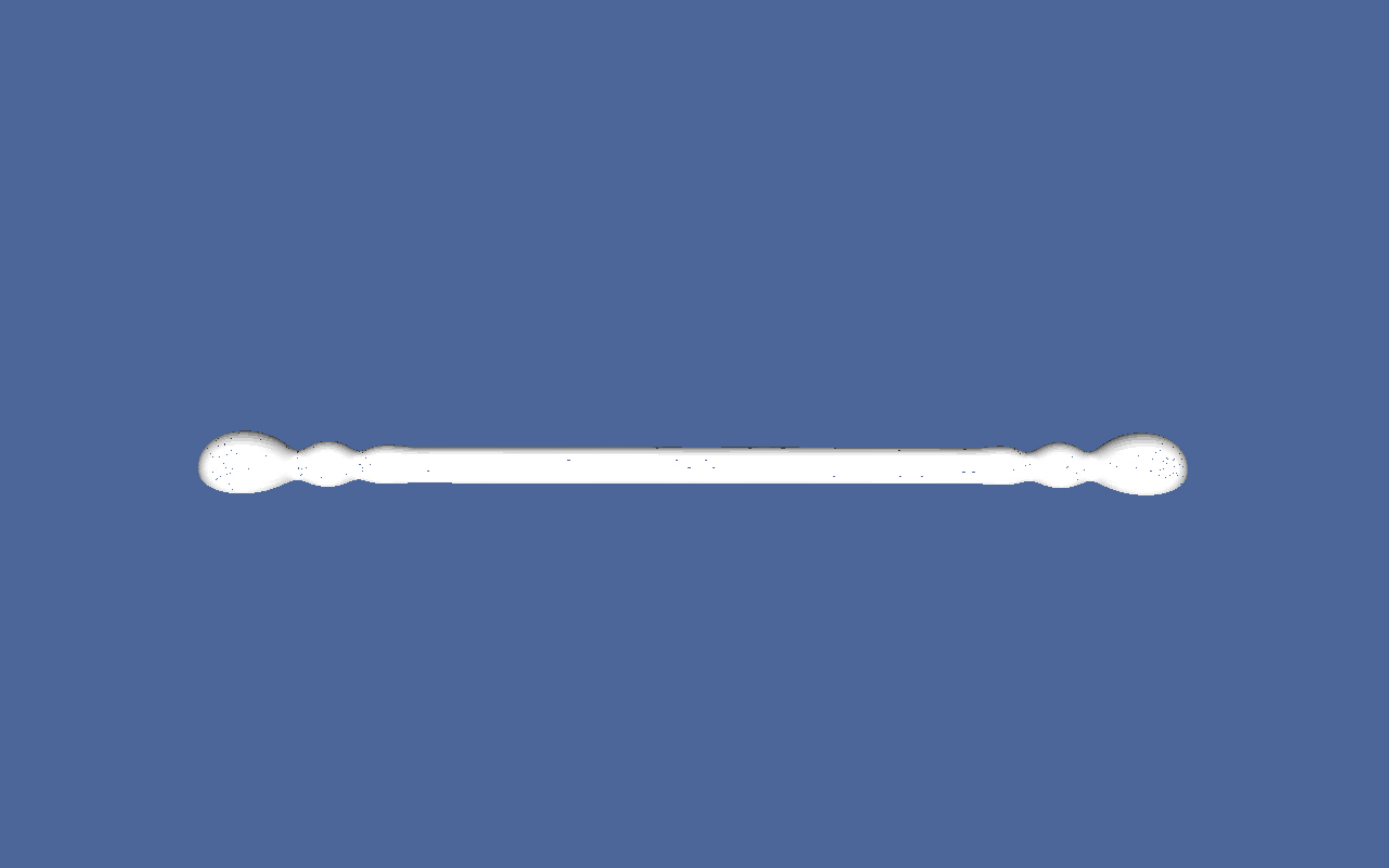}&
  \includegraphics[width=.25\textwidth]{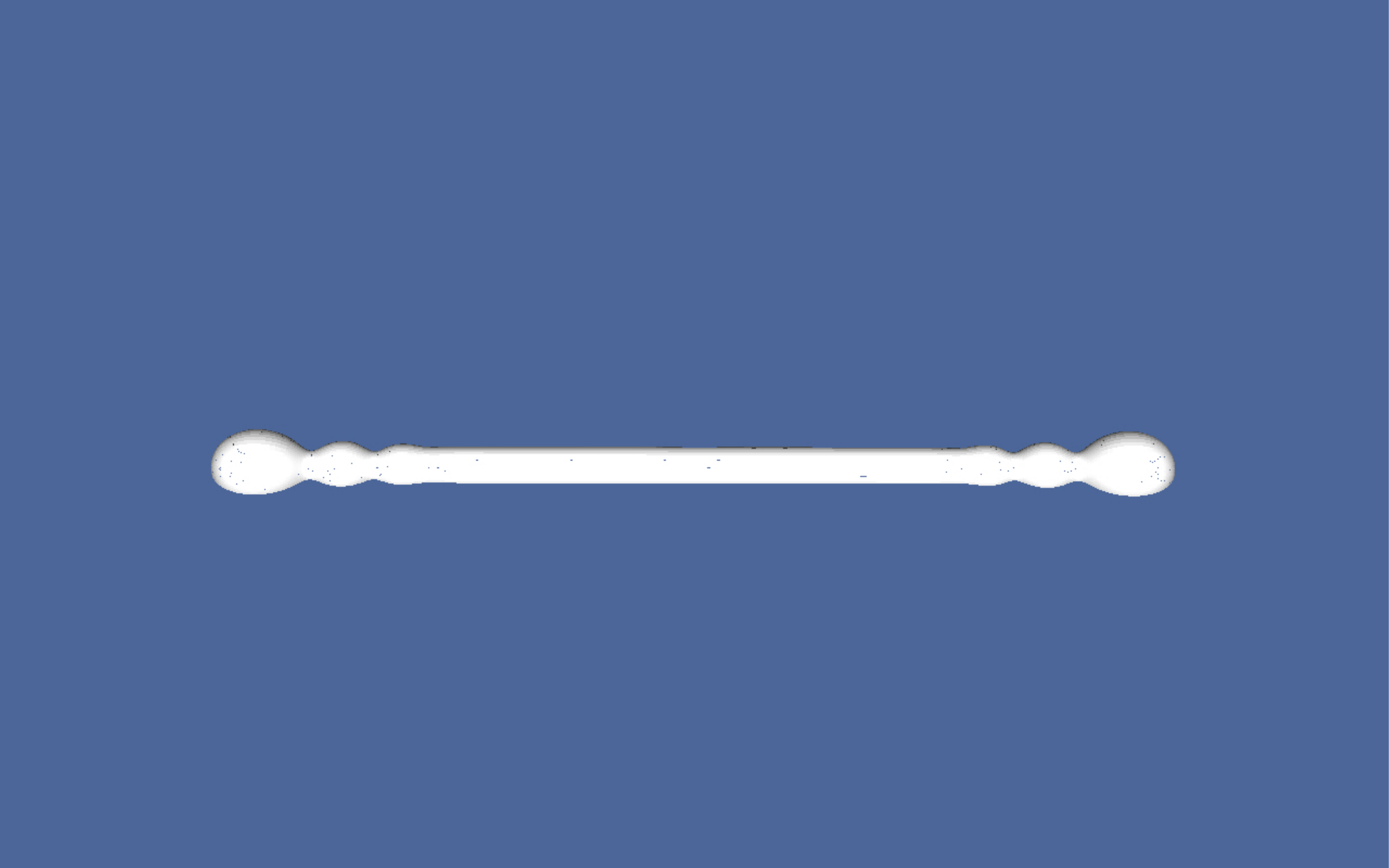}\\
  \includegraphics[width=.25\textwidth]{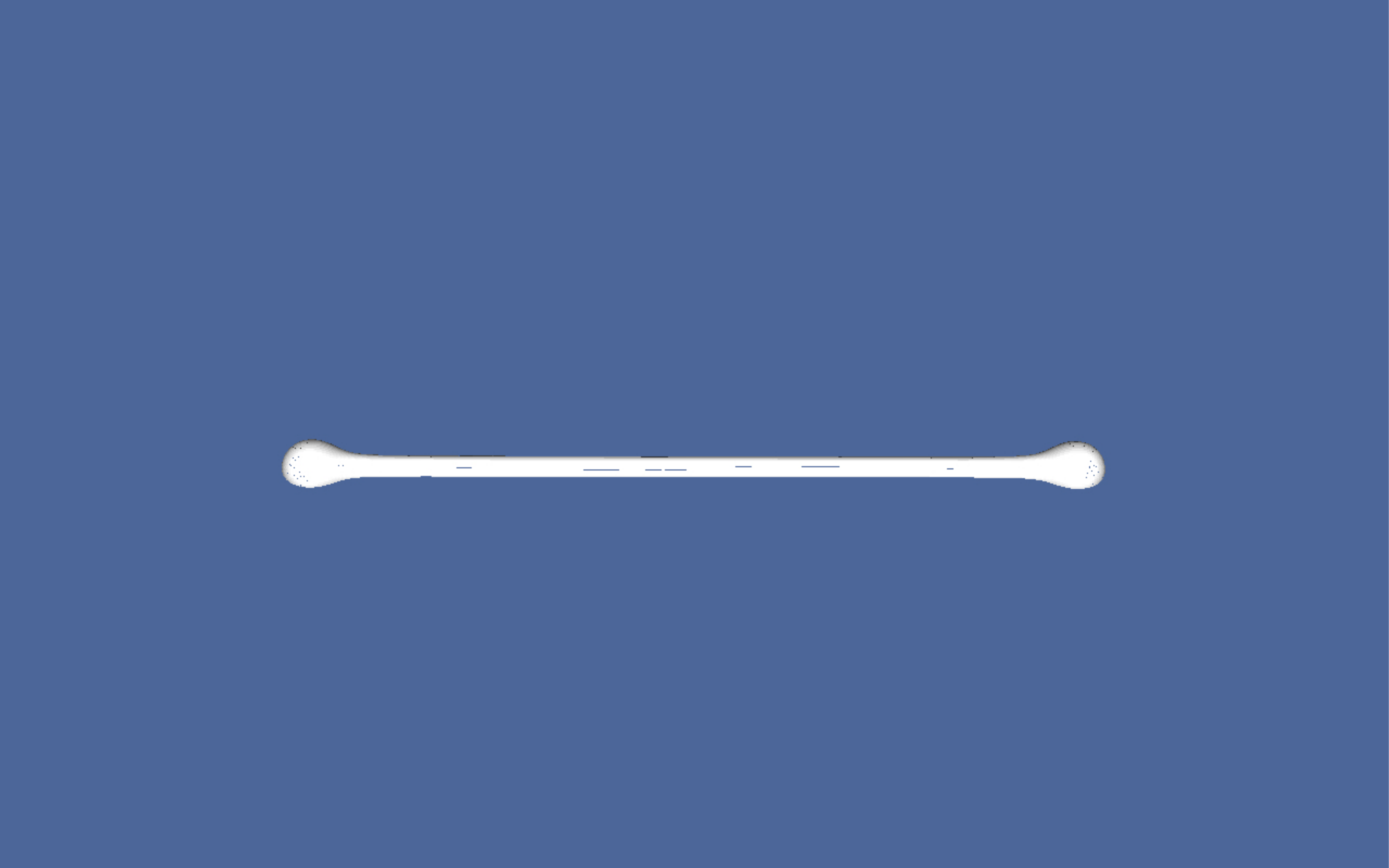}&
  \includegraphics[width=.25\textwidth]{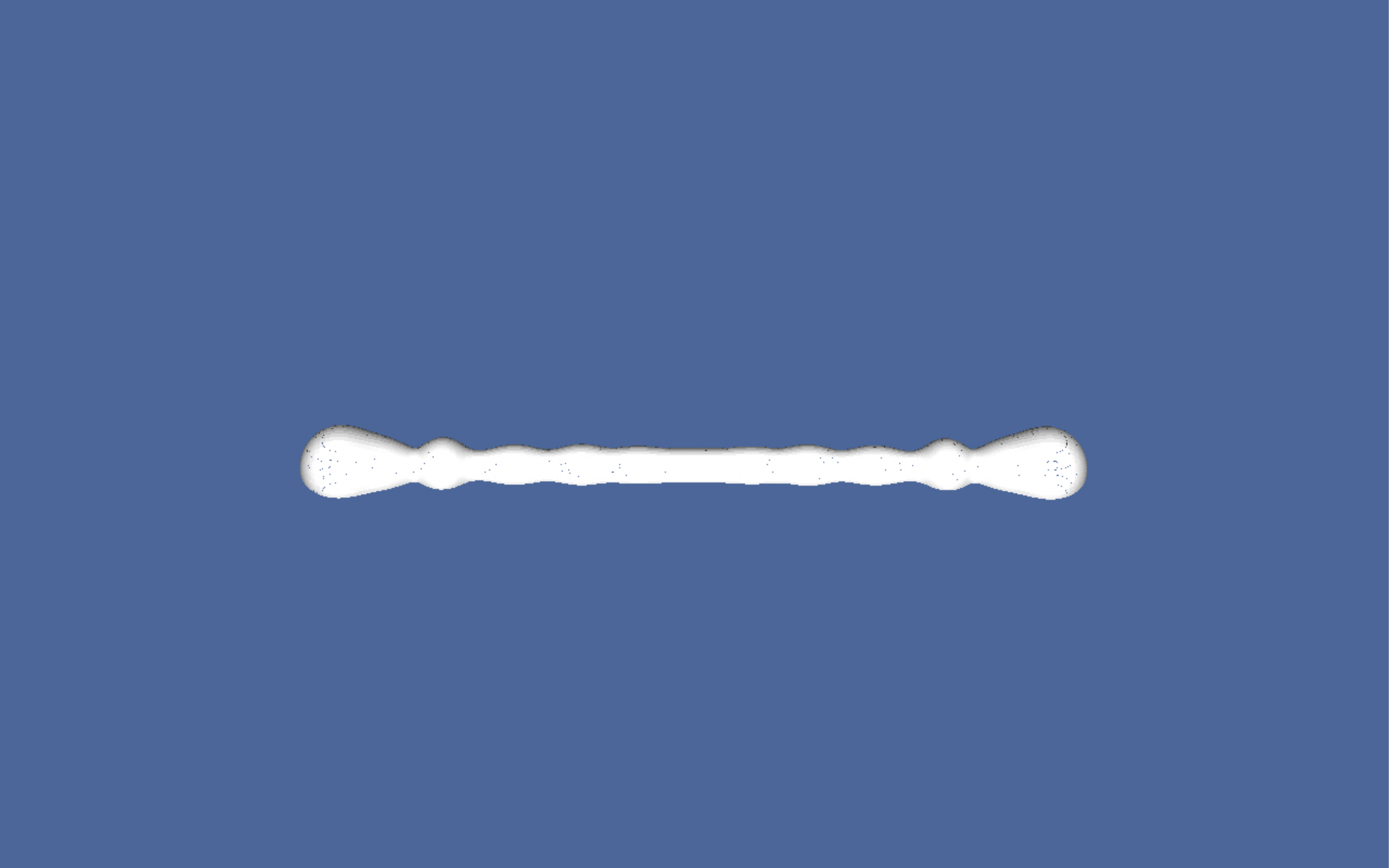}&
  \includegraphics[width=.25\textwidth]{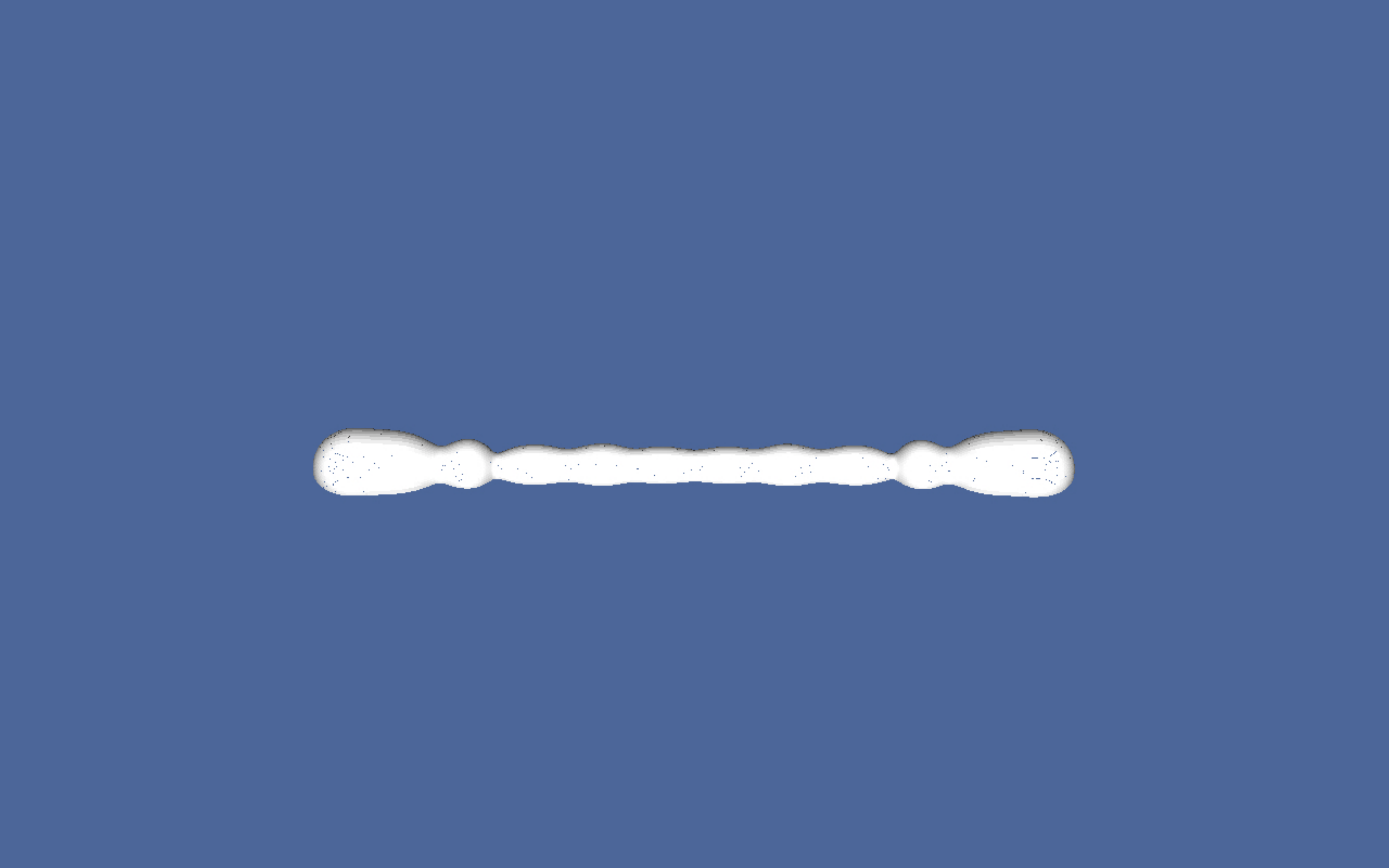}\\
  \includegraphics[width=.25\textwidth]{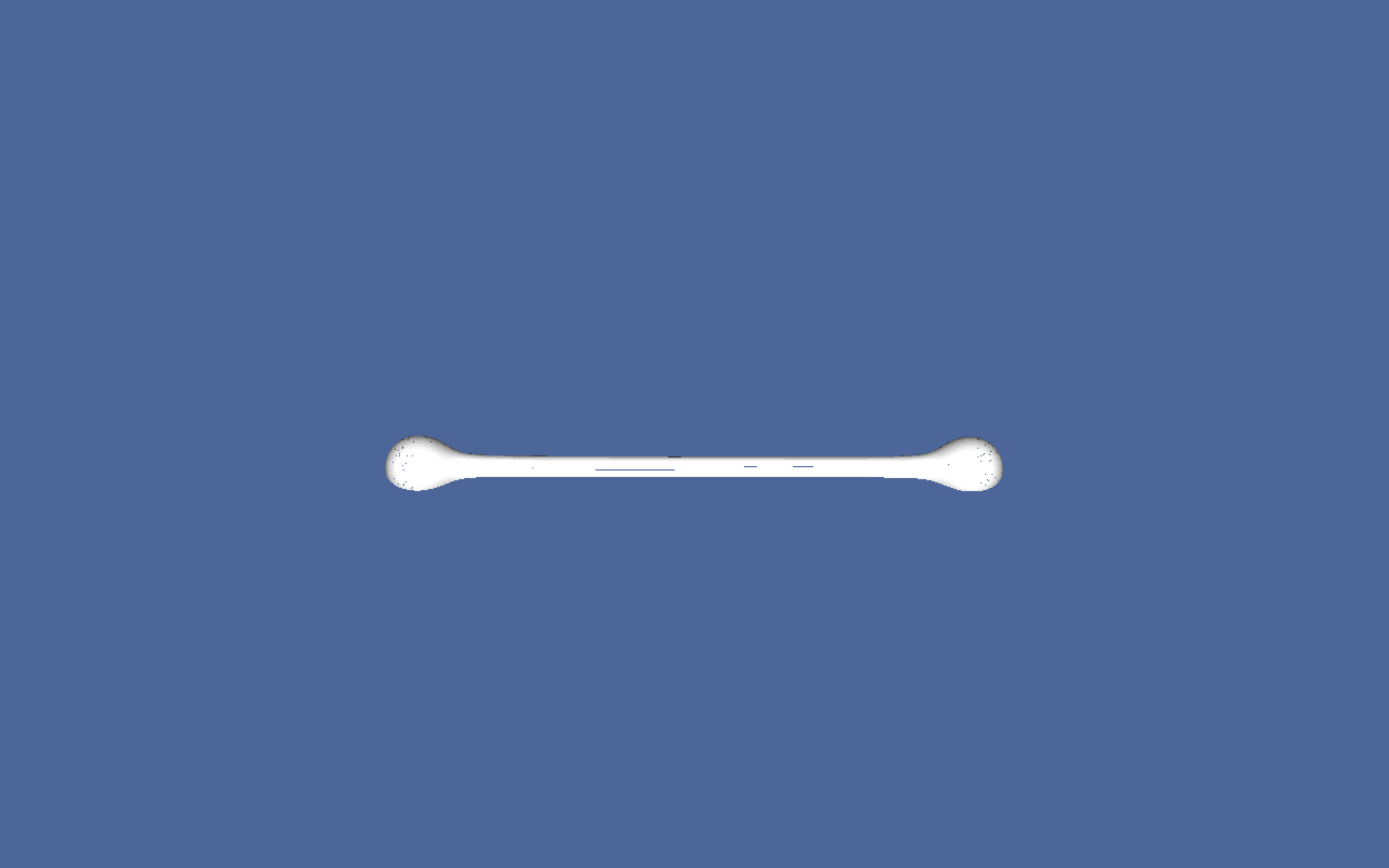}&
  \includegraphics[width=.25\textwidth]{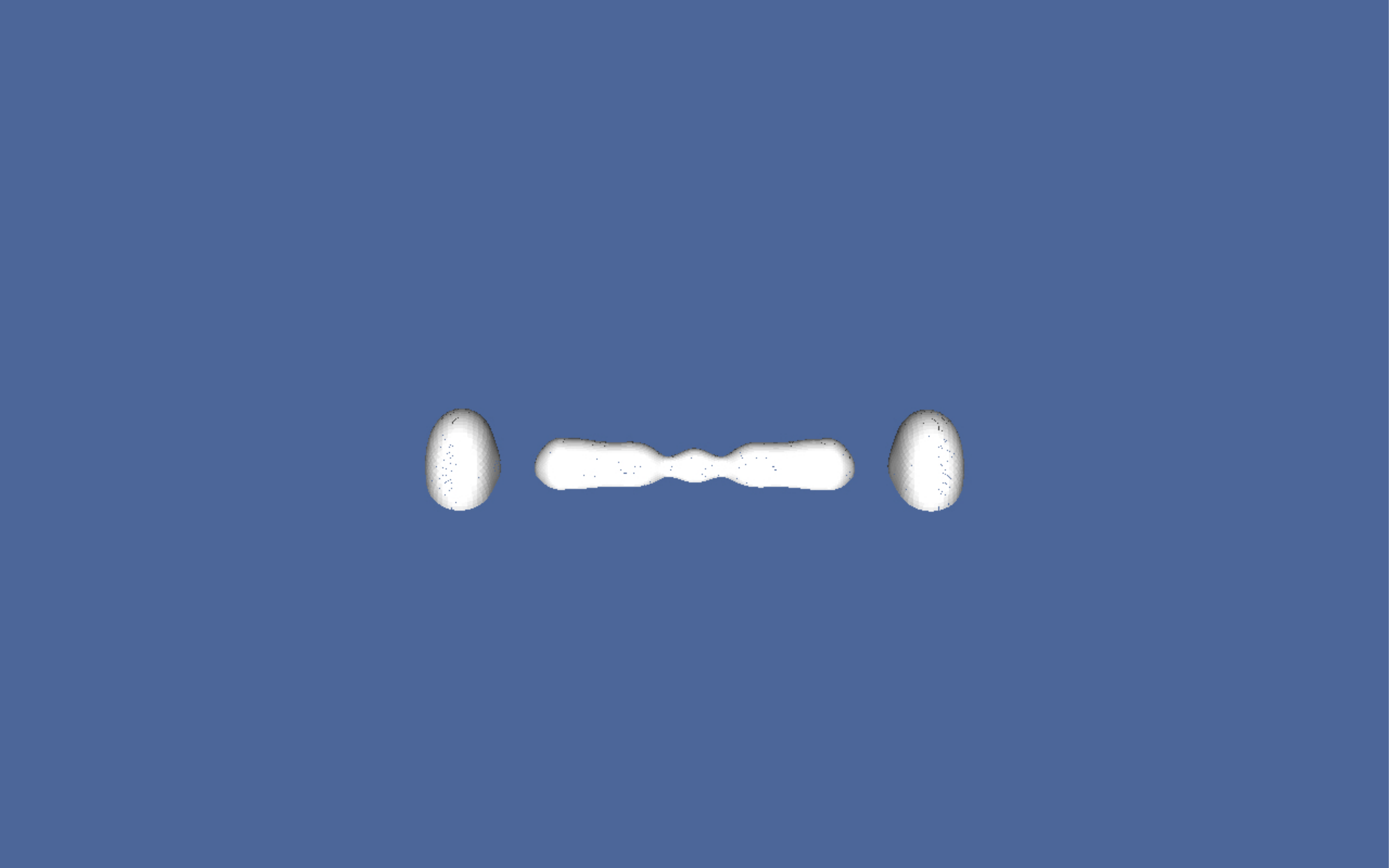}&
  \includegraphics[width=.25\textwidth]{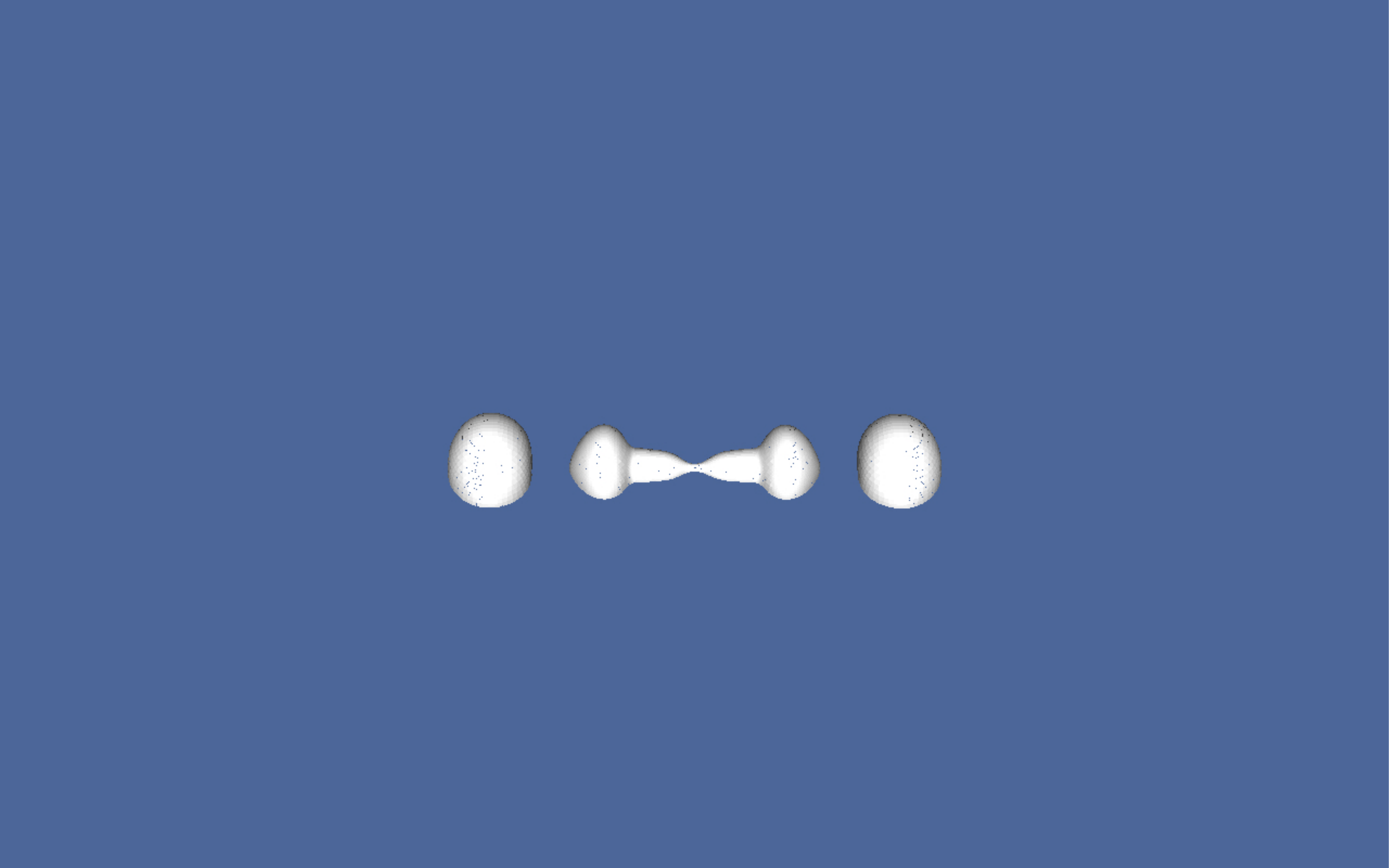}\\
  \includegraphics[width=.25\textwidth]{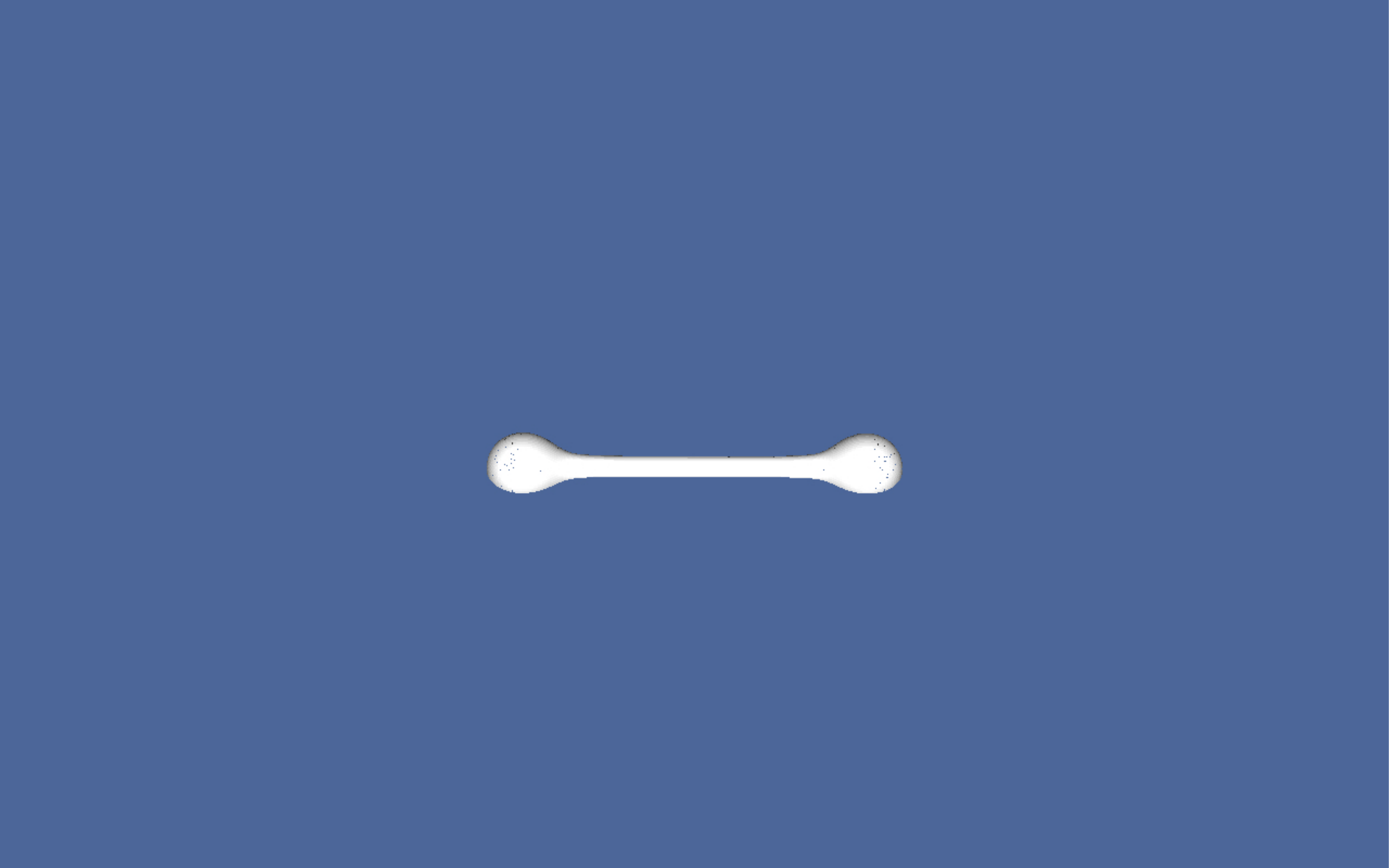}&
  \includegraphics[width=.25\textwidth]{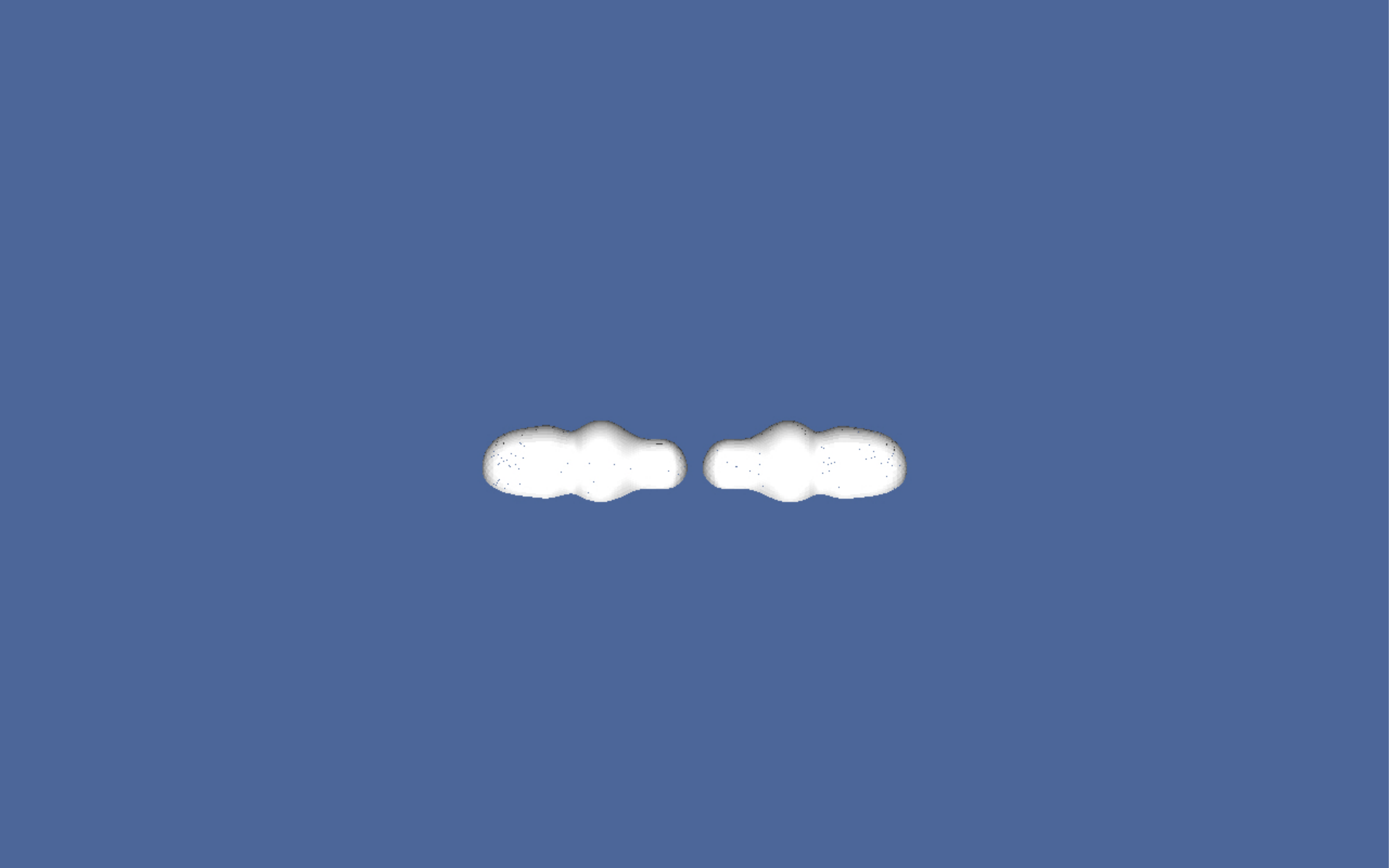}&
  \includegraphics[width=.25\textwidth]{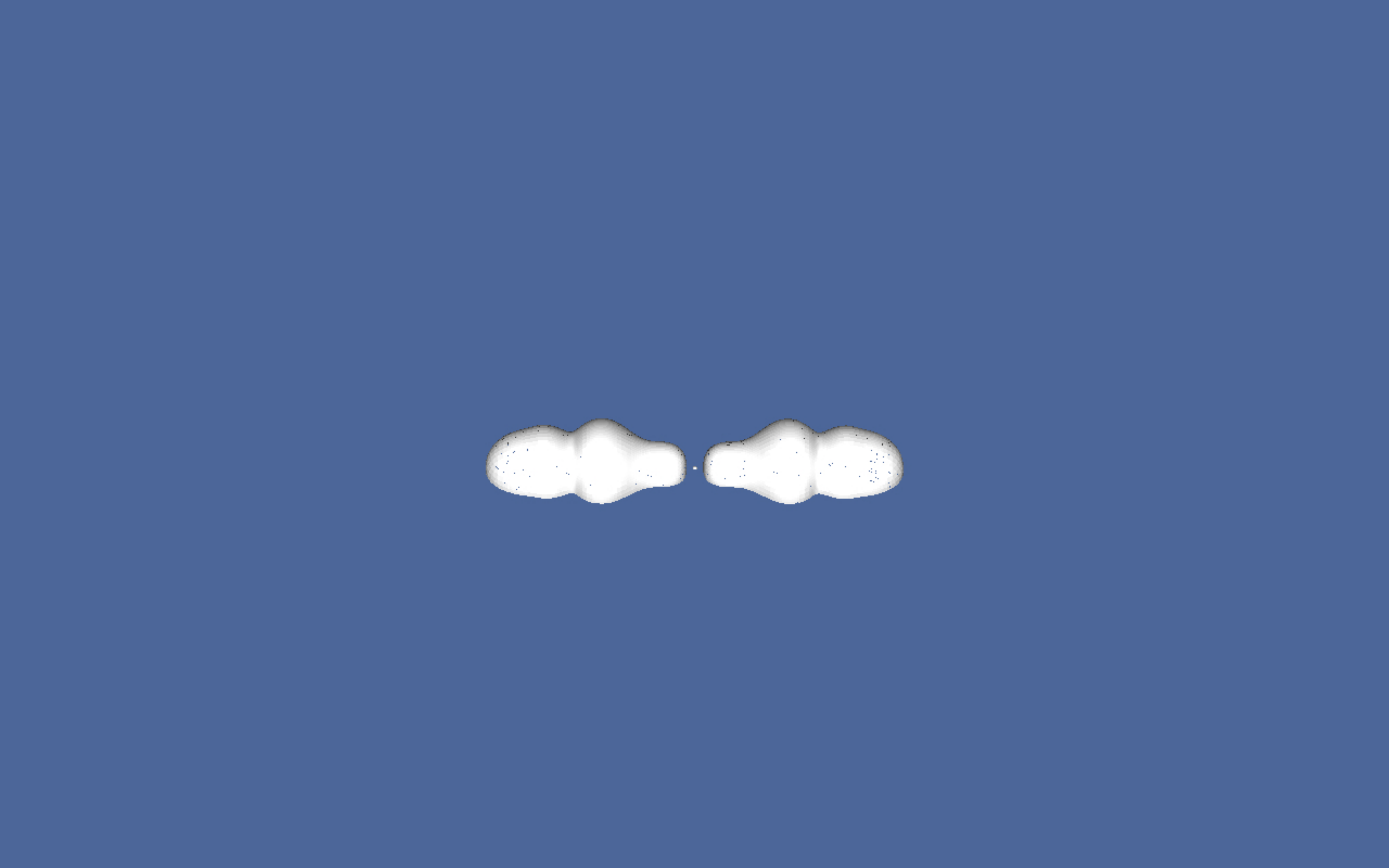}\\
  \includegraphics[width=.25\textwidth]{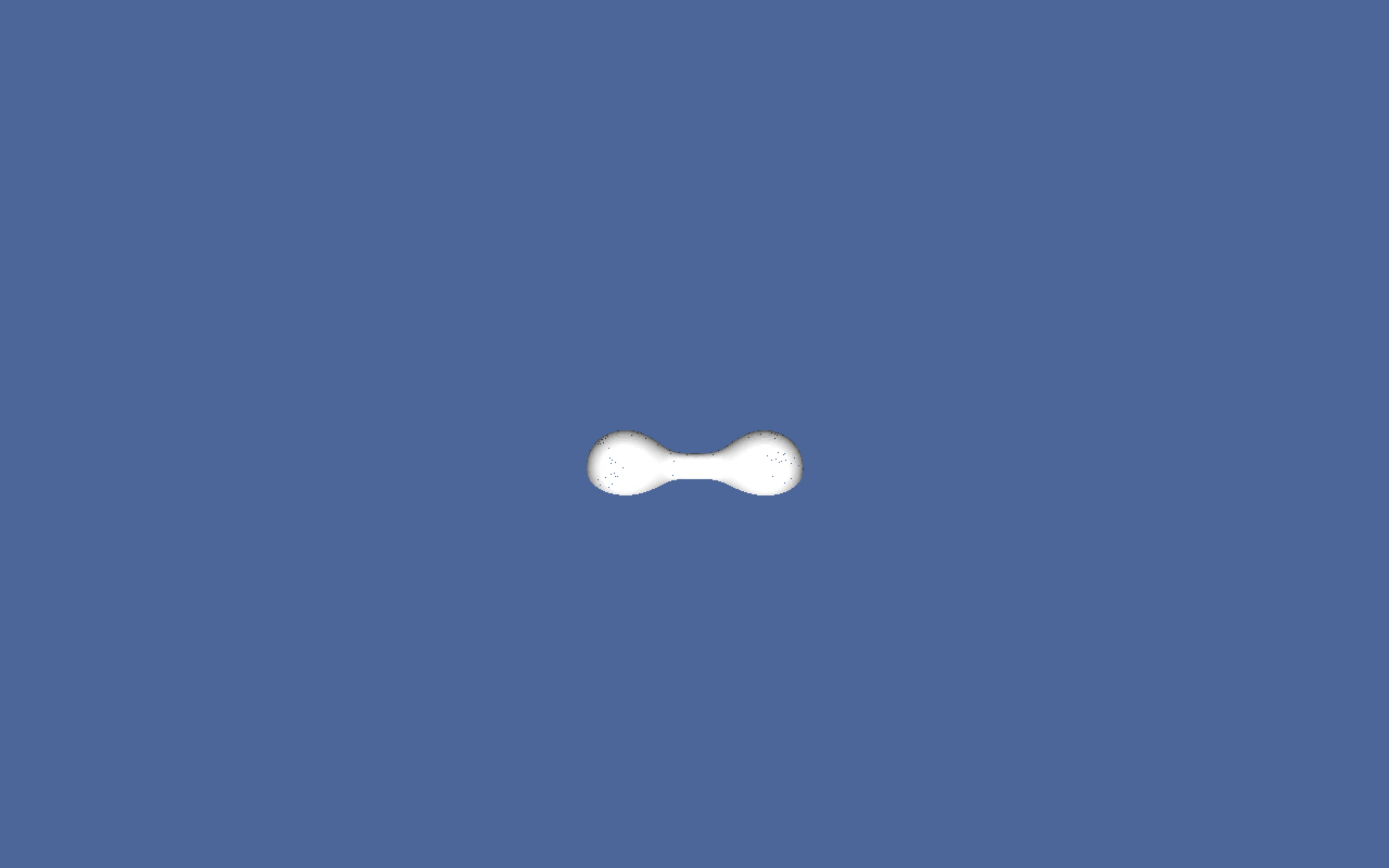}&
  \includegraphics[width=.25\textwidth]{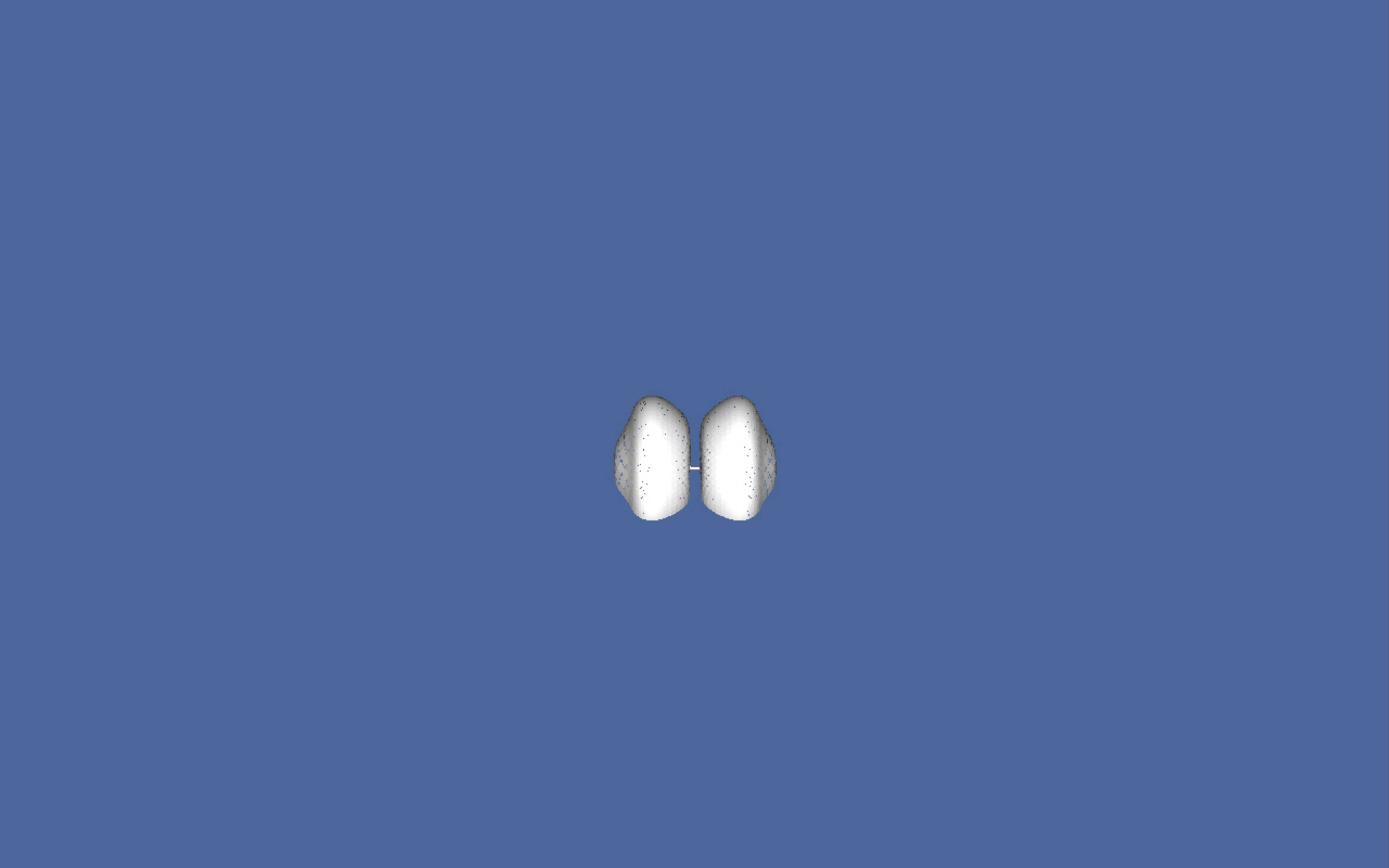}&
  \includegraphics[width=.25\textwidth]{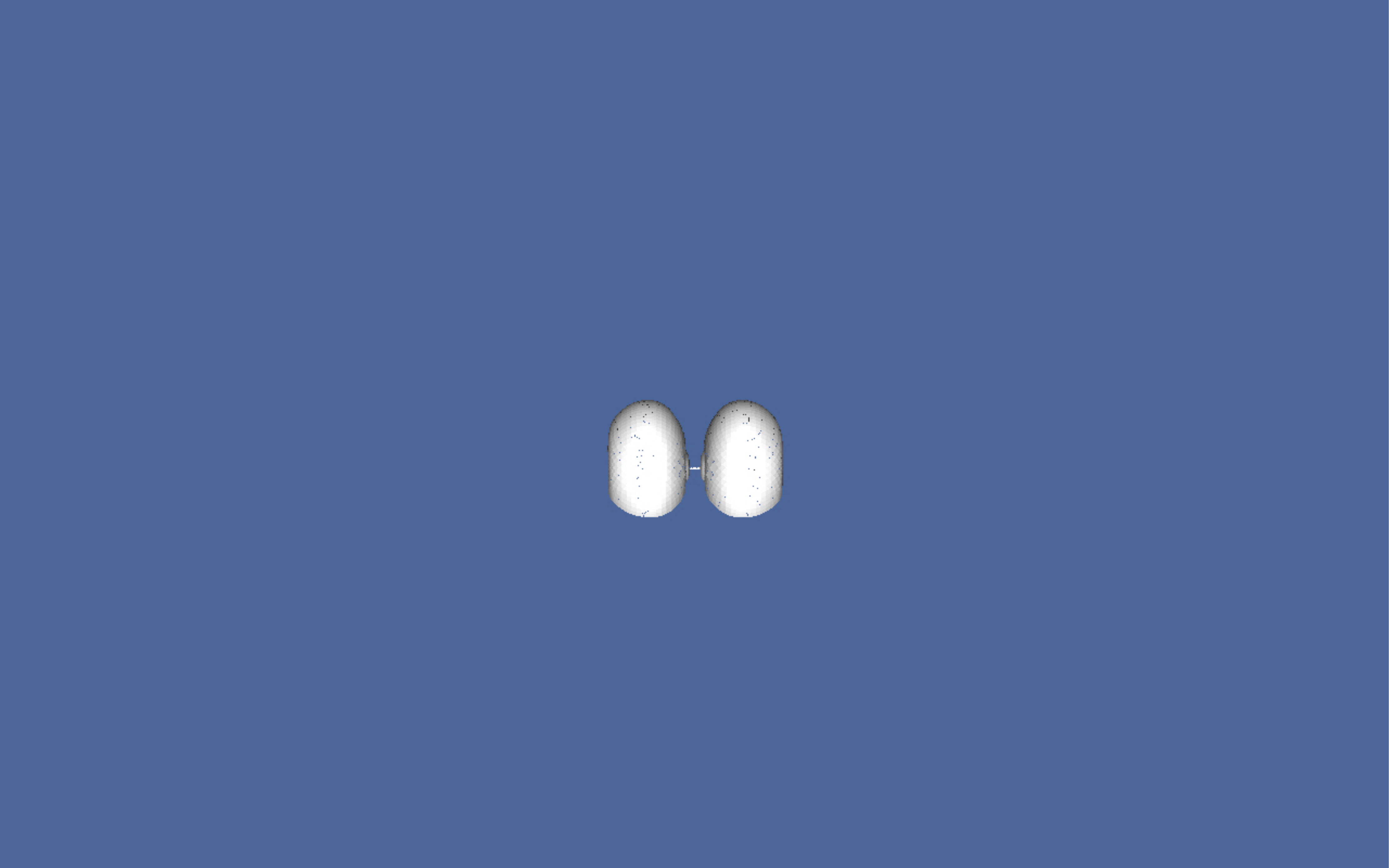}\\
  \includegraphics[width=.25\textwidth]{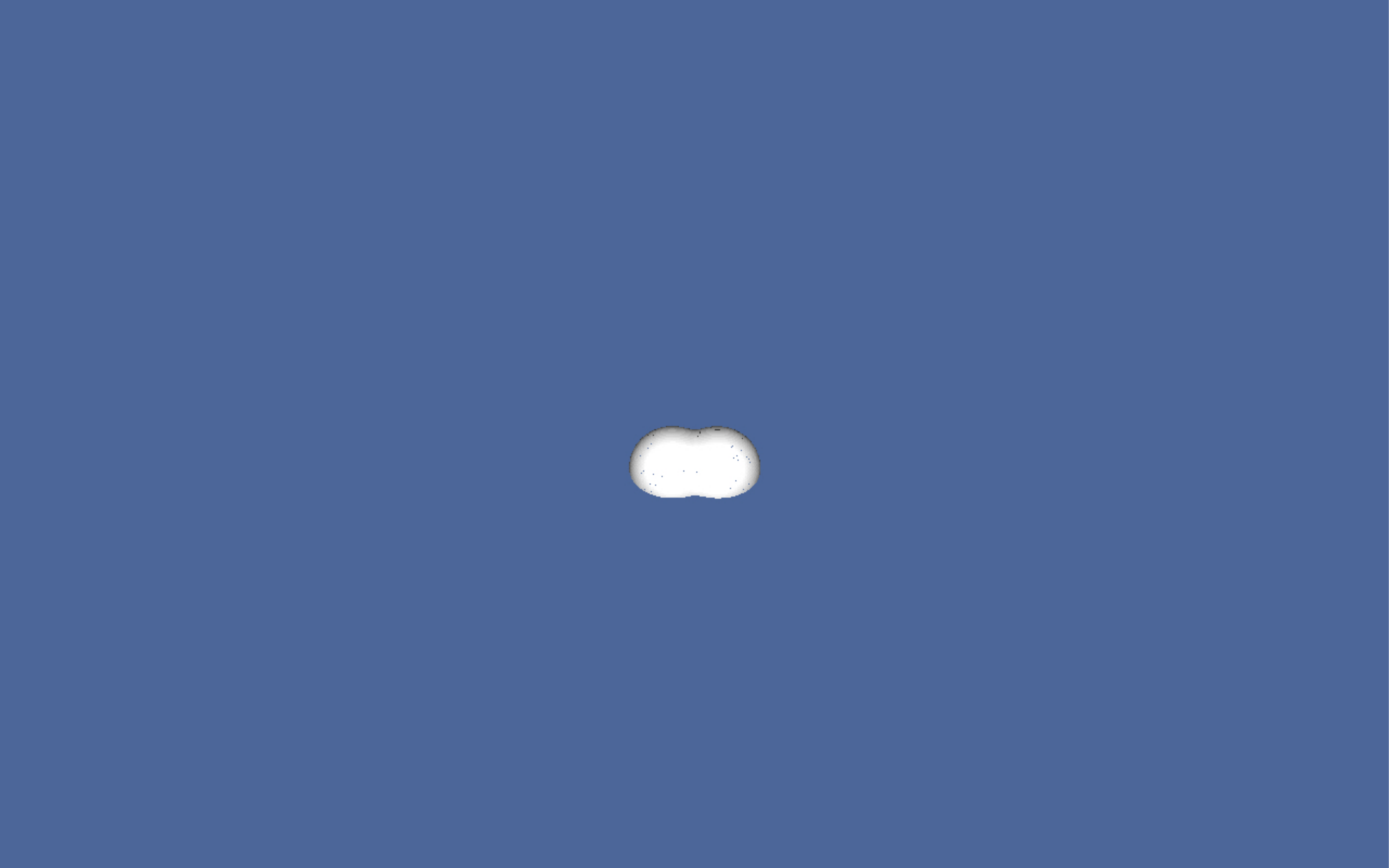}&
  \includegraphics[width=.25\textwidth]{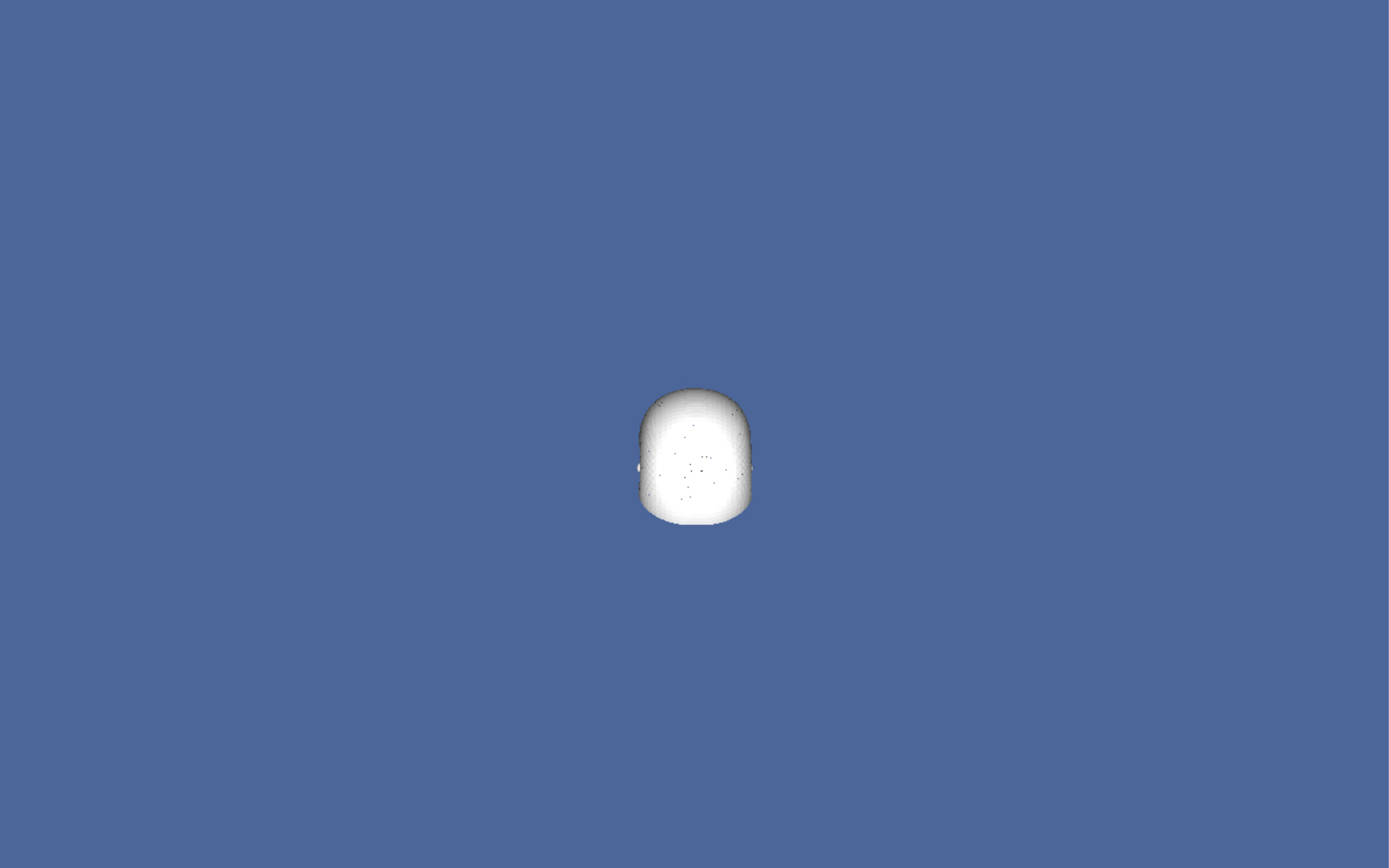}&
  \includegraphics[width=.25\textwidth]{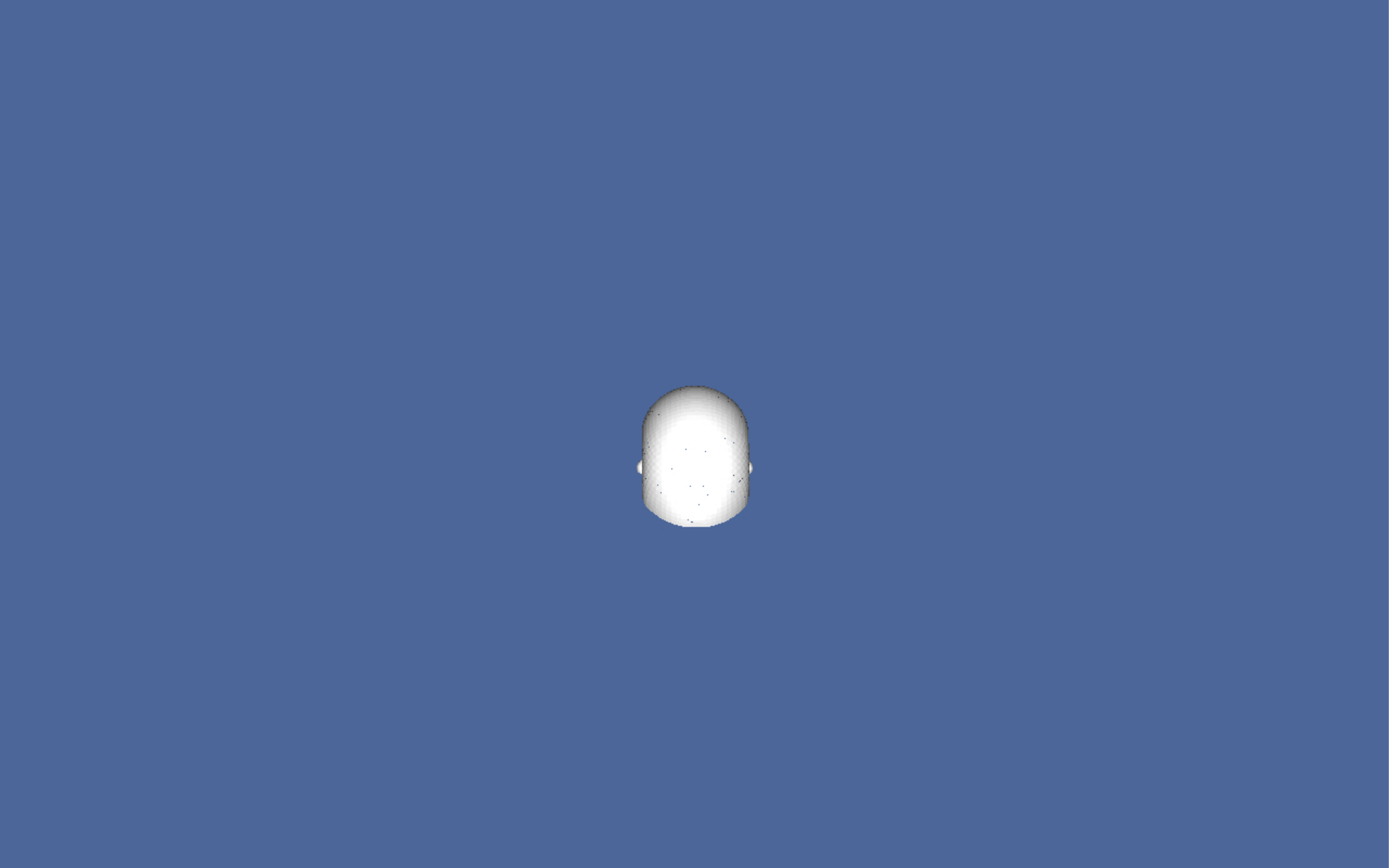}\\
  \includegraphics[width=.25\textwidth]{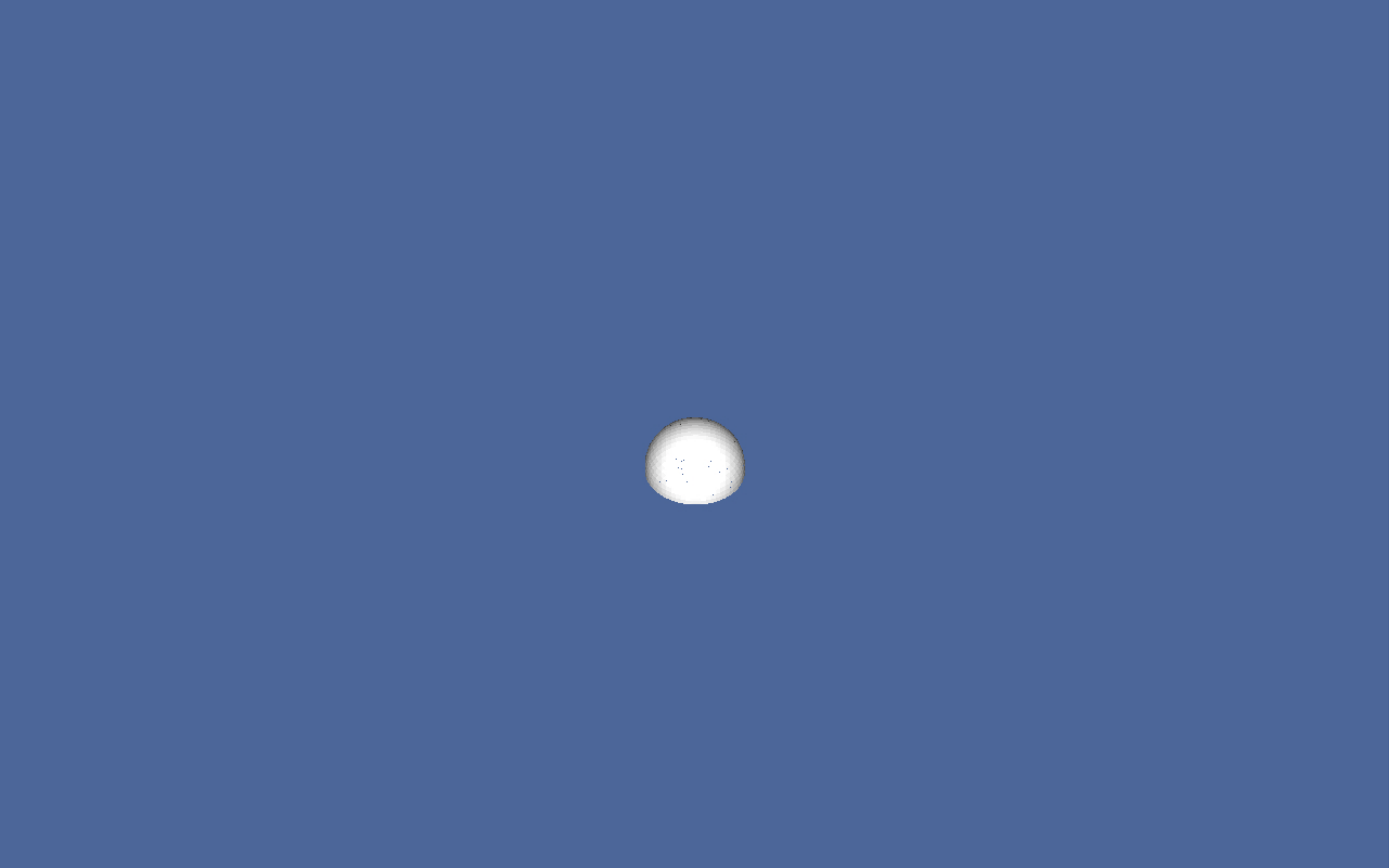}&
  \includegraphics[width=.25\textwidth]{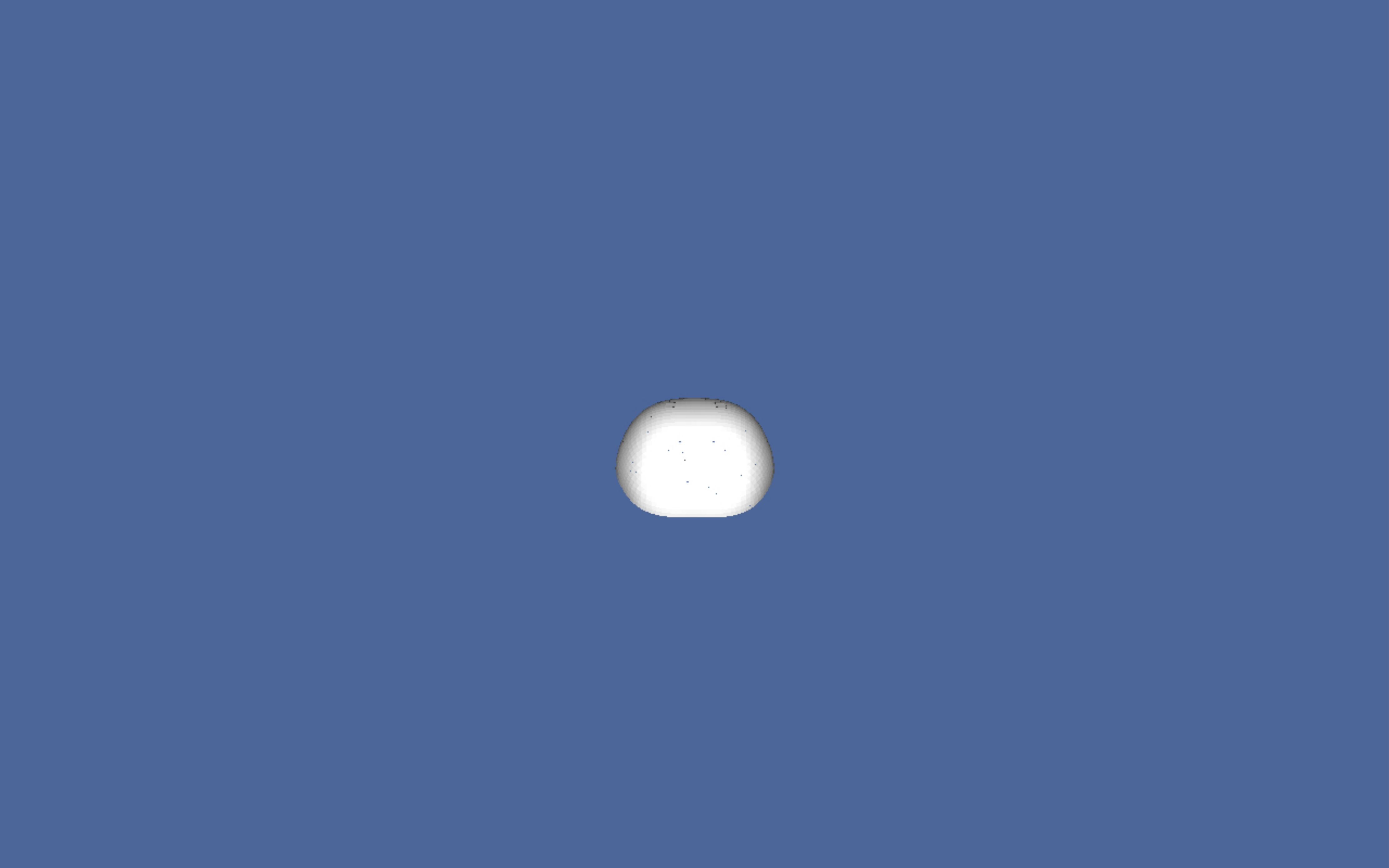}&
  \includegraphics[width=.25\textwidth]{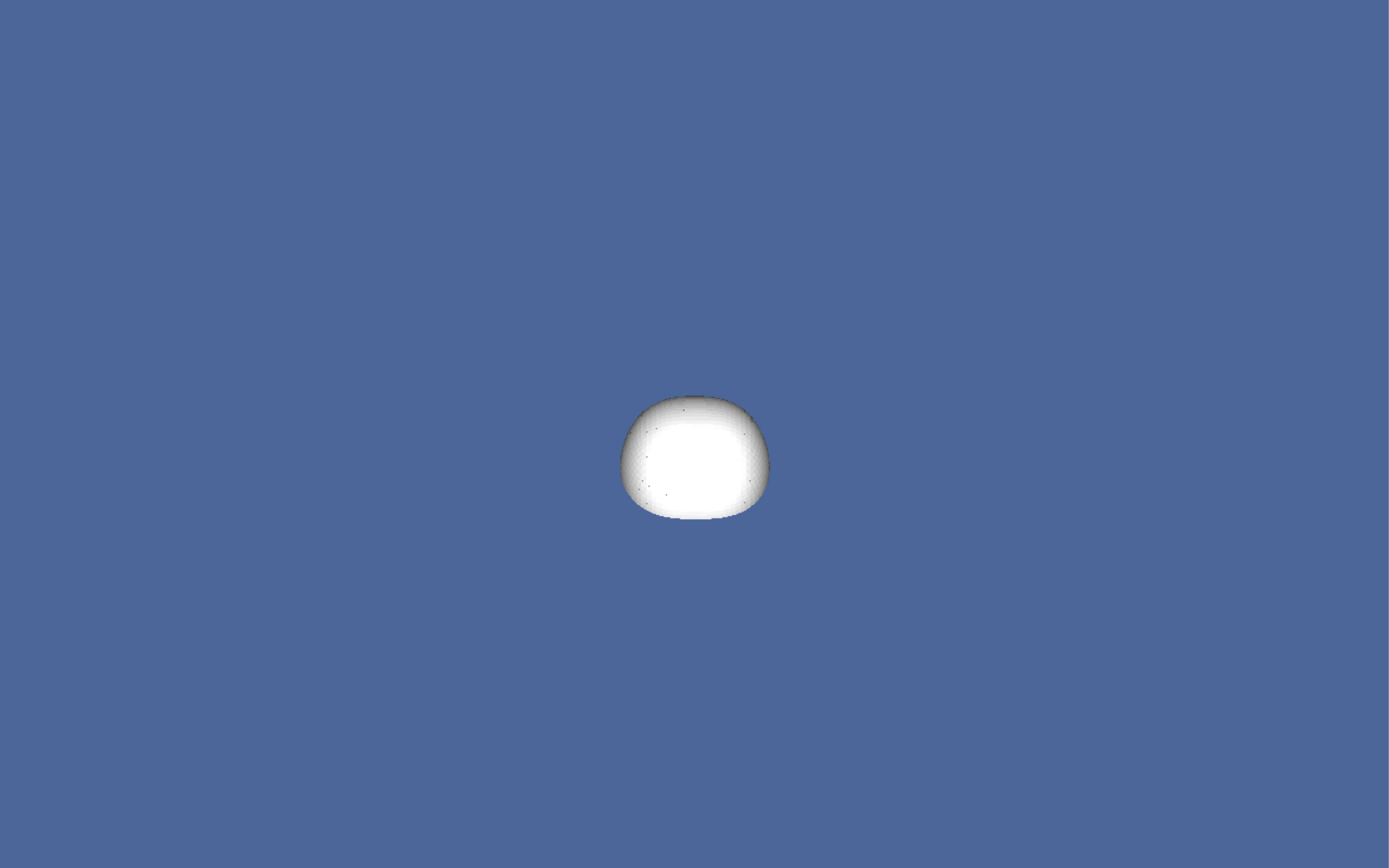}\\
  a &b &c 
\end{tabular}
\caption{Simulations for free-standing jets: (a) (Oh,AR) = (0.8,49) and (b,c) (Oh,AR) = (0.015,28) for (8,9) levels of refinement.
Only top half of the filament is shown.}
\vspace{-0.1in}
\label{fig:scenarios1}
\end{figure}

\begin{figure}[thb]
\centering
\begin{tabular}{cccc}
  \includegraphics[width=.25\textwidth]{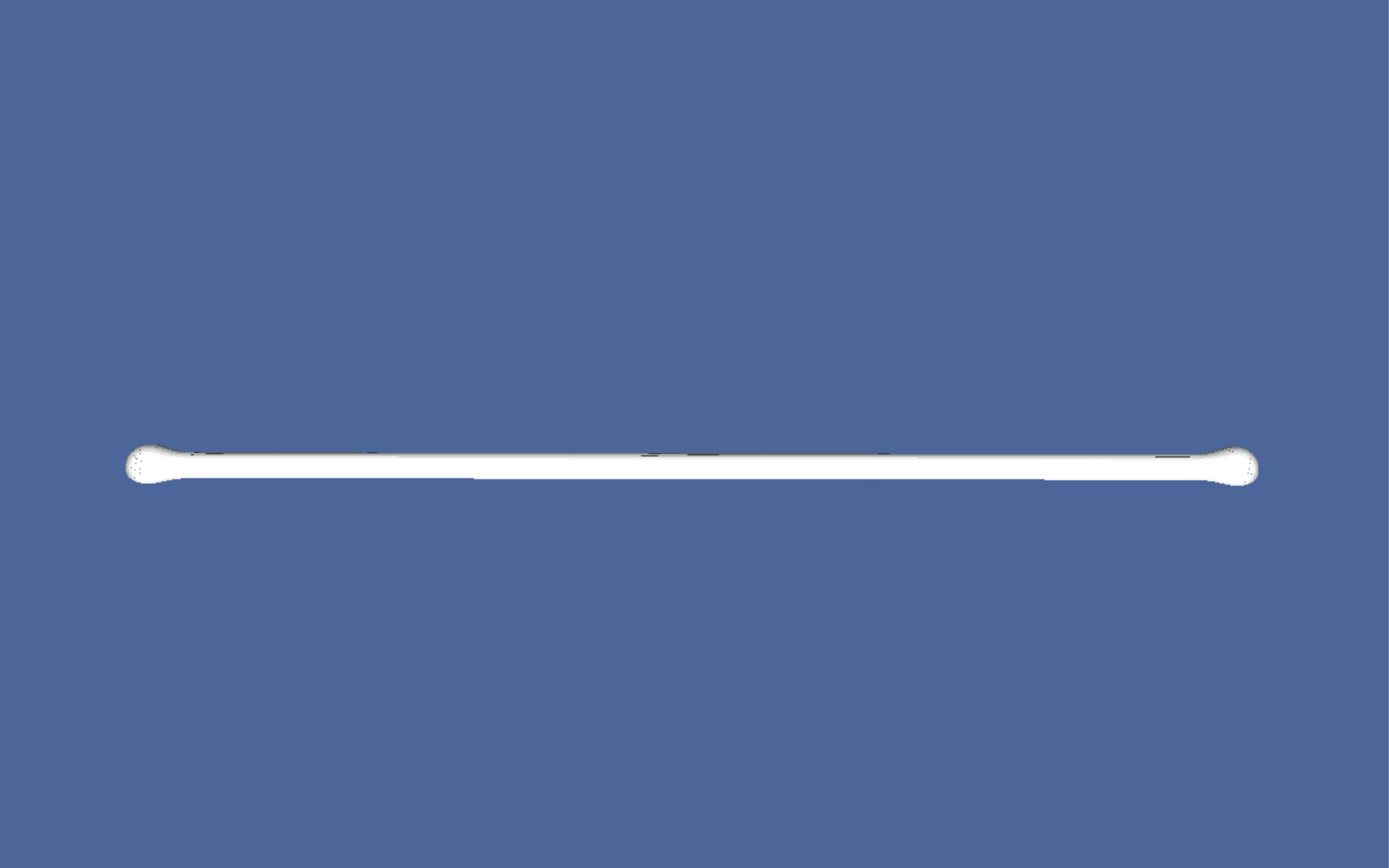}&
  \includegraphics[width=.25\textwidth]{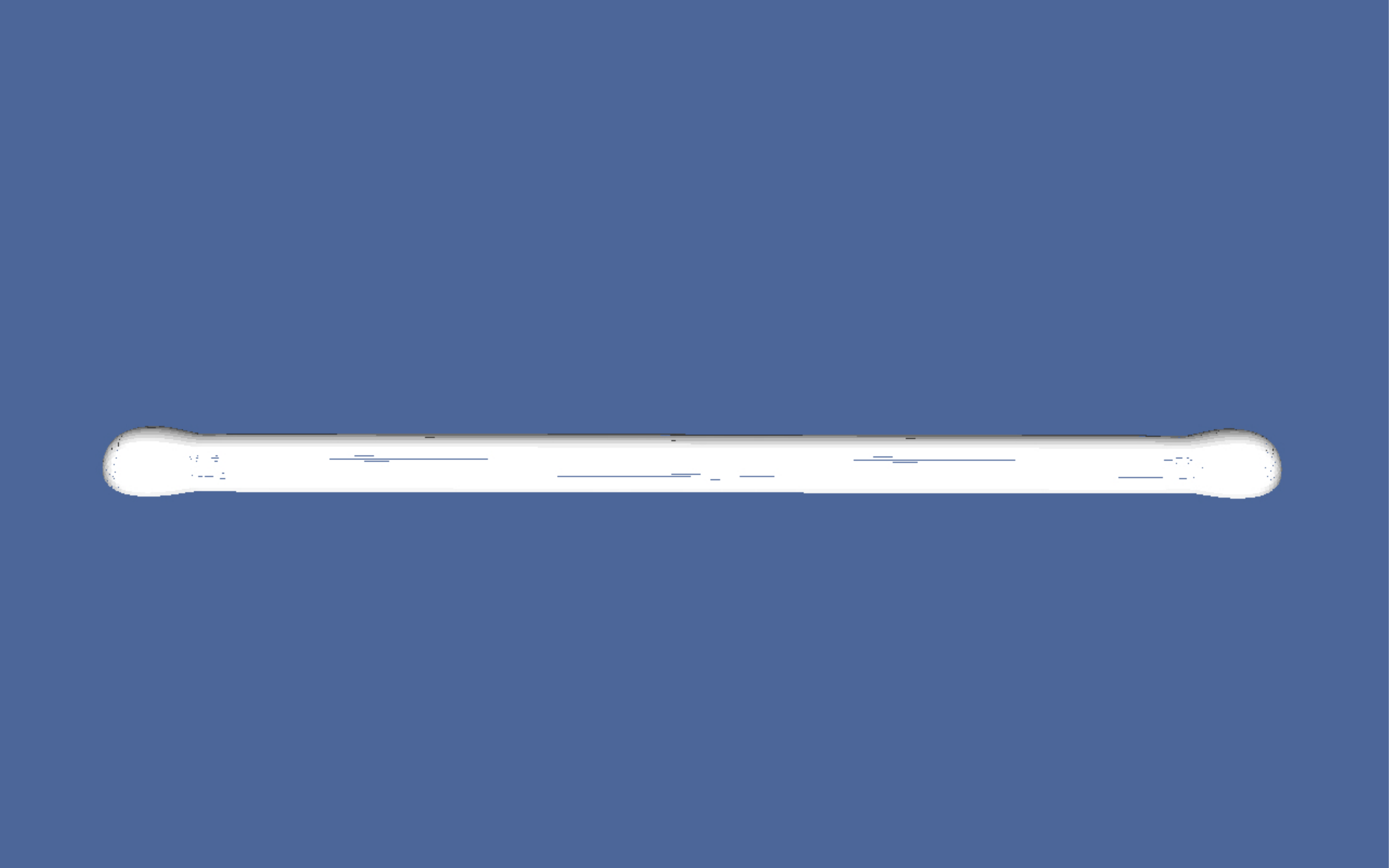}&
  \includegraphics[width=.25\textwidth]{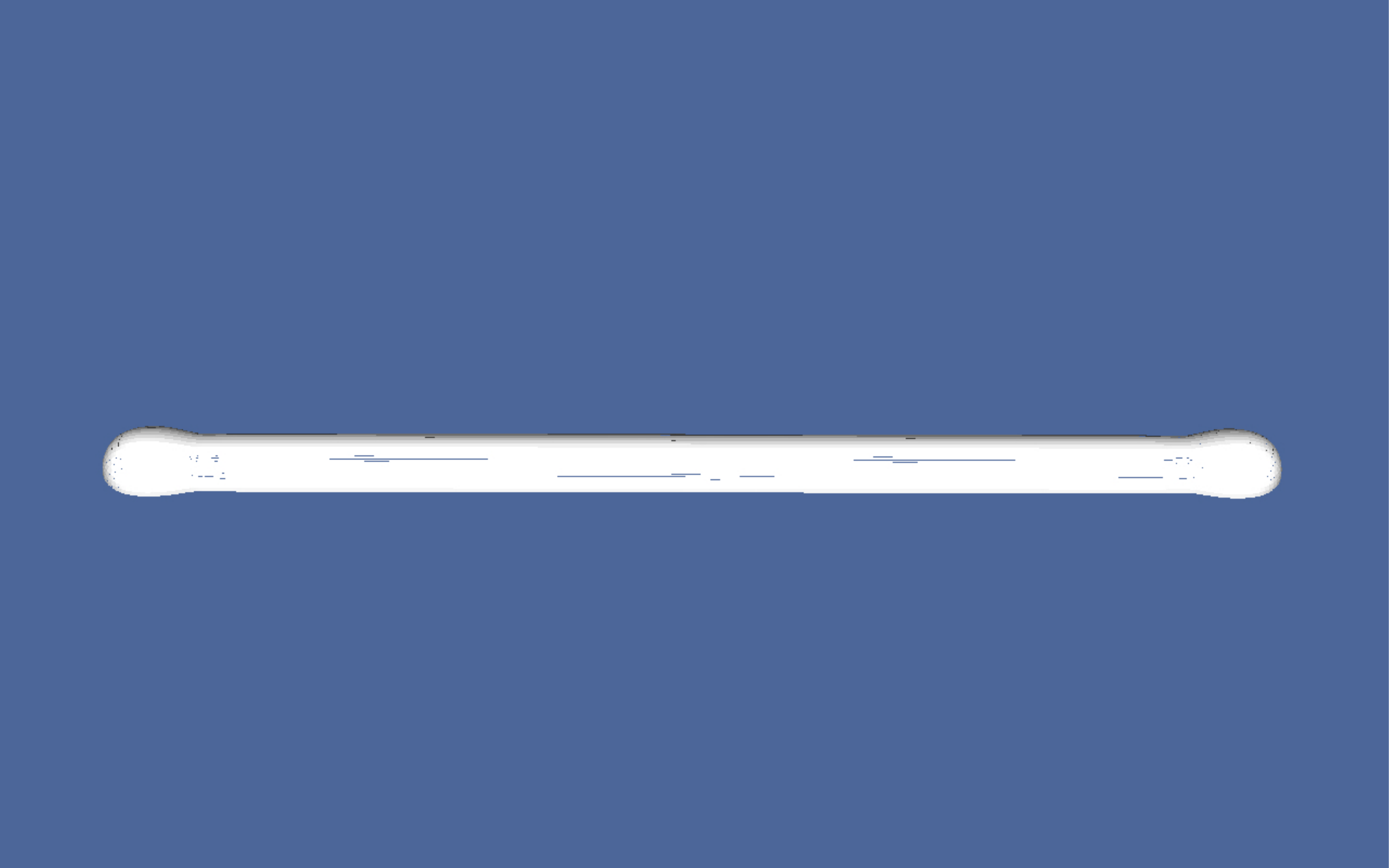}&
  \includegraphics[width=.25\textwidth]{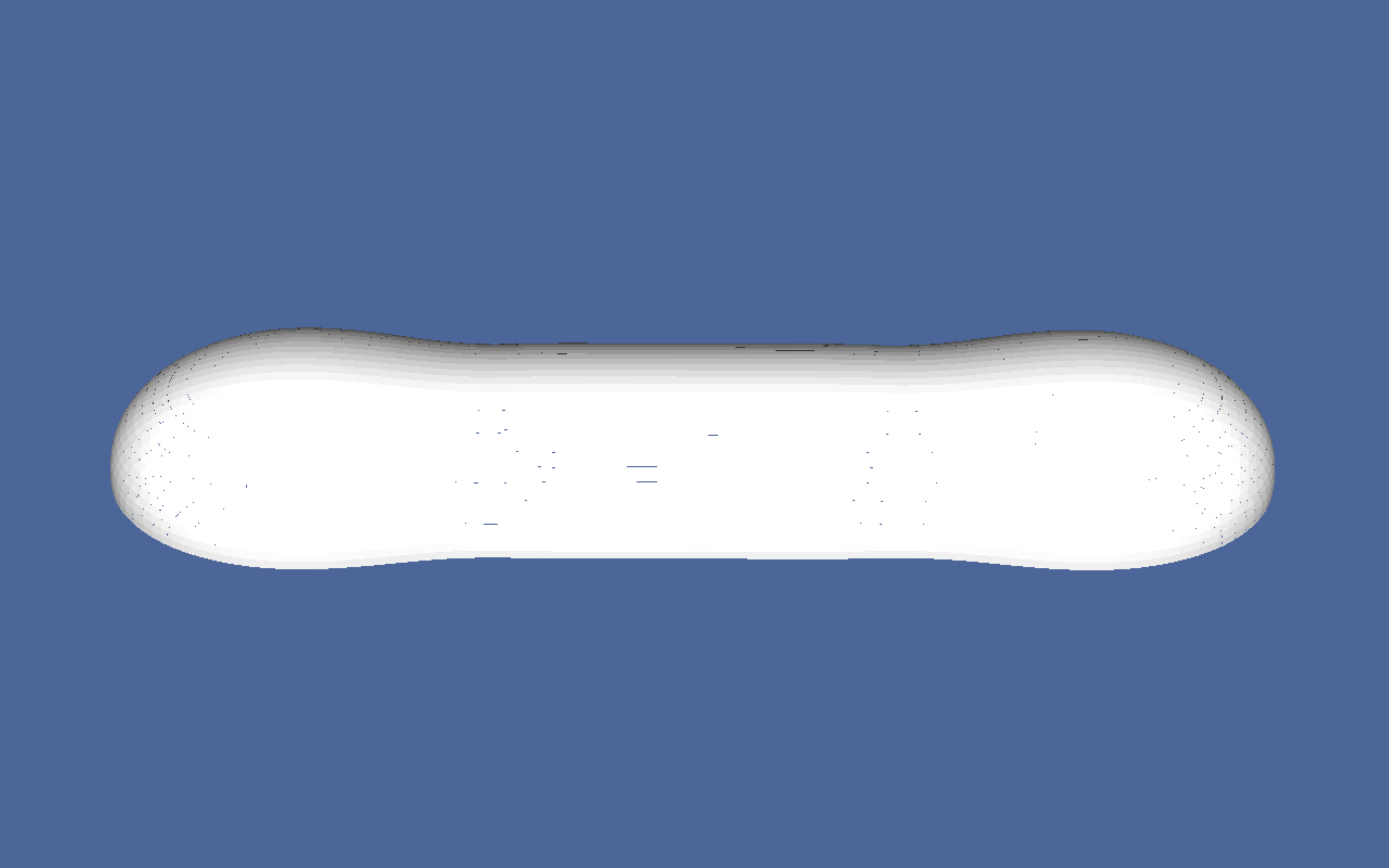}\\
  \includegraphics[width=.25\textwidth]{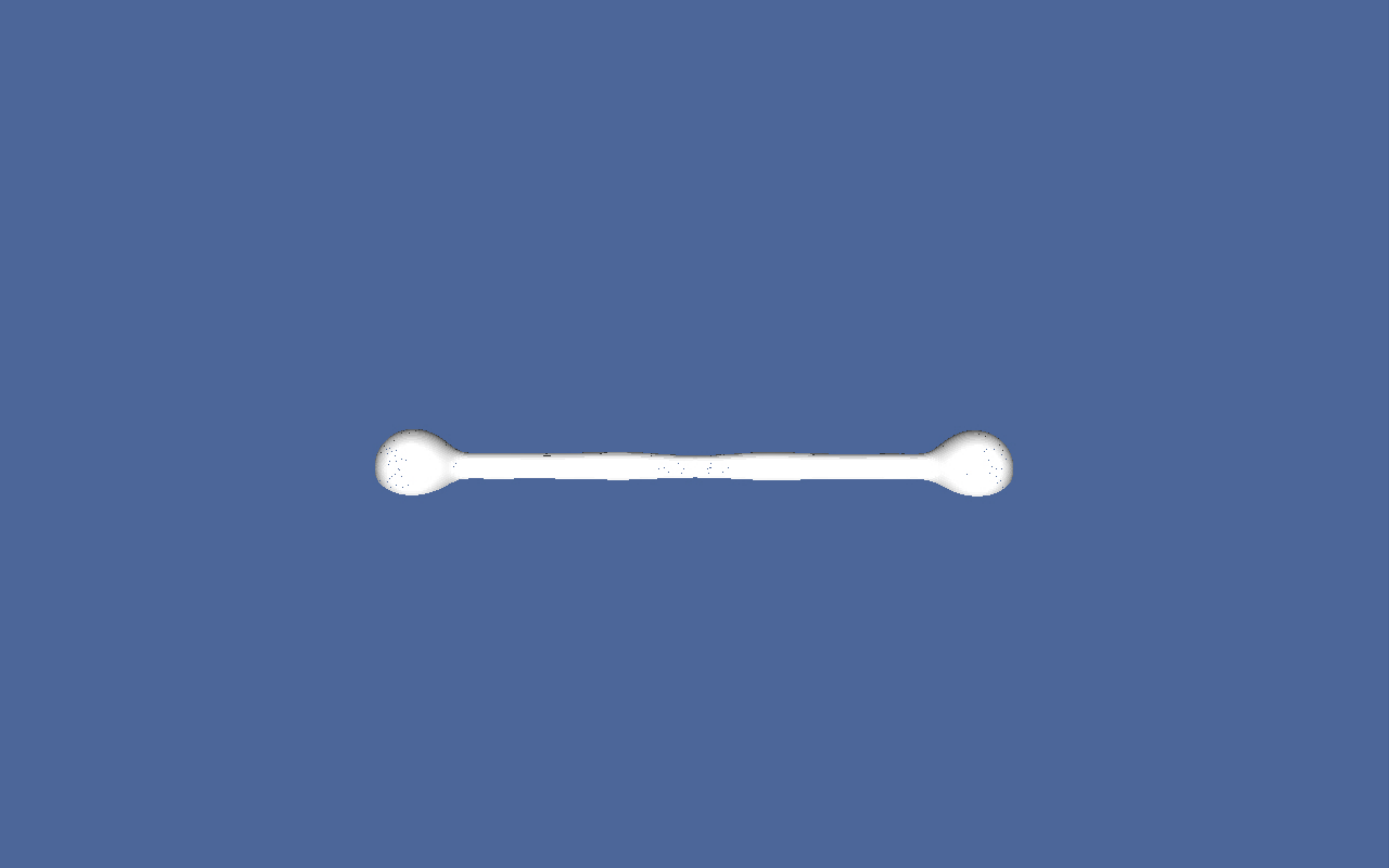}&
  \includegraphics[width=.25\textwidth]{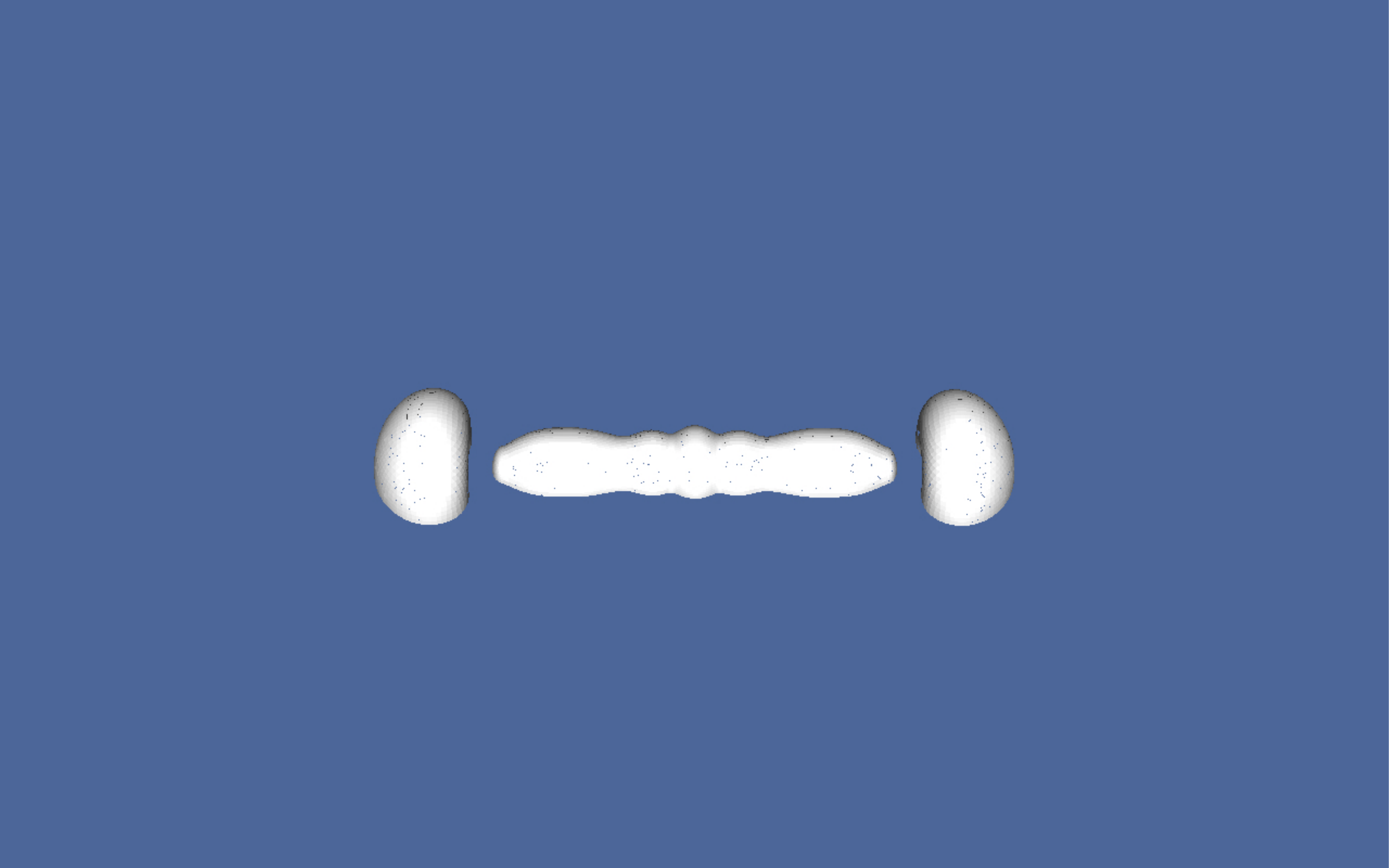}&
  \includegraphics[width=.25\textwidth]{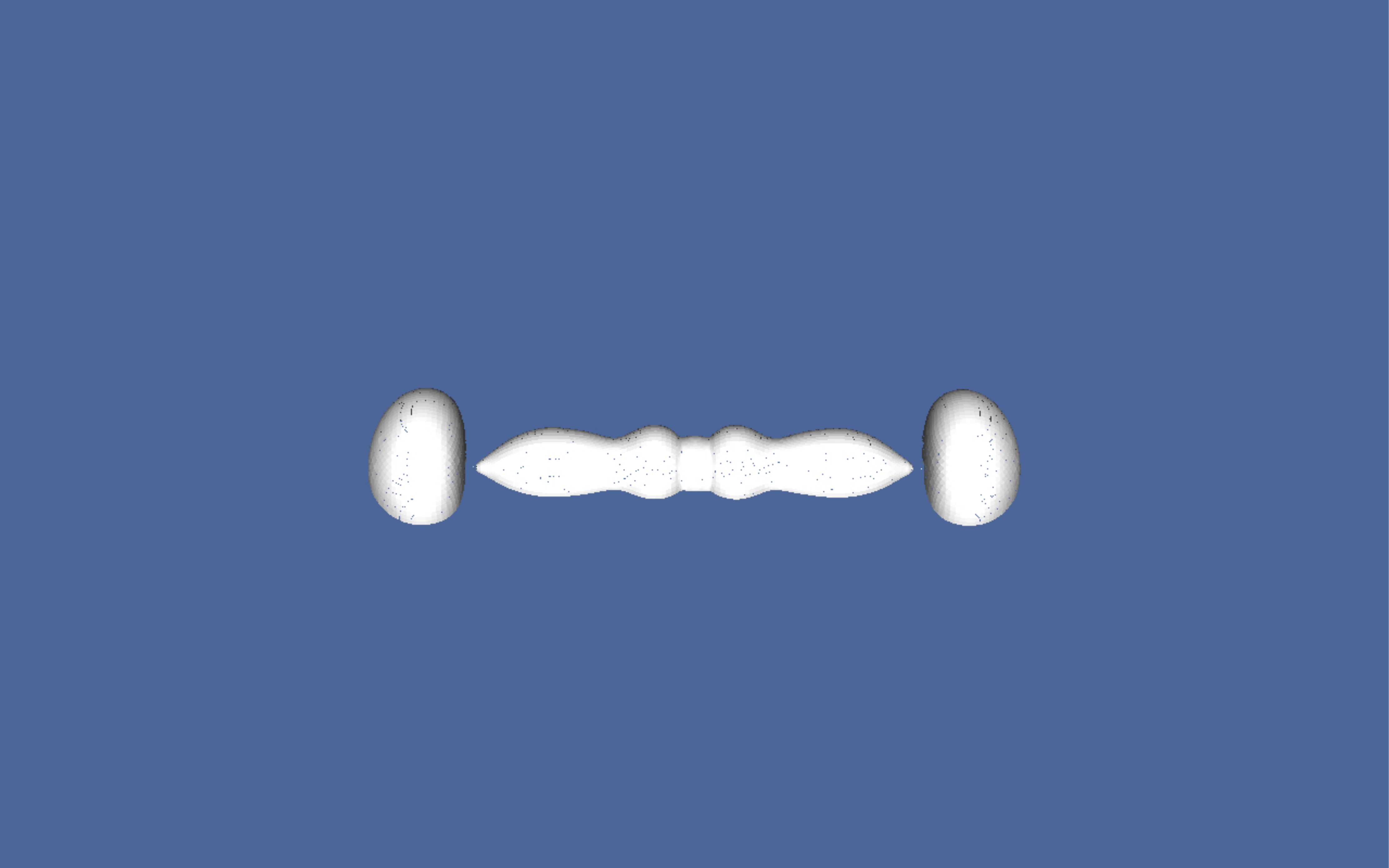}&
  \includegraphics[width=.25\textwidth]{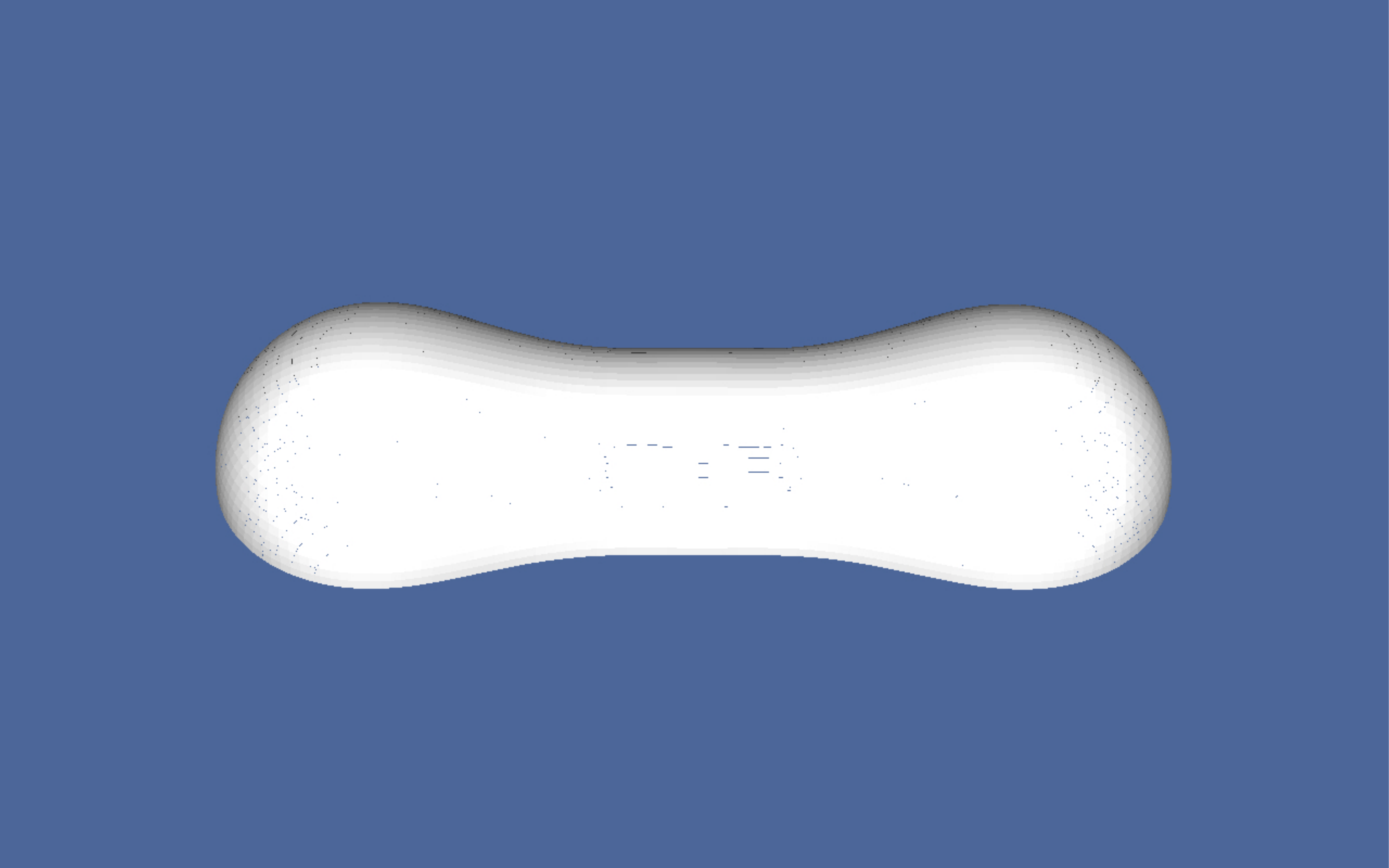}\\
  \includegraphics[width=.25\textwidth]{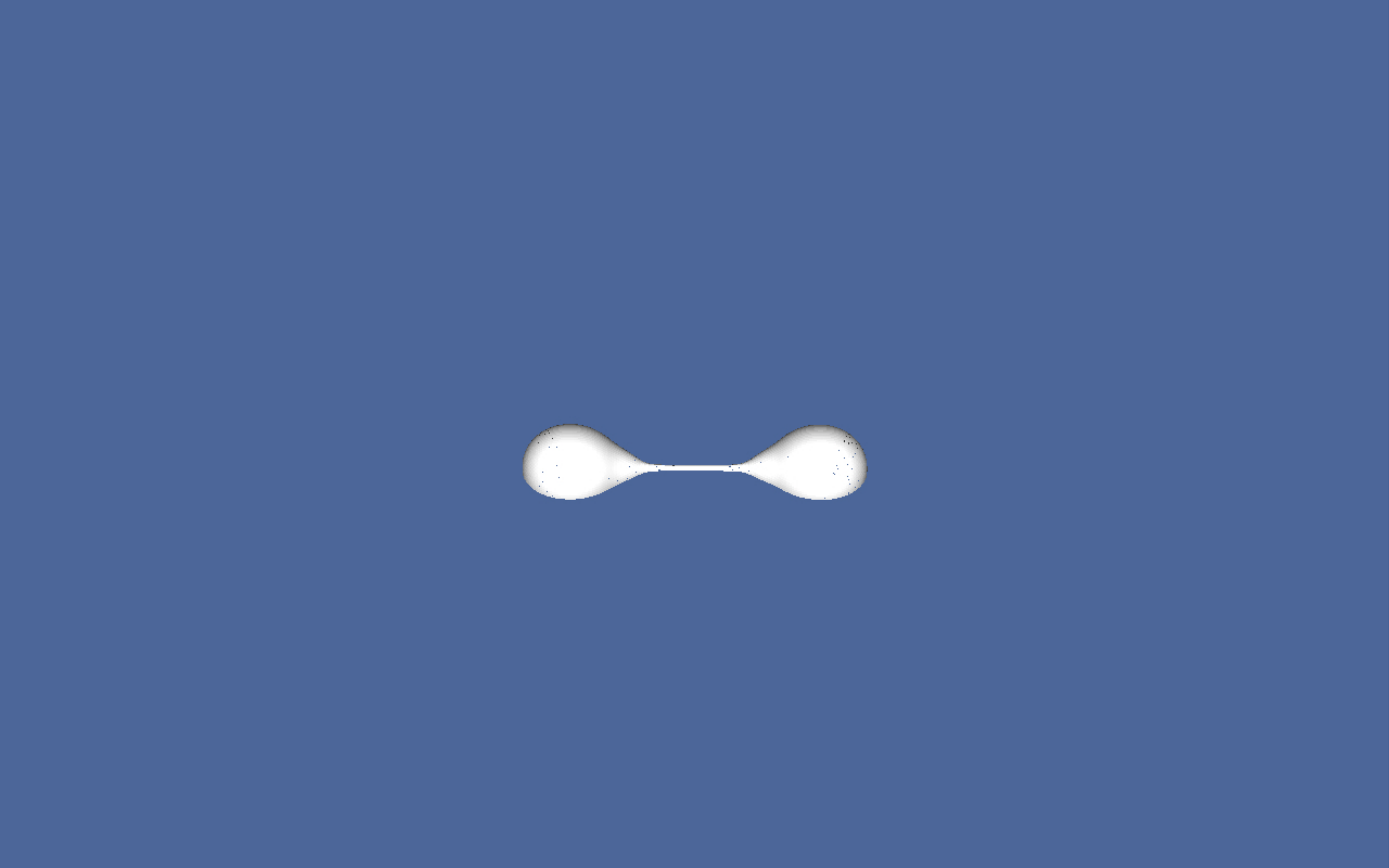}&
  \includegraphics[width=.25\textwidth]{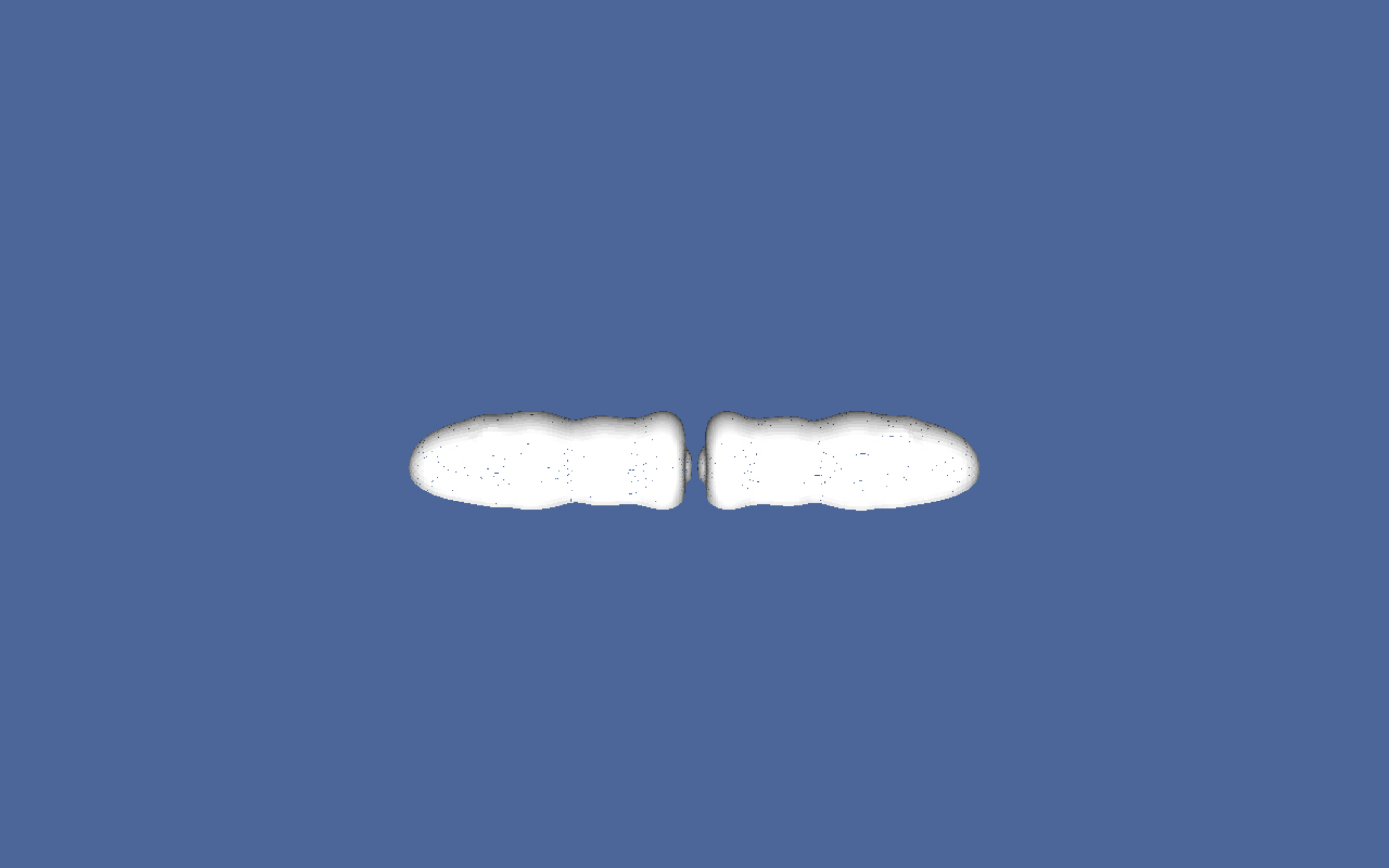}&
  \includegraphics[width=.25\textwidth]{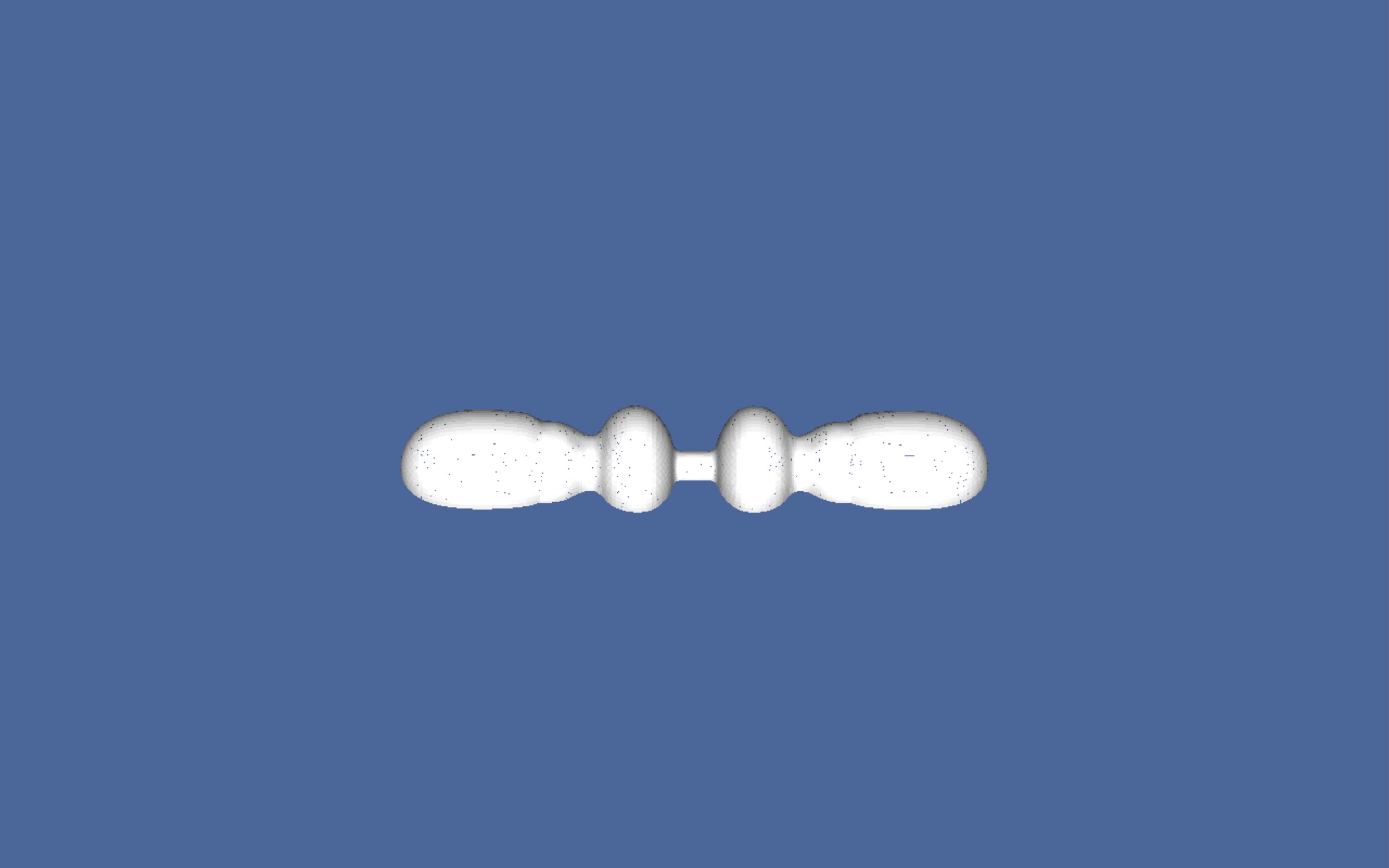}&
  \includegraphics[width=.25\textwidth]{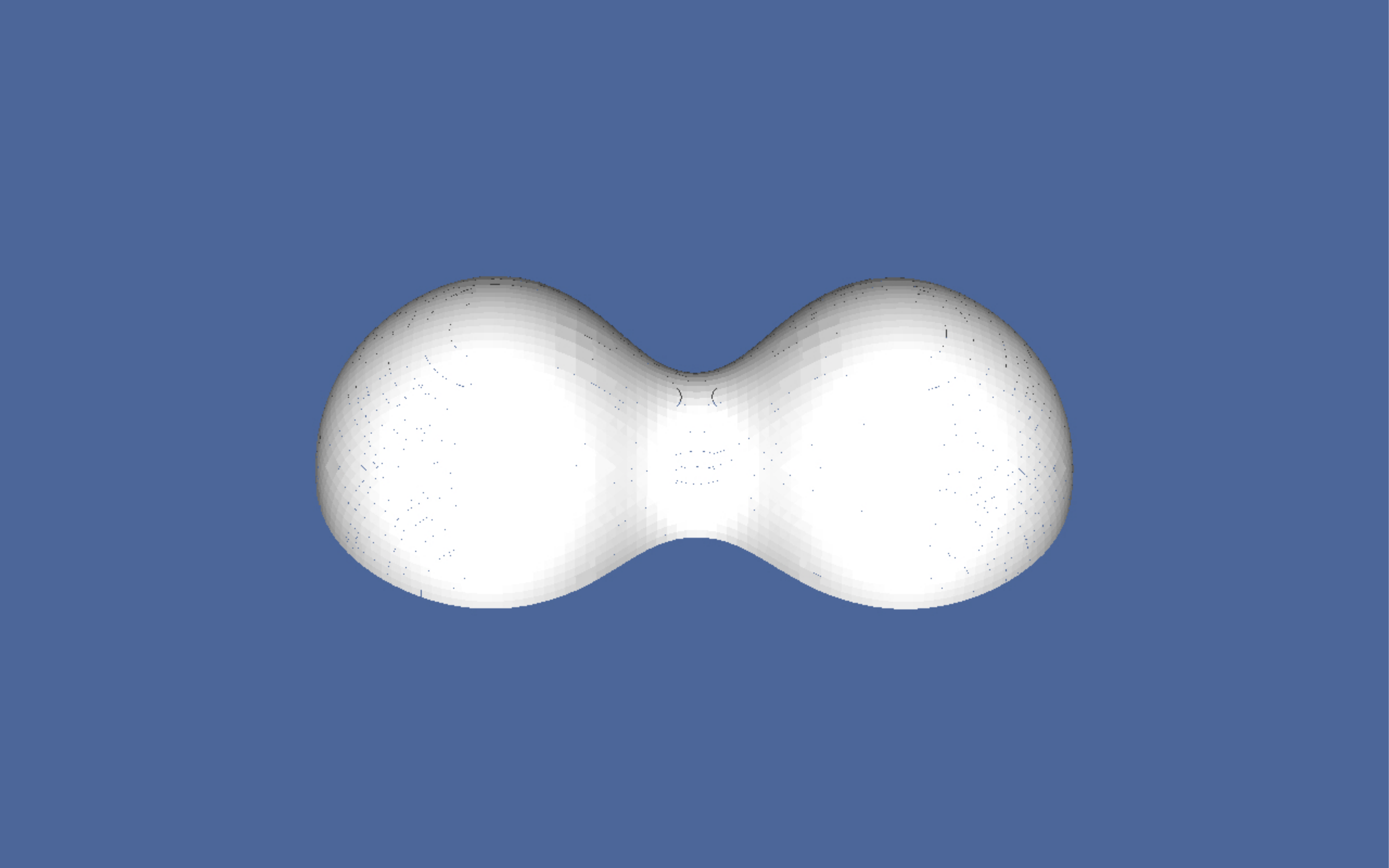}\\
  \includegraphics[width=.25\textwidth]{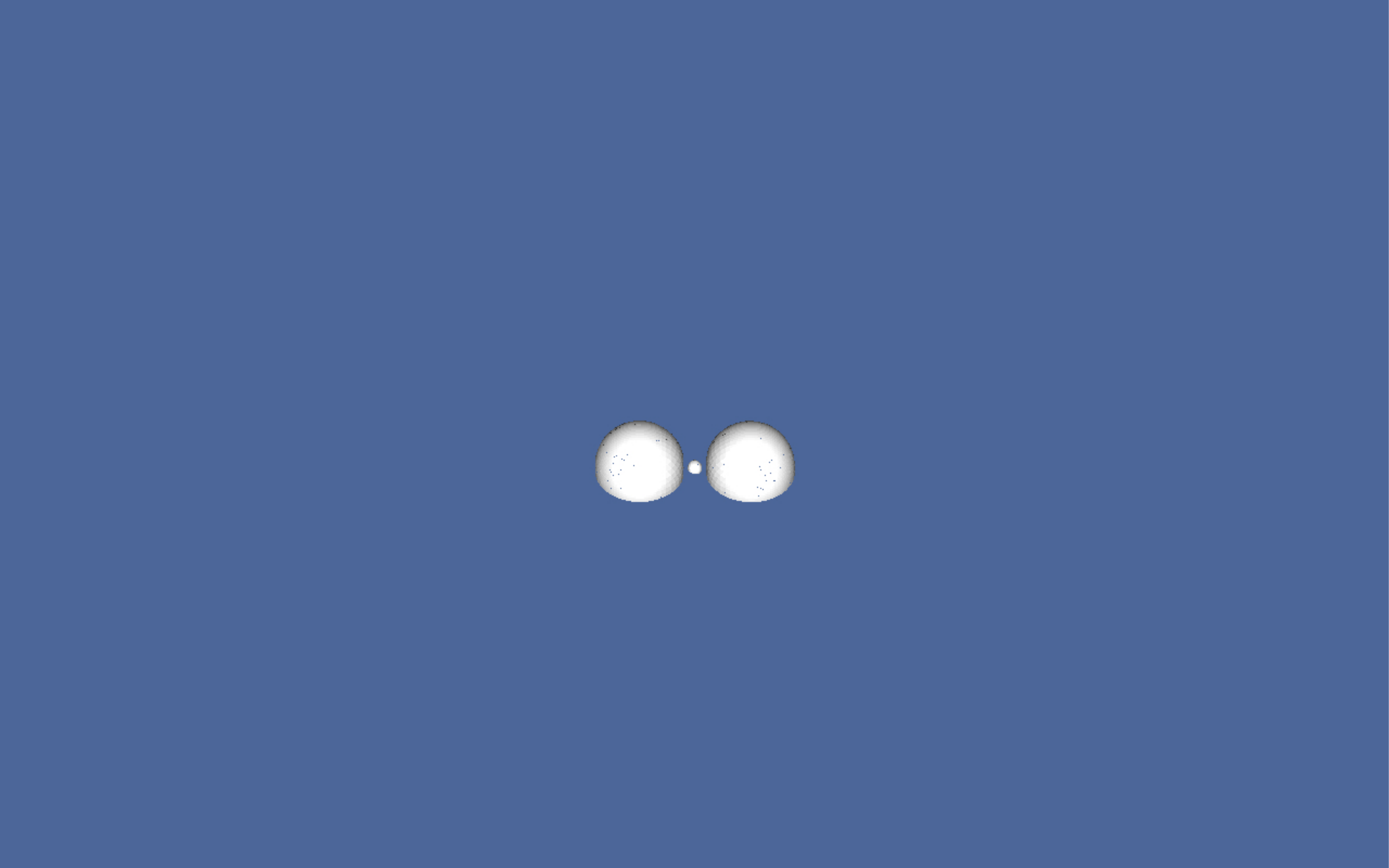}&
  \includegraphics[width=.25\textwidth]{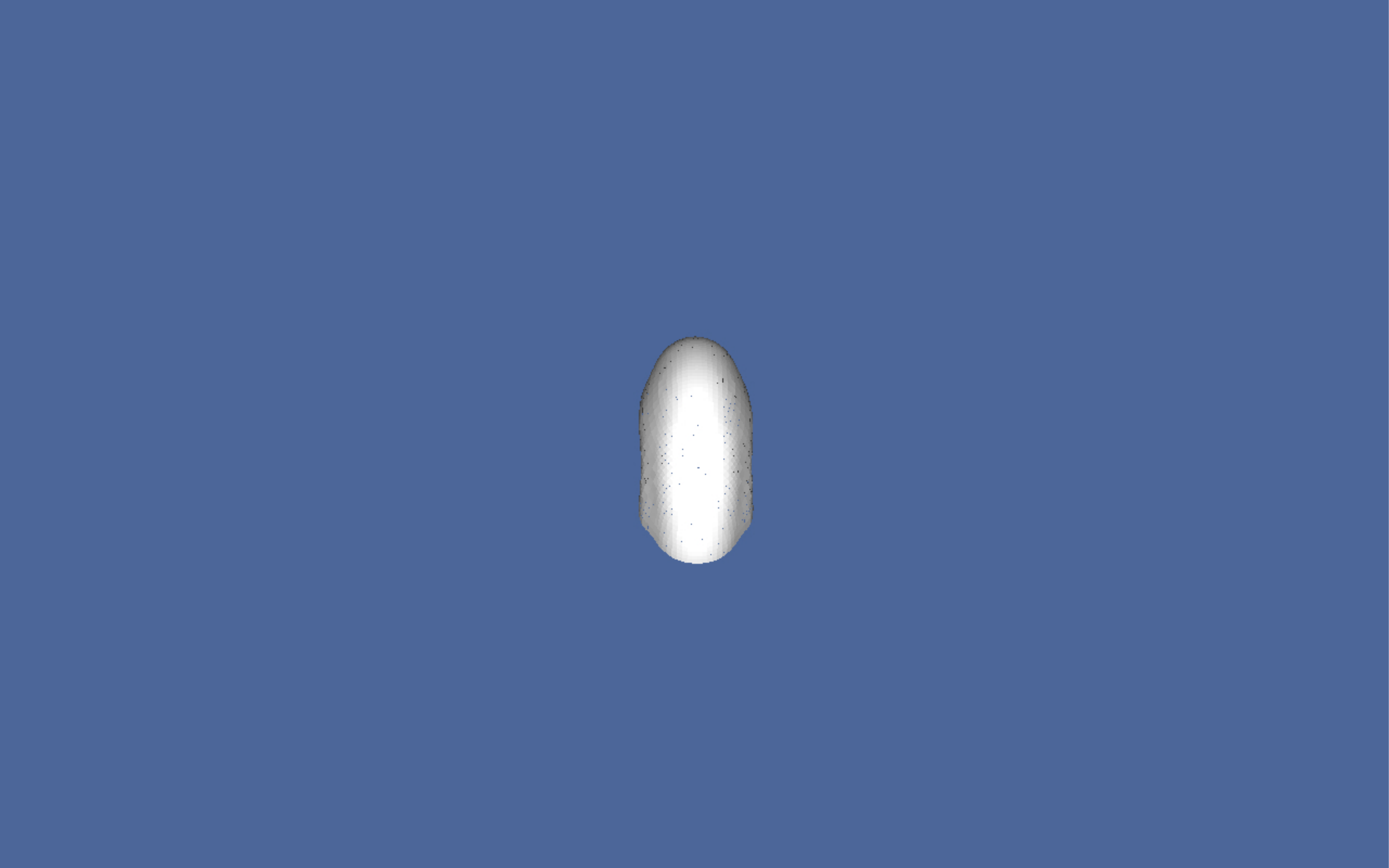}&
  \includegraphics[width=.25\textwidth]{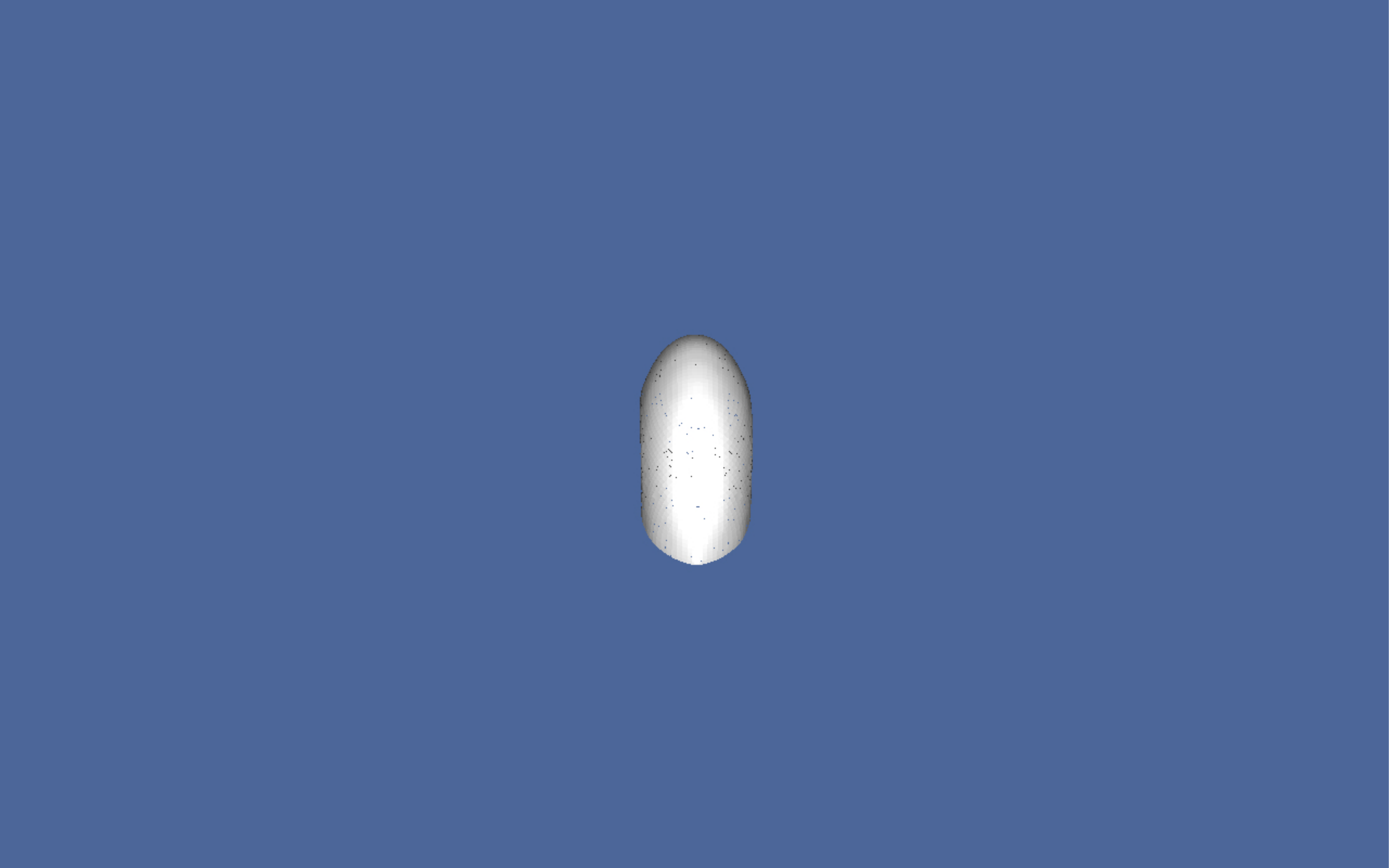}&
  \includegraphics[width=.25\textwidth]{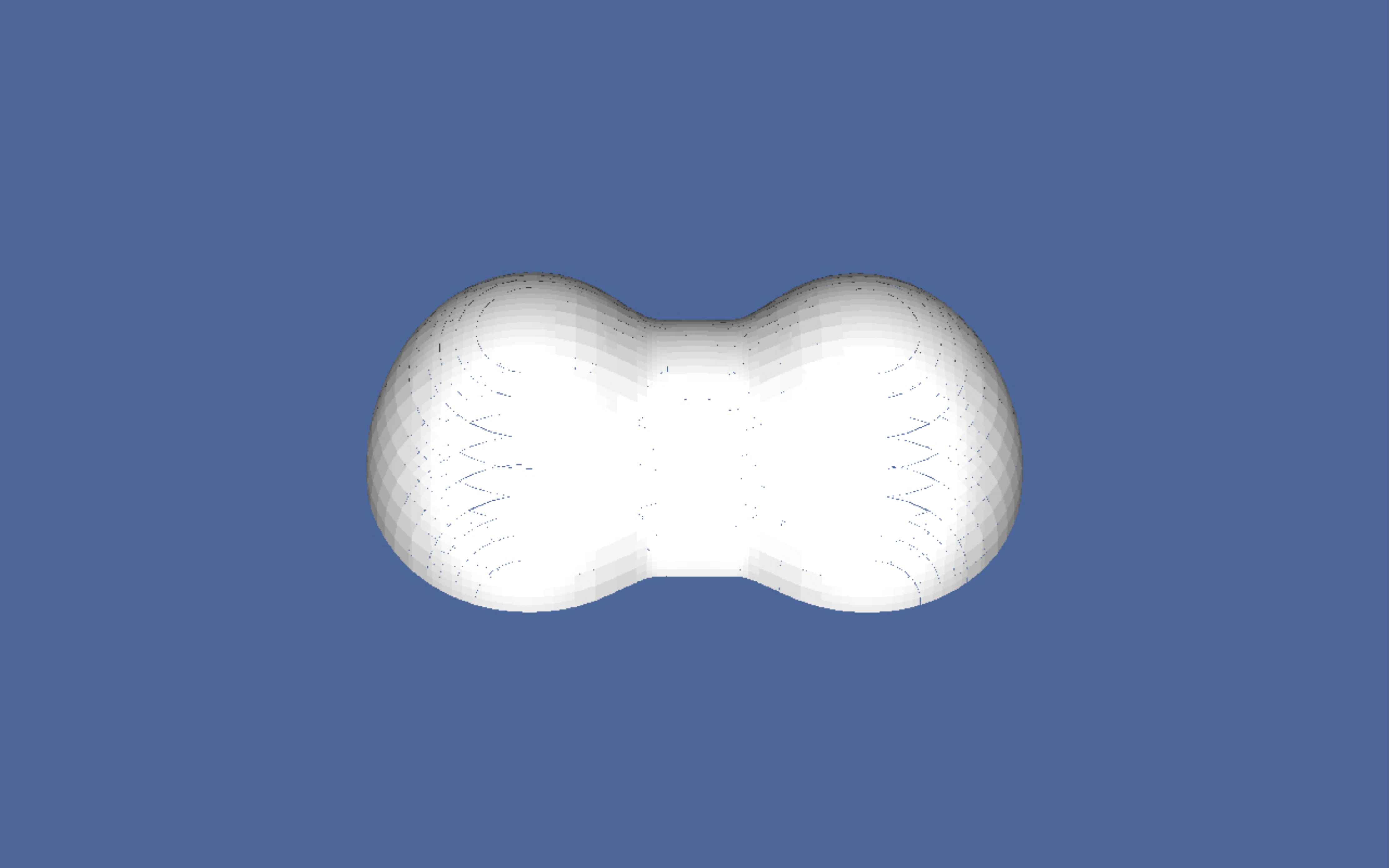}\\
  \includegraphics[width=.25\textwidth]{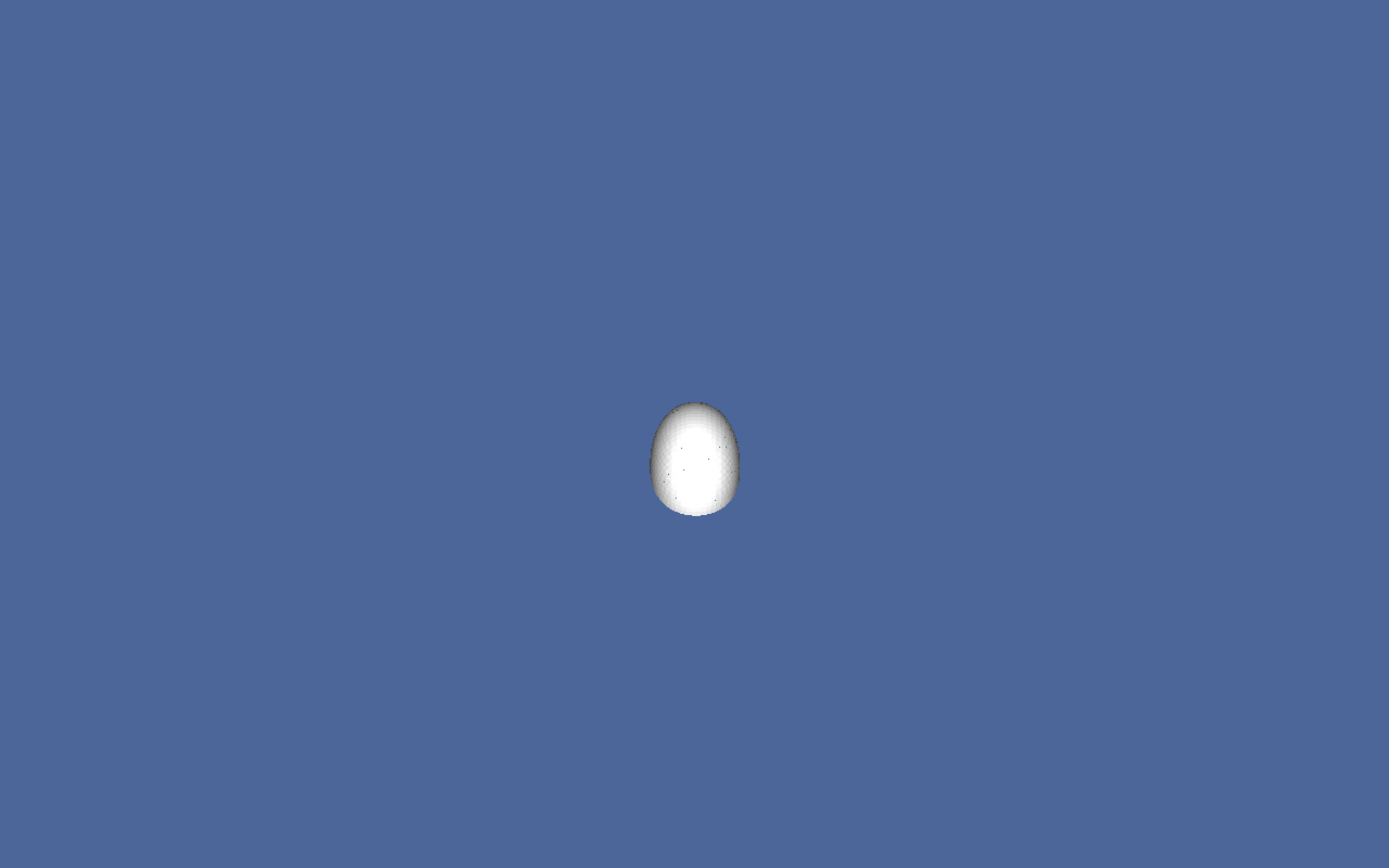}&
  \includegraphics[width=.25\textwidth]{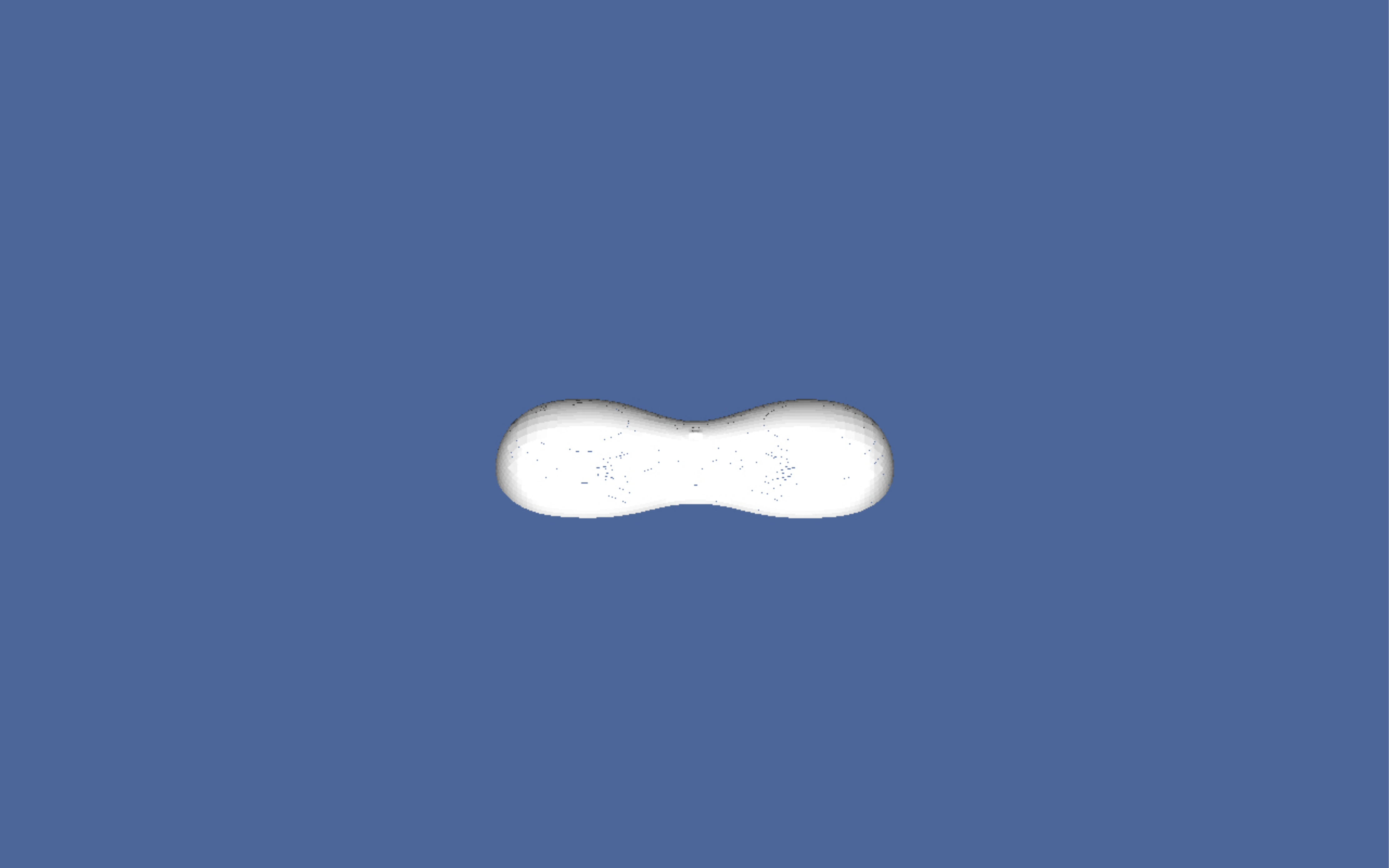}&
  \includegraphics[width=.25\textwidth]{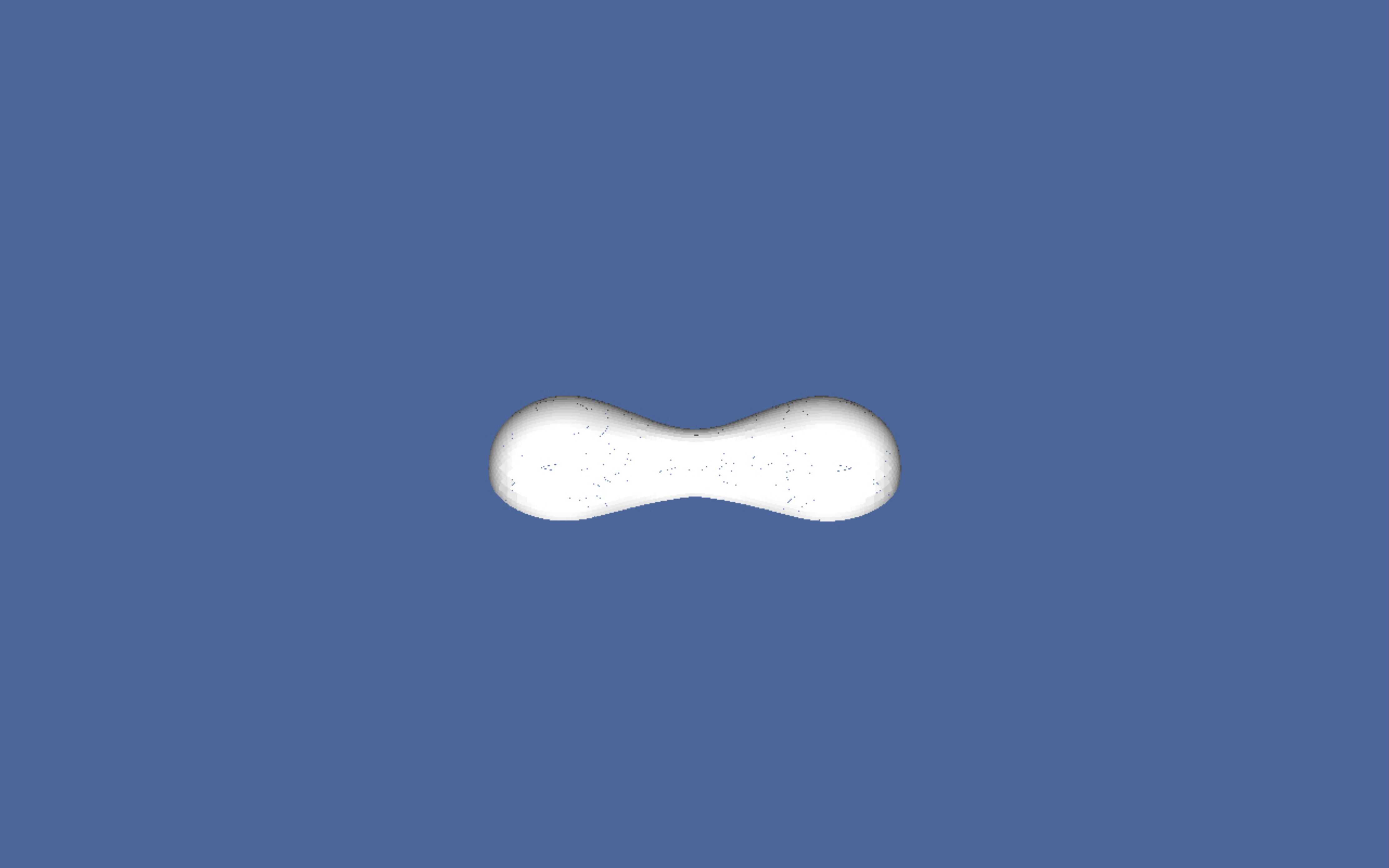}&
  \includegraphics[width=.25\textwidth]{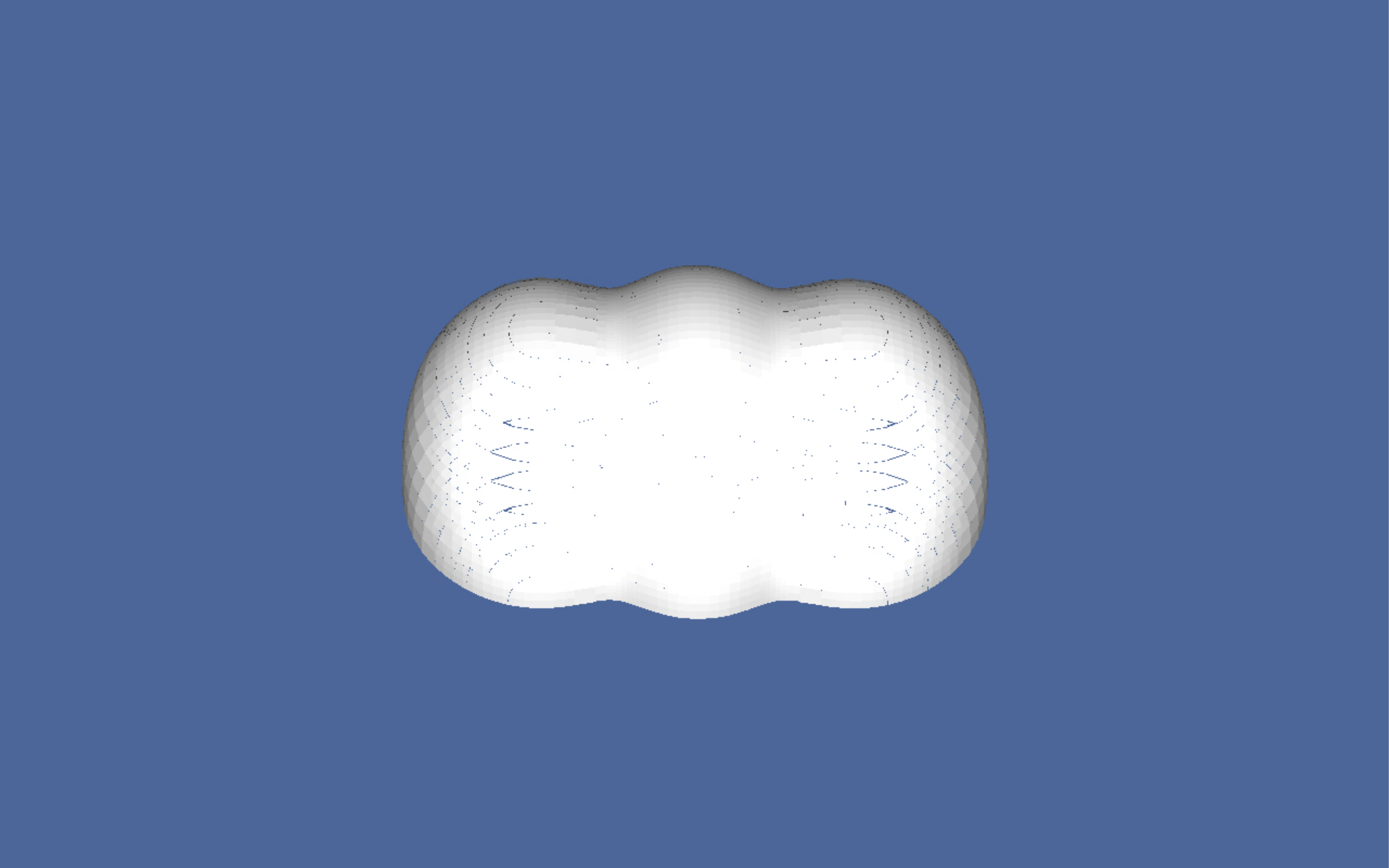}\\
  \includegraphics[width=.25\textwidth]{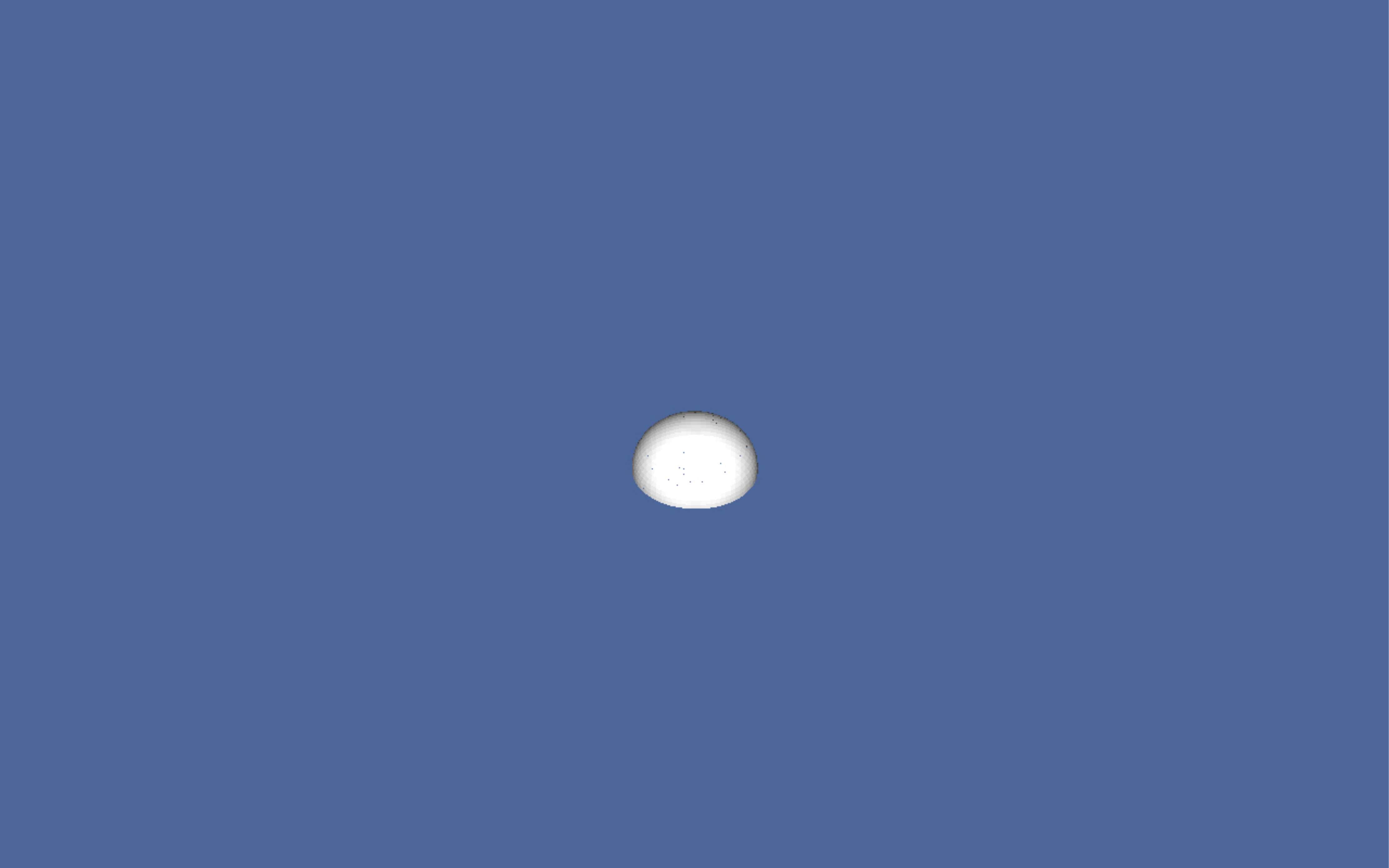}&
  \includegraphics[width=.25\textwidth]{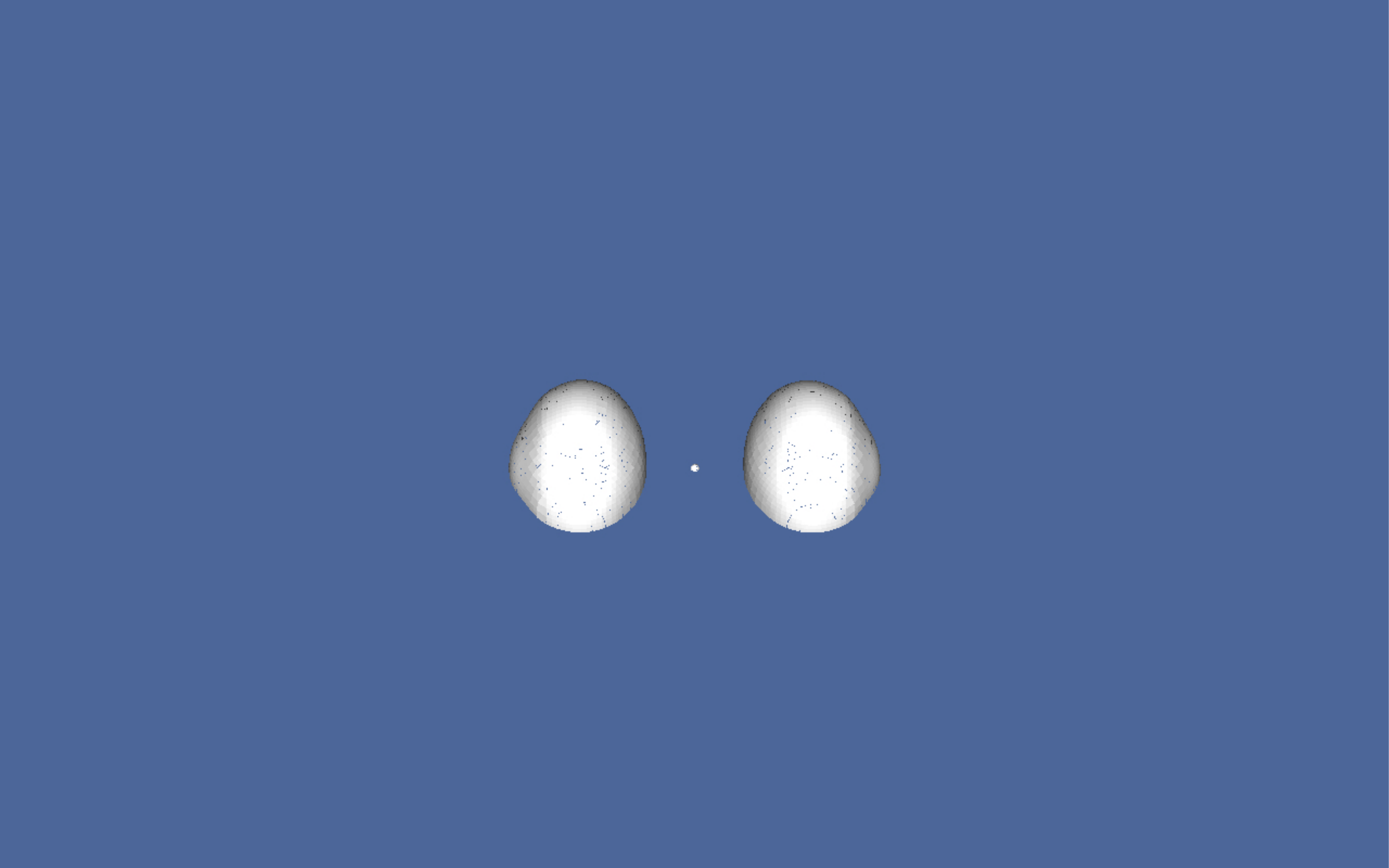}&
  \includegraphics[width=.25\textwidth]{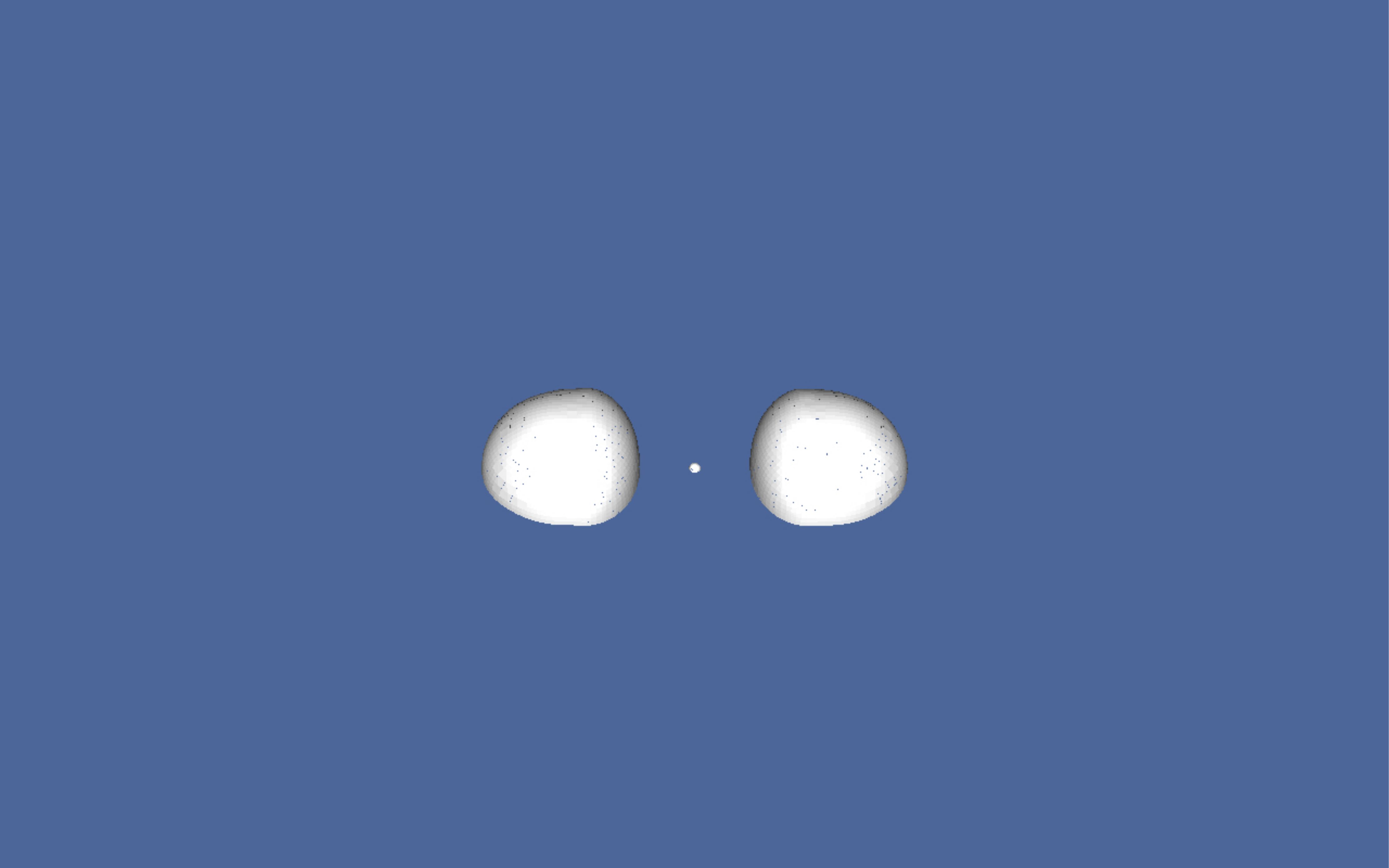}&
  \includegraphics[width=.25\textwidth]{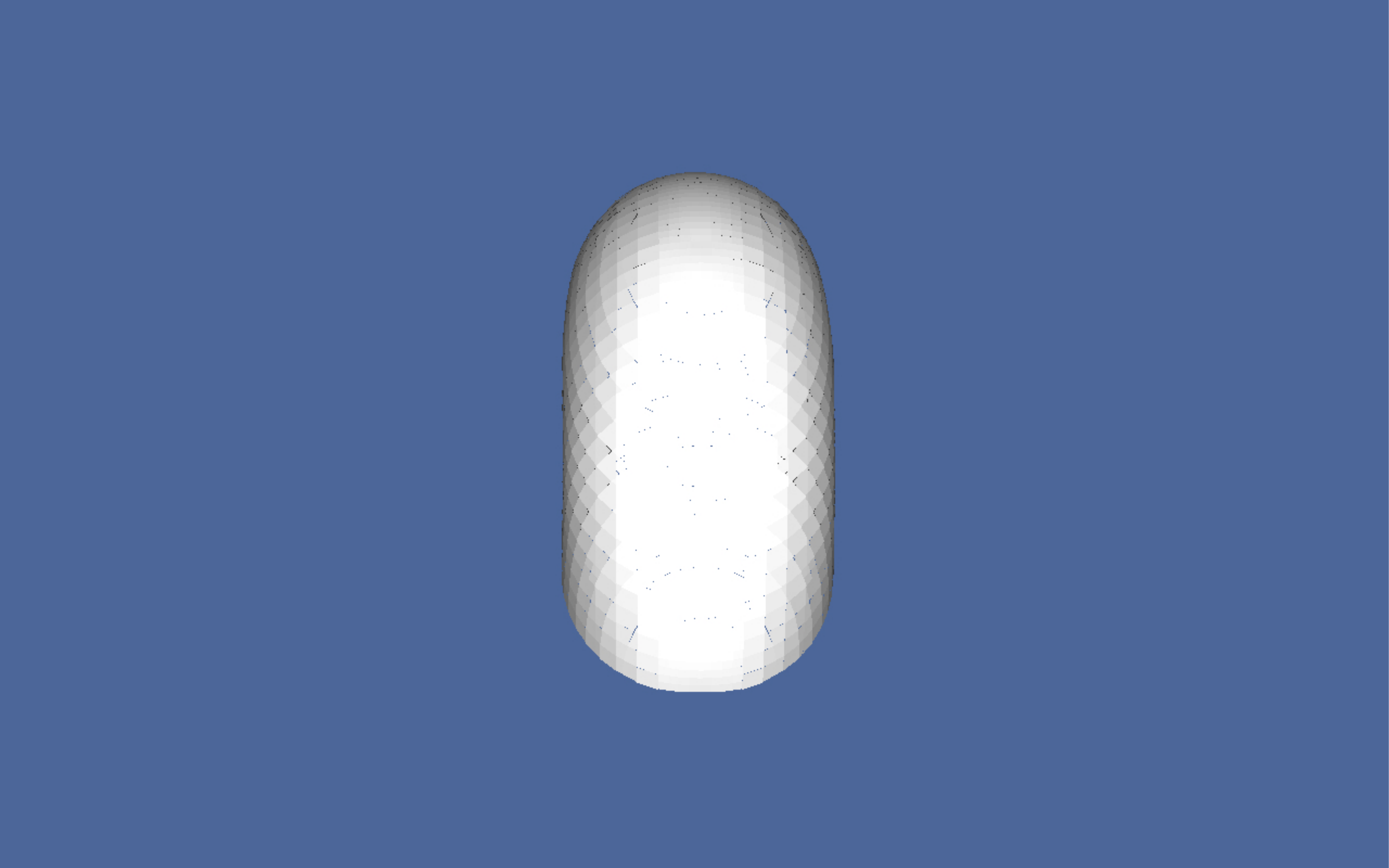}\\
  \includegraphics[width=.25\textwidth]{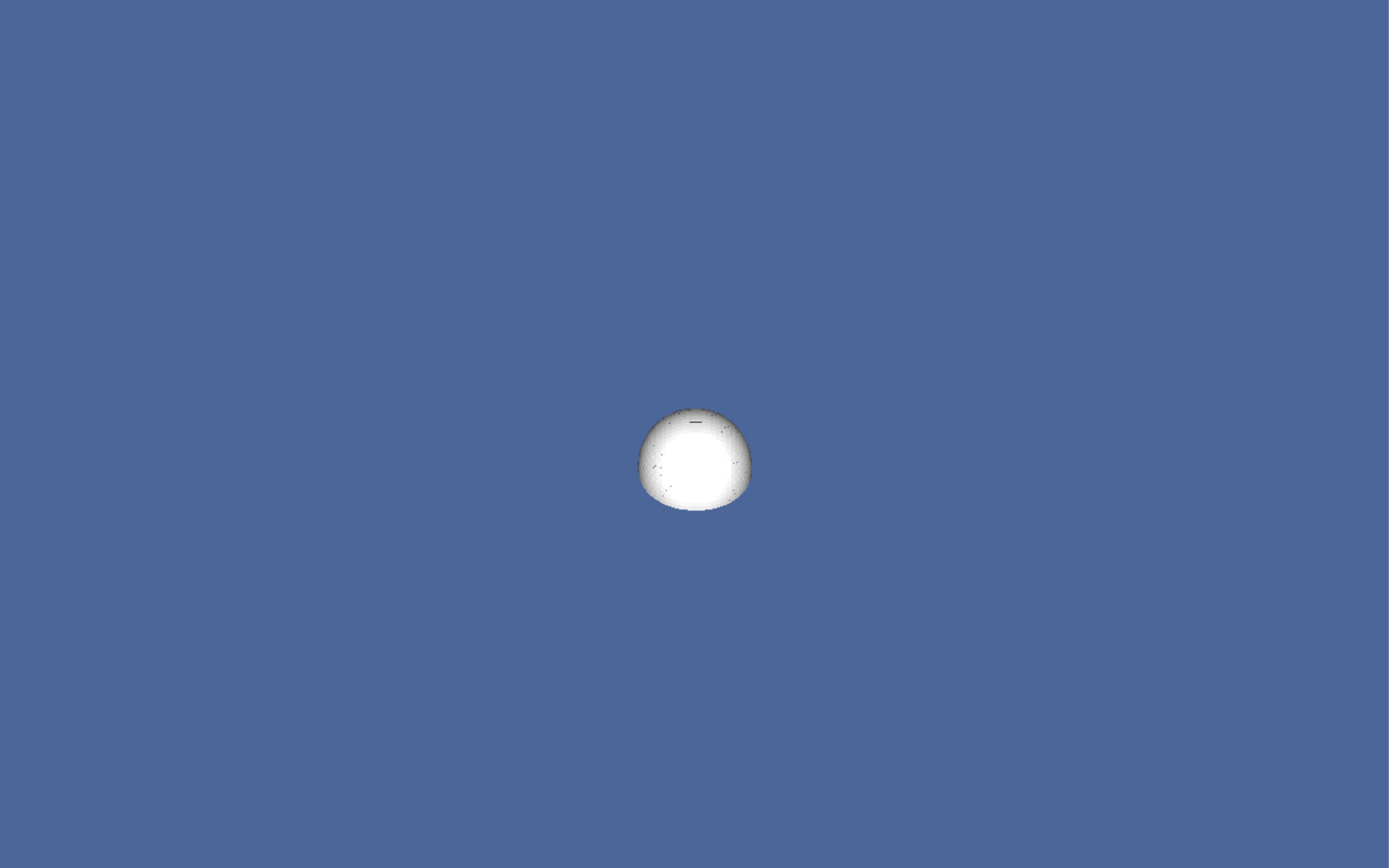}&
  \includegraphics[width=.25\textwidth]{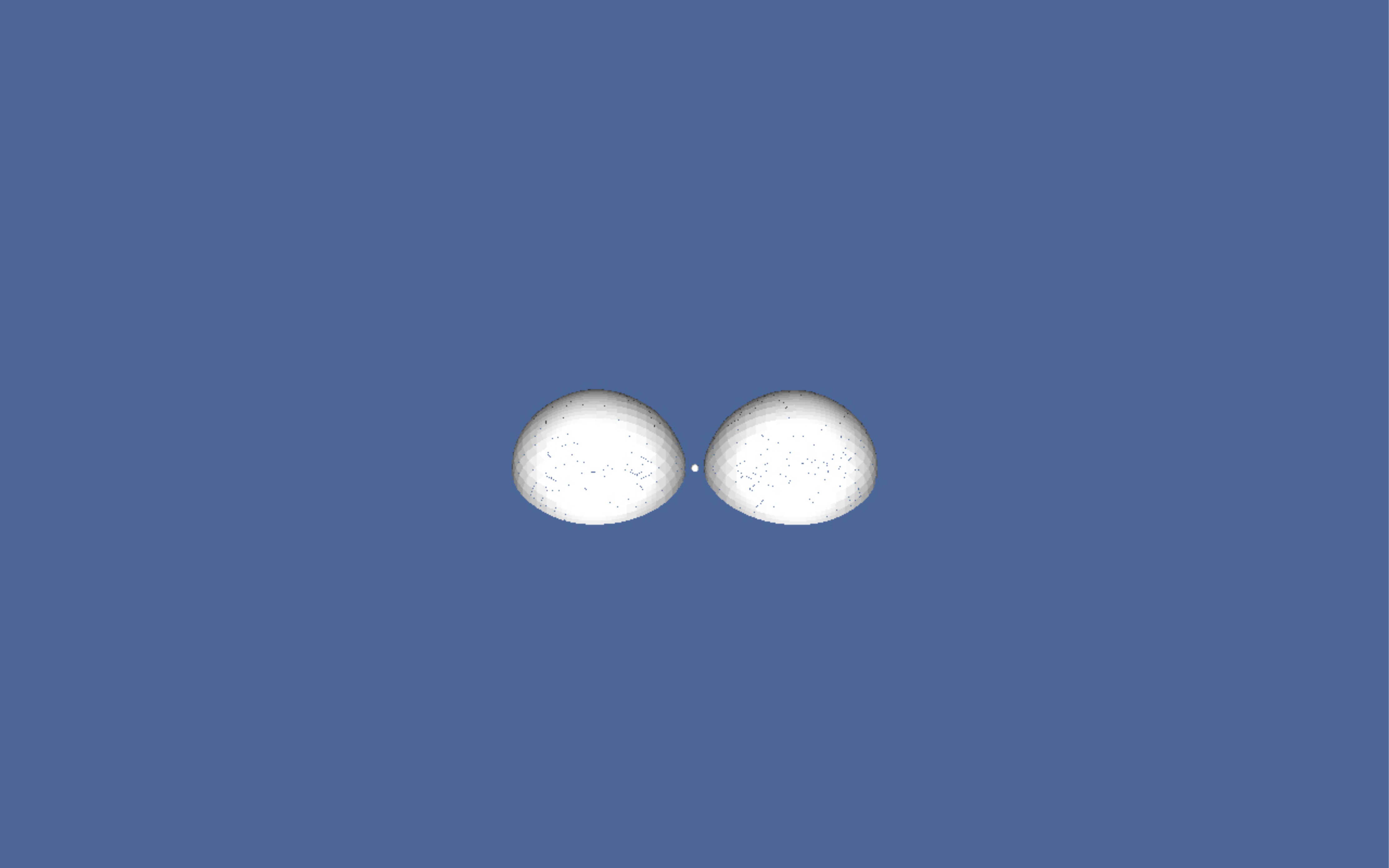}&
  \includegraphics[width=.25\textwidth]{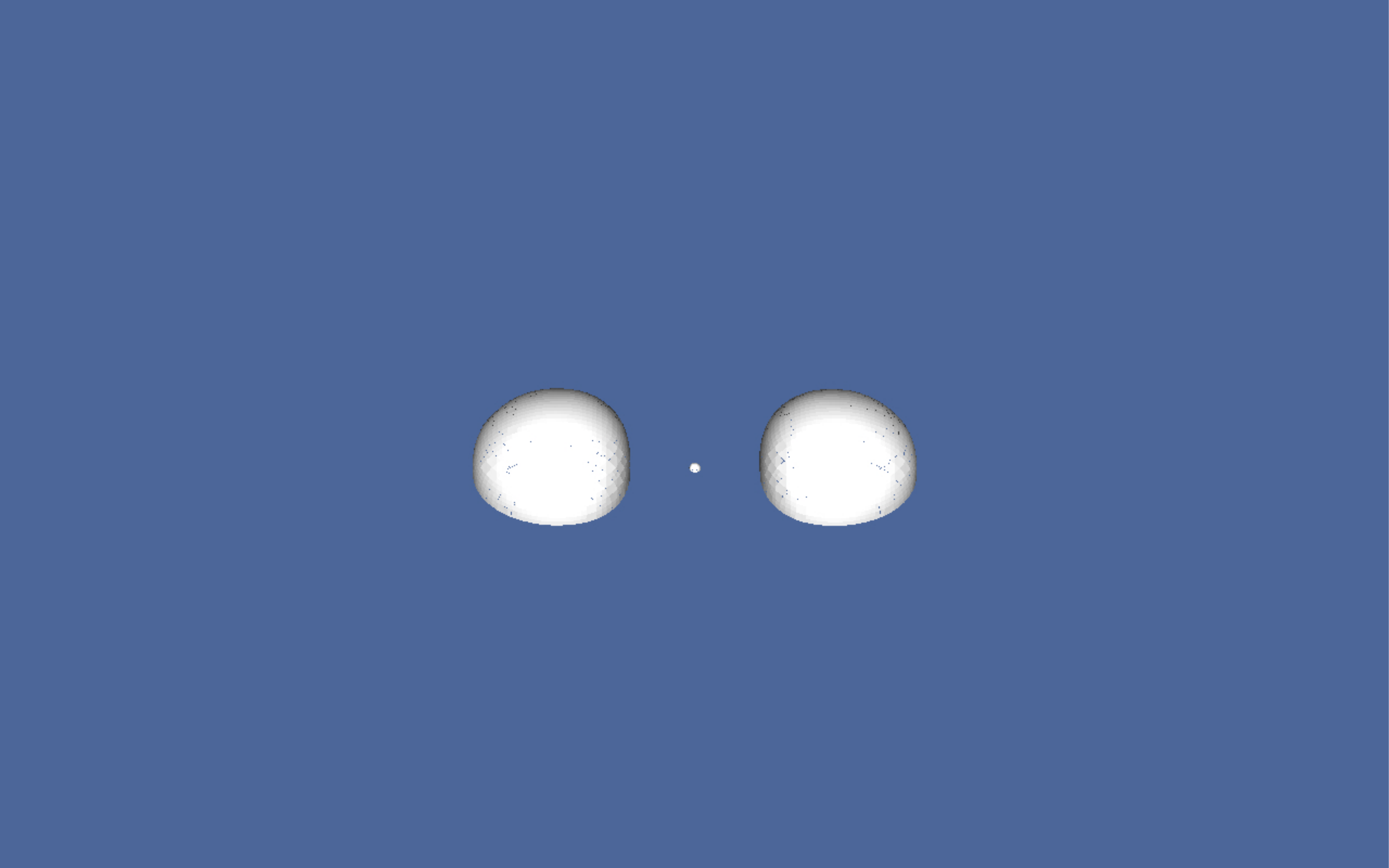}&
  \includegraphics[width=.25\textwidth]{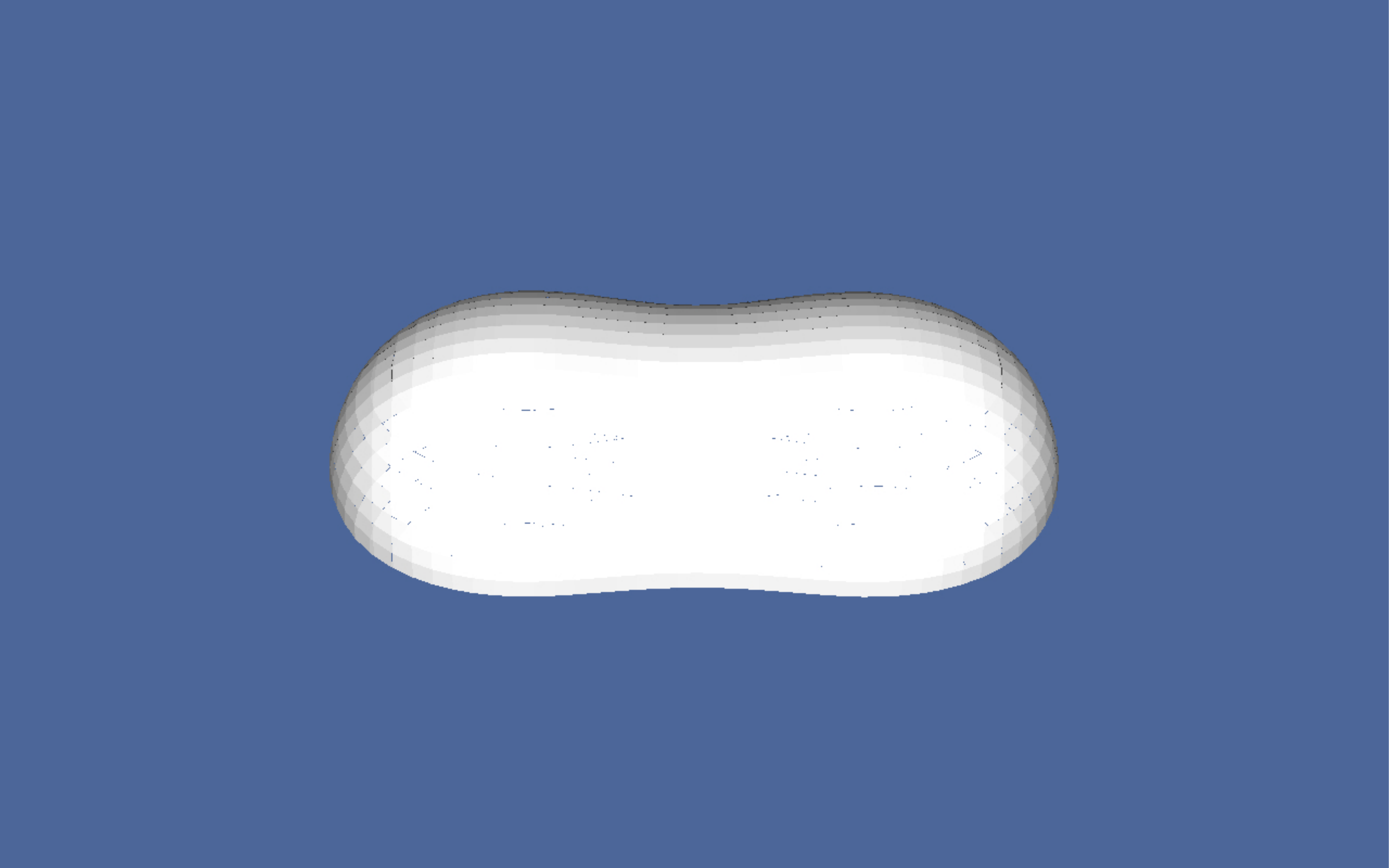}\\
  \includegraphics[width=.25\textwidth]{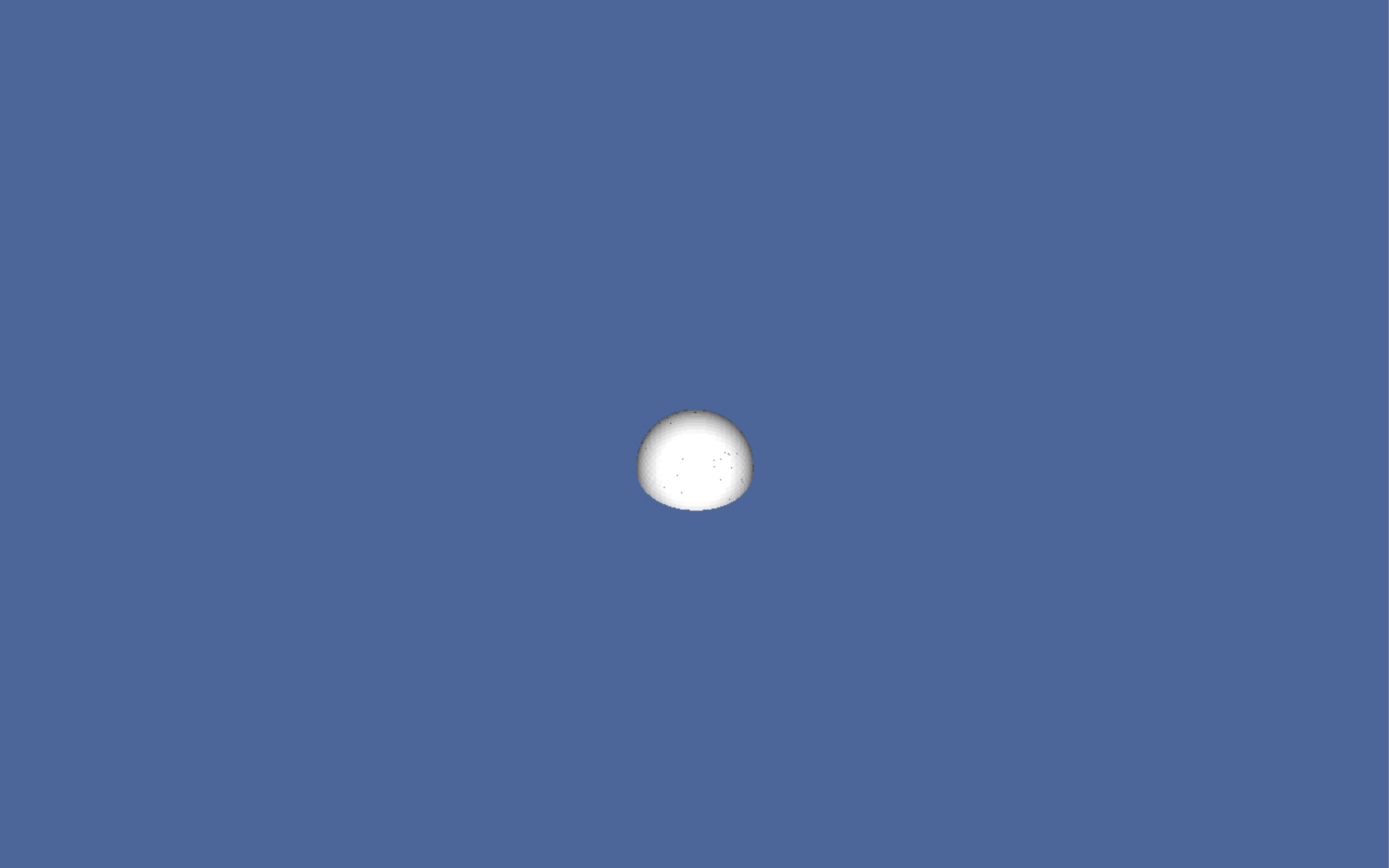}&
  \includegraphics[width=.25\textwidth]{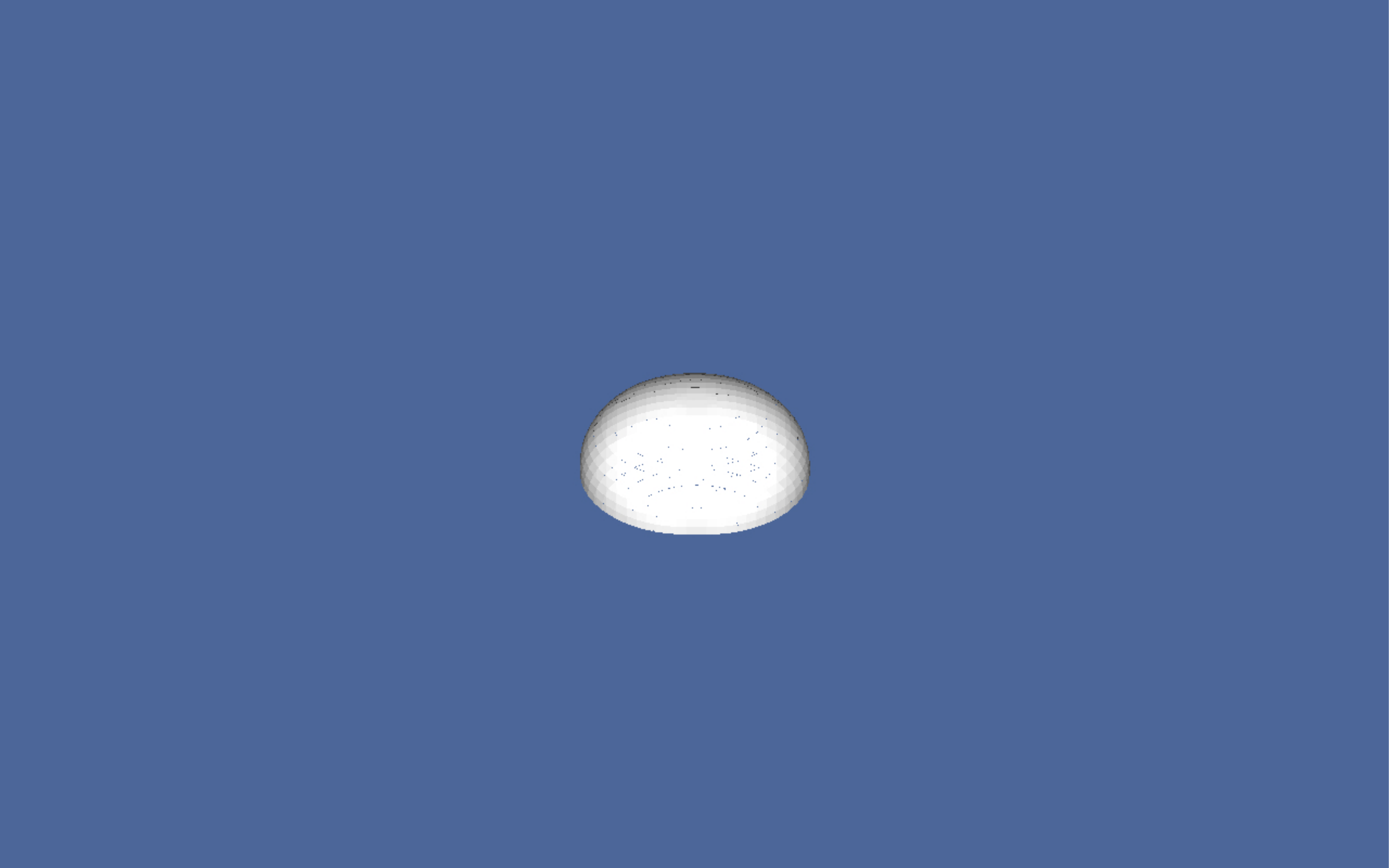}&
  \includegraphics[width=.25\textwidth]{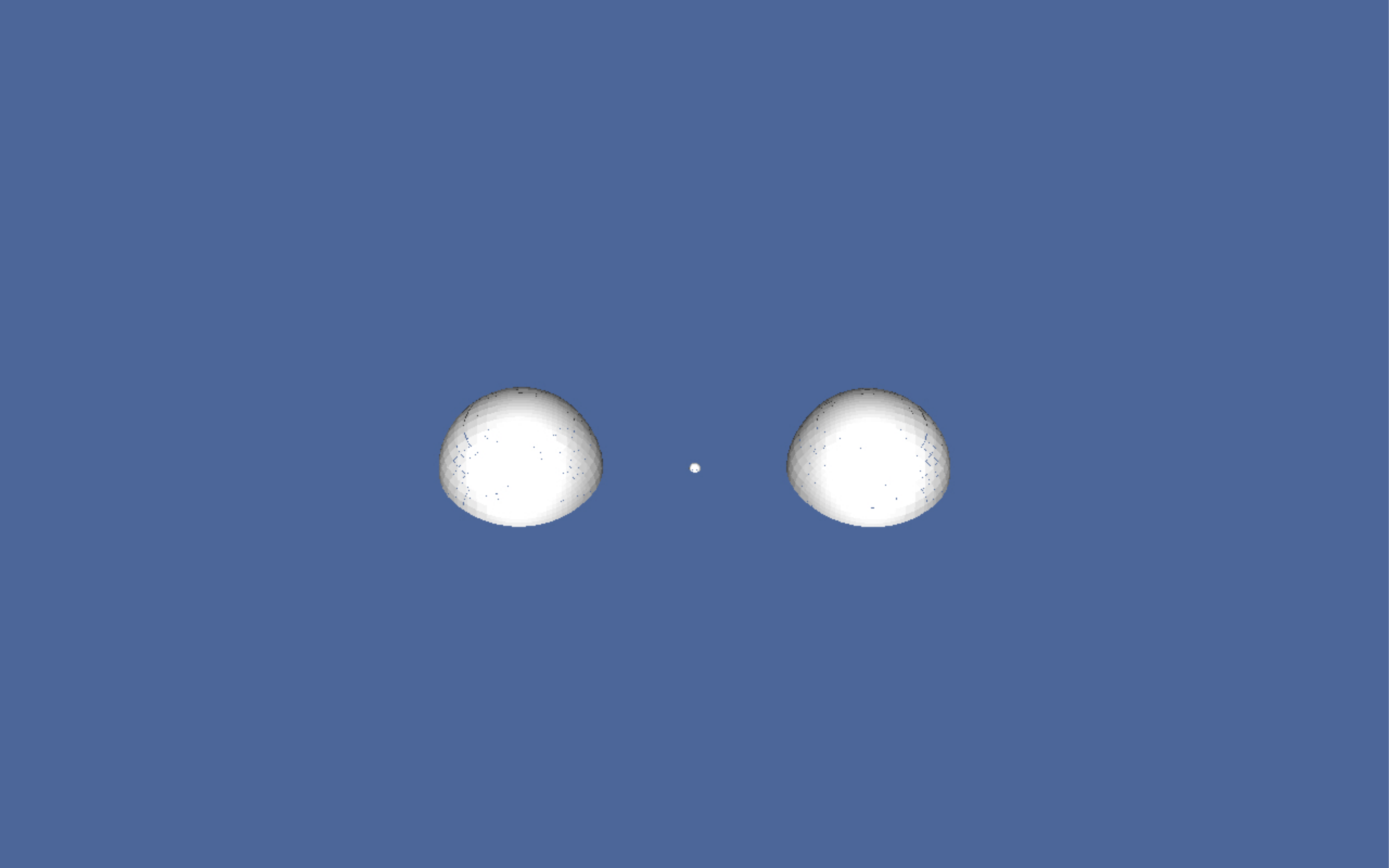}&
  \includegraphics[width=.25\textwidth]{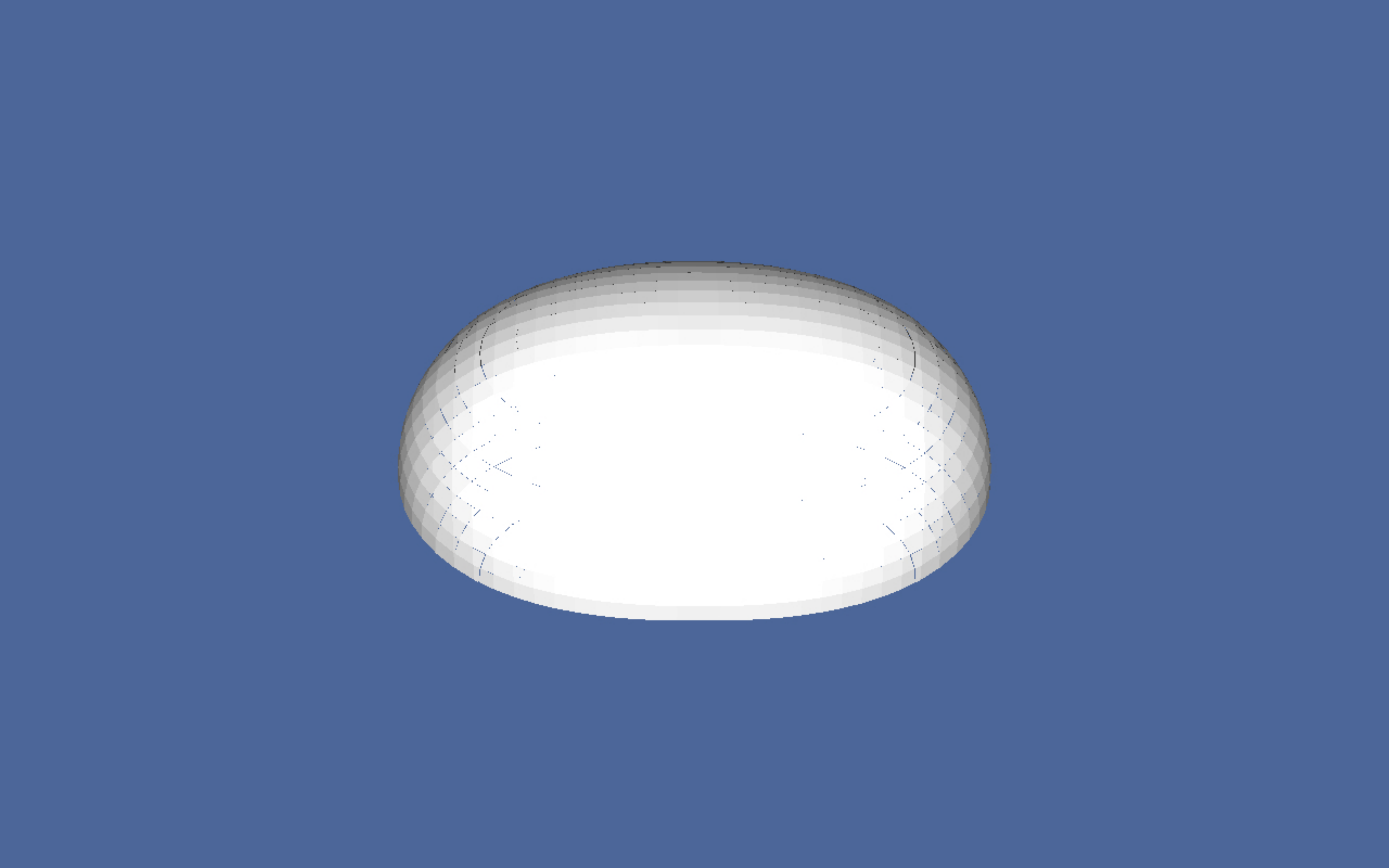}\\
  a &b &c &d
\end{tabular}
\caption{Simulations for free-standing jets: (a) (Oh,AR) = (0.4,49), (b) (Oh,AR) = (0.00125,18), 
(c) (Oh,AR) = (0.001,18), and (d) (Oh,AR) = (0.015,5).
Only top half of the filament is shown.}
\vspace{-0.1in}
\label{fig:scenarios2}
\end{figure}

We define the transition between breakup and no-breakup as whether the filament breaks up  
at anytime during its evolution toward a final equilibrium state. During the evolution, 
two main scenarios can be identified:
(i) no breakup and collapse into a single droplet, see Fig.~\ref{fig:scenarios1}(a), or (ii) the breakup of the filament, 
see Fig.~\ref{fig:scenarios1}(b). We have checked and made sure this characterization is independent
of the mesh resolution; see Fig.~\ref{fig:scenarios1}(c) that shows the breakup scenario for a higher
resolution compared to the case shown in Fig.~\ref{fig:scenarios1}(b). Although the results show quantitative differences,
whether the breakup occurs or not is independent of the gird size - we have also checked this for
other simulations. Interestingly, for free
standing jets, at the transition AR, the broken up filament always condenses into a single droplet, regardless of the 
Oh number, see for example, Fig.~\ref{fig:scenarios1}(b). This is quite different from the 
filaments that are supported by a substrate; we will show later that in that case, the final configuration 
can also be two (or multiple) distinguished droplets (without recoalescence). For scenario (ii) above,
we can also identify a number of cases: (ii-a) the filament breaks up into two droplets that eventually recoalesce
into one single drop, see Fig.~\ref{fig:scenarios2}(a); (ii-b) the filament breaks up into two droplets, then condenses 
into a single drop that breaks up again due to significant inertia, and eventually recoalesces into a single droplet
because of strong oscillations,
see Fig.~\ref{fig:scenarios2}(b); and (ii-c) the same as case (ii-b) but without the final recoalescence, so two 
distinct drops form at the end,
see Fig.~\ref{fig:scenarios2}(c).
    
\section {Results and Discussions} 
We first validate our numerical results by comparing them against the
experimental observations in \cite{pita} for standing liquid jets. 
We carry out full 3D simulations despite the fact that
due to axial symmetry in this case, axisymmetric simulations suffice.  
However, apart from the consistency with the results of the substrate-supported filaments,
for visualization purposes, we also perform 3D computations
for free-standing filaments.
As discussed in \cite{Driessen2013}, for viscous liquid filaments, 
the transition from collapse to breakup can be
described as a competition between the capillary driven end retraction 
and the Rayleigh--Plateau \cite{Rayleigh1878} type instability mechanism. Filaments with 
a small aspect ratio do not break up irrespective of the Oh number; 
in this case, end pinching does not have time to develop and the whole
filament collapses into a single drop; see e.g.~Fig.~\ref{fig:scenarios2}(d). 
Very viscous filaments (Oh $>1$) are also
always stable regardless of AR; in this case, the Rayleigh--Plateau mechanism is not
operative to cause the filament to break up; see e.g.~Fig.~\ref{fig:scenarios1}(a). 
The breakup occurs when 
the breakup time due to Rayleigh--Plateau instability is comparable to 
the time required for the end pinching to happen; see e.g.~Fig.~\ref{fig:scenarios2}(a). 

\begin{figure}[tb]
\includegraphics[width=1\textwidth]{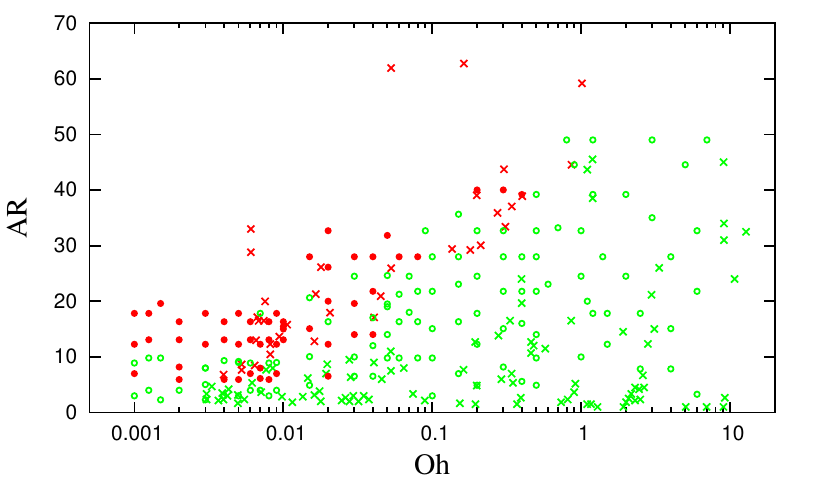}
\caption{Breakup and no-breakup results plotted as a function of Oh versus AR: 
breakup ({\textcolor{red}{$\bullet$}}) and no-breakup ({\textcolor{green}{$\circ$}}).
Experimental results in \cite{pita}:  breakup ({\textcolor{red}{$\times$}}) and no-breakup ({\textcolor{green}{$\times$}}).}
\label{fig:pita}
\end{figure}

Figure \ref{fig:pita} presents the comparison of our numerical results
with the experimental data in \cite{pita}. As shown, there is a clear agreement
between our results and the experimental ones. The breakup transition predicted
numerically is also shown to agree with the observations in  \cite{pita}. Additionally,
we extend the parameter space to smaller values of Oh ($O(0.001)$),
where prior results are not available in the literature for that range. 
In this range, the dynamics is very fast and surface capillary waves do not have enough
time to grow. However, it appears from the numerical results that the
boundary between the breakup and no-breakup remains unchanged when decreasing
Oh below about $0.01$. This is the region which was not explored in the experiments in \cite{pita},
suggesting that there is an AR of about $10$, above which the breakup always occurs for Oh$<0.01$.
 
\begin{figure}[tb]
	\centering
	\begin{tabular}{ccc}
		\includegraphics[width=.25\textwidth]{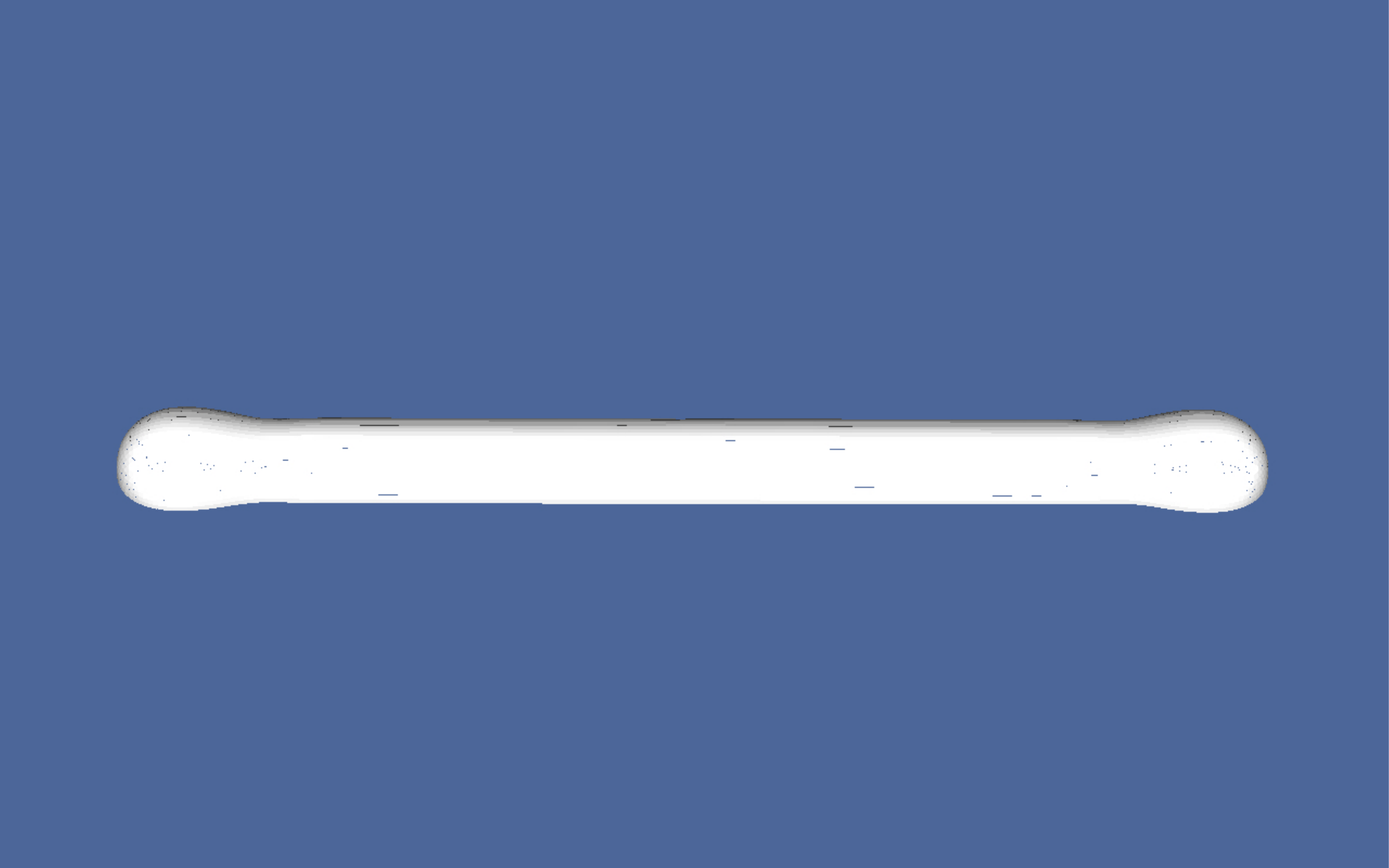}&
		\includegraphics[width=.25\textwidth]{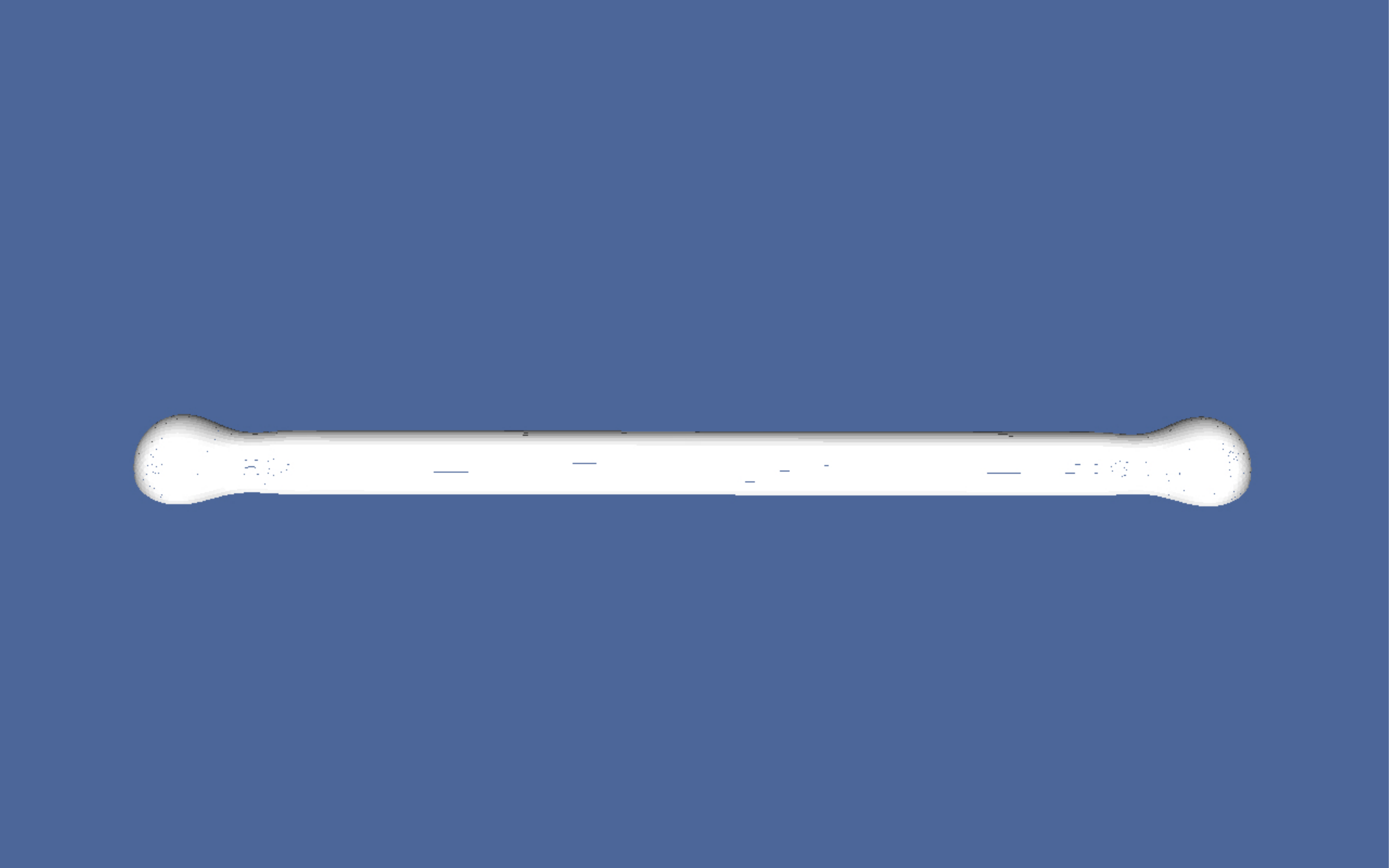}&
		\includegraphics[width=.25\textwidth]{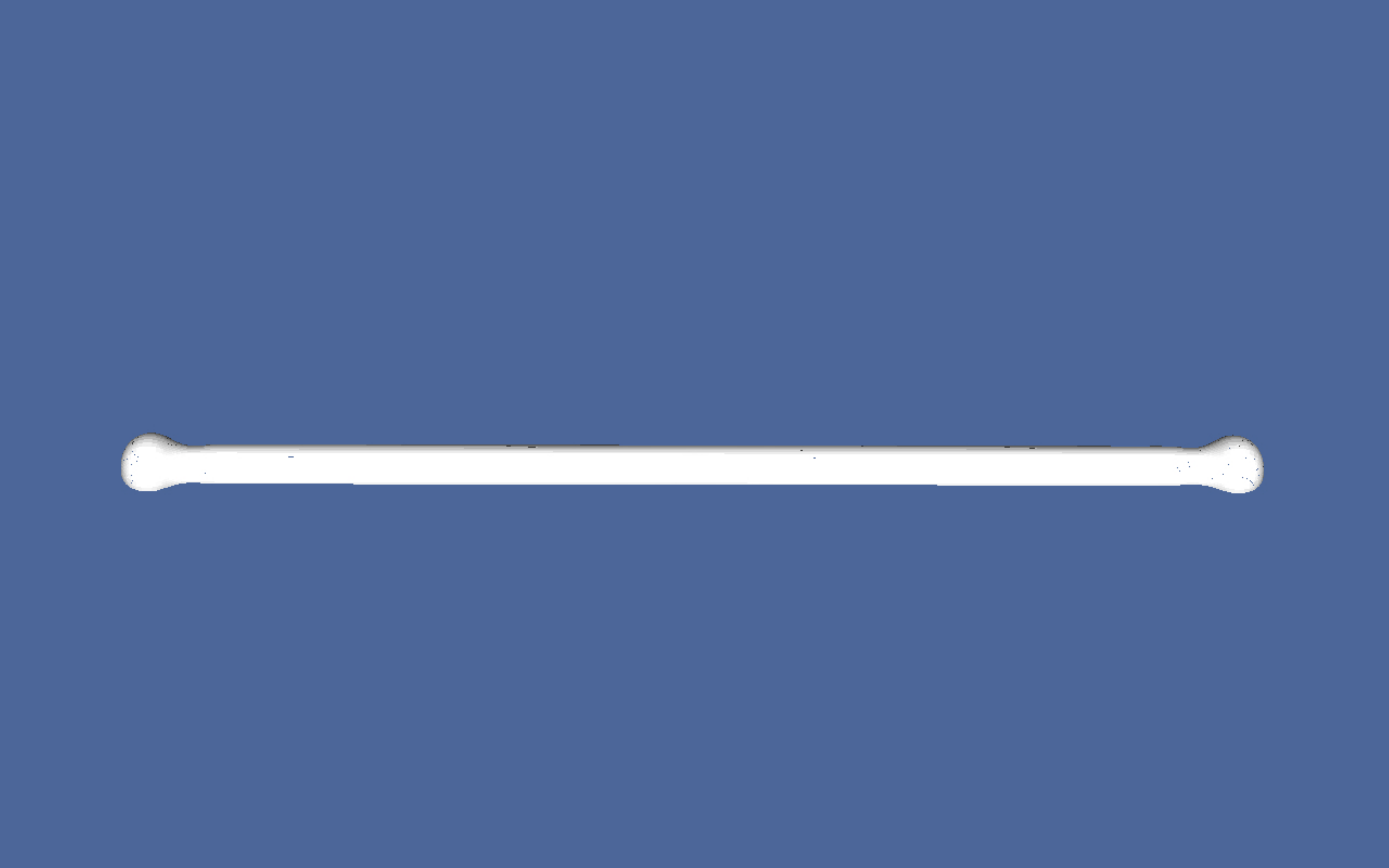}\\
		\includegraphics[width=.25\textwidth]{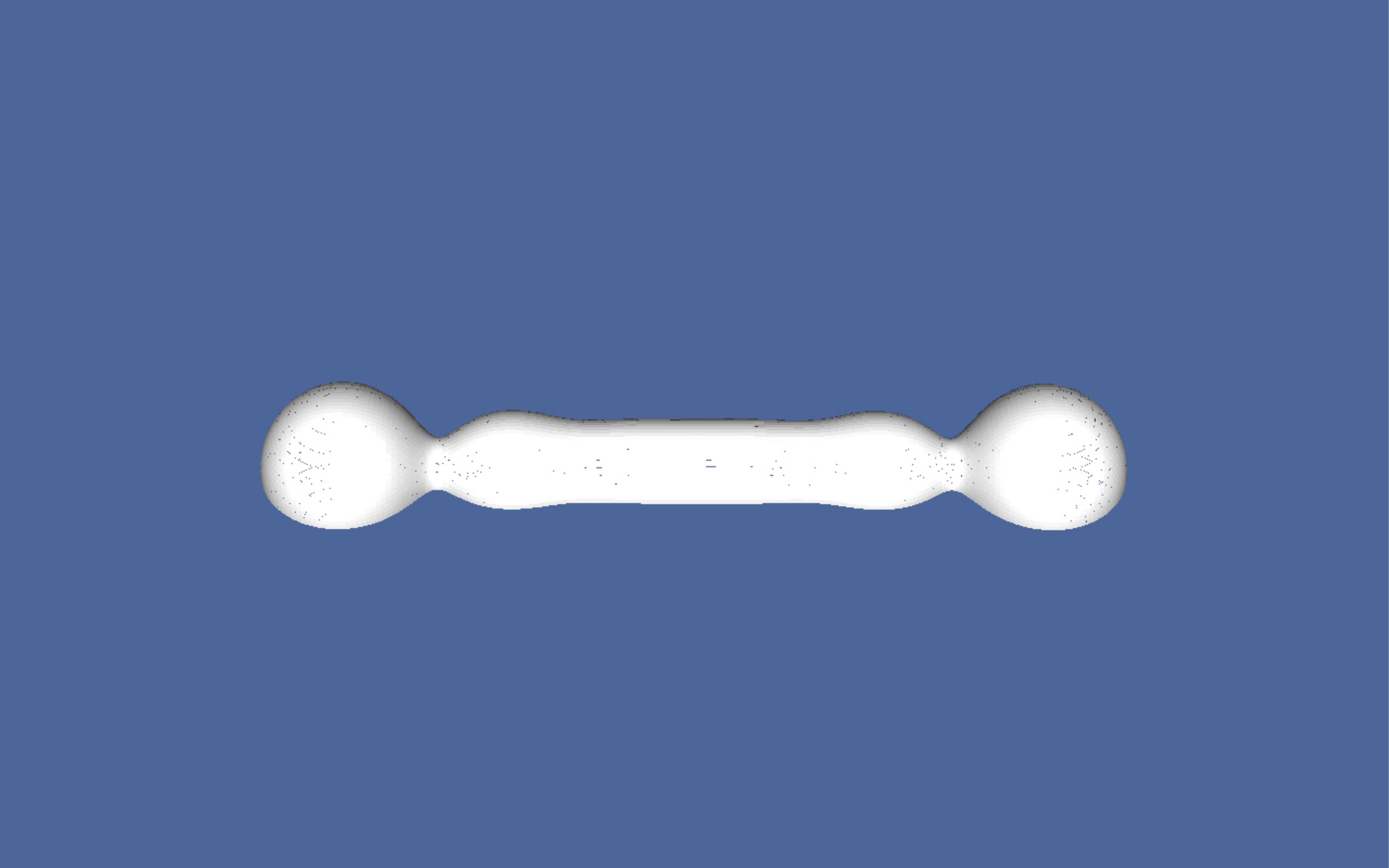}&
		\includegraphics[width=.25\textwidth]{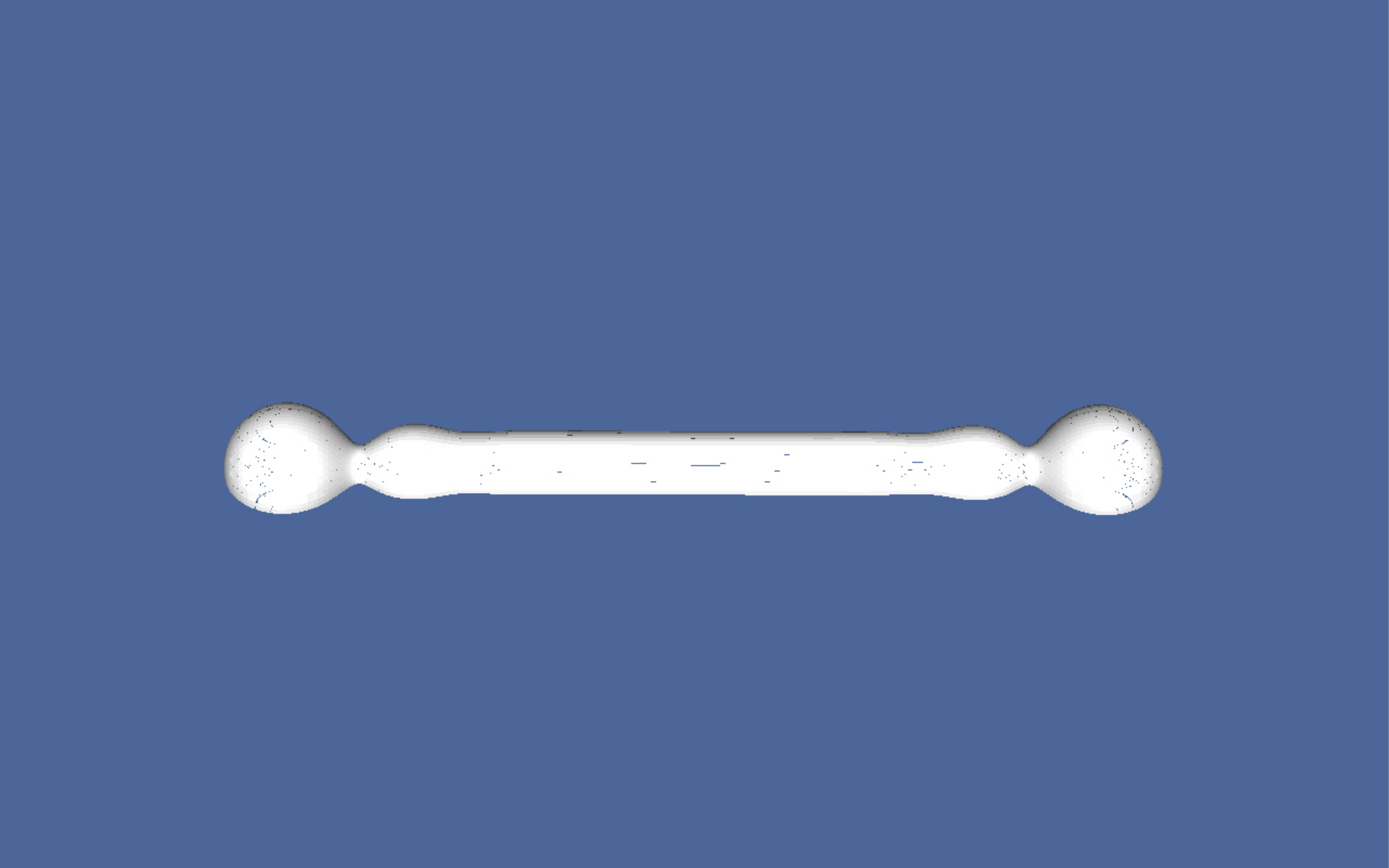}&
		\includegraphics[width=.25\textwidth]{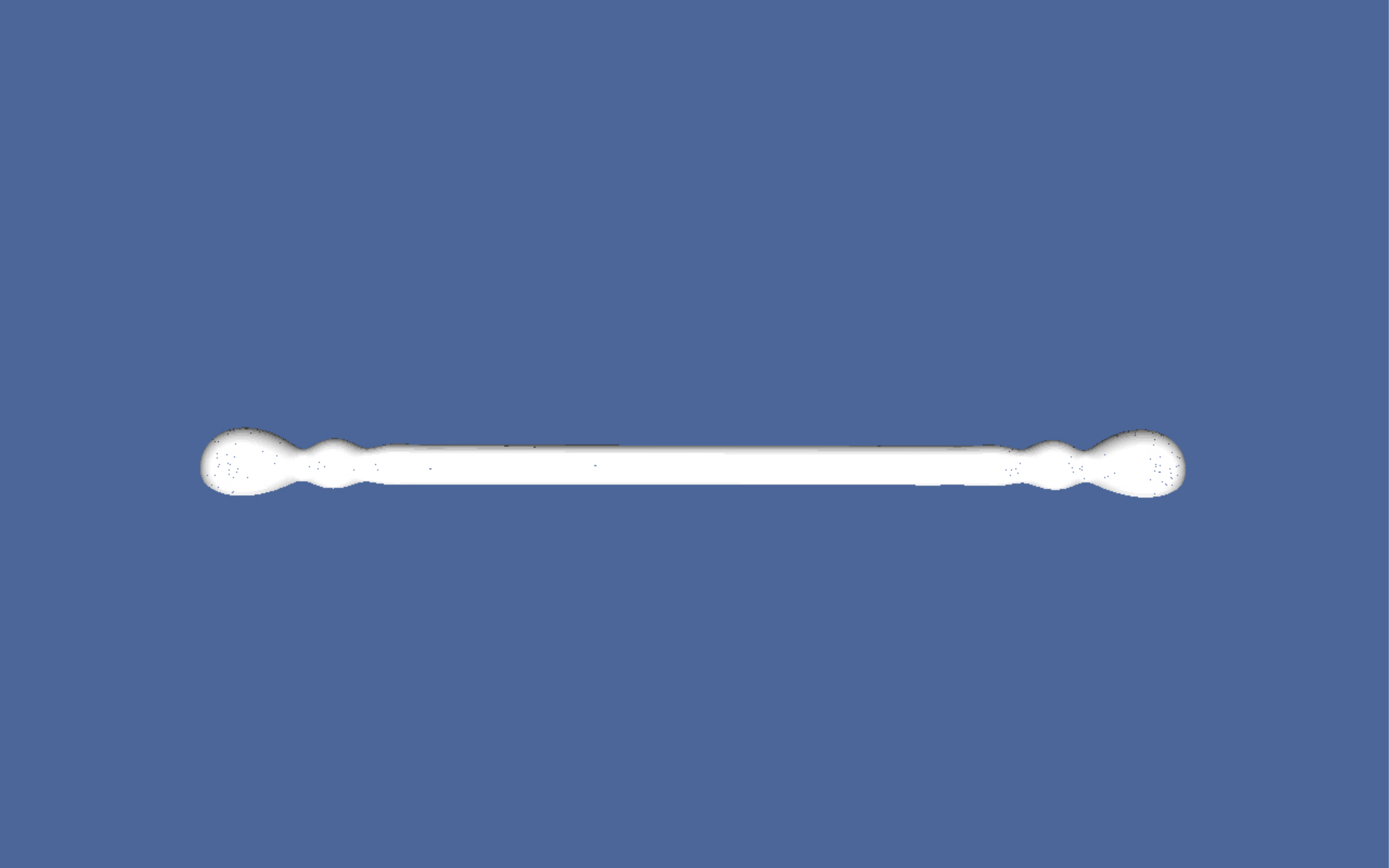}\\
		\includegraphics[width=.25\textwidth]{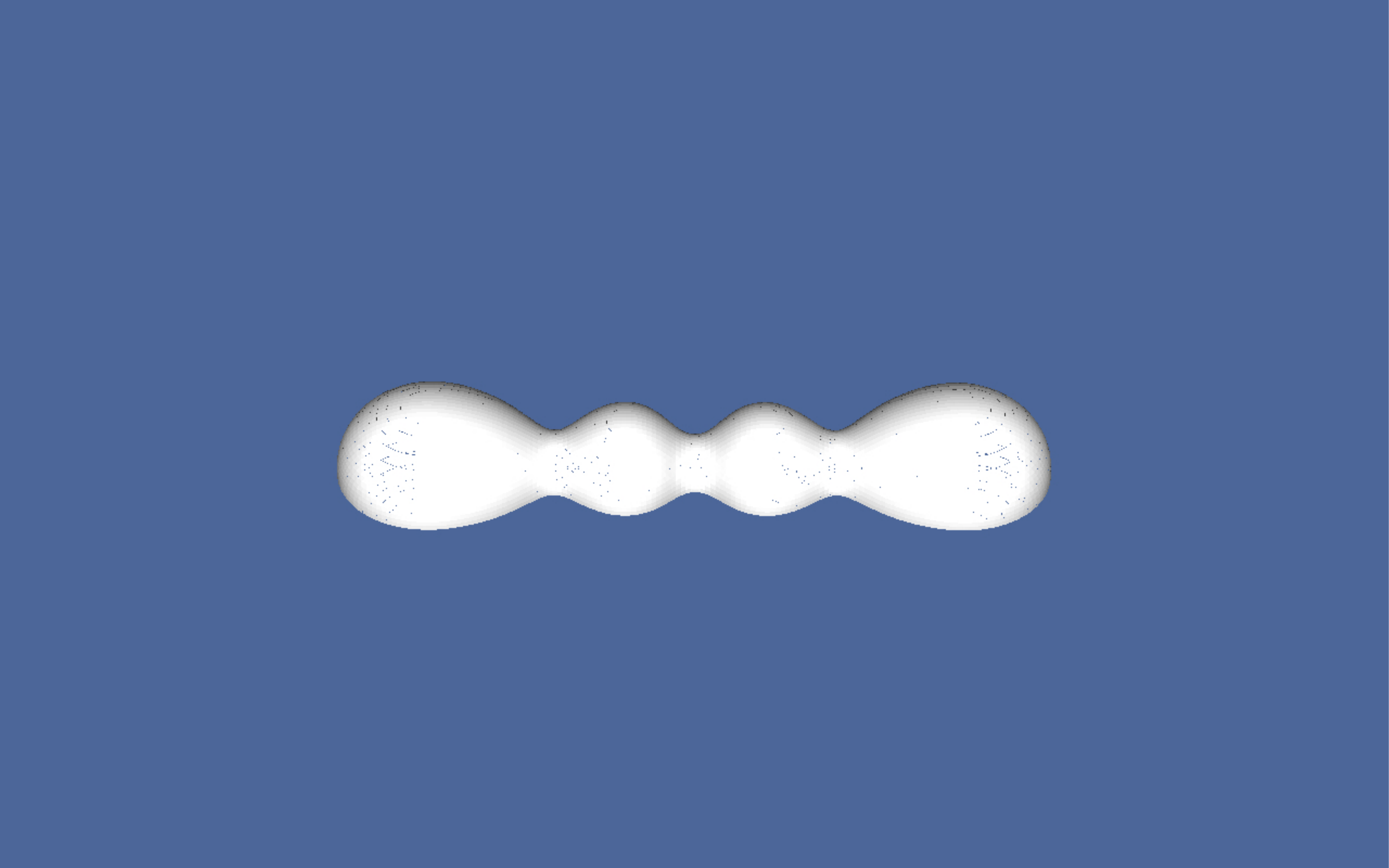}&
		\includegraphics[width=.25\textwidth]{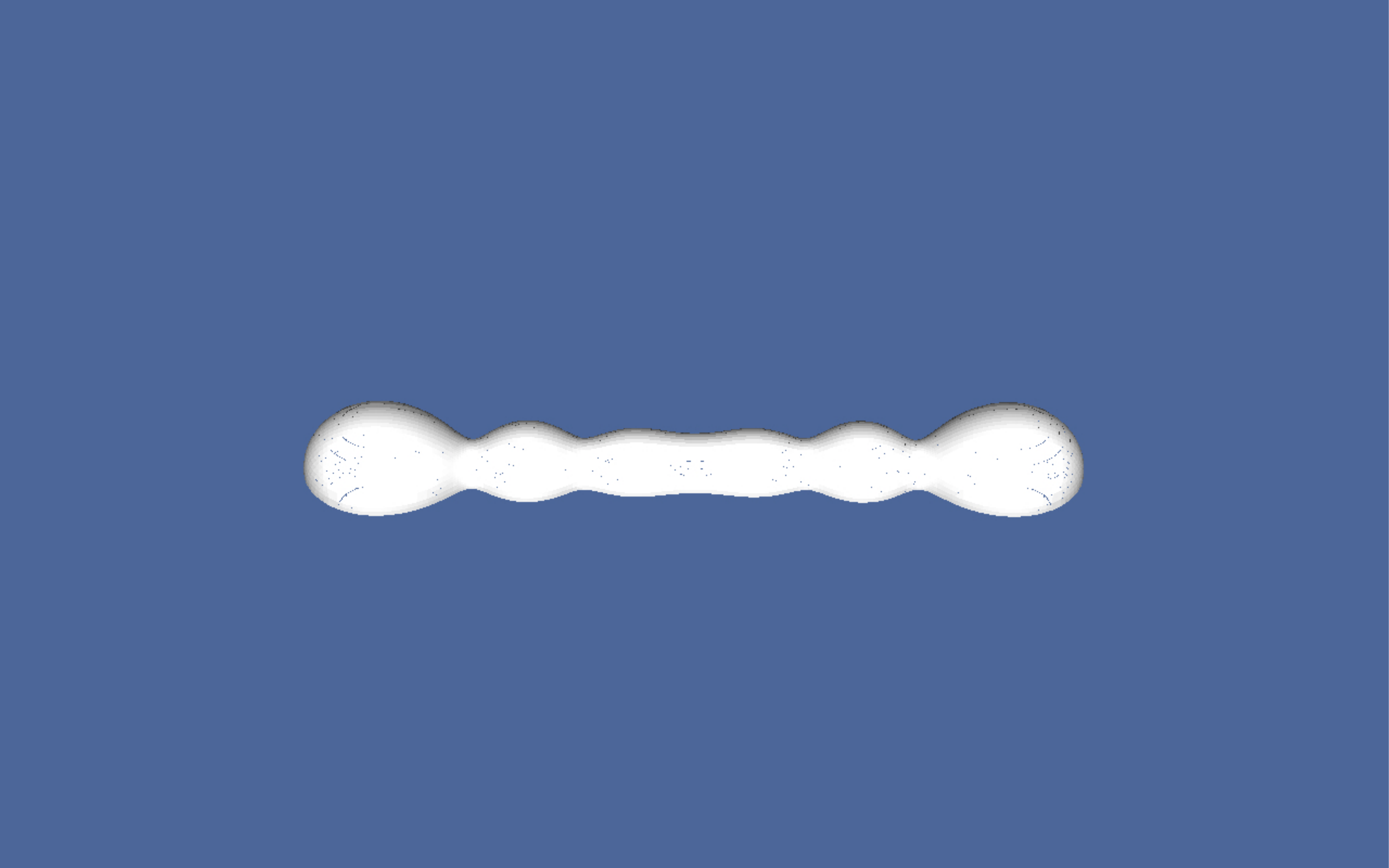}&
		\includegraphics[width=.25\textwidth]{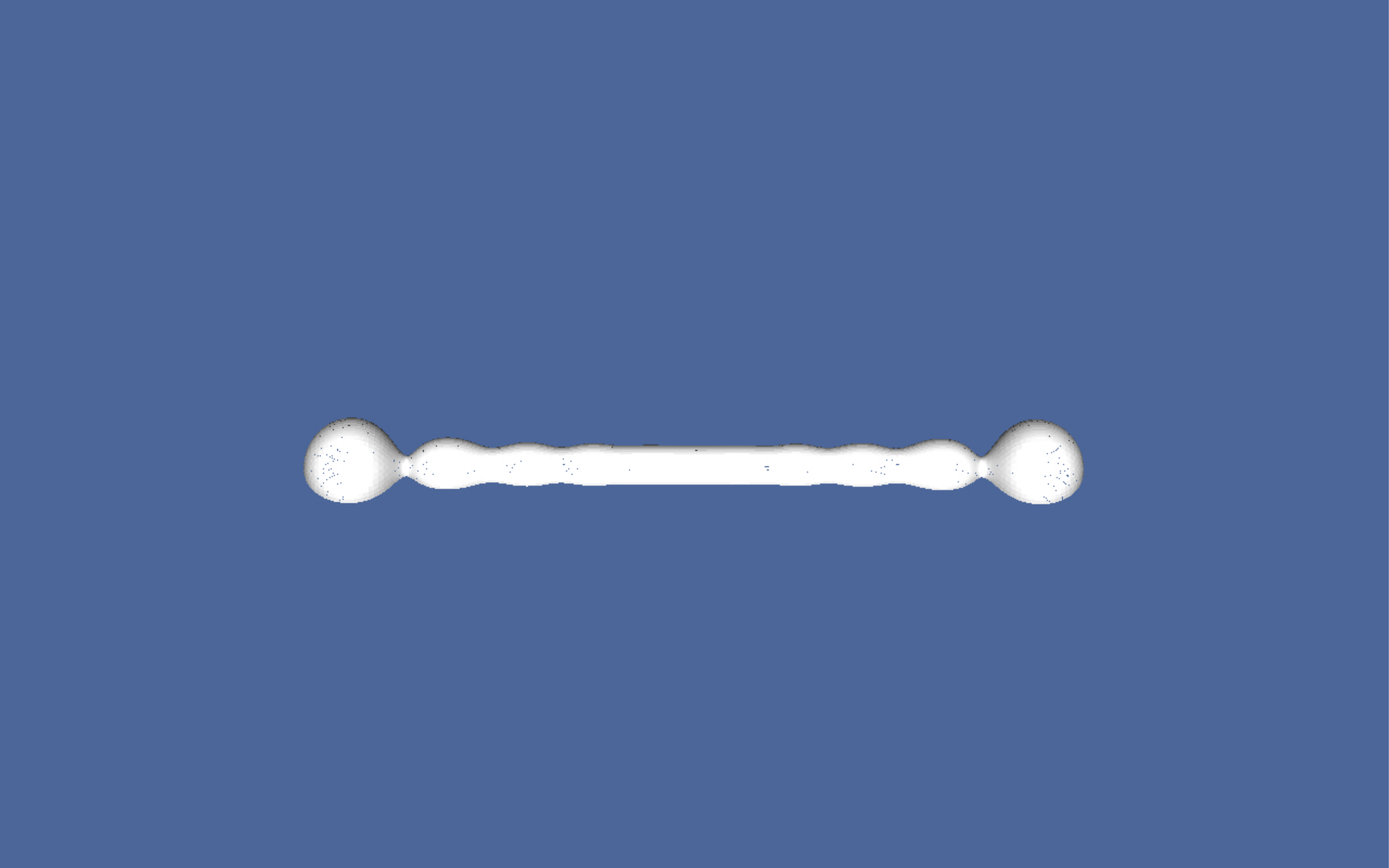}\\
		\includegraphics[width=.25\textwidth]{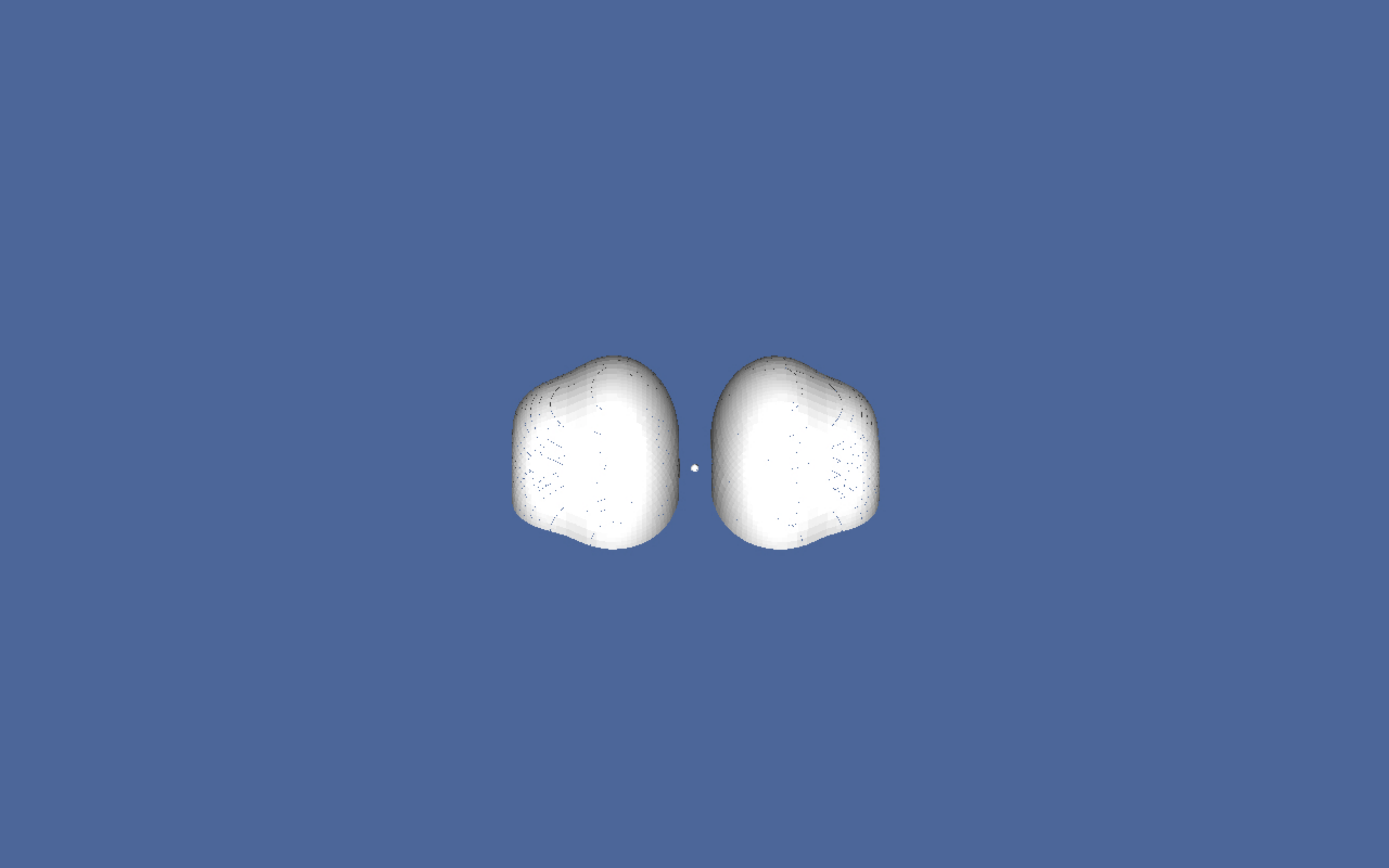}&
		\includegraphics[width=.25\textwidth]{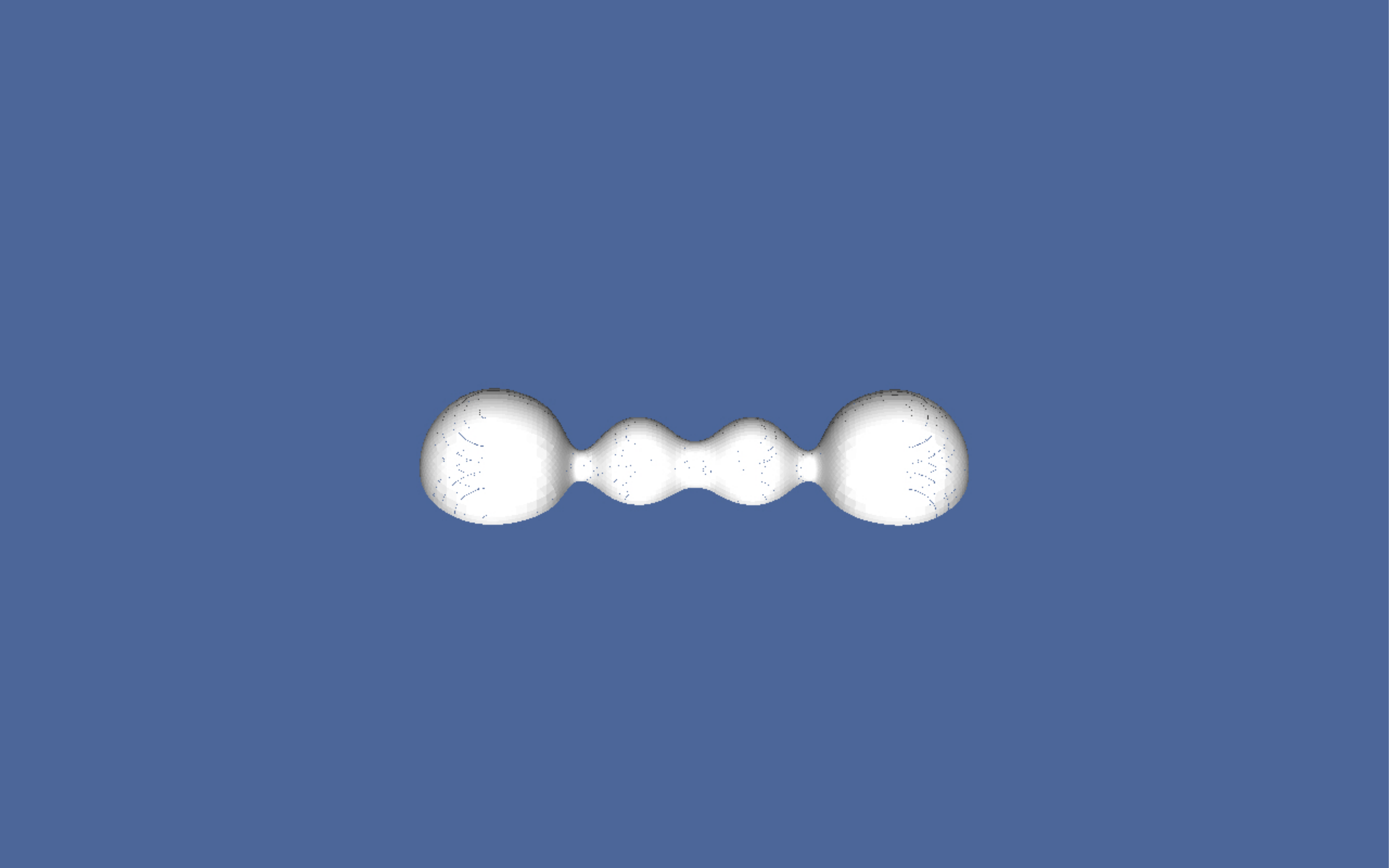}&
		\includegraphics[width=.25\textwidth]{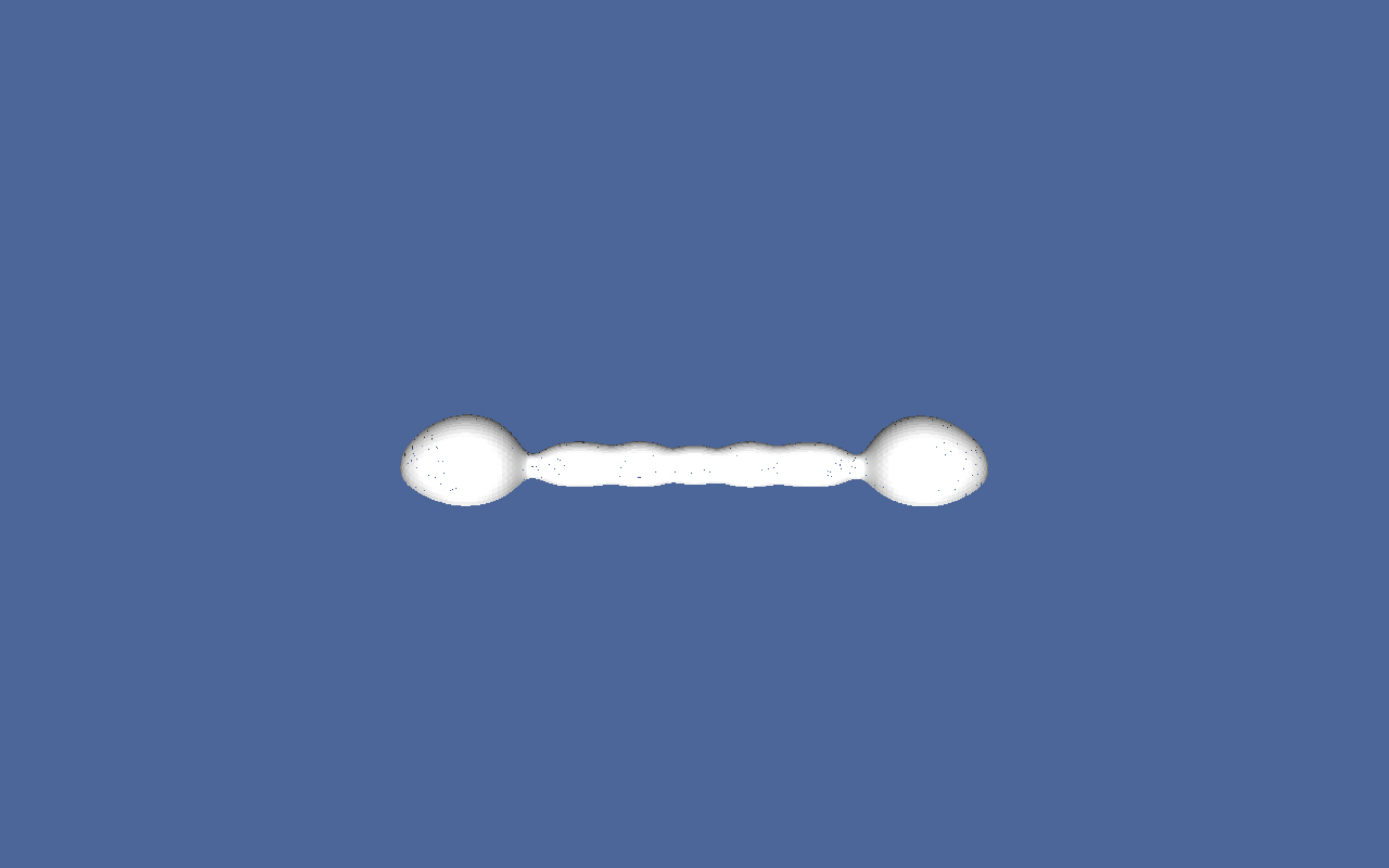}\\
		\includegraphics[width=.25\textwidth]{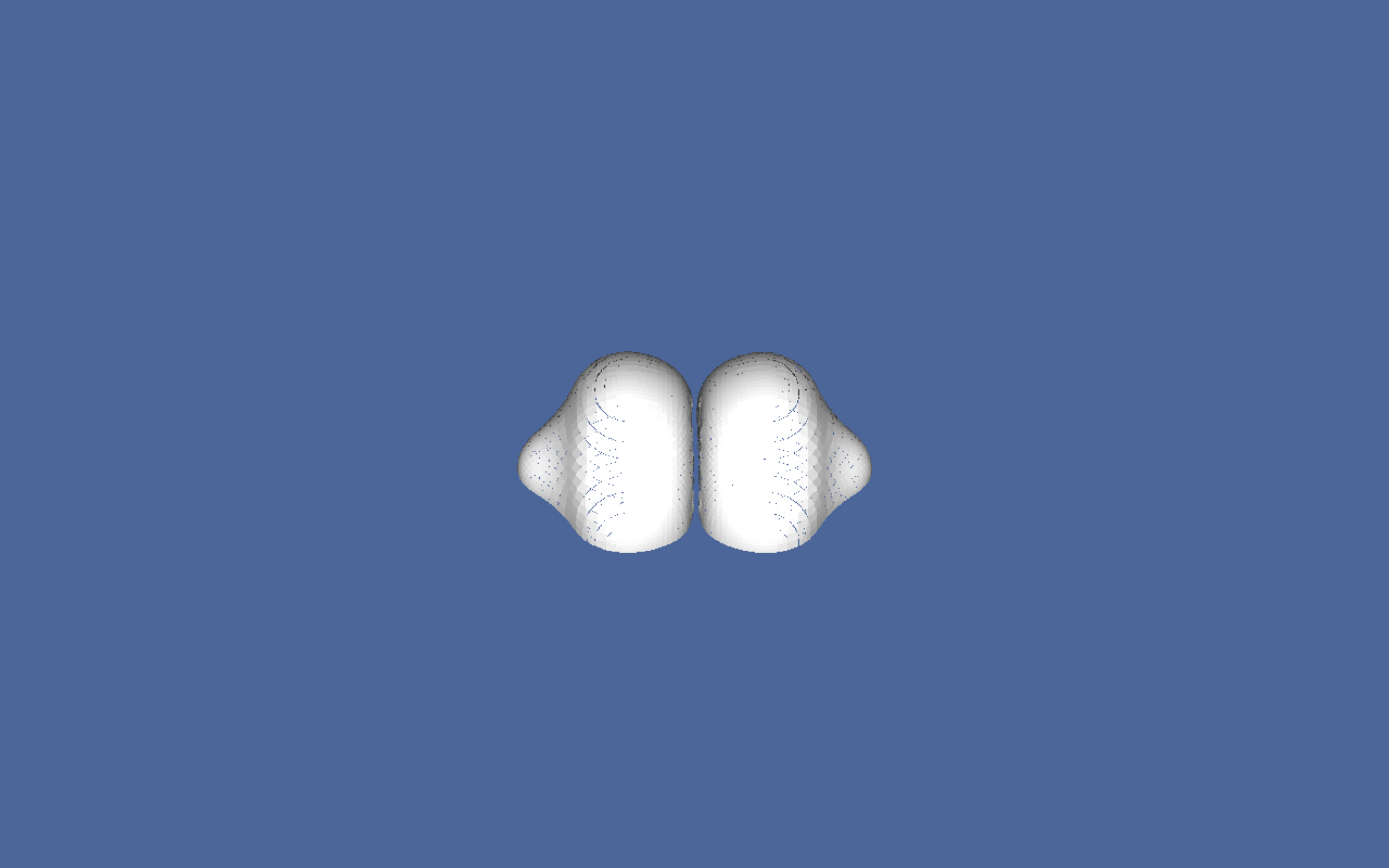}&
		\includegraphics[width=.25\textwidth]{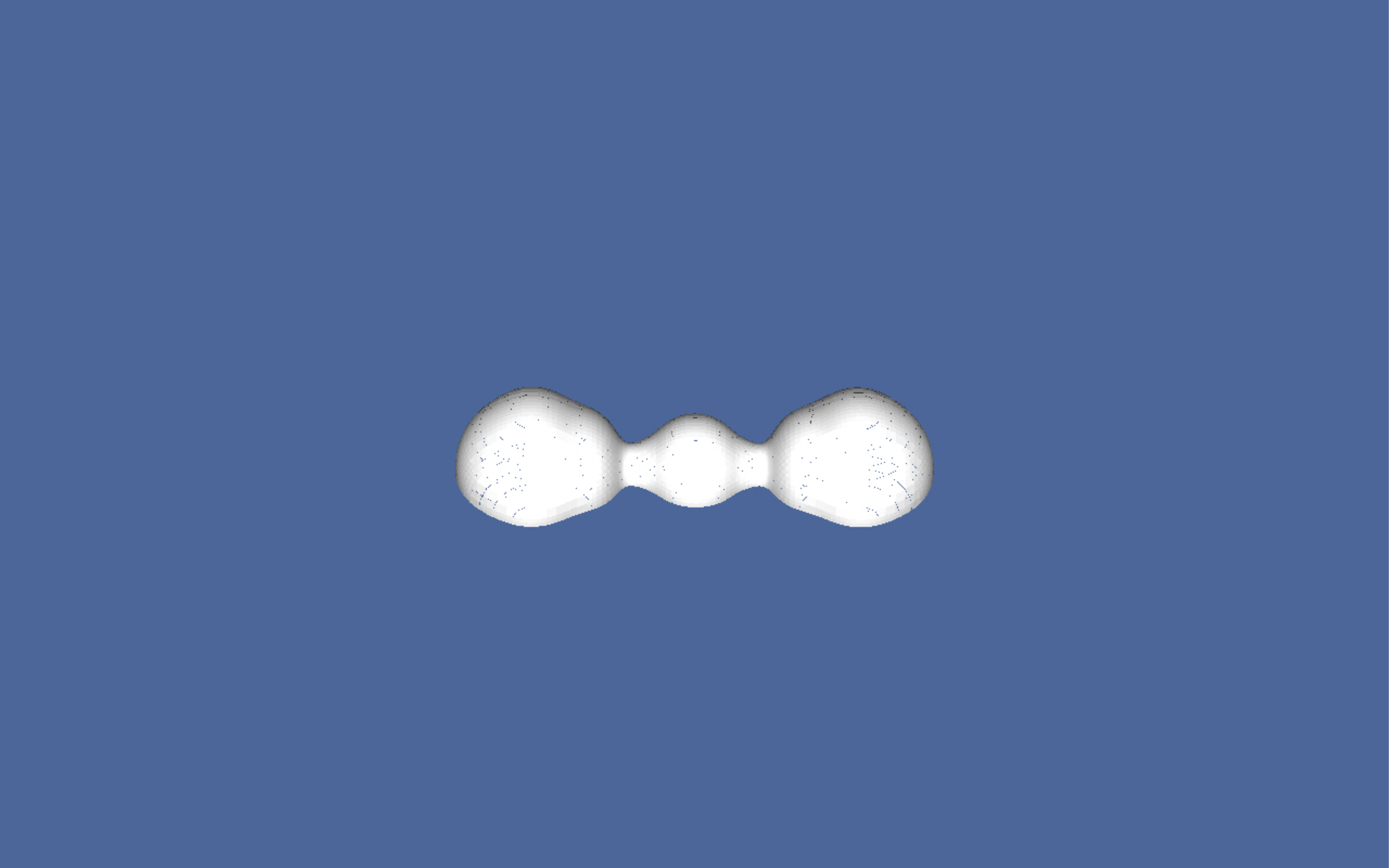}&
		\includegraphics[width=.25\textwidth]{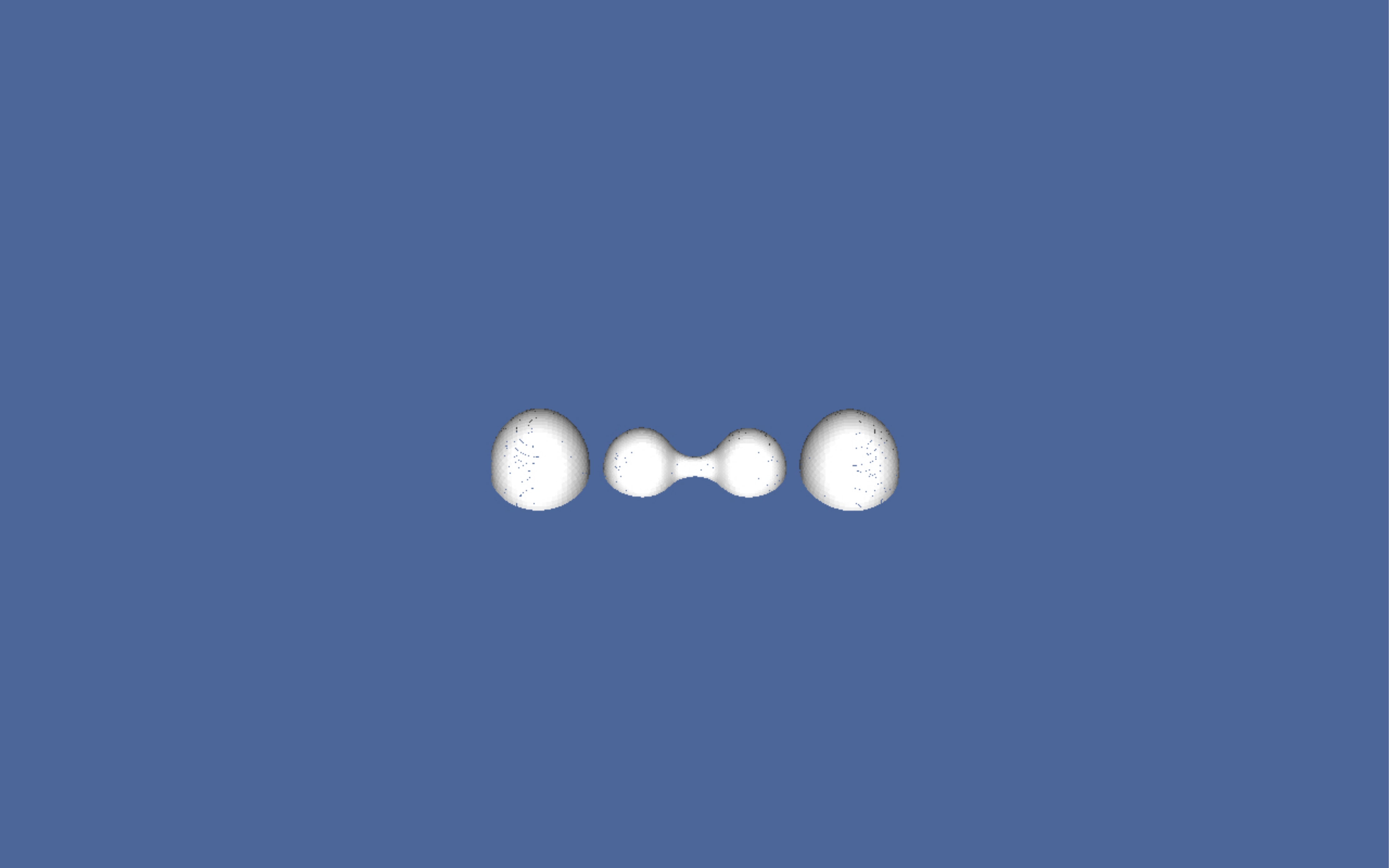}\\   
		\includegraphics[width=.25\textwidth]{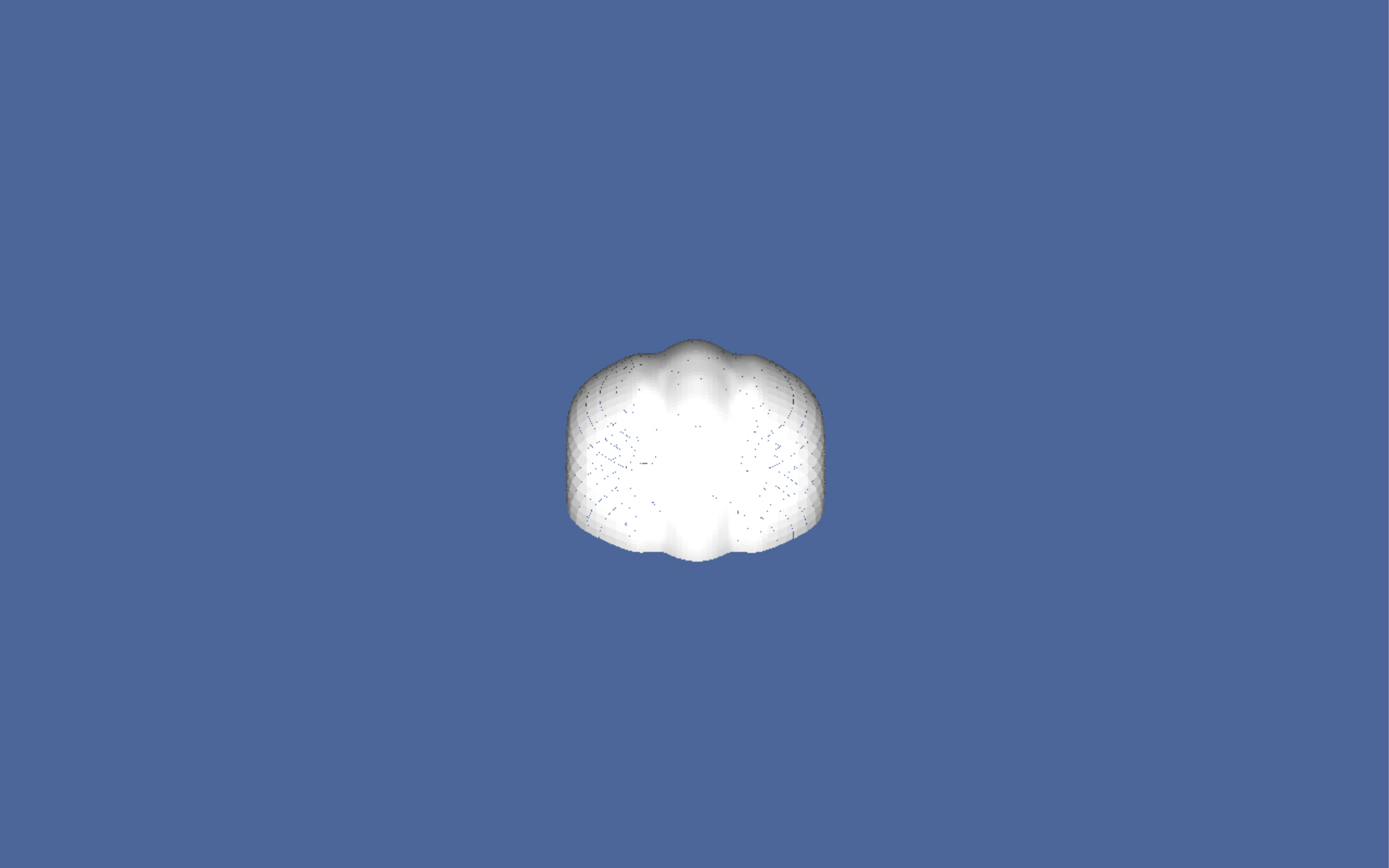}&
		\includegraphics[width=.25\textwidth]{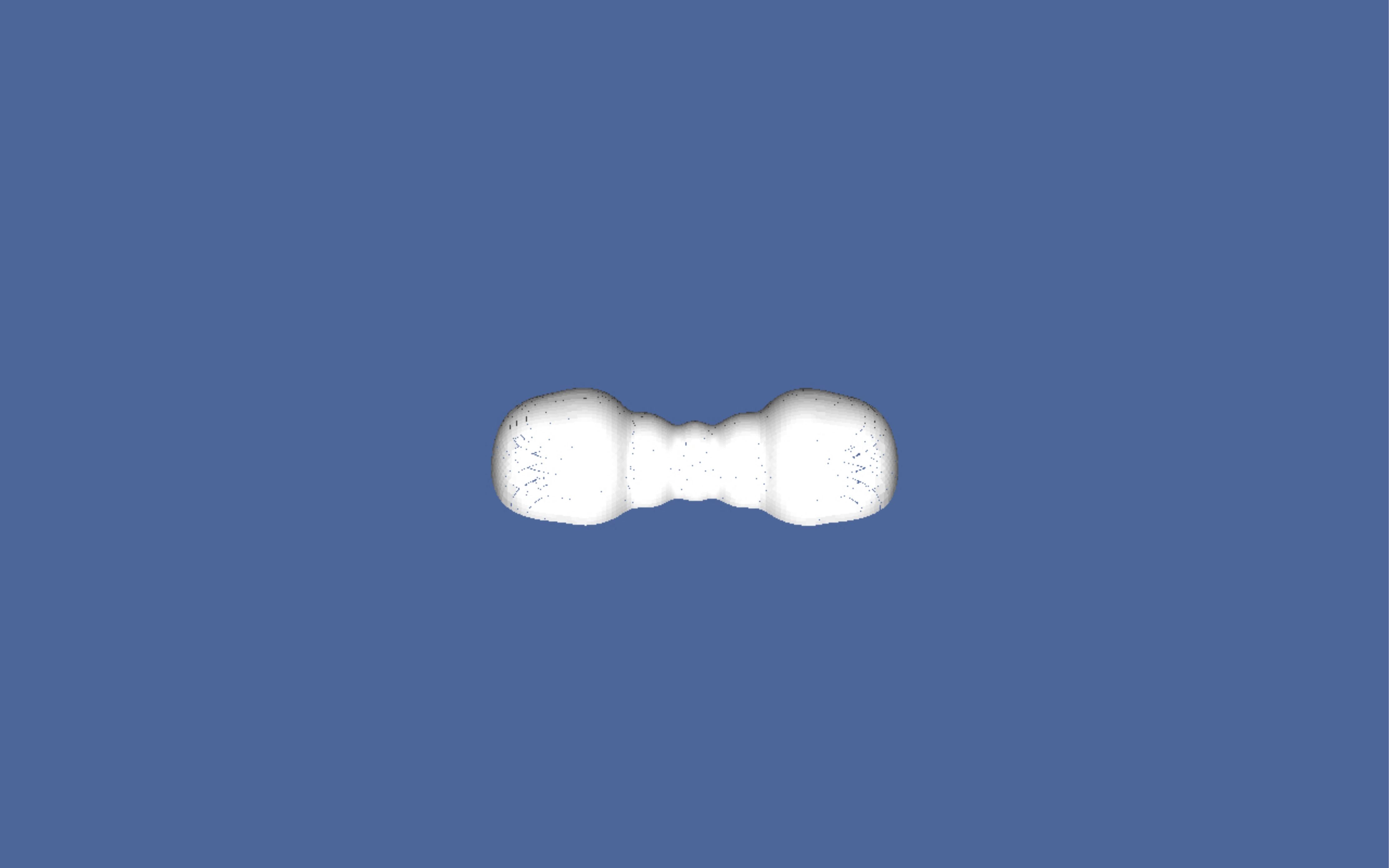}&
		\includegraphics[width=.25\textwidth]{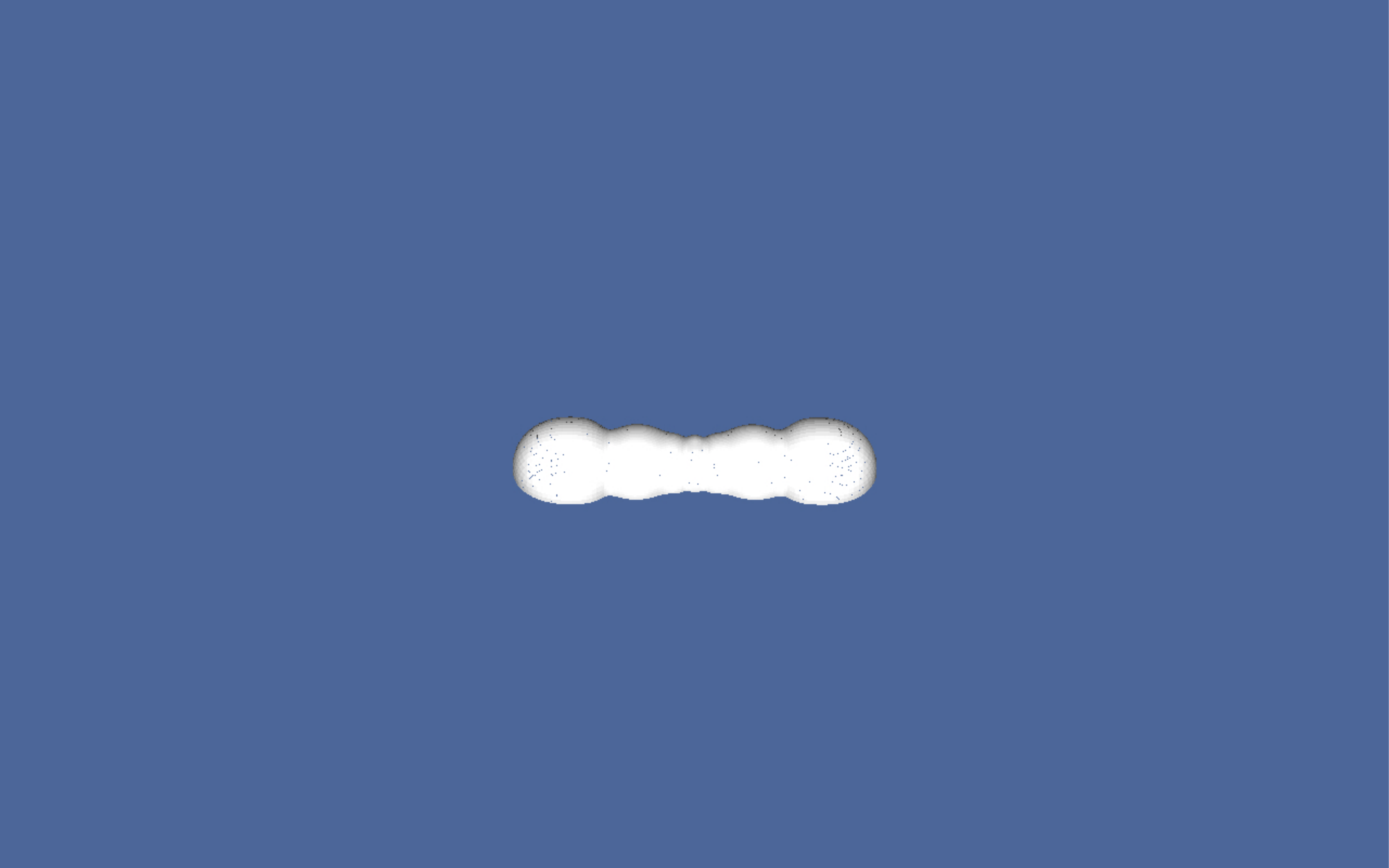}\\  
		\includegraphics[width=.25\textwidth]{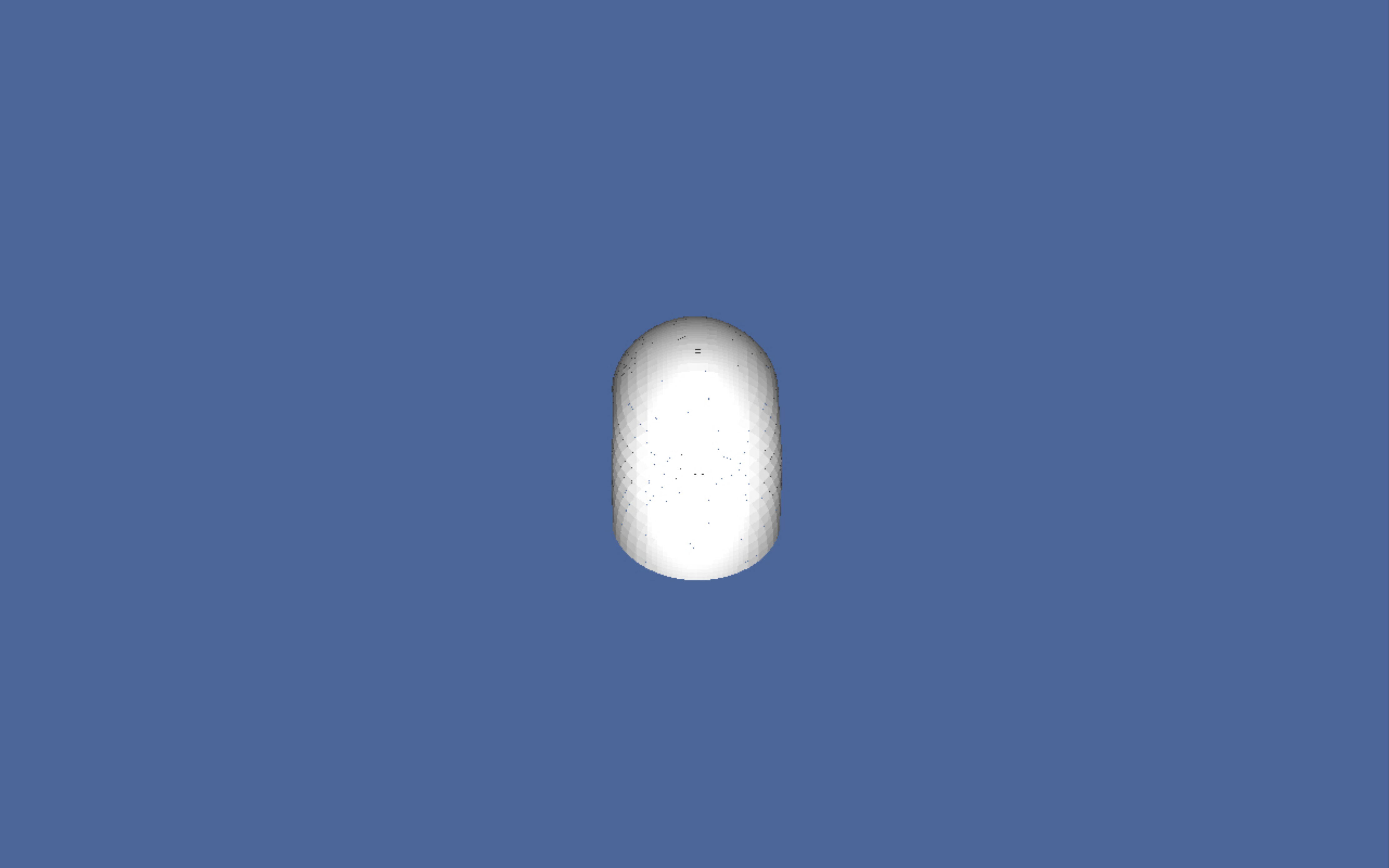}&
		\includegraphics[width=.25\textwidth]{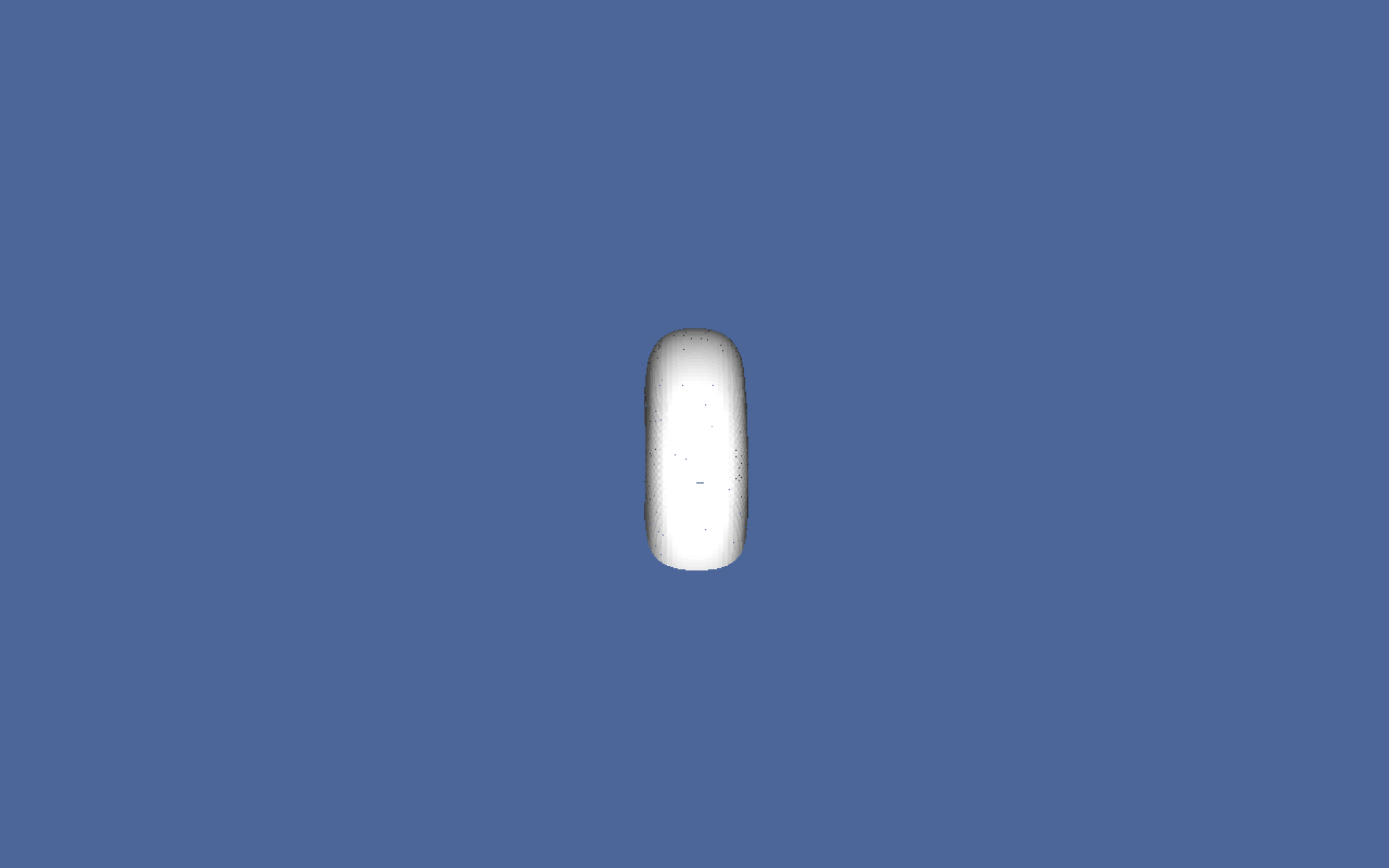}&
		\includegraphics[width=.25\textwidth]{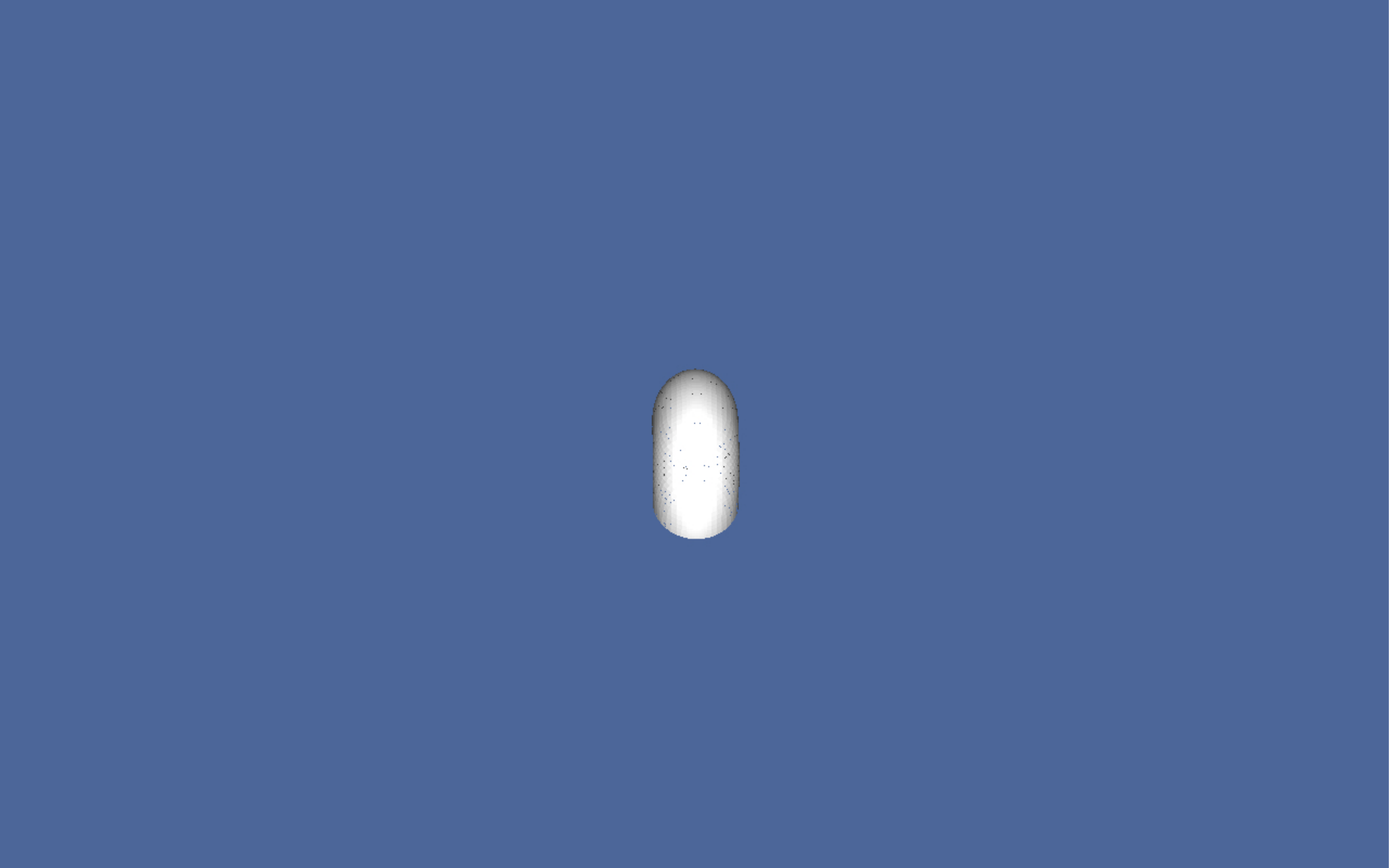}\\   
		a &b &c 
	\end{tabular}
	\caption{Simulations for free-standing jets for Oh  $=0.02$, (a) AR  $= 12$, (b)  AR  $= 16$ and (c) AR  $= 26$.
		Only top half of the filament is shown.}
	\vspace{-0.1in}
	\label{fig:scenarios3}
\end{figure}

Interestingly, the simulations reveal  a non-continuous behavior in the breakup and no-breakup
regime when AR is varied for a fixed Oh number and when the Oh number is
sufficiently small. For example, consider Oh $=0.02$, where 
the simulations show breakup for AR $=12$, no-breakup for AR $=16$, and breakup for AR $=26$;
see Fig.~\ref{fig:scenarios3}. (We note that this behavior is not presented in 
the experimental data in \cite{pita}, mainly because in the
region that the numerical results show non-continuity in breakup to no-breakup, there are either
limited or no experimental observations; this can be due to the fact that the breakup occurs on a very fast time scale and therefore high framing speeds are needed to capture fast breakups -
the numerical simulations can indeed be very useful here. As authors in \cite{pita} also report,
there are other factors such as surface contamination, 
surface vibration, filaments not at initial rest, temperature driven effects, 
as well as generated filaments not being completely symmetrical, that can lead to possible
differences between the results.)

We explore the above non-continuous behavior in more details; as shown 
in Fig.~\ref{fig:scenarios3}, a subtle competition 
between inertial effects,  oscillating modes, and capillarity, 
can lead to this peculiar dynamics: 
for sufficiently small Oh numbers, the time scale over which the filament retracts is comparable to
the time scale over which the surface instabilities grow. Meanwhile, due to inertial effects,
capillary waves propagate on the surface, which are dampened by the viscous effects. 
From looking at Fig.~\ref{fig:scenarios3}(b), it is clear that 
the capillary waves develop on the surface; if the oscillation of the capillary waves 
happens at the time of the breakup, it can eventually prevent the necking of the 
filament and the consequent breakup. For a small Oh number, internal flows
are important, unlike for highly viscous filaments. Additionally, the time
scale of collapse is comparable with the typical time scale on which capillary waves
propagate. The transition from no-breakup to breakup can therefore become sensitive
to when all  these effects  are  in balance, showing a non-continuous behavior. 
Future work should consider quantifying each effect to provide a more
in depth analysis of the transition mechanism.
 
In summary, our study of free-standing filaments reveal a remarkable agreement with
the experimental results obtained in \cite{pita}. We also observer that a liquid filament 
with Oh number greater than 1 never breaks up regardless of the filament aspect ratio.
Additionally, we extend the range of parameter space to very small Oh numbers where
there is a rich spectrum of dynamics present. We show how subtle interplay of the dynamical
effects can lead to unexpected outcomes.  We next focus on the transition from no-breakup to breakup
including substrate effects. We also study the influence of slip on the breakup transition and
discuss the specific regions where a remarkably different behavior occurs when compared
with the free-standing jets.

%In the final Section we compare the experimental results with the ones obtained by fully
%nonlinear simulations based on a VoF method. By solving the three-dimensional
%Navier-Stokes (N-S) equations, these simulations provide additional insight regarding the
%competing instability mechanisms. It is assumed that the fluid flow can be modeled as an
%isothermal Newtonian fluid obeying the N-S equations.
%
%\begin{equation}
%  \rho\frac{D\textbf{u}}{Dt}=-\nabla p+\nabla\cdot(\mu(\nabla\textbf{u} + \nabla\textbf{u}^T))+\sigma\kappa\delta_S\textbf{n}
%\end{equation}
%\begin{equation}
%  \nabla\cdot\textbf{u}=0
%\end{equation}
%
%where $\textbf{u}=(u,v,w)$ is the velocity field, $p$ is the pressure, $\kappa$ is the
%curvature of the fluid-vapor interface, and $\delta n$ denotes the normal vector of the
%liquid interface. A contact angle of $90^0$ is imposed during the simulations.
%Experimental results could not be reproduced without taking contact line slip and laser
%melting/solidification effects into account. The contact line slip is represented by a
%slip condition,
%
%\begin{equation}
%  (u,w)|_{y=0}=\lambda\partial_y(u,w)_{y=0}
%\end{equation}
%
%where $\lambda$ denotes the slip length, and $(u,w)$ are the in-plane compoenents of
%velocity. 

\subsection{Transition from no-breakup to breakup
including substrate effects}
\begin{figure}[tb]
\begin{tabular}{c}
\centering
  \includegraphics[width=0.5\textwidth,trim=0 0mm 31mm 0,clip=true]{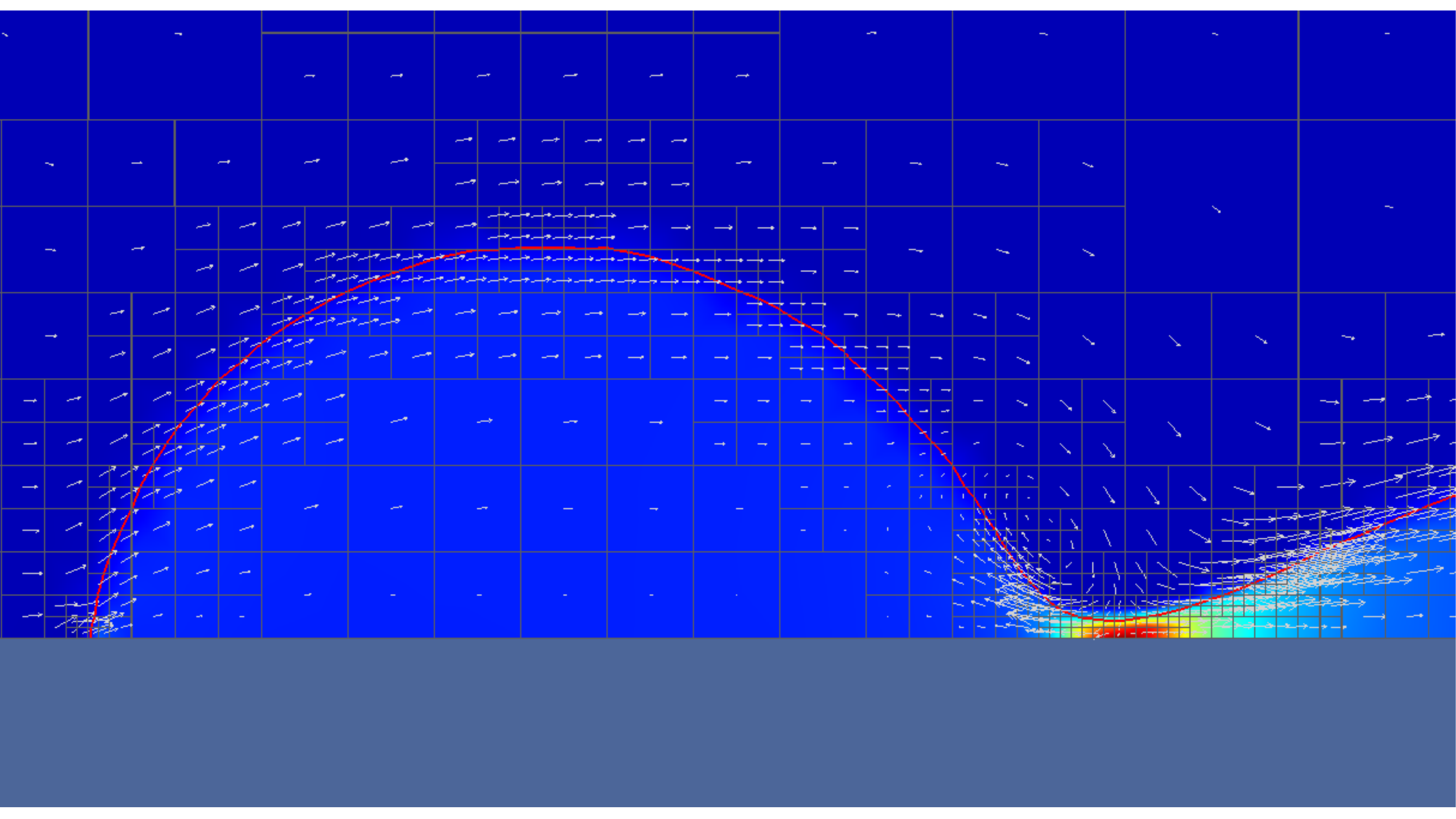}
\end{tabular}
\caption{Zoom of a cross section ($z=0$) of the bulge and the neck region prior to breakup, showing 
the details of the flow field (white arrows) and the pressure distribution 
(red(blue) shows the areas of high(low) pressure);
the background shows the adaptive mesh. Oh$ = 0.2$, AR $=33$, and 
$\lambda = 0$.}
\label{fig:neck}
\end{figure}
Here we consider finite-size filaments deposited on a substrate that retract axially
and either break up during the retraction or collapse into a single droplet.  
As discussed earlier, for viscous filaments, we also observe in this case that
as the tails of the filament recede, two bulges form at the end of the filament.
As the bulges grow, a neck that connects the bulges to the filament forms. 
Figure \ref{fig:neck} shows a typical simulation image of the necking process
just prior to the breakup. From inspecting the velocity field, 
we see how the fluid is squeezed out from the neck to the bulge,
while the filament is retracting. 
If the time that the bulges have to travel before collapsing in the middle is smaller
than the time required for the neck to pinch off, then the filament will not break up during
the traction process, creating a strong vortical flow at the neck. The color map
also shows an area of high pressure at the ridge connecting the bulge to the 
filament in the middle.  Our aim is to investigate this process and the dynamics resulting  
the breakup.  
We extend the results discussed above for a free-standing filament to include the 
substrate effects. The introduction of the substrate is expected to profoundly change
the dynamics by delaying the retraction of the tails of the filament and therefore allowing
for the instabilities to have sufficient time to grow, promoting breakup.  For small to moderate Oh
values, where the breakup occurs due to either the end-pinching or the Rayleigh-Plateau 
instability developing on the filaments connecting the end bulges, the inclusion of the substrate results 
in smaller critical aspect ratio than the one for the free-standing filament. 
A striking conclusion of our study is that for very viscous filaments, i.e. for Oh $>1$,
the filament still breaks up due to the Rayleigh-Plateau instability, in contrast to the results
of simulations and observations described in the previous section for free-standing filaments. 
One experimental example of this effect was shown in  Fig.~\ref{fig:Experiments2}.

As noted above, the breakup of the filament on the substrate can be largely influenced
by the retraction velocity of the tails of the filament. Providing a theoretical explanation 
of the critical AR that leads to breakup is however difficult due to the fact that  
the typical velocity of tails of the filament cannot be found analytically, mainly
as a result of the contact line singularity and issues related to that; see e.g.~\cite{MCL}.
It is the aim of the following study to provide a numerical database to characterize the
no-break to breakup regions based on the extent of the slip on the substrate.
We consider various levels of slip, defined by $\lambda$ in Eq.~\ref{eq:slip},  
and repeat the numerical simulations for a range of Oh and AR values. We note
that the case of a perfect slip condition at the substrate corresponds to the free-standing
filaments. In our numerical model, we impose a Navier slip through  Eq.~\ref{eq:slip}; 
when $\lambda=0$, this condition amounts to a no-slip condition on the velocity field. 
Here, we consider $\lambda=0,0.01,0.1,1$ and carry out an exhaustive study of the effect 
of slip on the breakup transition. 

\begin{figure}[tb]
\centering
\begin{tabular}{cc}
  \includegraphics[width=.515\textwidth]{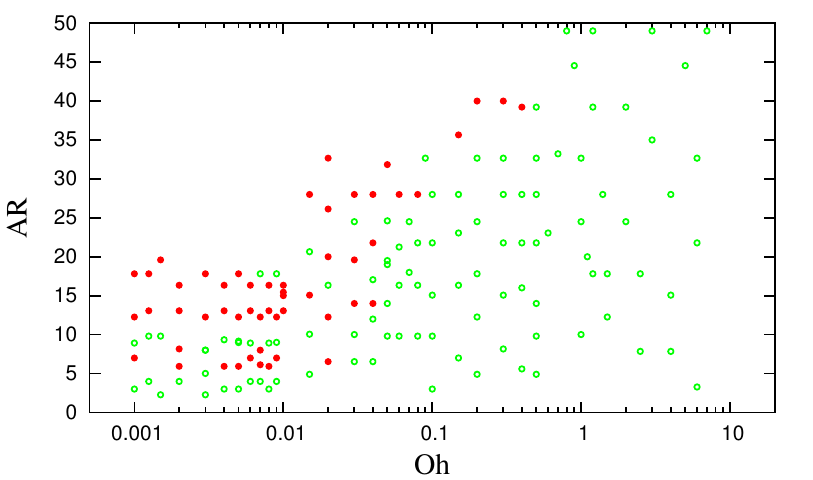}&
  \includegraphics[width=.515\textwidth]{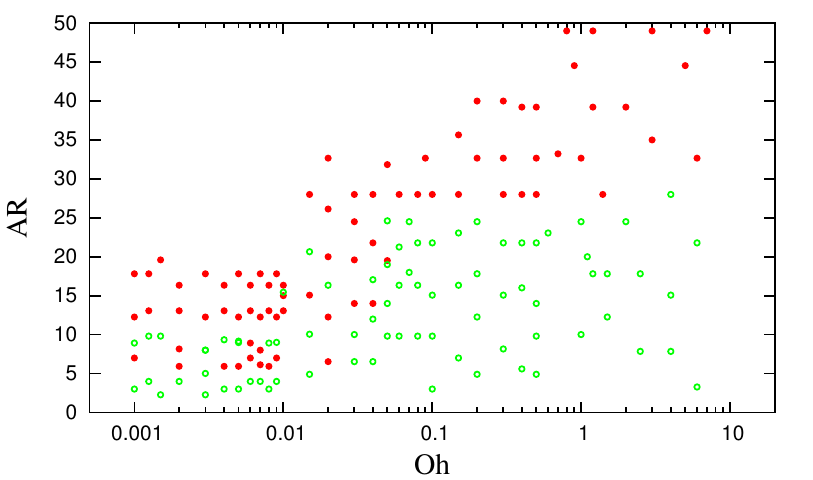}\\
   (a) & (b)\\
  \includegraphics[width=.515\textwidth]{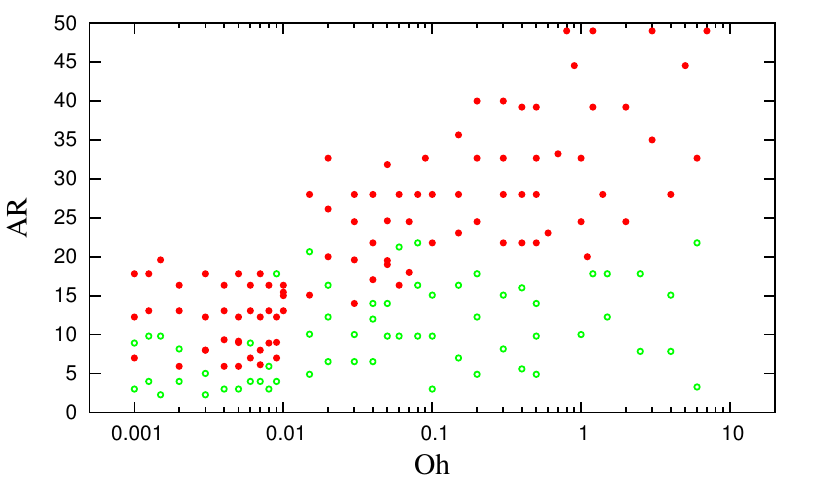}&
  \includegraphics[width=.515\textwidth]{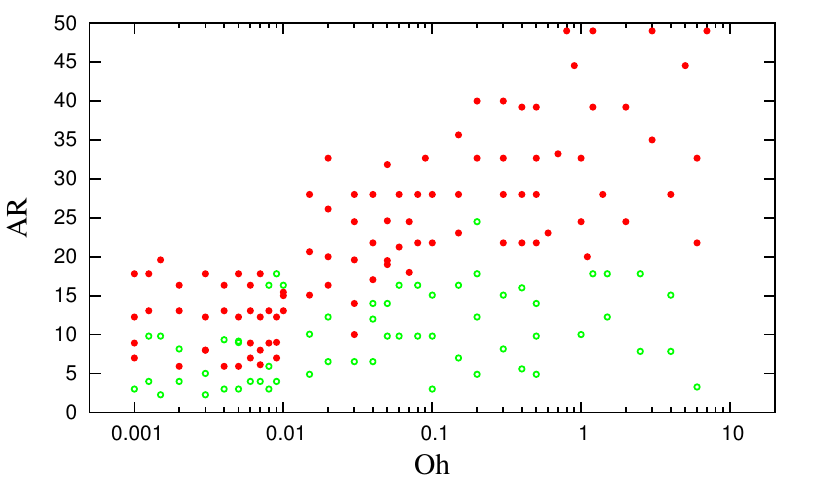}\\\
   (c) & (d)
\end{tabular}
\caption{Breakup and no-breakup results plotted as a function of Oh versus AR: 
breakup ({\textcolor{red}{$\bullet$}}) and no-breakup ({\textcolor{green}{$\circ$}}),
for $\lambda=1$ (a), $0.1$ (b), $0.01$ (c), and $0$ (d).}
\label{fig:slip}
\end{figure}

Figures \ref{fig:slip} (a-d) show how the breakup transition depends on the slip length:
in general, the critical AR of the breakup decreases  with a decrease of $\lambda$. This 
can be understood based on the fact that  the retraction of the filament ends 
is decelerated by decreasing the amount of slip (i.e.~by increasing the viscous 
dissipation at the contact line), which leaves sufficient  
time for the necking and/or the Rayleigh-Plateau instability to occur. 
Clearly, as the slip length is decreased, the transition from no-breakup to breakup 
is shifted down, including more data points that represent a breakup. For small Oh($<0.01$) however 
there are anomalies where the breakup transitions to no-breakup (and again to breakup) when 
increasing the amount of slip; this is mainly due to a complex and subtle interaction with inertial effects. Next, we elaborate on these competing effects.

\begin{figure}[thb]
\centering
\begin{tabular}{cccc}
  \includegraphics[width=.25\textwidth]{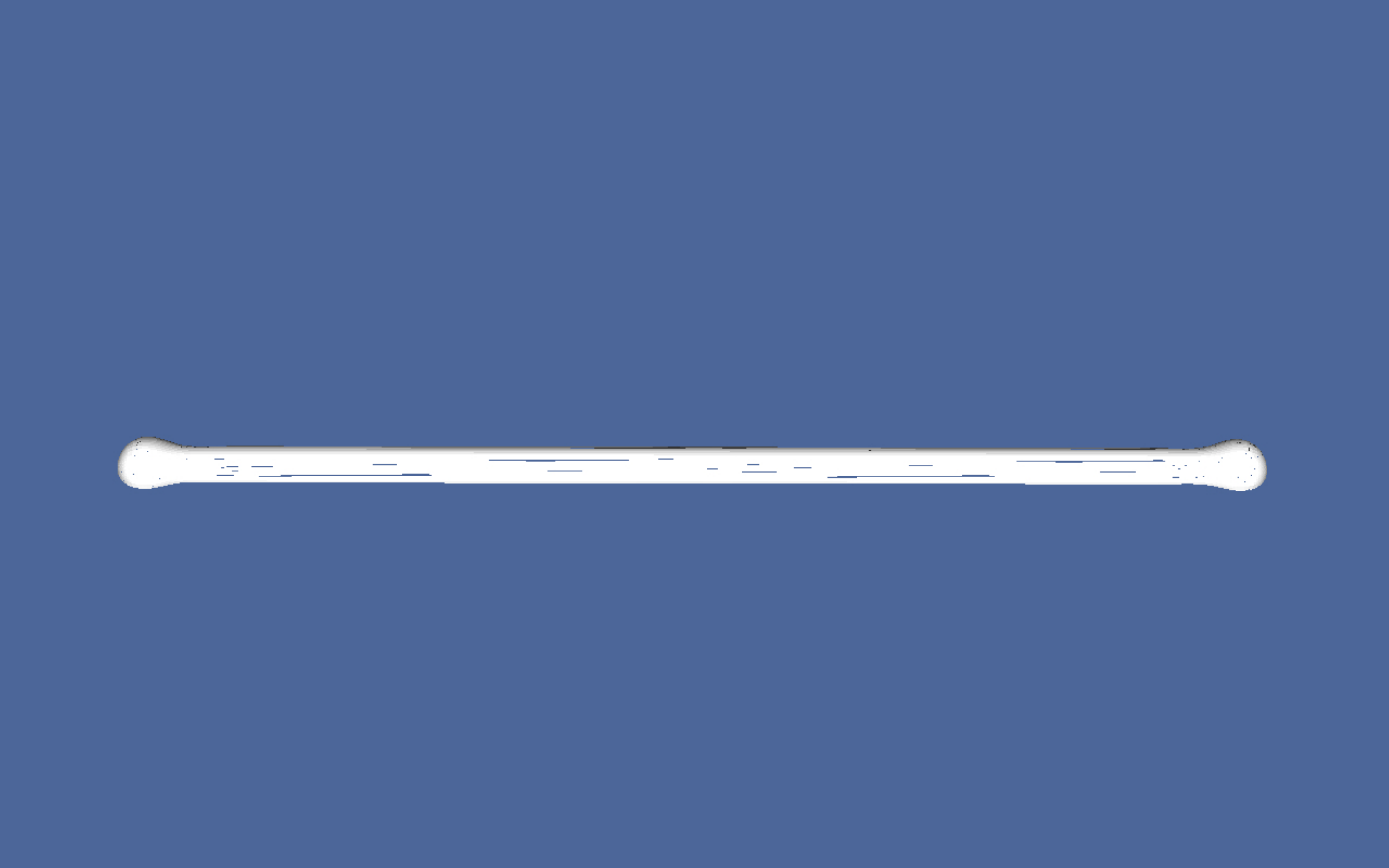}&
  \includegraphics[width=.25\textwidth]{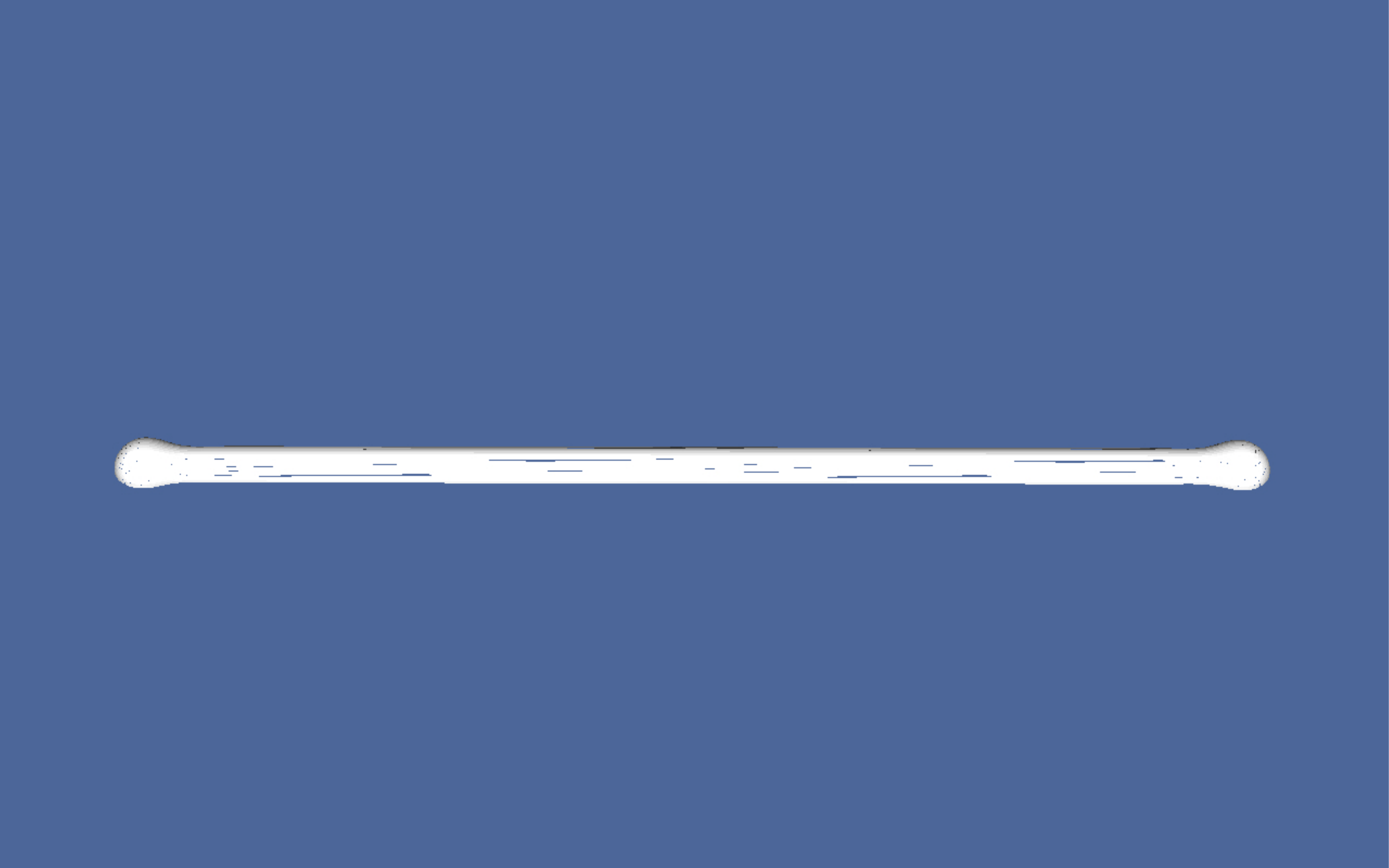}&
  \includegraphics[width=.25\textwidth]{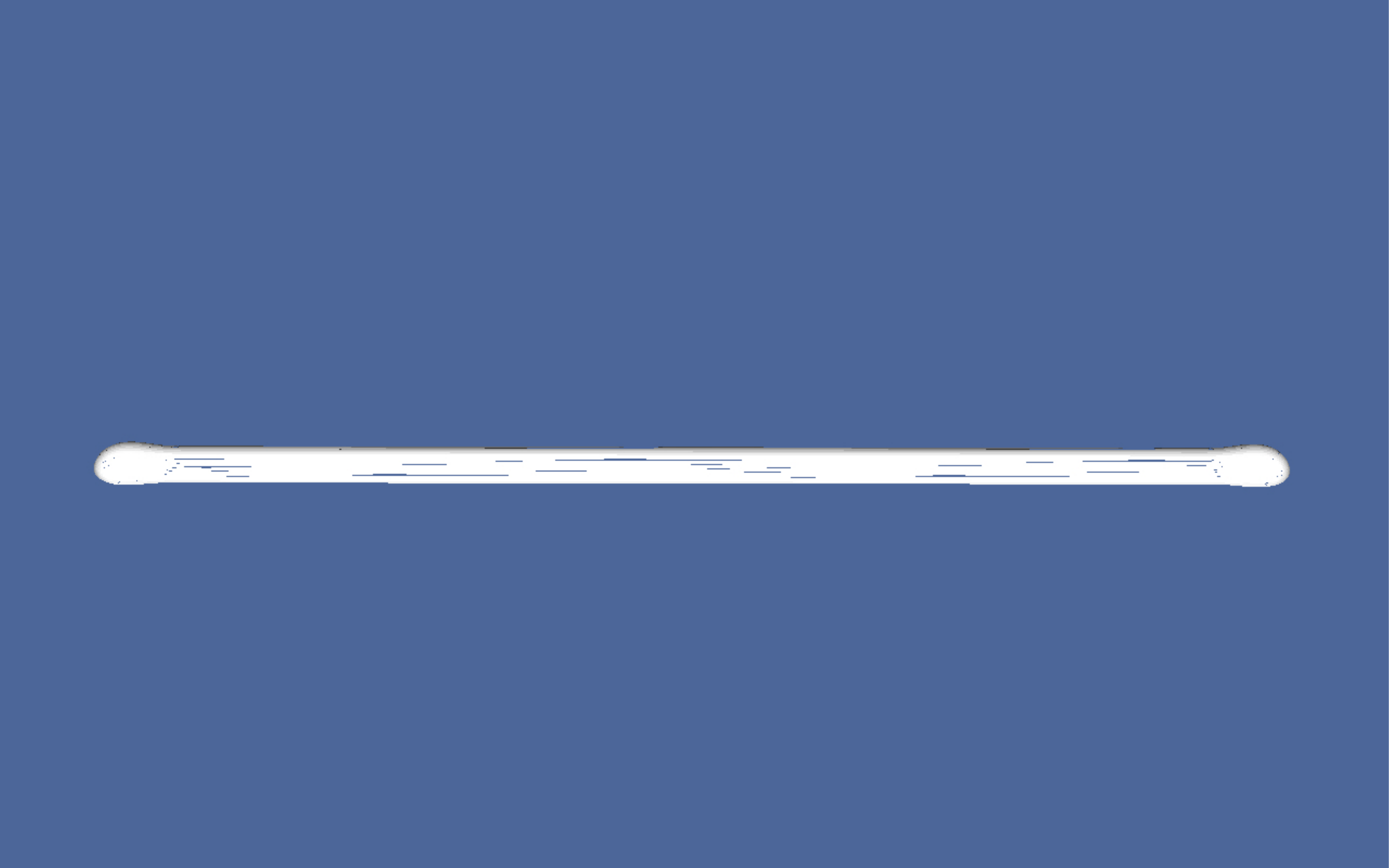}&  
  \includegraphics[width=.25\textwidth]{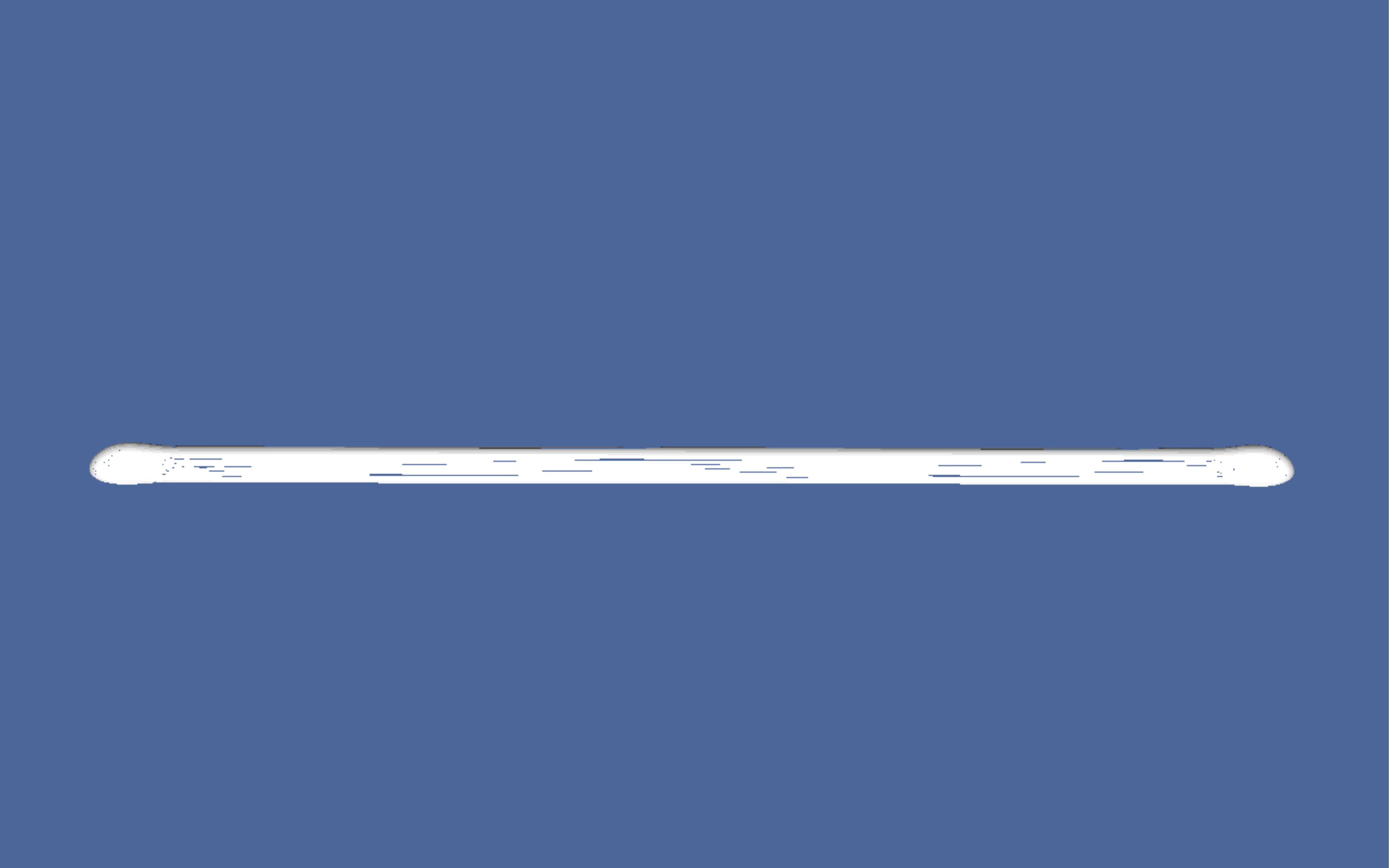}\\
  \includegraphics[width=.25\textwidth]{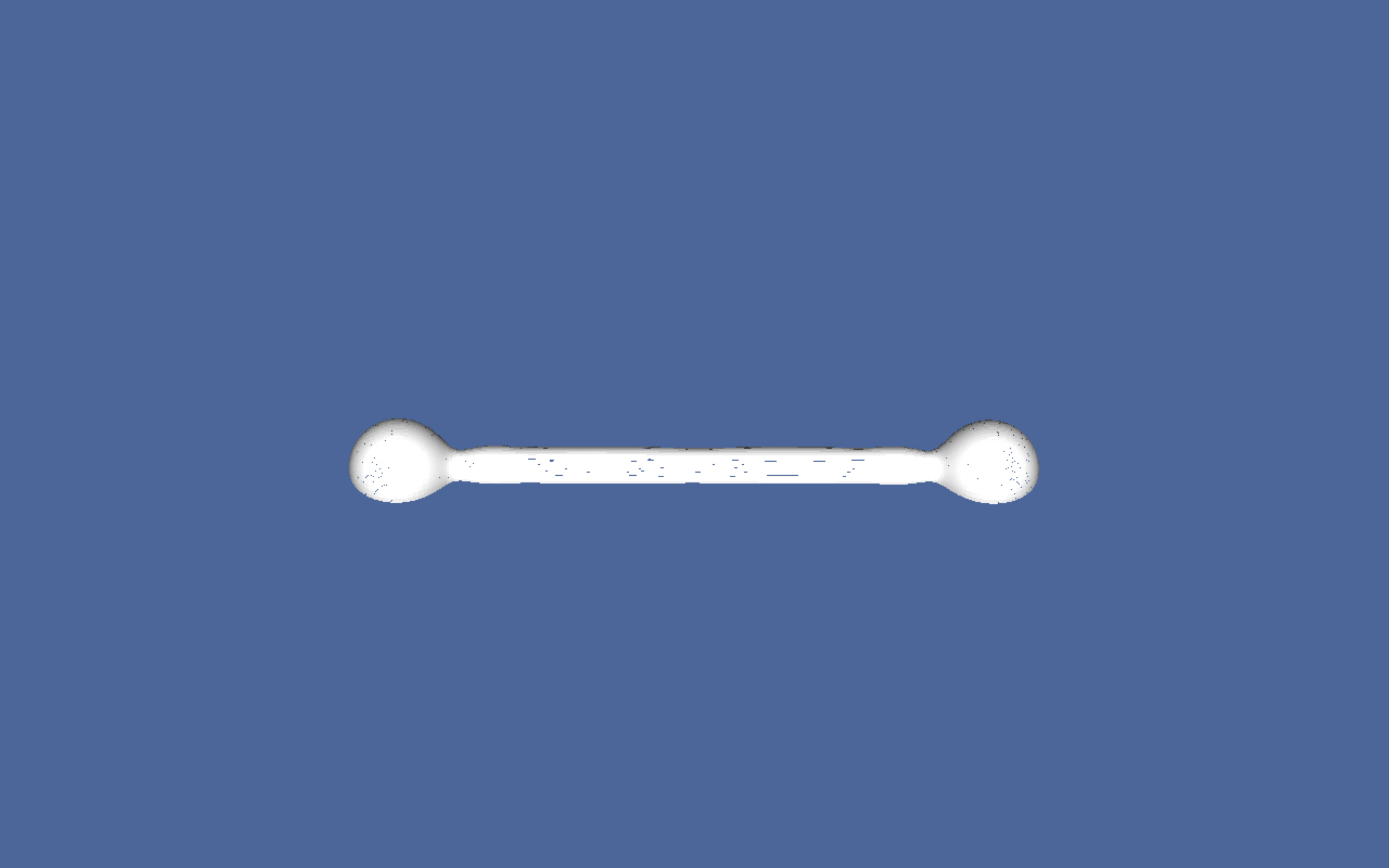}&
  \includegraphics[width=.25\textwidth]{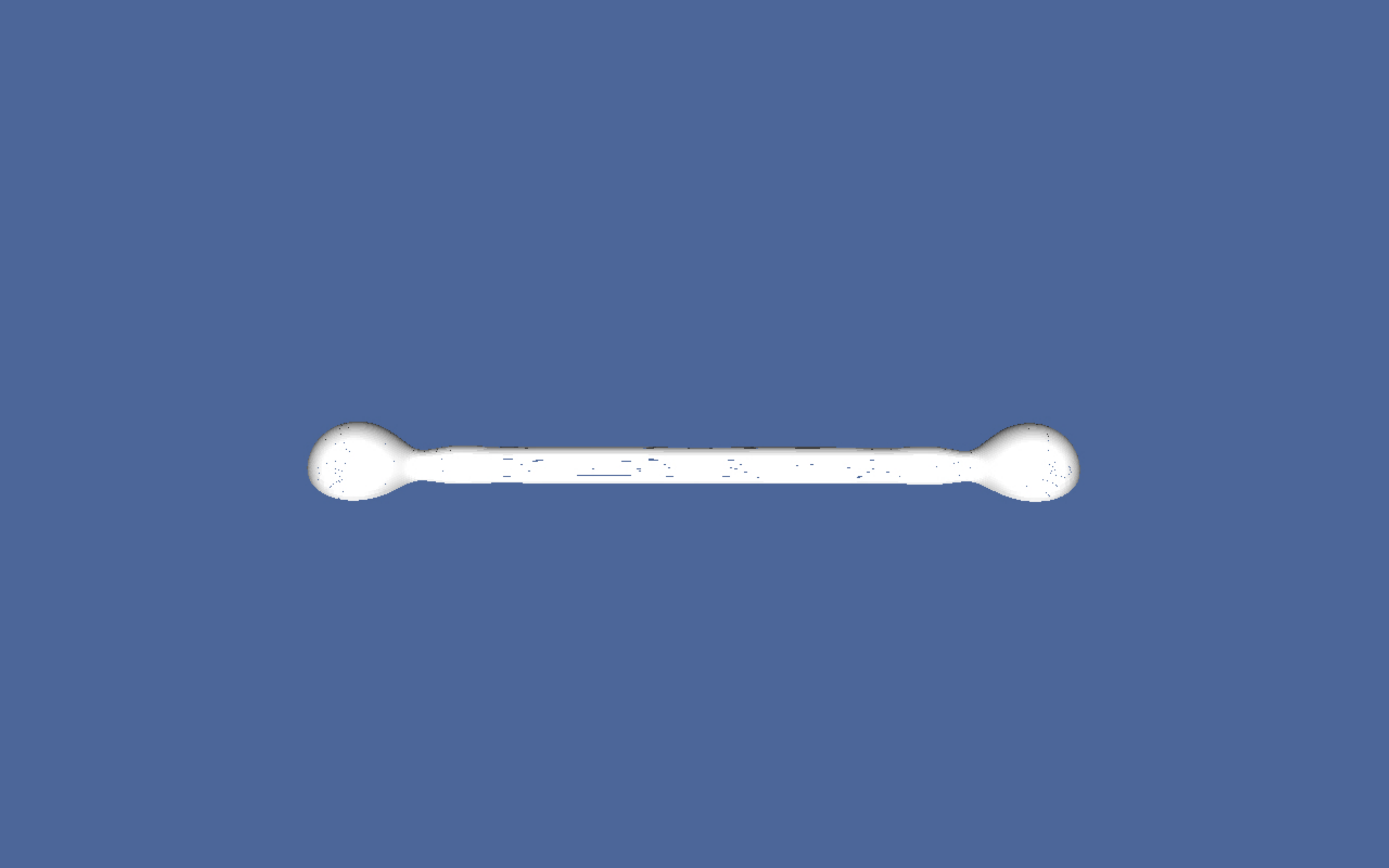}&
  \includegraphics[width=.25\textwidth]{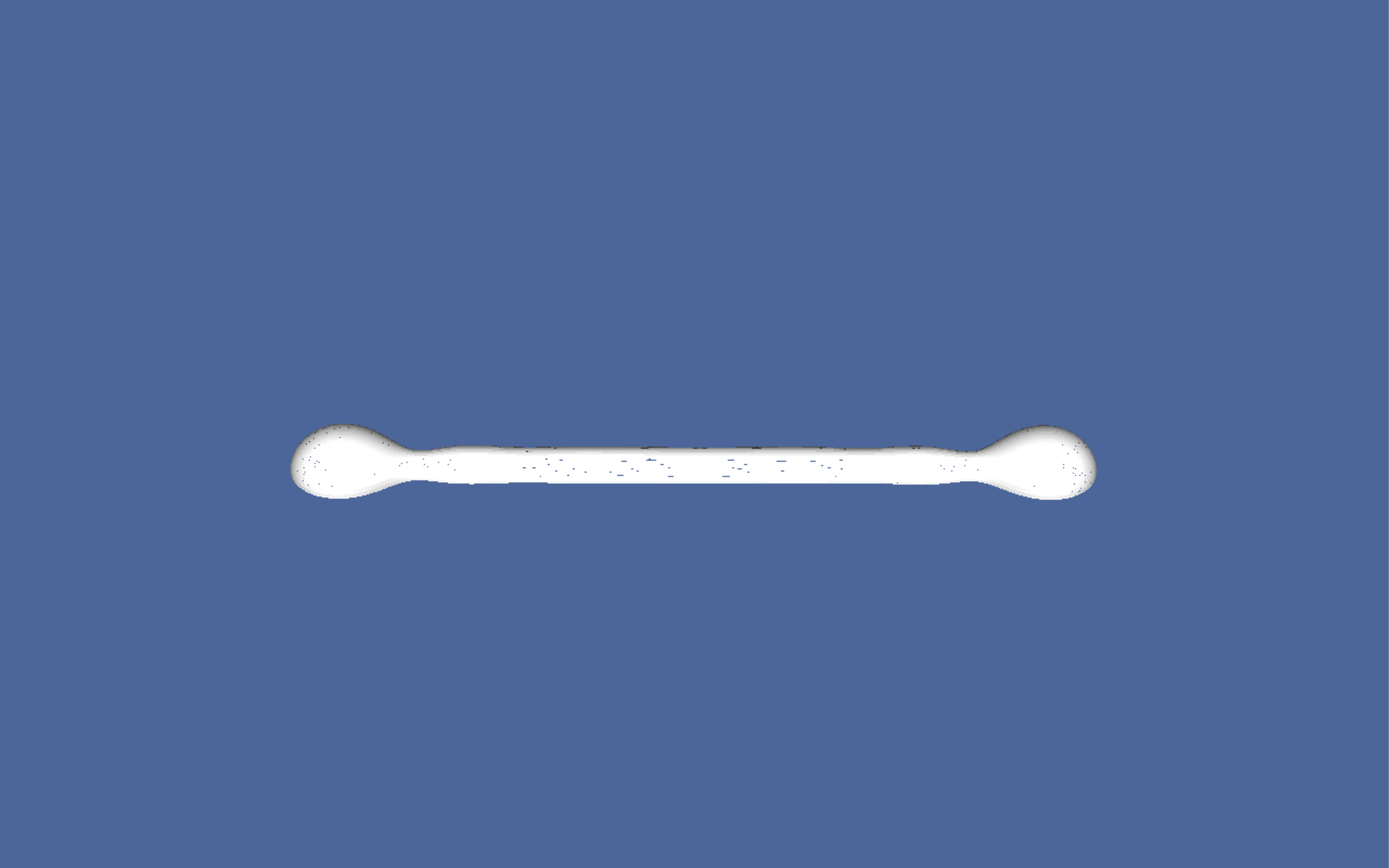}&  
  \includegraphics[width=.25\textwidth]{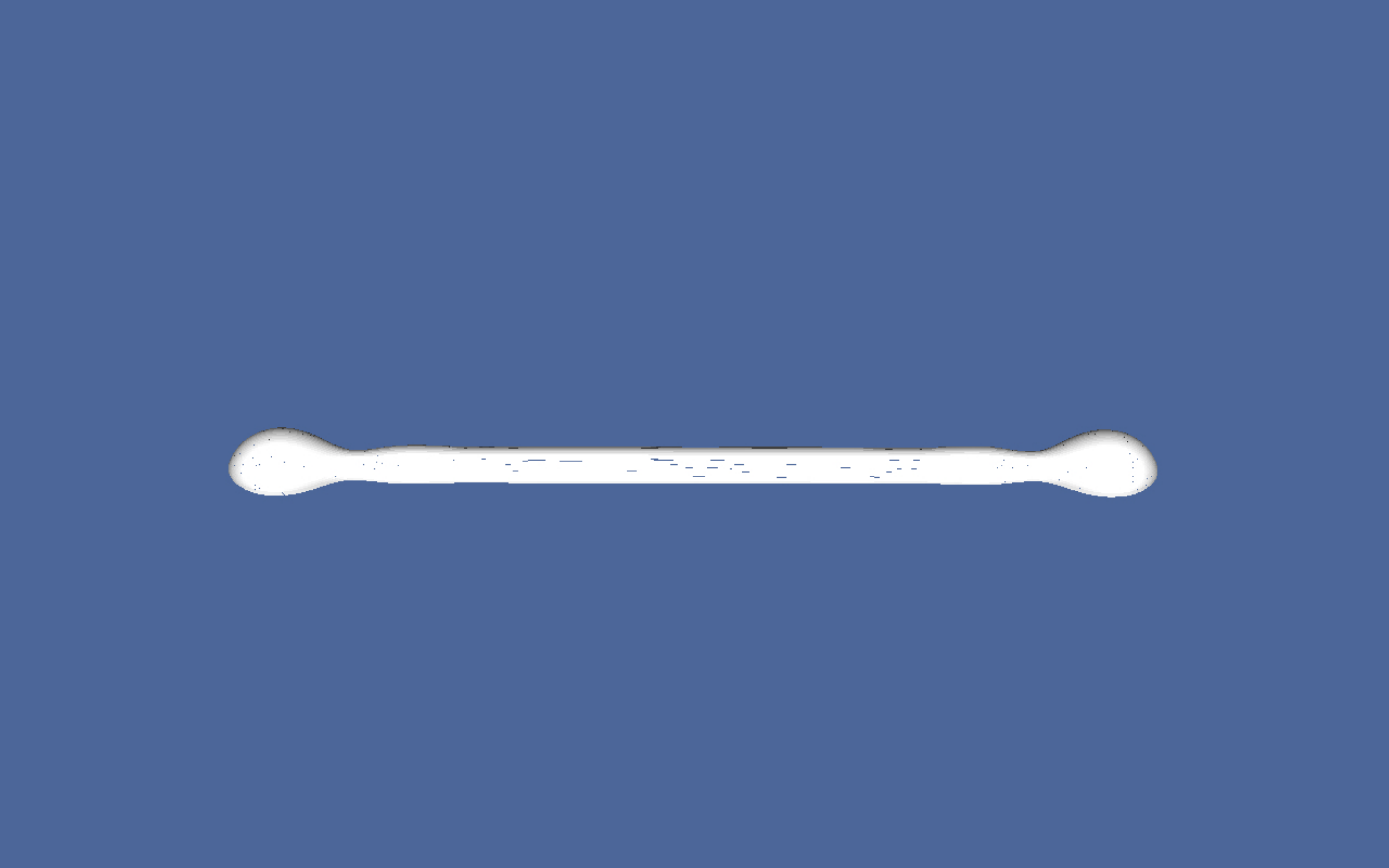}\\
   \includegraphics[width=.25\textwidth]{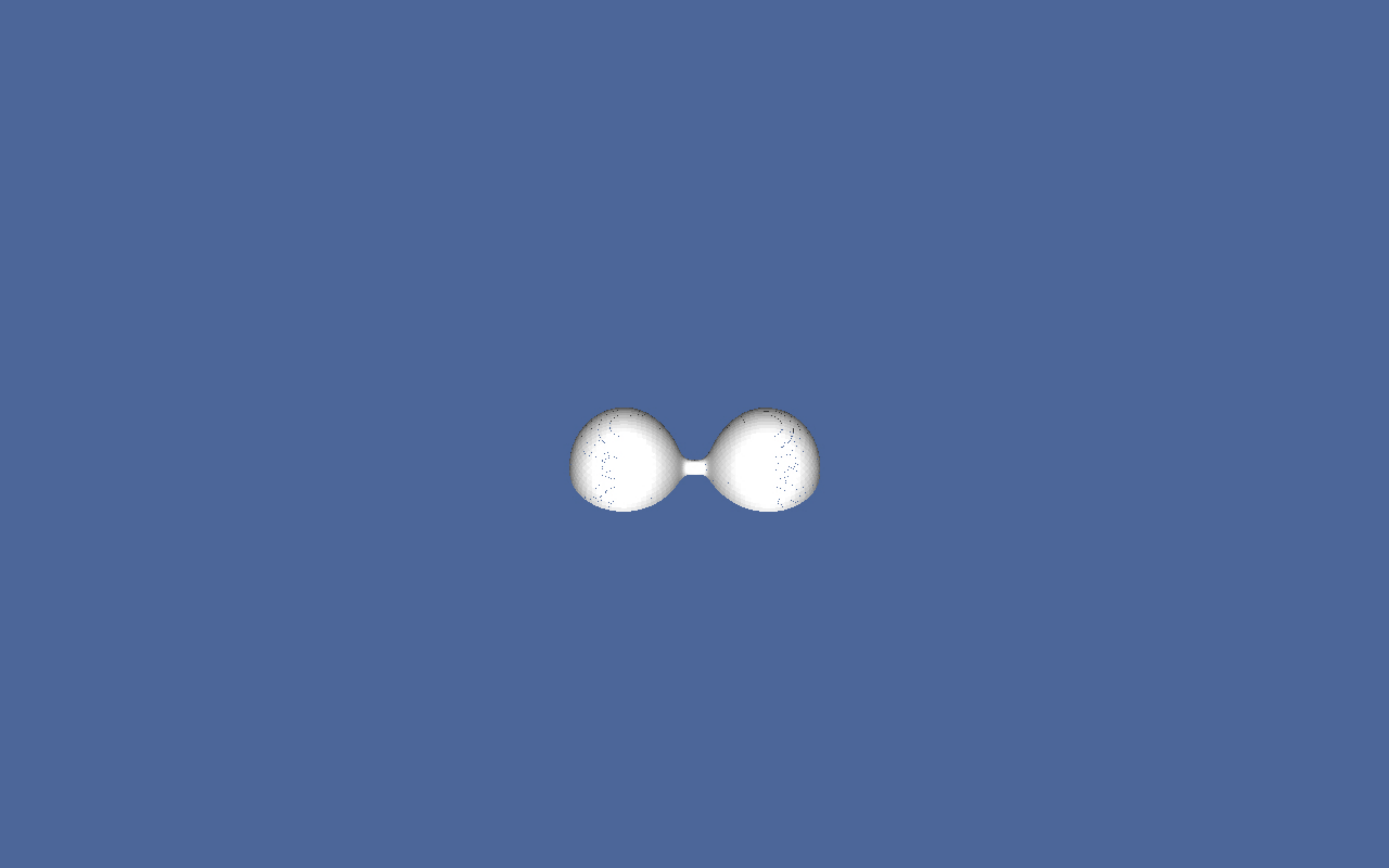}&
  \includegraphics[width=.25\textwidth]{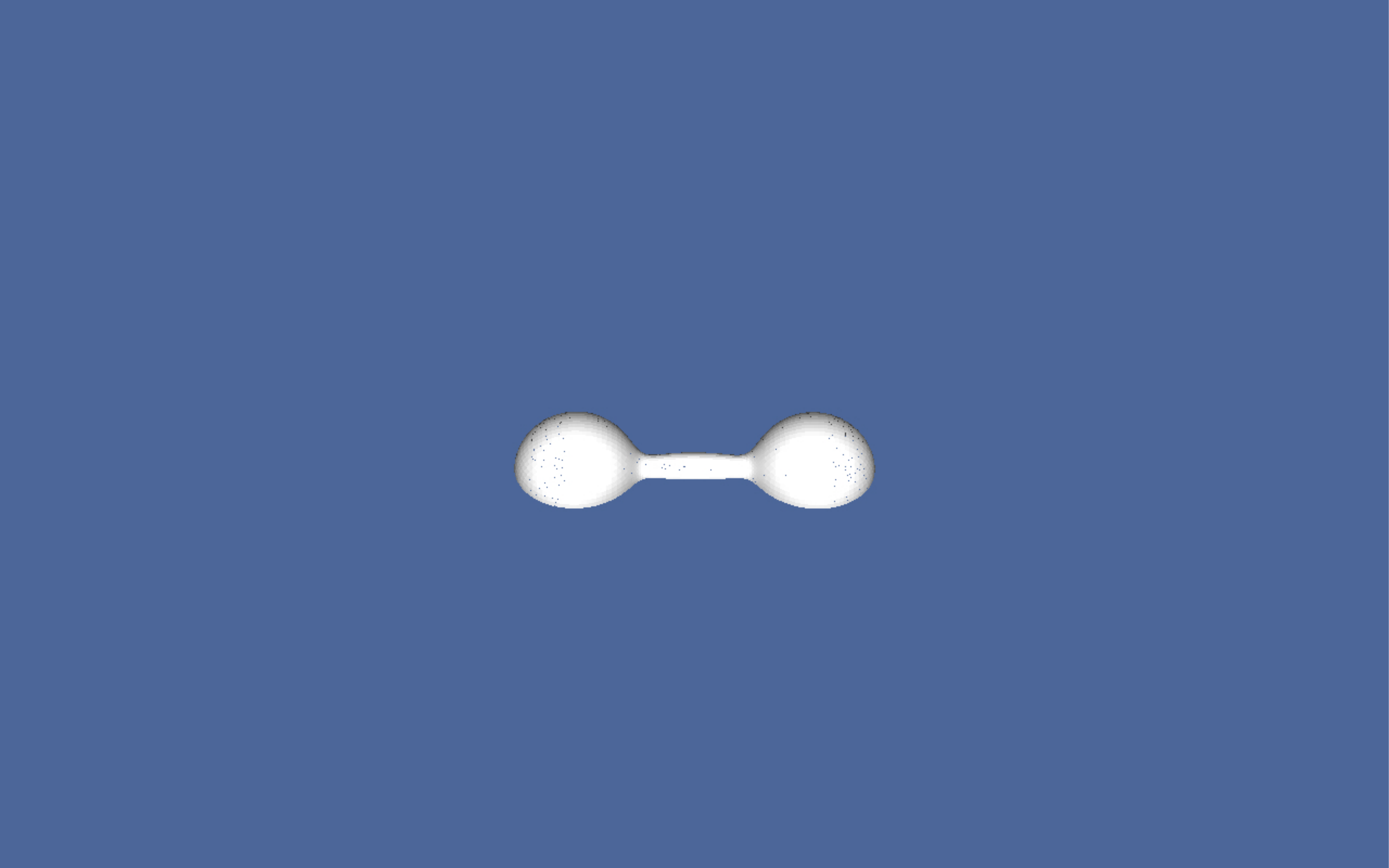}&
  \includegraphics[width=.25\textwidth]{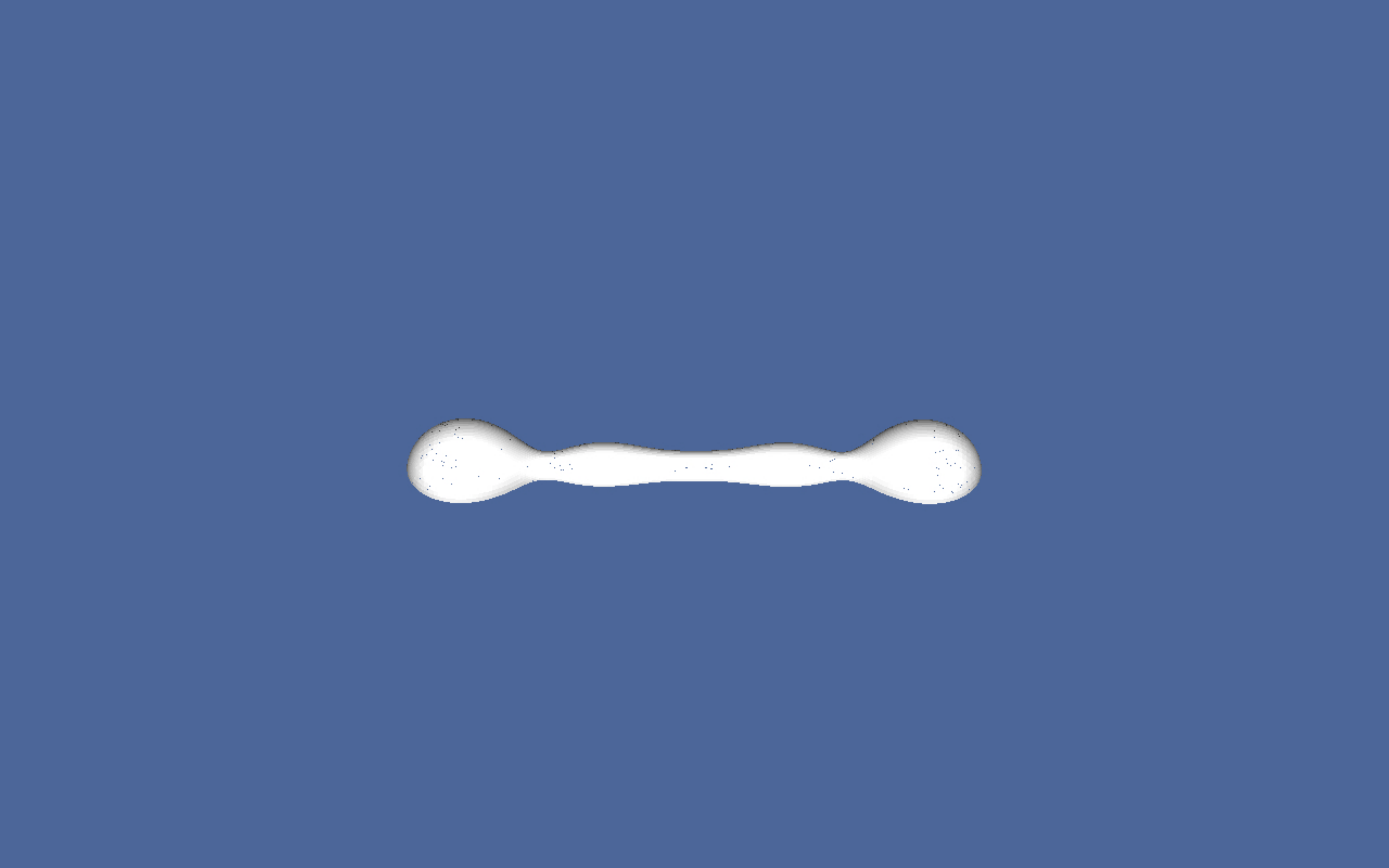}&  
  \includegraphics[width=.25\textwidth]{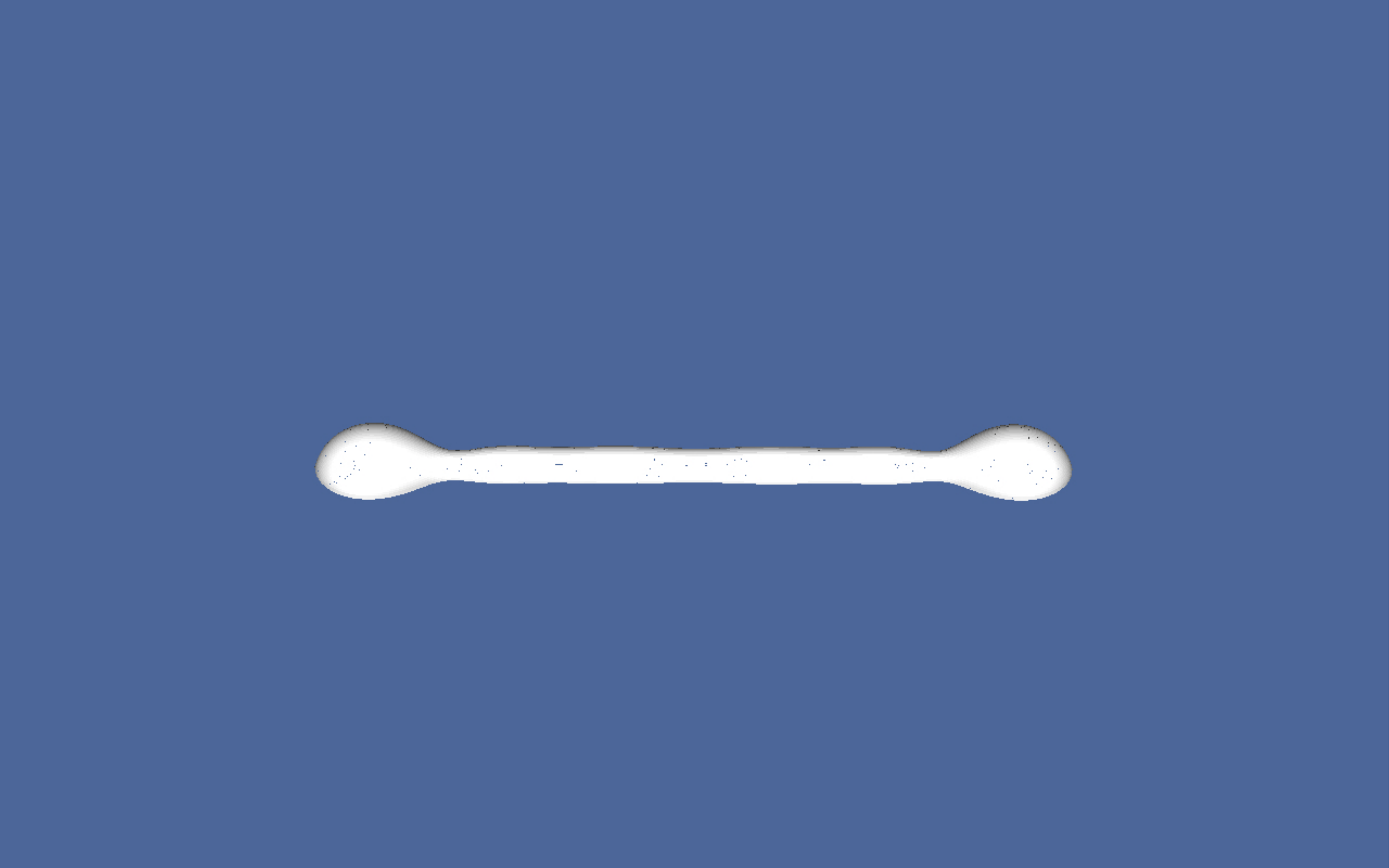}\\
  \includegraphics[width=.25\textwidth]{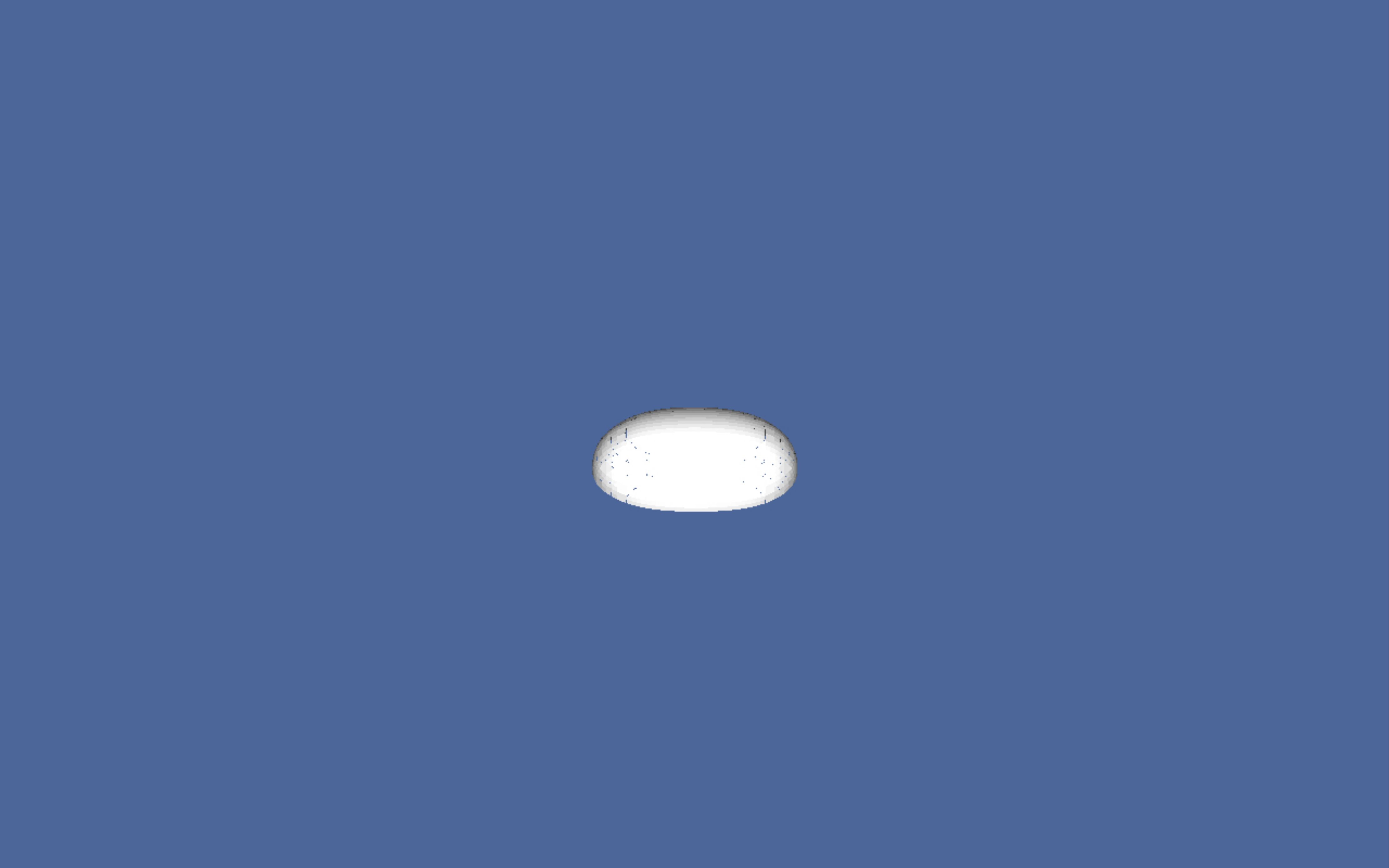}&
  \includegraphics[width=.25\textwidth]{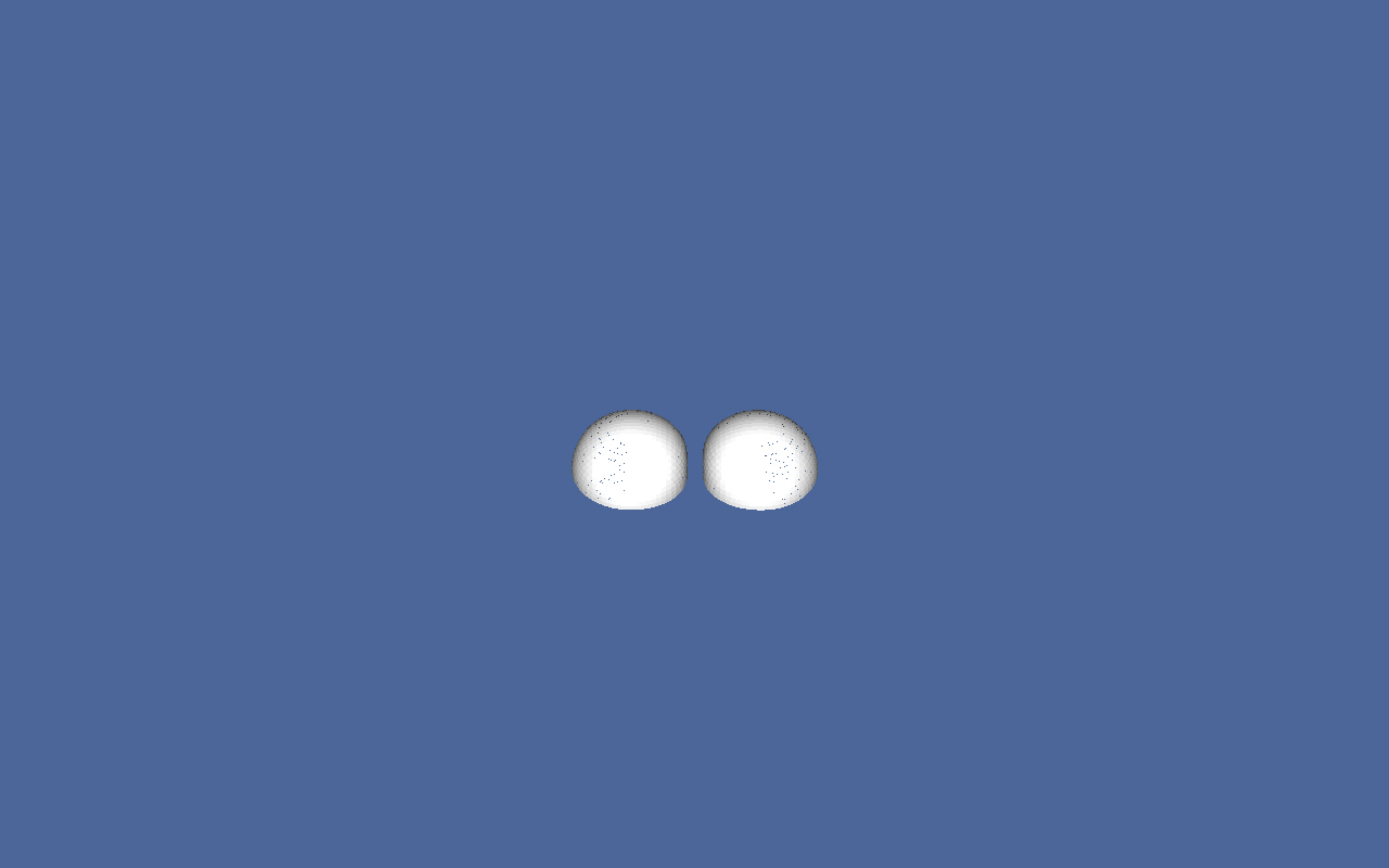}&
  \includegraphics[width=.25\textwidth]{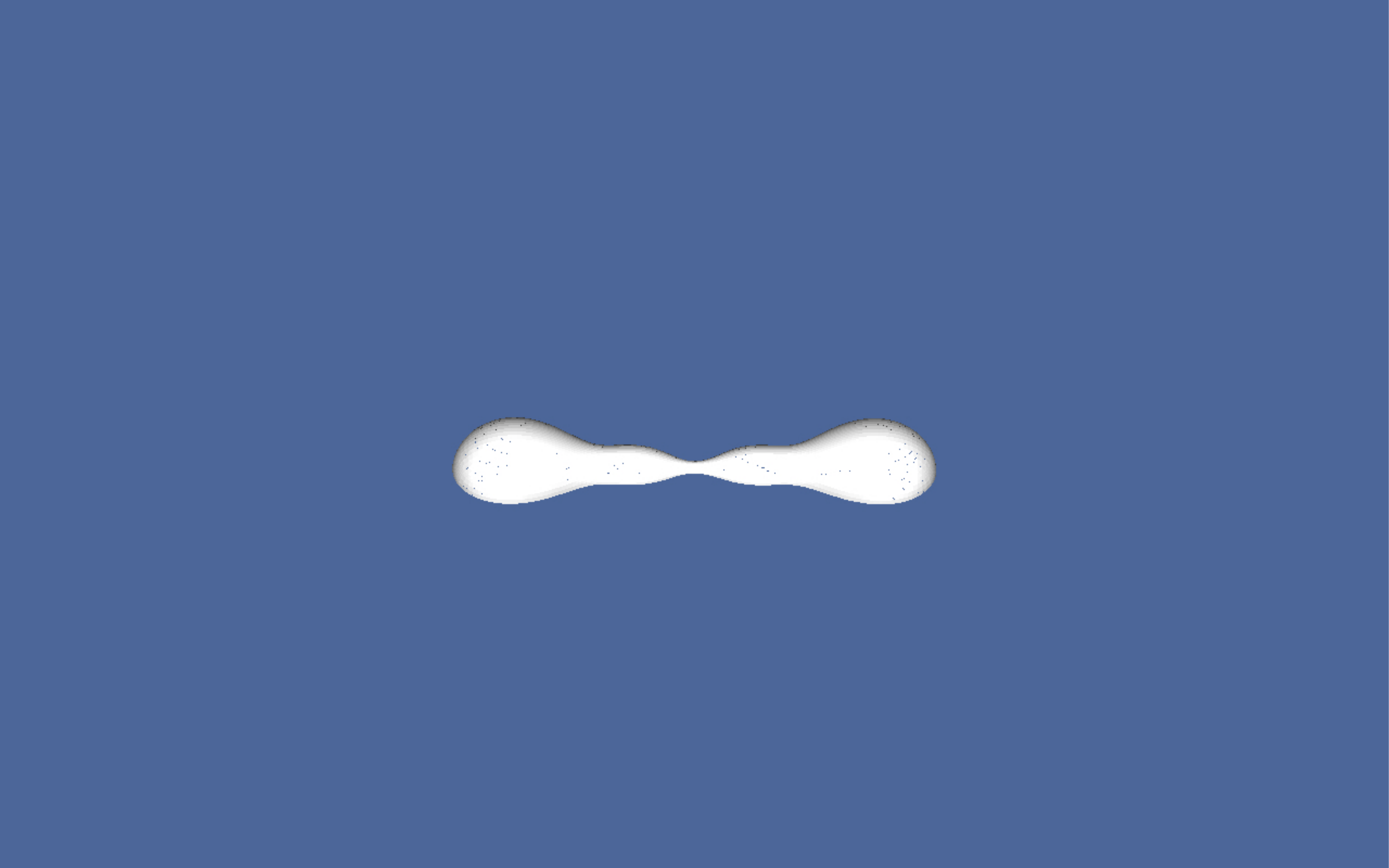}&  
  \includegraphics[width=.25\textwidth]{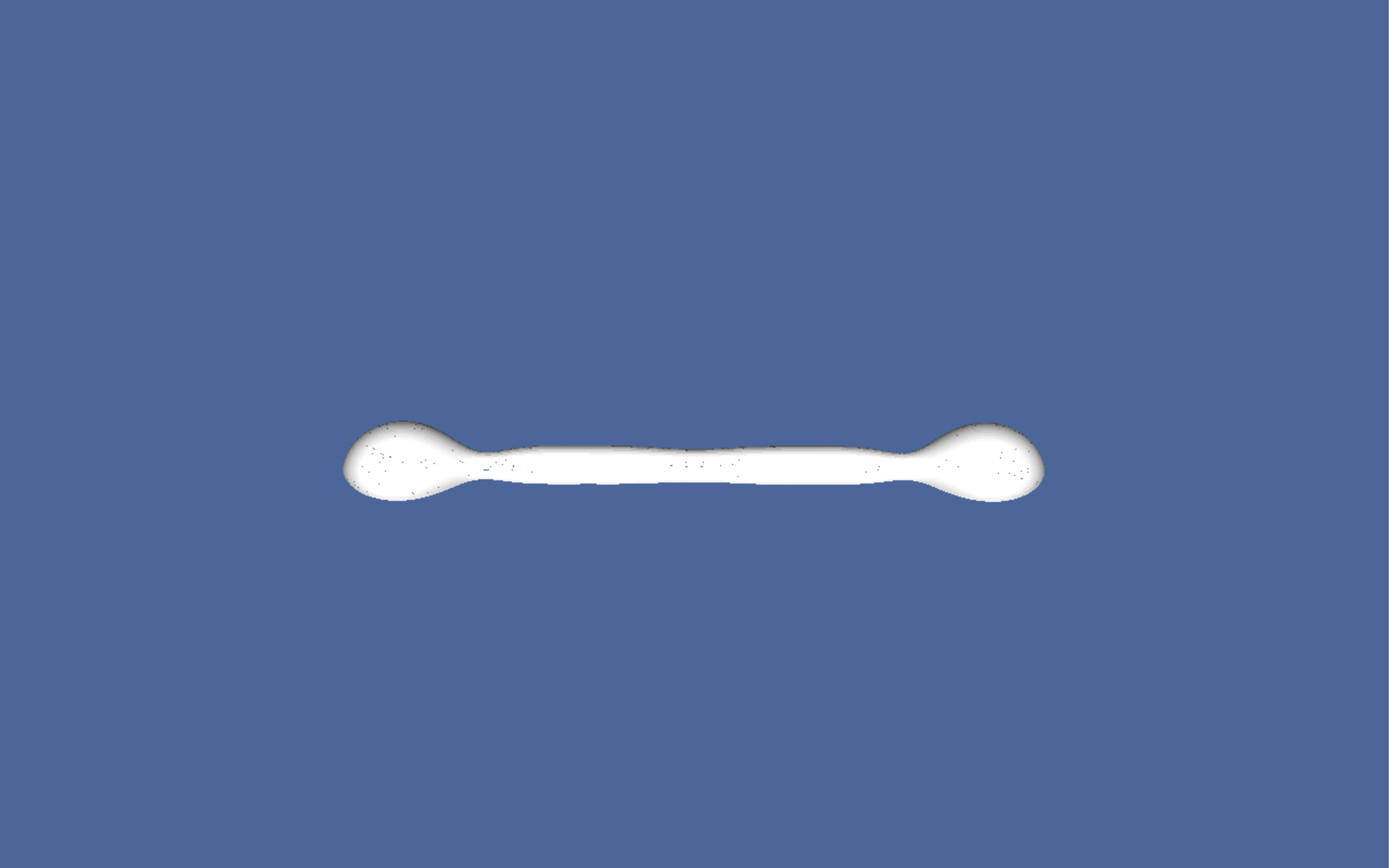}\\
  \includegraphics[width=.25\textwidth]{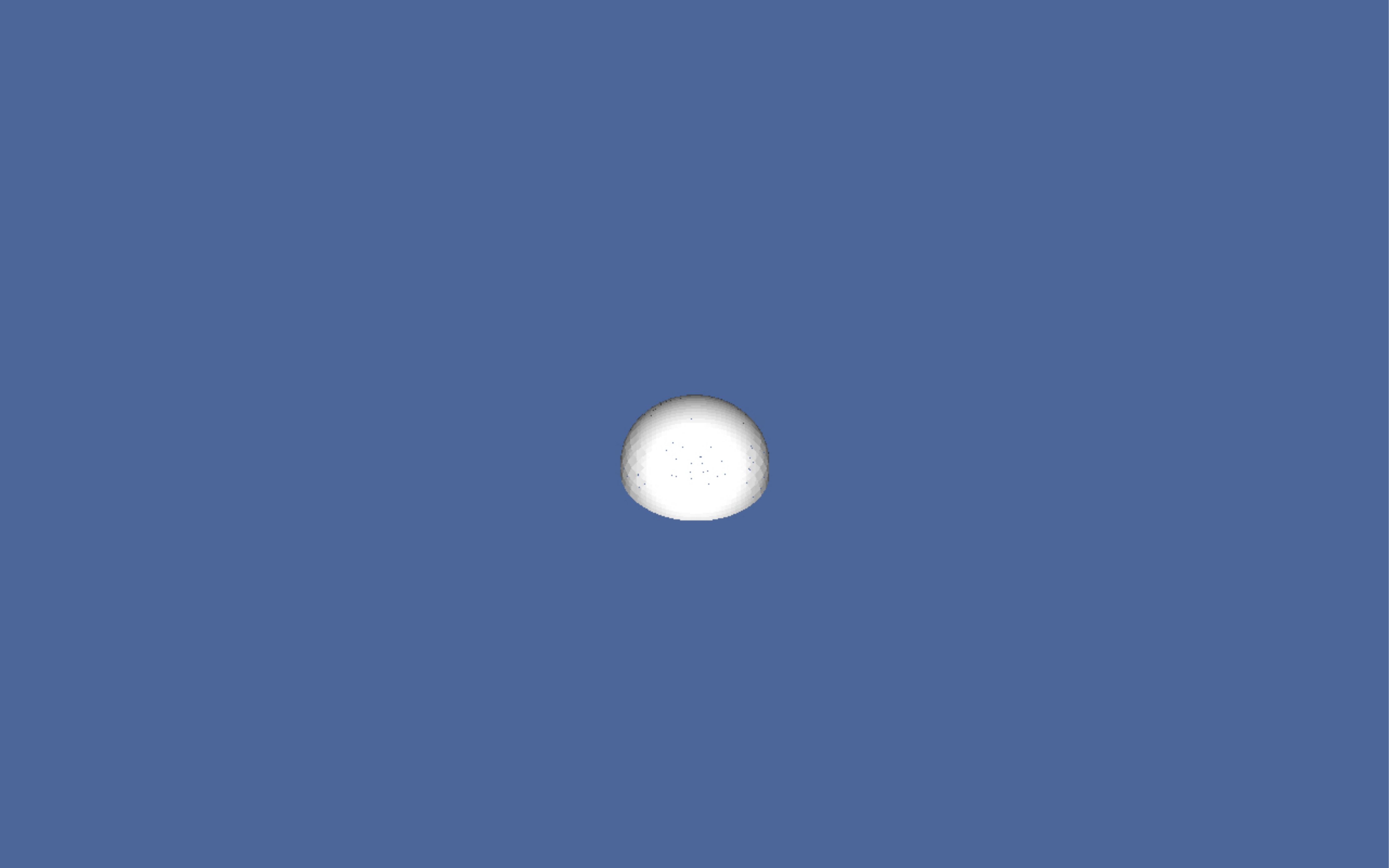}&
  \includegraphics[width=.25\textwidth]{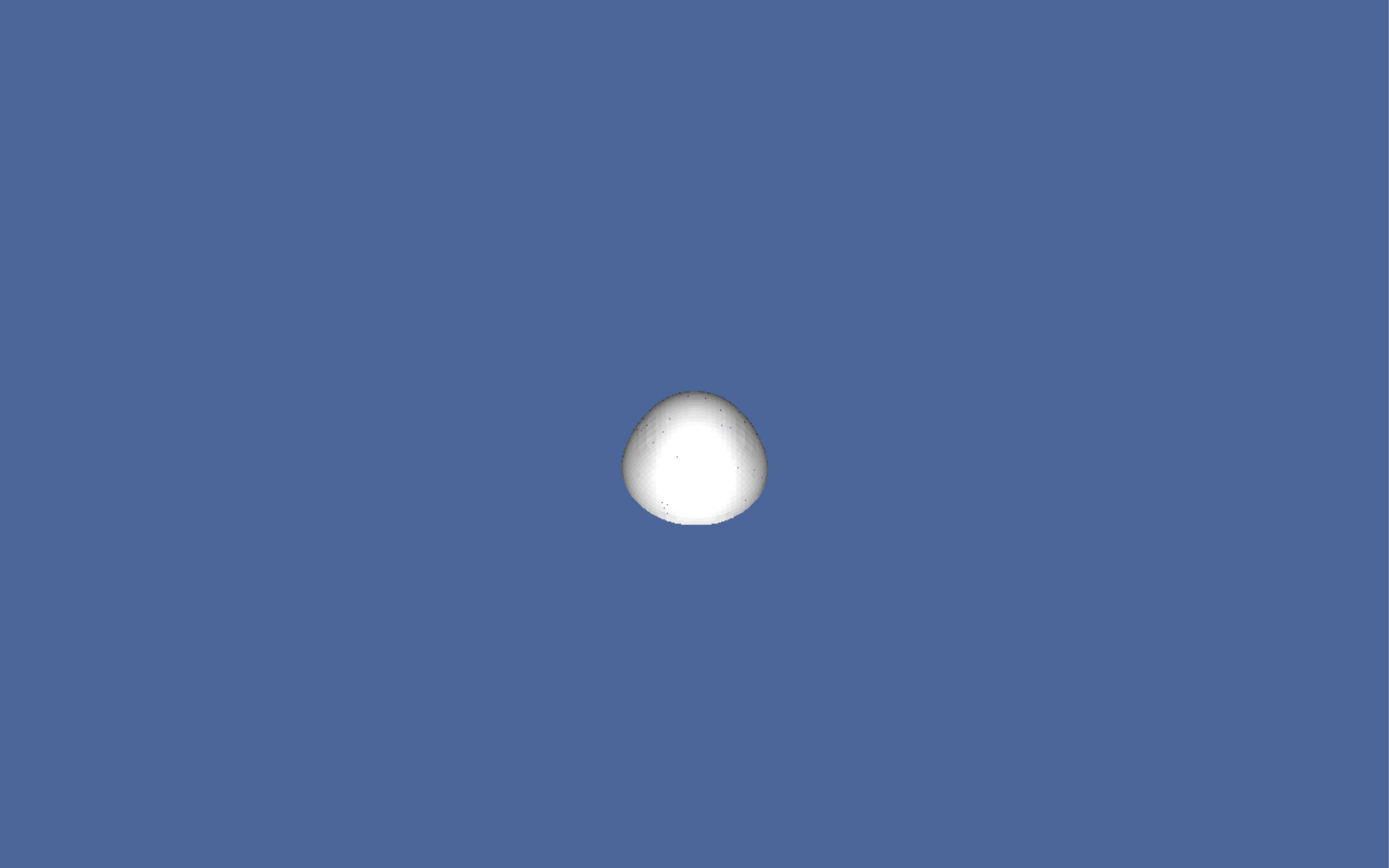}&
  \includegraphics[width=.25\textwidth]{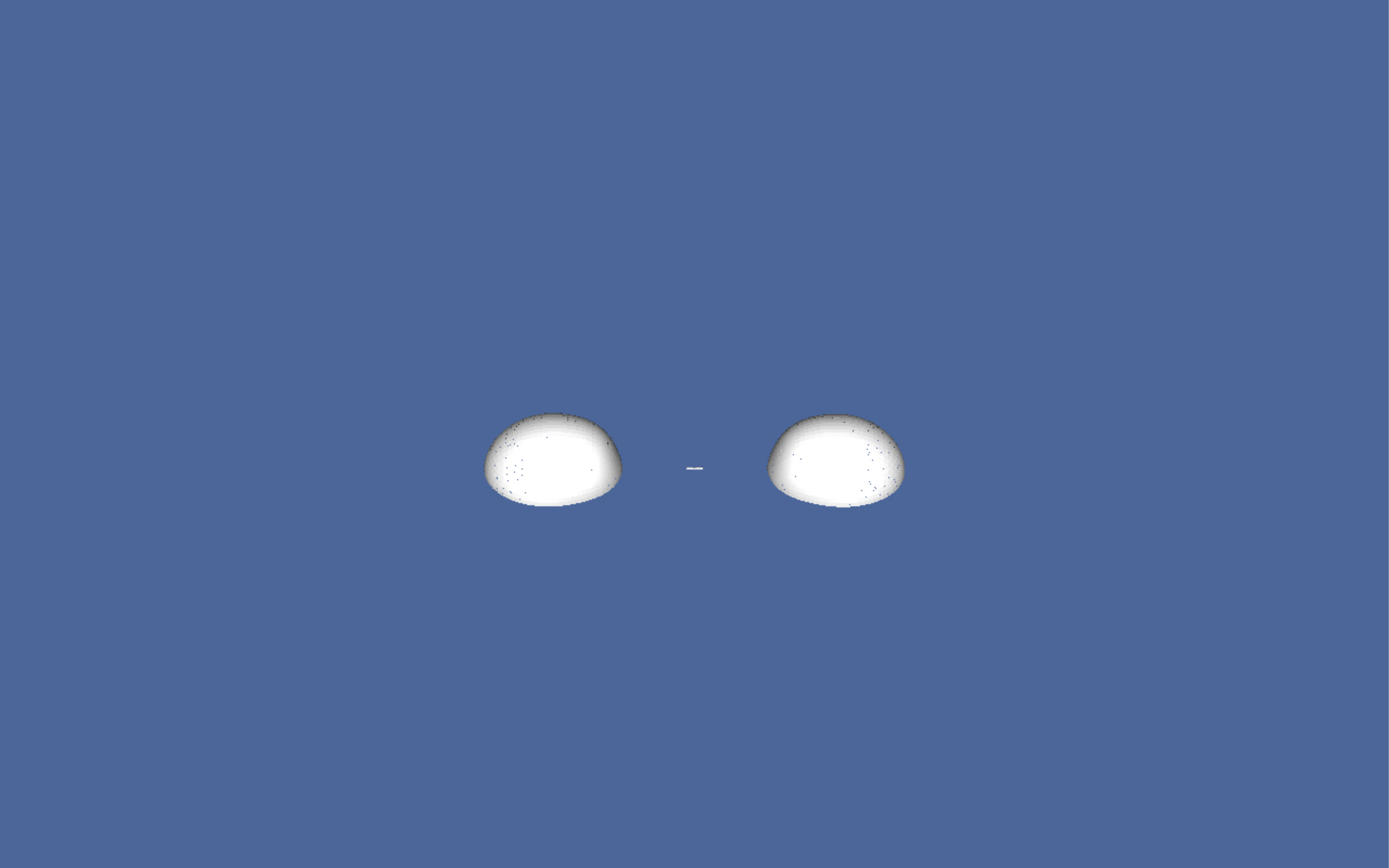}&  
  \includegraphics[width=.25\textwidth]{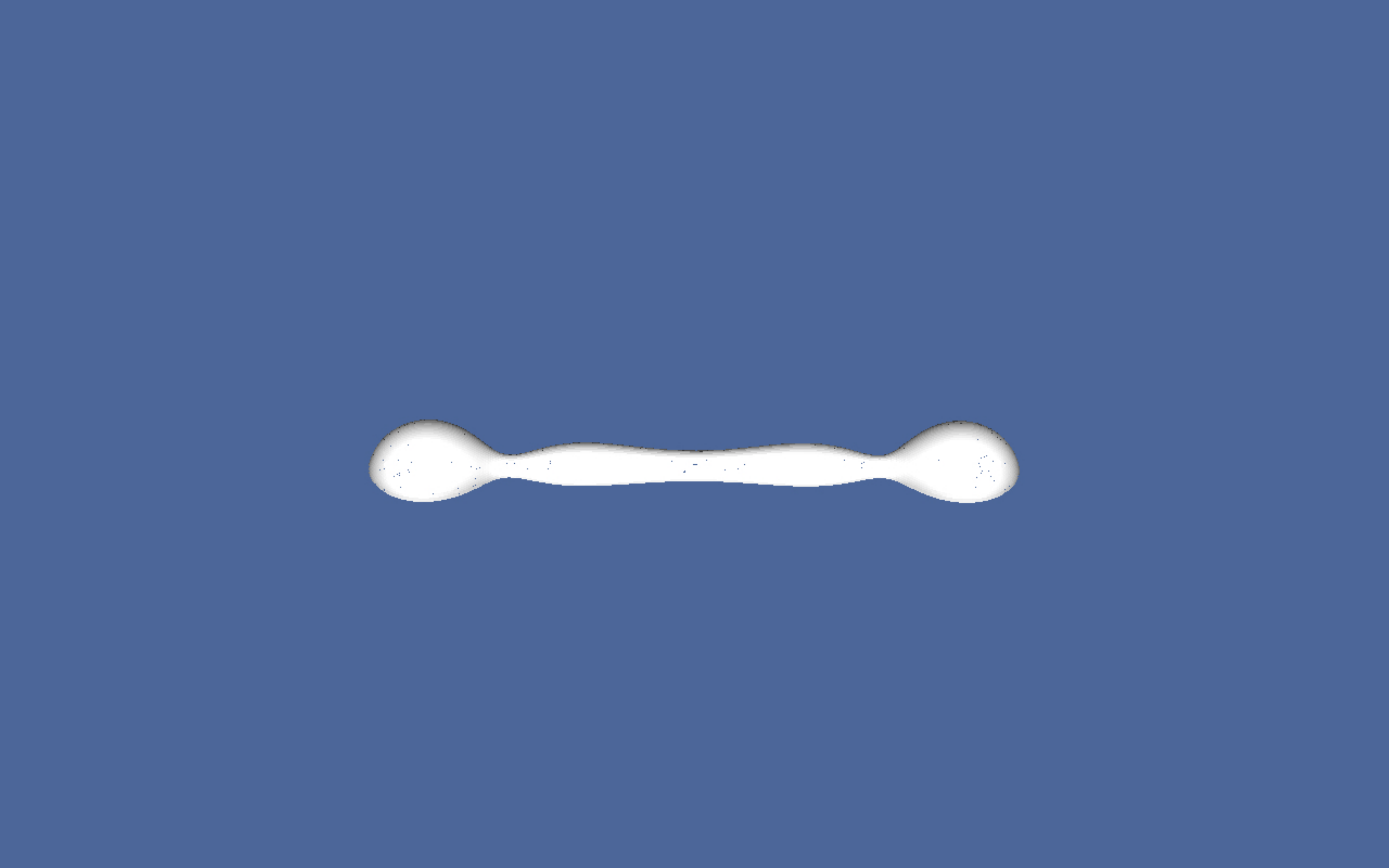}\\
  \includegraphics[width=.25\textwidth]{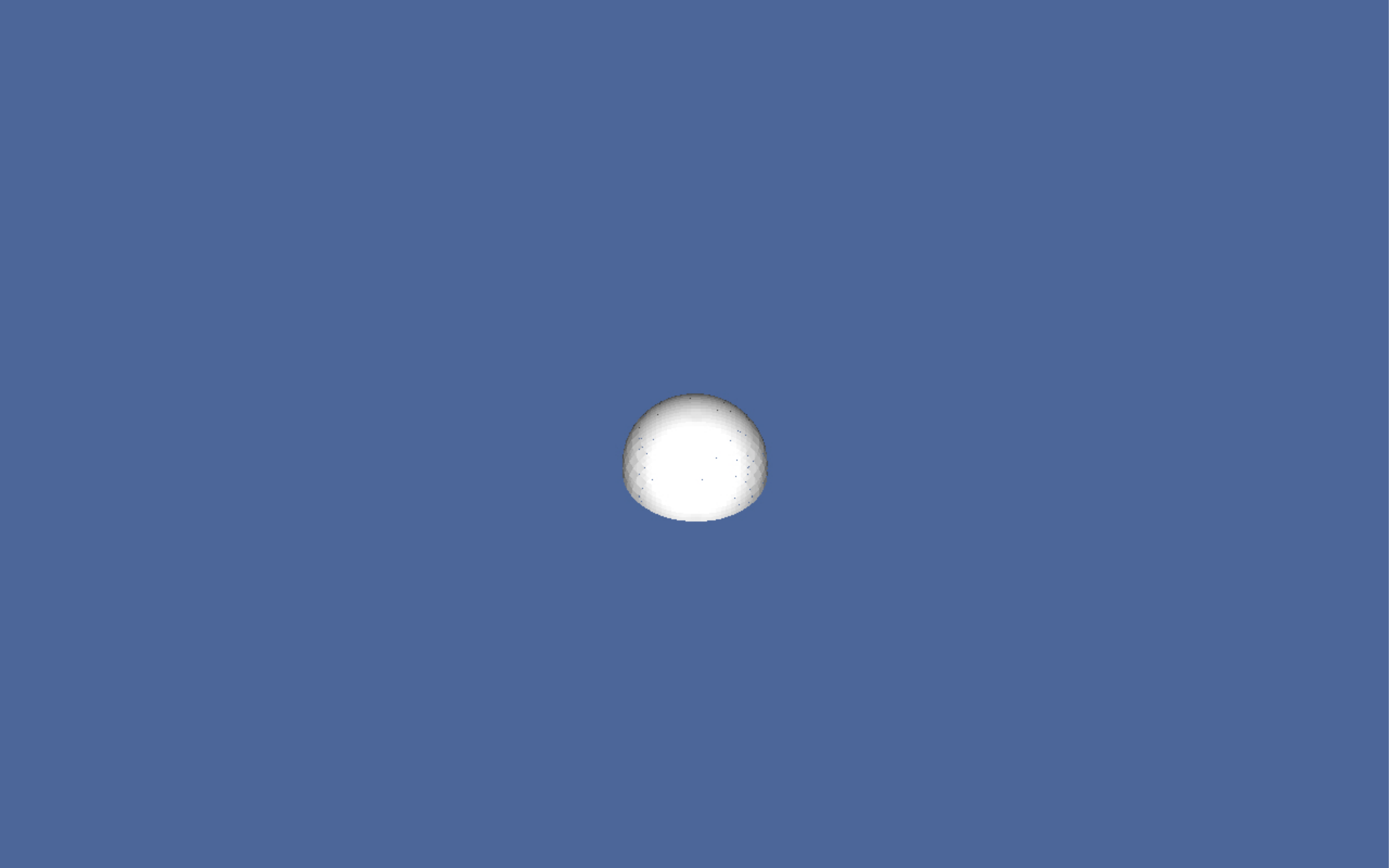}&
  \includegraphics[width=.25\textwidth]{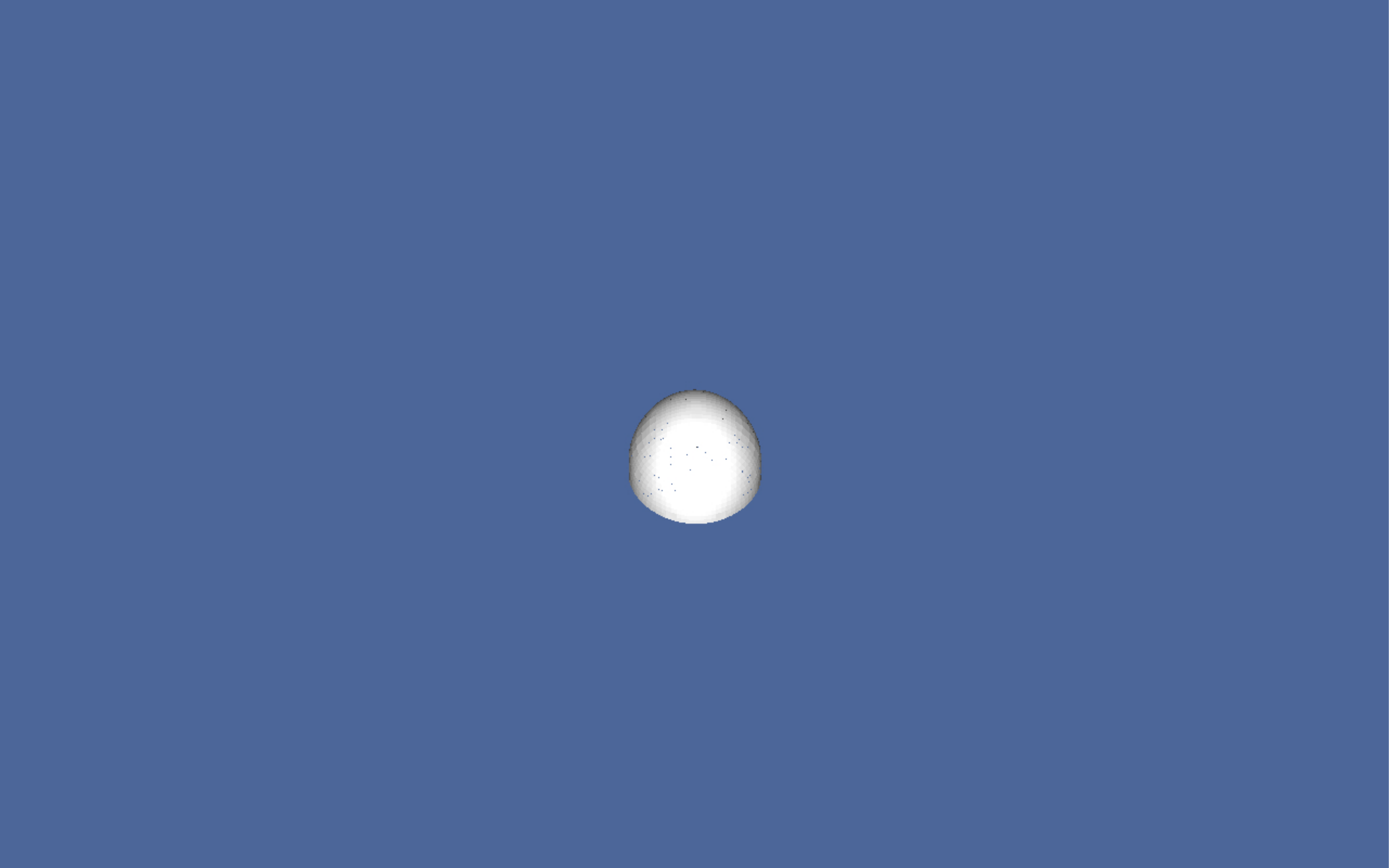}&
  \includegraphics[width=.25\textwidth]{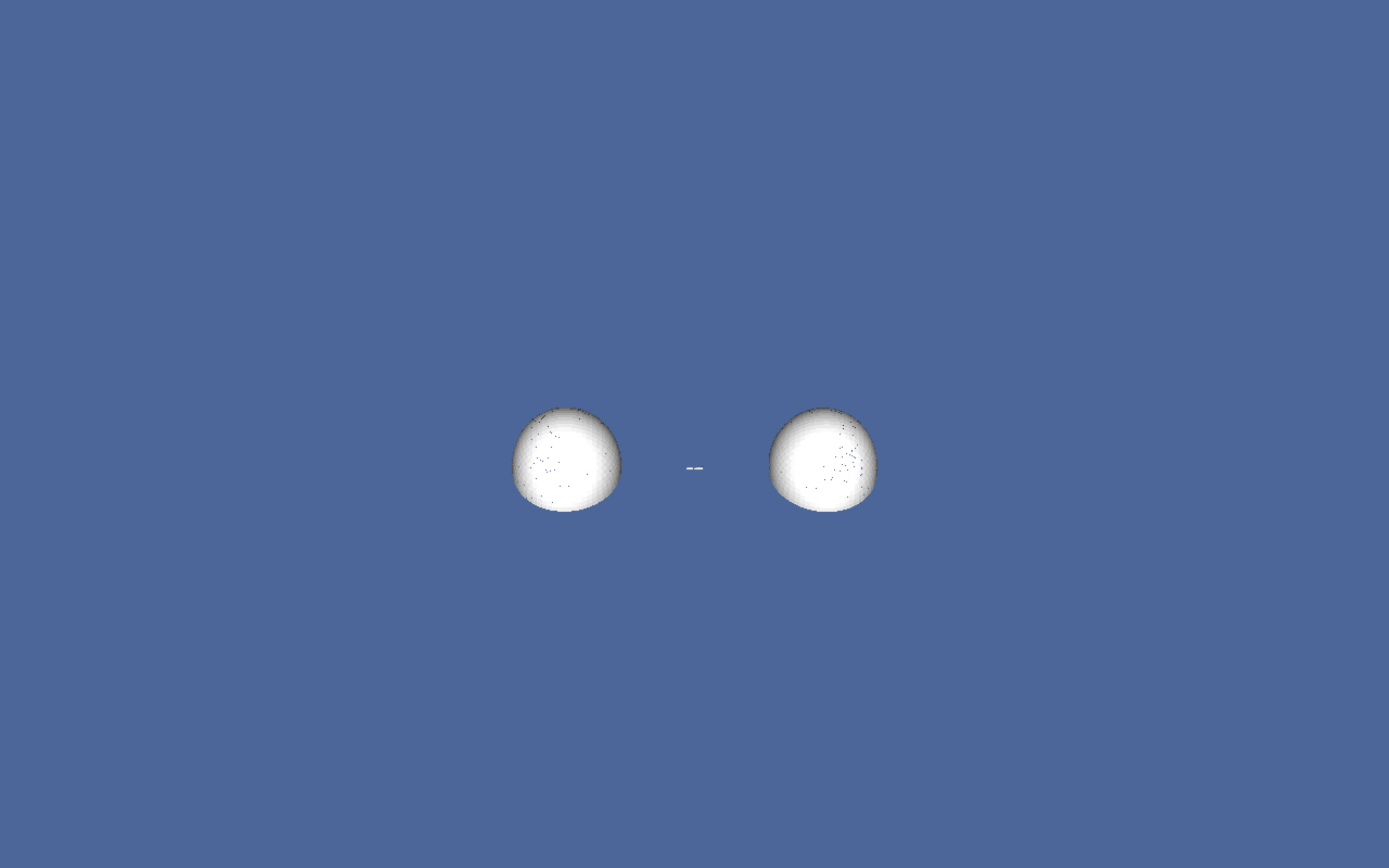}&  
  \includegraphics[width=.25\textwidth]{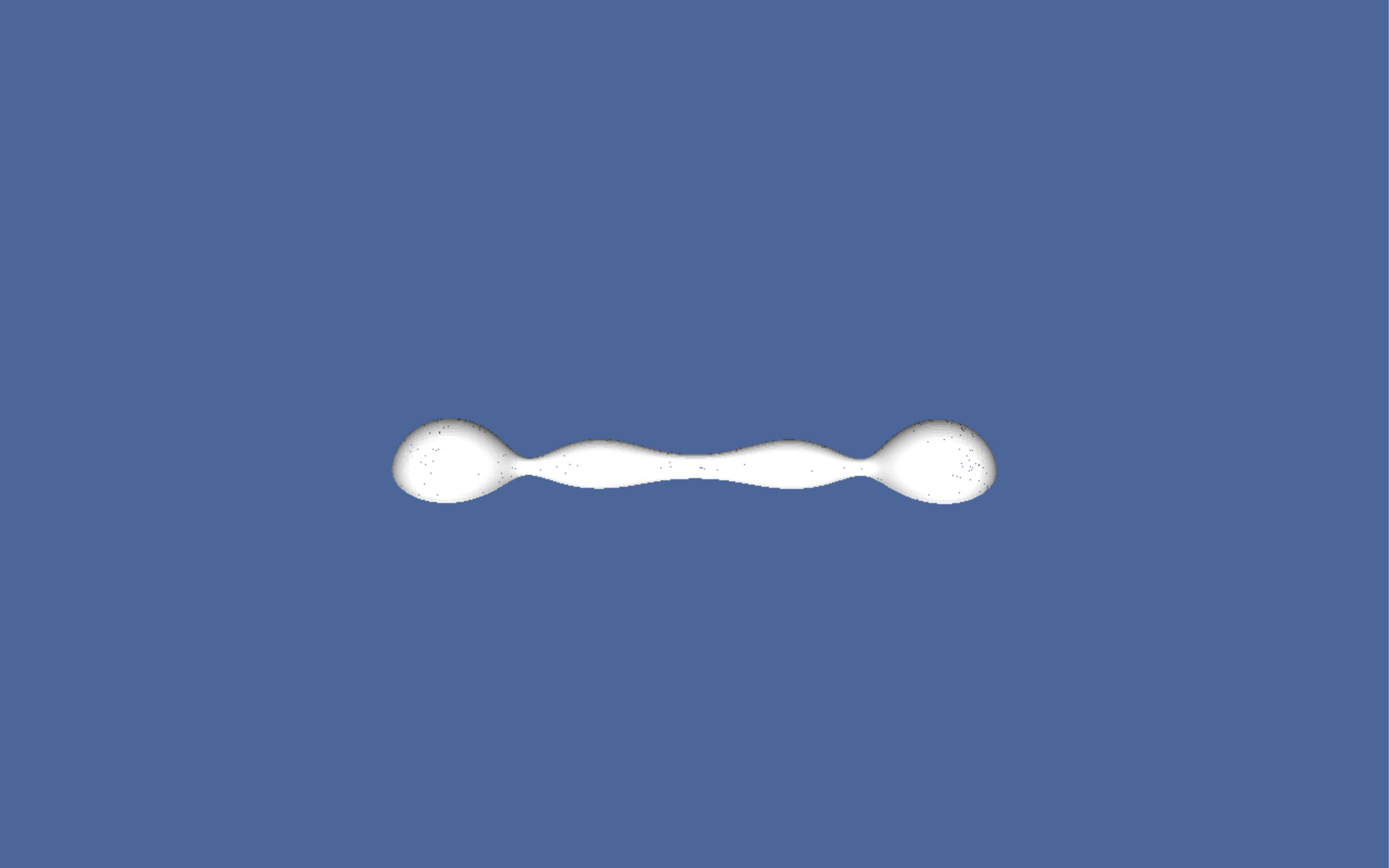}\\
  \includegraphics[width=.25\textwidth]{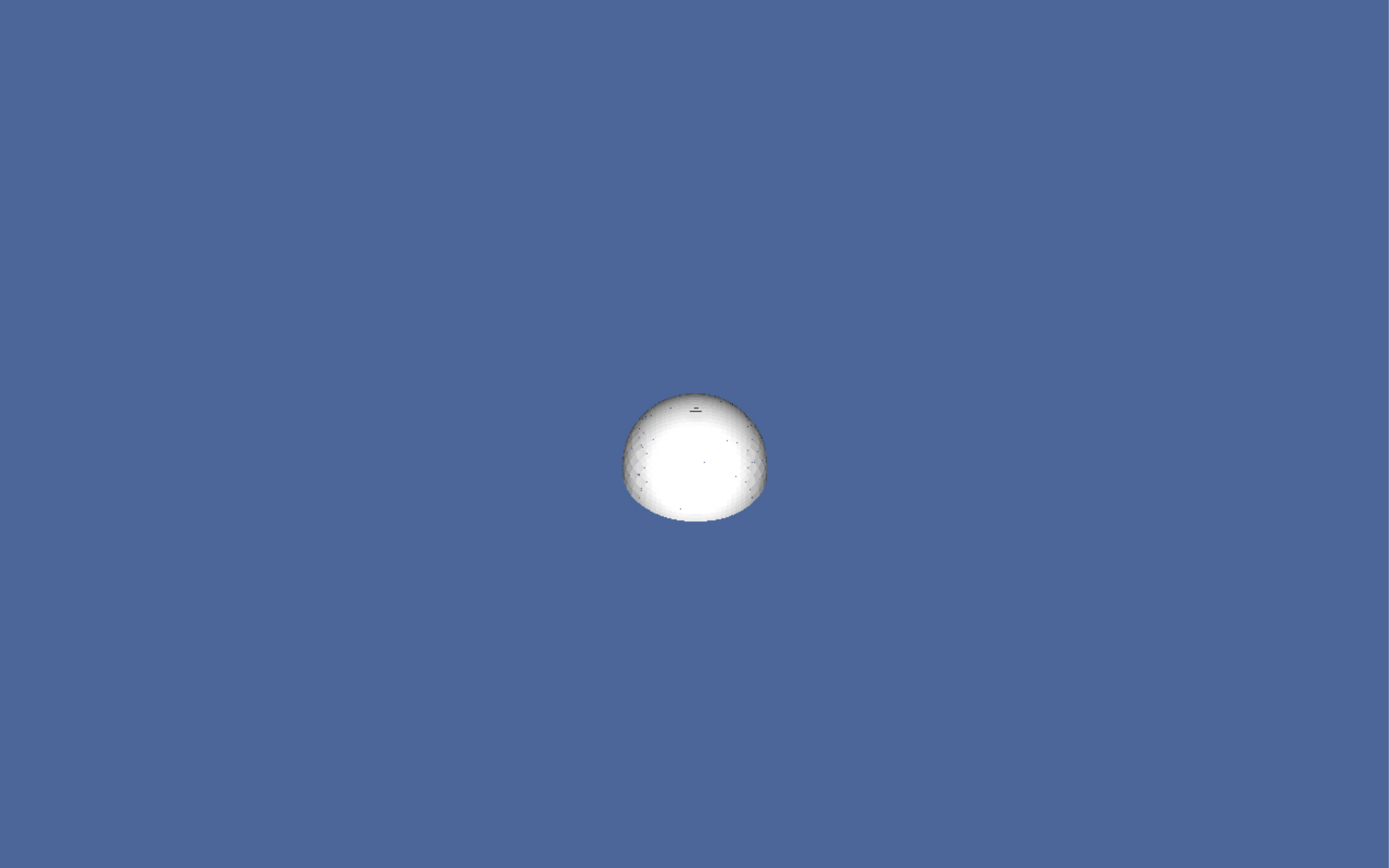}&
  \includegraphics[width=.25\textwidth]{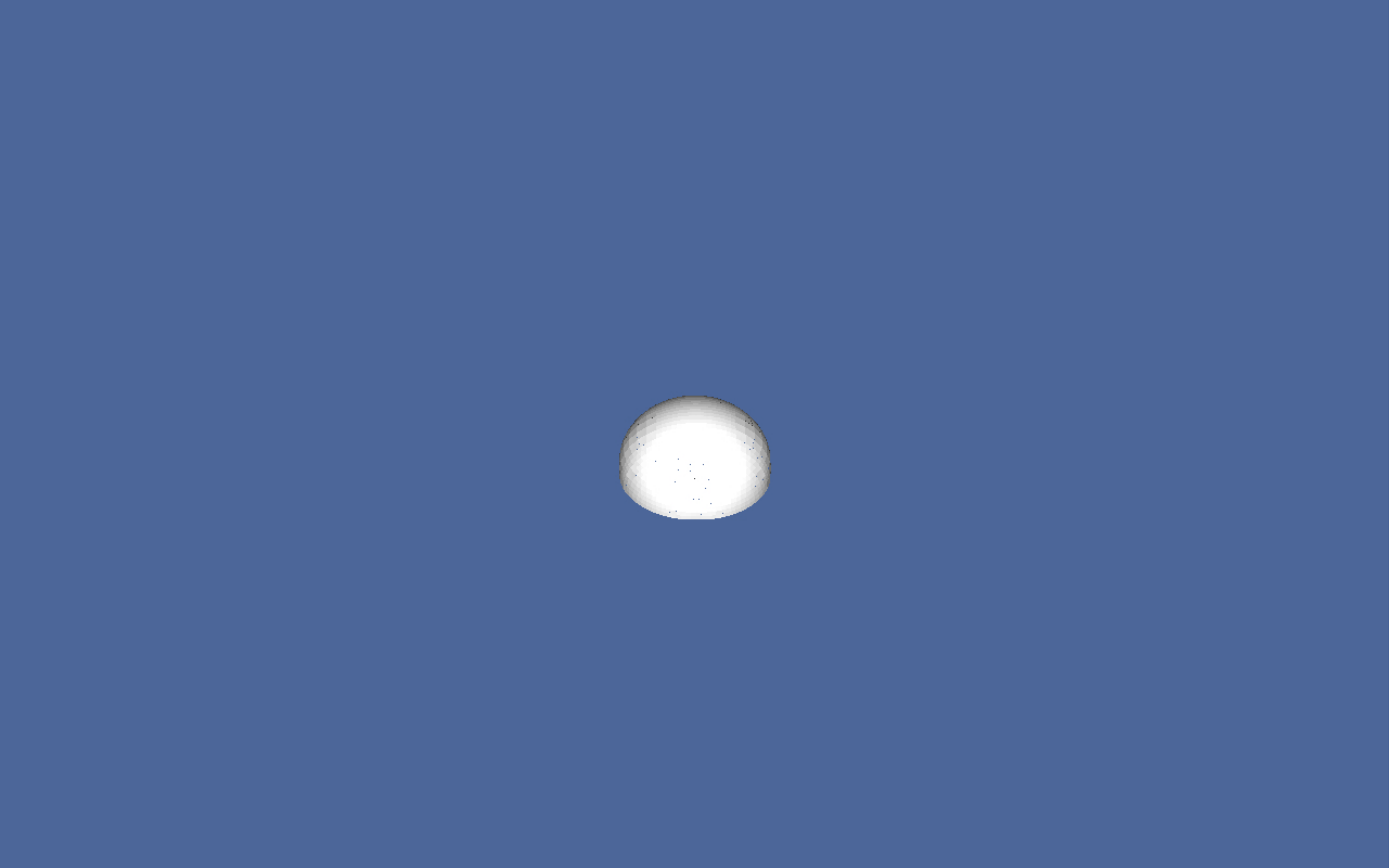}&
  \includegraphics[width=.25\textwidth]{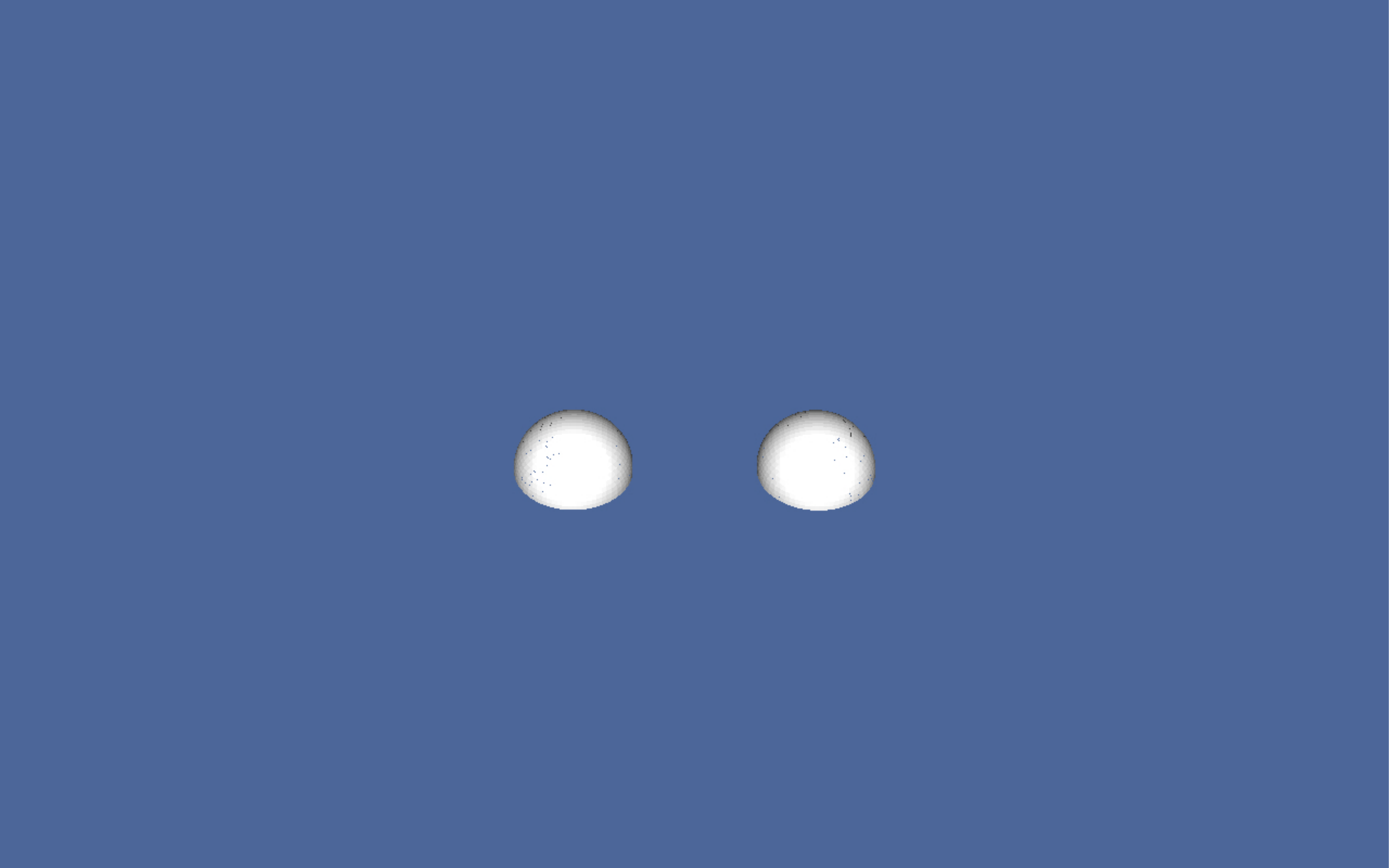}&  
  \includegraphics[width=.25\textwidth]{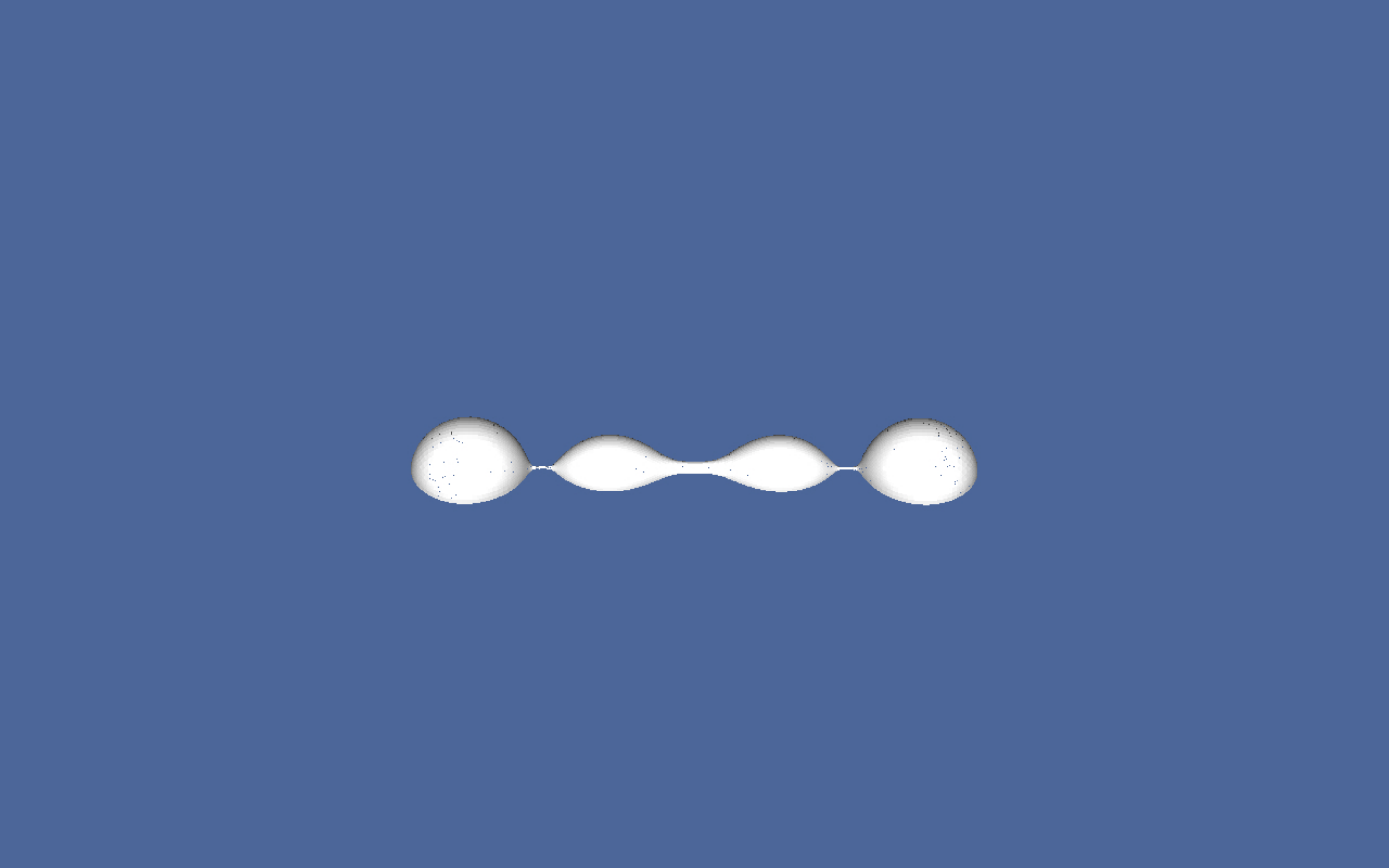}\\
  \includegraphics[width=.25\textwidth]{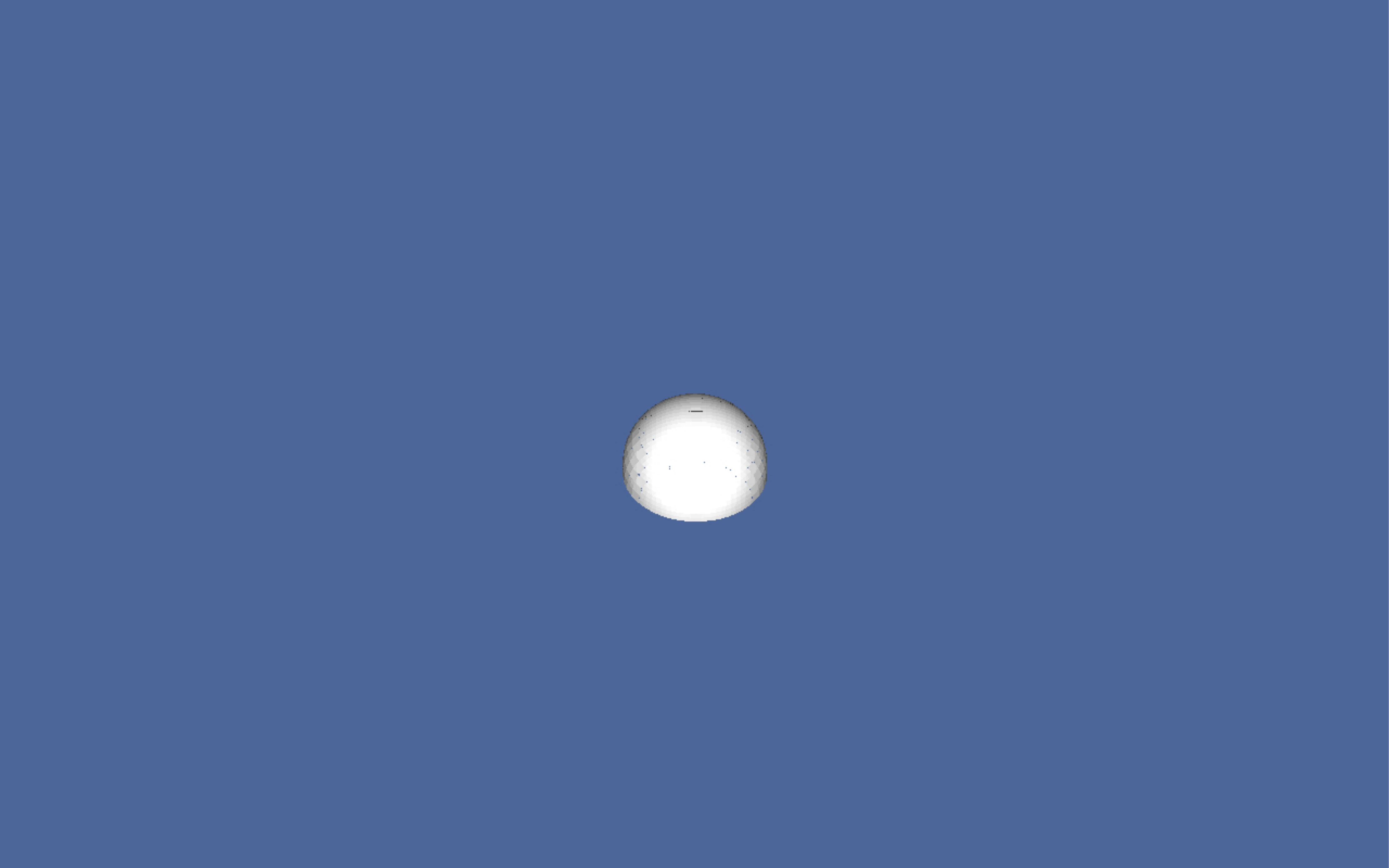}&
  \includegraphics[width=.25\textwidth]{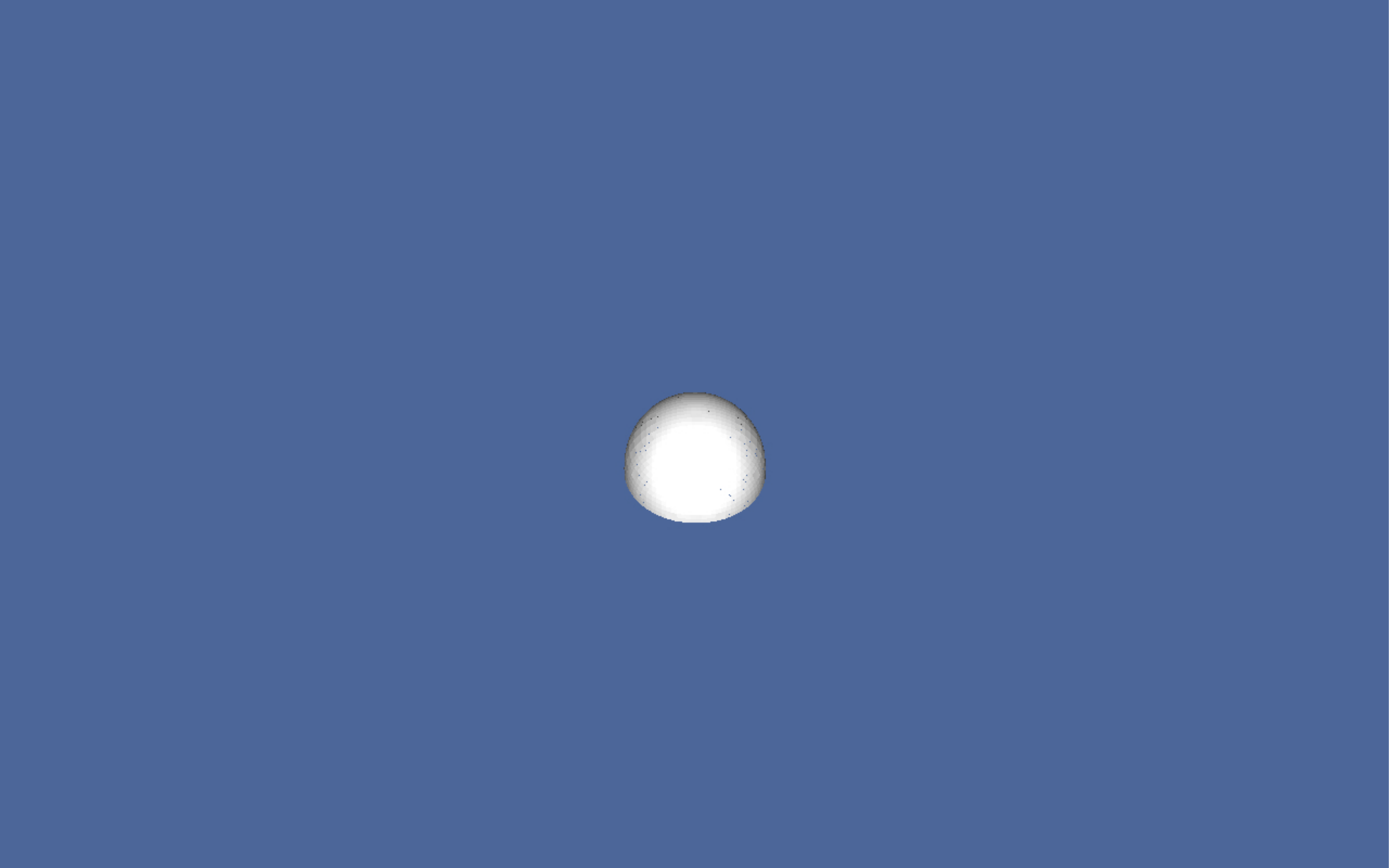}&
  \includegraphics[width=.25\textwidth]{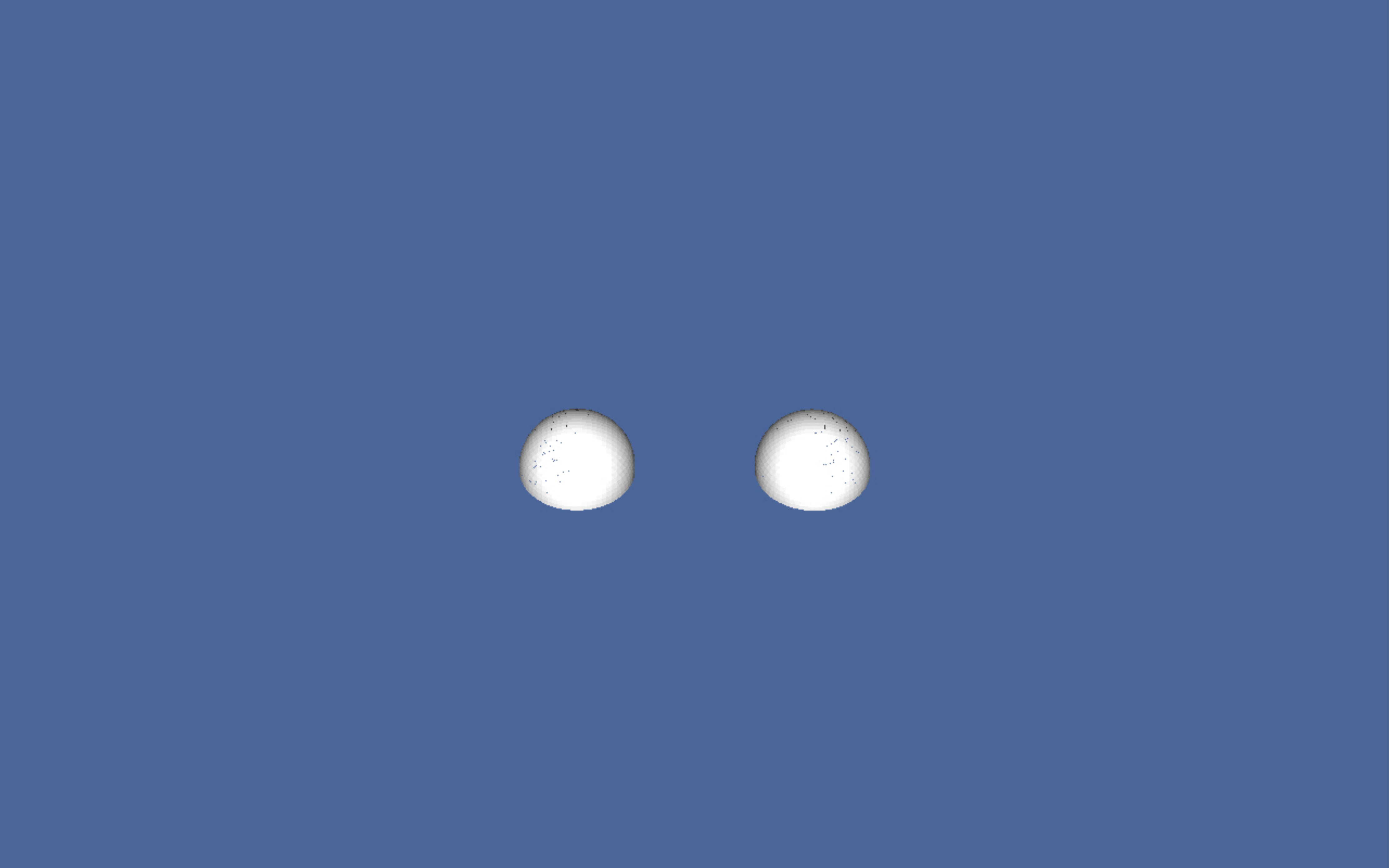}&  
  \includegraphics[width=.25\textwidth]{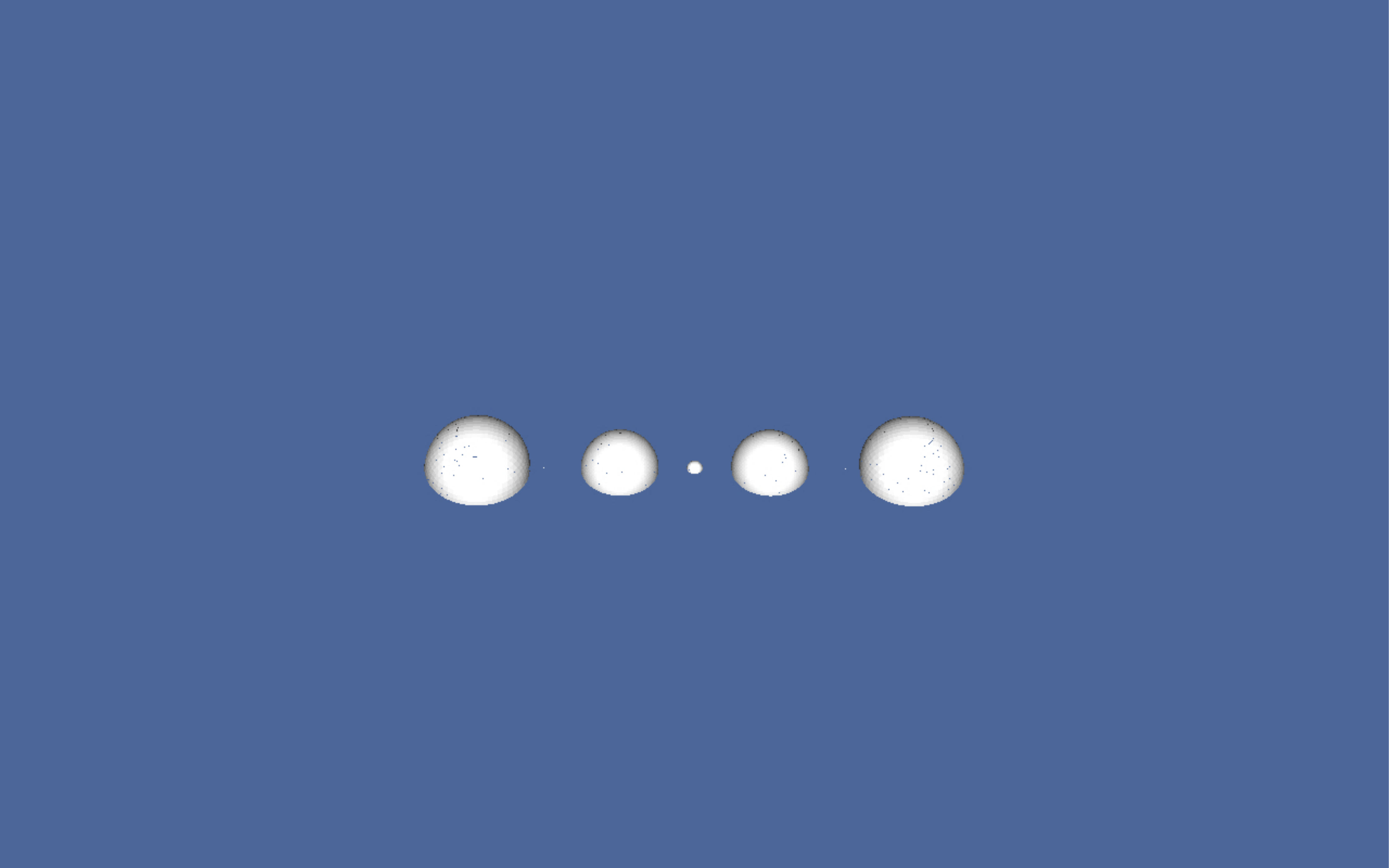}\\
  (a) & (b) & (c)& (d)
\end{tabular}
\caption{Snapshots of the simulations revealing various scenarios for Oh$ = 0.15$ and AR $=28$, and for
$\lambda = 1$ (a), $0.1$ (b), $0.01$ (c), and $0$ (d).}
\label{fig:lambda}
\end{figure}

To provide more insight into the effect of slip on the breakup transition,
we show in Fig.~\ref{fig:lambda}(a-d), a sequence of simulations for a moderate Oh $=0.15$ and relatively large
AR $=28$, for $\lambda=1,0.1,0.01,0$. As seen, the transition from no-breakup to breakup
on a substrate depends on subtleties in the problem arising from certain situations at the moment of the breakup:
(a) breakup may not occur for a large slip $\lambda=1$ as the retraction speed is fast so the 
thinning of the neck happens as the end bulges get too close to each other preventing further
necking and the pinch-off; (b) for a moderate slip, $\lambda=0.1$, breakup occurs due to end-pinching  
which happens at the middle of the filament - the resulting droplets however merge due to inertial effects;
(c) for a small slip, $\lambda=0.01$, both necking at end points and the Rayleigh-Plateau instability in the middle
are operative at the same time - however, the instability in the middle wins over the end-pinching leading
to the formation of two stable droplets; and (d) for zero slip, $\lambda=0$, the filament breaks up due to both
mechanisms -  end-pinching and the instabilities growing in the middle due to the development of capillary waves,
resulting four primary droplets in this case.  

\section{Conclusions}

This work presents an extensive computational study of the breakup of finite-size liquid filaments, 
including substrate effect, using direct numerical simulations.
We carry out detailed computations to provide an enhanced understanding of the
sublet and complex competition between the end-pinching and the Rayleigh-Plateau instability that
leads to the transition from no-breakup to breakup region in the parameter space that
we consider. We focus on the effects of Ohnesorge number, the filament
aspect ratio, and slip length; we did not consider the effect of the 
the contact angle due to the limitation existed in the numerical framework and leave
the study of this effect to future work. We however note that a prior work in the literature
has considered such effect \cite{Feng2013} and the interested reader is referred to 
this study for further discussions. Additionally,  we refer to the work carried out
by two of the co-authors on the issues related to
the computations of moving contact line problem \cite{MCL}.
Nevertheless, our study provides a most complete picture, 
that has not been available in the literature so far, of the breakup 
of finite-size filaments, when also considering the substrate effect. 

Our study also provides detail insight into various scenarios, namely,
a collapse into a single droplet, the breakup into one or multiple droplets,
and recoalescence into a single droplet after the breakup (or even possibly 
another breakup after recoalescence), and finally the formation of distinct droplets
when there is the substrate effect.  We also show that when there is a slip
effect, the liquid filament can break up for large Oh ($>1$), considerably different
from free-standing filaments that show no breakup for this case regardless of 
the AR value.  Our comprehensive results significantly extend the available parameter space
over which the transition from no-breakup to break occurs, when also including the substrate effect.
We observe how the retraction velocity of the end points on the substrate can strongly influence
this transition. A careful comparison of the retraction speed with experimental results will allow  
the calibration of the slip length, whose understanding clearly requires future work.

\section*{Acknowledgements}

This research was partially supported by the NSF grant No.~CBET-1604351. 
We gratefully acknowledge Kim Lozarito, Michel Castor, Matthew Marner,
Elise Burkhardt, and Brandon DeGraw for carrying out the experiments.
S.~A.~gratefully acknowledges helpful discussions with
S.~Zaleski. 

\section*{Author Contributions}
A. Dziedzic, M. Nakrani, B. Ezra, and M. Syed carried out all the simulations,
performed the analysis, and generated the visualizations. S. Popinet created 
the simulations example and authored the  numerical code.  S. Afkhami  
conceived the project, planned the simulations, and wrote the first draft of the manuscript. 
All authors discussed, edited, and revised the manuscript.

%\section*{References}

\end{document}